\documentclass[iop,revtex4,numberedappendix,openany]{emulateapj}
\usepackage{float}
\usepackage{rotating}
\usepackage{graphicx}
\usepackage{chngpage}
\usepackage{amsmath}
\usepackage[caption=false]{subfig}

\shorttitle{Groups in Strong Gravitational Lens Fields}
\shortauthors{Wilson et al.}

\begin{document}

\title{A Spectroscopic Survey of the Fields of 28 Strong Gravitational Lenses: The Group Catalog}

\author{Michelle L. Wilson\altaffilmark{1}, Ann I. Zabludoff\altaffilmark{1}, S. Mark Ammons\altaffilmark{2}, Ivelina G. Momcheva\altaffilmark{3}, Kurtis A. Williams\altaffilmark{4}, and Charles R. Keeton\altaffilmark{5}}

\altaffiltext{1}{Steward Observatory, University of Arizona, 933 North Cherry Avenue, Tucson, AZ 85721, USA}
\altaffiltext{2}{Lawrence Livermore National Laboratory, Physics Division L-210, 7000 East Ave., Livermore CA 94550}
\altaffiltext{3}{Space Telescope Science Institute, 3700 San Martin Drive, Baltimore, MD 21218, USA}
\altaffiltext{4}{Department of Physics and Astronomy, Texas A\&M University-Commerce, Commerce, TX, 75428, USA}
\altaffiltext{5}{Department of Physics and Astronomy, Rutgers University, 136 Frelinghuysen Road, Piscataway, NJ 08854, USA}

\begin{abstract} 
With a large, unique spectroscopic survey in the fields of 28 galaxy-scale strong gravitational lenses, we identify groups of galaxies in the 26 adequately-sampled fields.  Using a group finding algorithm, we find 210 groups with at least five member galaxies; the median number of members is eight. Our sample spans redshifts of 0.04 $\le z_{grp} \le$ 0.76 with a median of 0.31, including 174 groups with $0.1 < z_{grp} < 0.6$.  Groups have radial velocity dispersions of 60 $\le \sigma_{grp} \le$ 1200 km s$^{-1}$ with a median of 350 km s$^{-1}$.  We also discover a supergroup in field B0712+472 at $z =$ 0.29 consisting of three main groups.  We recover groups similar to $\sim$ 85\% of those previously reported in these fields within our redshift range of sensitivity and find 187 new groups with at least five members.  The properties of our group catalog, specifically 1) the distribution of $\sigma_{grp}$, 2) the fraction of all sample galaxies that are group members, and 3) the fraction of groups with significant substructure, are consistent with those for other catalogs.  The distribution of group virial masses agrees well with theoretical expectations.  Of the lens galaxies, 12 of 26 (46\%) (B1422+231, B1600+434, B2114+022, FBQS J0951+2635, HE0435-1223, HST J14113+5211, MG0751+2716, MGJ1654+1346, PG 1115+080, Q ER 0047-2808, RXJ1131-1231, and WFI J2033-4723) are members of groups with at least five galaxies, and one more (B0712+472) belongs to an additional, visually identified group candidate. There are groups not associated with the lens that still are likely to affect the lens model; in six of 25 (24\%) fields (excluding the supergroup), there is at least one massive ($\sigma_{grp} \ge$ 500 km s$^{-1}$) group or group candidate projected within 2$^{\prime}$ of the lens.
  
\end{abstract}


\section{Introduction}

Gravitational lensing is an attractive method to measure cosmological parameters \citep[e.g.,][]{refsdal64,oguri07,suyu13}, because the lensing phenomenon only depends on the geometry of the universe and the mass distribution along the line-of-sight.  However, determining that mass can be difficult.  Many of the properties of lens systems (e.g., shears or lensed image positions) are not well described by a model with only one galaxy contributing to the lensing potential, showing that groups or clusters of galaxies at the lens and even along the lens line-of-sight are important to the lensing \citep[e.g.,][]{keeton97,lehar97,keeton04,wambsganss05,jaroszynski12,collett13}.  

Galaxy groups are interesting apart from their possible impact on gravitational lensing due to their importance to galaxy evolution.  Groups are the most common environment for galaxies, with even our own Milky Way residing in the Local Group.  Their lower velocity dispersions foster stronger galaxy-galaxy interactions than do clusters \citep[e.g.,][]{zabludoff98}.  Also, cluster galaxy properties, such as mass or color, may first develop through ``pre-processing'' in the less dense group environment \citep[e.g.,][]{zabludoff96,zabludoff98,mcgee09,just15}.

To produce a new group sample with which to investigate the environments of lens systems and line-of-sight structures, as well as the effect of environment on galaxy properties, we employ a spectroscopic survey in the fields of 28 galaxy-scale strong gravitational lenses \citep{momcheva15}.  In 26 of those fields, we construct a group finding algorithm that works well for spectroscopic samples of variable completeness, field size, and sampling footprint, modifying the iterative method of \citet{ammons14}.  Our method also is designed to find a range of group masses without favoring the most dynamically evolved systems (i.e., those with a significant quiescent galaxy population and/or detectable hot intracluster gas), unlike methods that use galaxy color, galaxy luminosity, or X-ray halo detections to identify groups or determine group properties.  The resulting catalog includes groups ranging from the very poor ($\sigma_{grp} \sim 100$ km s$^{-1}$) to rich clusters ($\sim 1000$ km s$^{-1}$) at $0.04 \le z \le 0.76$.

In this paper, we describe the data (Section \ref{sec:data}), our group finding algorithm (Section \ref{sec:method}), and its performance (Section \ref{sec:evaluation}).  We present our group catalog and discuss its properties in Section \ref{sec:catalogsummary}, describe a supergroup we identify in Section \ref{sssec:supergroup}, and comment on the lens environments and line of sight structures in Sections \ref{ssec:lensenvs} and \ref{ssec:lenslos}, respectively.  We suggest ways of using our catalog and group quality flags for various science cases in Section \ref{ssec:usingthecatalog}.

Throughout this paper, we adopt the values of $H_0$ = 71 km~s$^{-1}$ Mpc$^{-1}$, $\Omega_m$ = 0.274, and $\Omega_{\Lambda}$ = 0.726 \citep{hinshaw09}.

\section{The Data}
\label{sec:data}

\subsection{Photometry}

Images of each lens field were collected using the Mosaic-1 imager on the Kitt Peak National Observatory (KPNO) Mayall 4 m telescope and the Mosaic II imager on the Cerro Tololo Inter-American Observatory (CTIO) Blanco 4 m telescope.  \citet{williams06} detail these data and their reduction, which we summarize here.  Each field was imaged using the ``nearly Mould'' $I$-band filter, as well as either the Harris $V$ filter or the Harris $R$ filter depending on the redshift at which the 4000\AA{} break of the lens galaxy enters the $R$ filter (z $\ge$ 0.35).  Photometry was obtained using SExtractor version 2.3.2's MAG\_AUTO \citep{bertin96}.  Galaxy colors were measured using aperture magnitudes after degrading one image to match the seeing of the other.  The photometry was then calibrated to the Kron-Cousins filter set using the imaging of the \citet{landolt92} standard star fields.  The photometry was used to select objects for spectroscopic followup and to calculate spectroscopic completeness.

\subsection{Spectroscopy}
\label{ssec:spectroscopy}

The spectroscopic follow-up is described in \citet{momcheva15}. We summarize the most pertinent details here.

A sample of objects from the photometry was selected for spectroscopic followup using Hectospec on the MMT 6.5 m telescope and LDSS-2, LDSS-3, and IMACS on the Magellan 6.5 m telescopes. 

Priorities were ranked in the following manner.  Objects identified as potential members of a red sequence \citep{williams06}, whether or not that sequence is at the lens redshift, were given the highest priority.  Objects outside potential red sequences but redder than models of starburst galaxies at the lowest redshift candidate red sequence, bluer than members of the highest redshift candidate red sequence, and within 5$^{\prime}$ of the lens in each field were given second highest priority.  Objects outside this color range but within 5$^{\prime}$ of the lens were given third highest priority.  Finally, objects farther than 5$^{\prime}$ from the lens were given the lowest priority.  The limiting magnitude for spectroscopic targeting was either 20.5 or 21.5 depending on the instrument used.

This target selection method results in a strong radial gradient in the spectroscopic completeness. Some fields have patchy completeness due to the multislit mask designs. Furthermore, the spectroscopic completeness varies greatly from field to field; within 2$^{\prime}$ of the lens, completenesses range from $\sim$20-90\% \citep[see Figures 5, 36, and 37 in][]{momcheva15}. Hectospec uses fibers with large holders that cannot be placed more closely on the sky than 20$^{\prime\prime}$, and the northern fields typically have fewer configurations.  Thus, fields observed with this instrument have sparser but smoothly varying radial completeness over our full field of view.  Those observed with LDSS-2 and -3 and IMACS, in many cases, display a rectangular and/or patchy completeness inconsistent with the total available aperture of the instrument (see Figure \ref{fig:grpskyplotsandvelhists}, e.g., fields b1422 and q1017).  These distributions were due to the mask designs.  Early LDSS-2 masks were all centered on the lens, resulting in higher completeness within 5$^{\prime}$ from the lens. Later observations were planned assuming a five mask configuration tiled in the shape of the pips on the five side of a die, with one mask in the middle and four masks that slightly overlapped each other arranged in a square.  The sampled field of view generally extends to at least 1$r_{vir}$ for groups with velocity dispersions of up to $\sim$ 500 km s$^{-1}$ at $z \gtrsim$ 0.1 and for richer clusters with velocity dispersions of up to $\sim$ 900 km s$^{-1}$ at $z \gtrsim$ 0.2.

Four fields (he0435, q0047, q0158, and rxj1131) were imaged while one of the eight CCDs in the Mosaic II camera was non-operational in late 2003.  This resulted in a dearth of imaging data over a $\sim 9\arcmin \times 18\arcmin$ region in the eastern portion of each field.  Since the spectroscopic targets were selected from our photometry, those regions have no spectroscopic coverage.  

For two of the 28 fields, b1608 and pmn2004, the spectroscopic completenesses are very low due to limited spectroscopic observations.  Thus, we do not include these fields in our group catalog.

Redshifts were calculated with a routine based on that of \citet{cool08}, which involves a $\chi^2$ fit of the measured spectra to templates.  The resulting redshifts of objects observed with IMACS and LDSS-3 were checked visually against the sky-subtracted, two-dimensional spectra.  The median uncertainty on the redshifts is $\sim$ 0.0002, or $\sim$ 60 km s$^{-1}$ at $z =$ 0.3.

Additional objects with published redshifts taken from NED were included as well.  In two fields, the NED additions were significant.  There were 240 NED additions to h12531, as it includes a previously studied supercluster at z = 1.237 \citep{demarco07}, and 133 to pg1115, which includes the z = 0.485 cluster RXJ1117.4+0743.  These additions are visible in the sky plots as small, densely sampled regions near the edges of our fields.  

We note that our algorithm (see Section \ref{ssec:method} and Appendix \ref{appendix:choosingparams}) is not optimized for the relatively overdense patches arising from adding many NED redshifts in the fields h12531 and pg1115.  In those two fields, before running the group algorithm, we thus discard the galaxies obtained from NED that have I-band magnitudes fainter than 21.5 mag (165 galaxies in field 12531 and 21 in field pg1115) to make the magnitude limits of those fields more comparable to the rest of our survey.  In h12531, this cut removes many galaxies at $z \sim 1.24$.  Since our work is optimized at $z <$ 1, we refer the reader to \citet{demarco07} and \citet{tanaka09} for further details.

In field sbs1520, the lens galaxy has two contradictory redshifts available in the literature: $z =$ 0.72 from \citet{chavushyan97} and $z =$ 0.761 from \citet{auger08}.  Thus, neither one appears in our redshift catalog, although both are marked in plots when relevant.

The final redshift catalog includes 9662 unique galaxies.  This catalog includes 9370 galaxies not in the preliminary catalog and analysis presented in \citet{momcheva06}, most of which are in 20 additional fields.

To determine the redshift interval over which each field is most sensitive, we consider the spectroscopic field size and how the spectroscopic completeness varies over our I-band magnitude range.  We assume a homogeneous universe, use the luminosity functions of \citet{faber07} for 0 $< z <$ 1, and implement K-corrections using an LRG spectral template, as in \citet{ammons14}.  Although the selection functions vary slightly from field to field, the peak is at z $\sim$ 0.2-0.25 with generally high sensitivity from 0.05 $< z <$ 0.5.  The survey's sensitivity decreases at $z \gtrsim$ 0.5, so it is more difficult for us to identify structures at these higher redshifts.

\section{Finding Groups}
\label{sec:method}

Several group and cluster identification methods exist in the literature.  Massive, dynamically evolved, low redshift structures can be readily detected in X-ray observations \citep[e.g.,][]{lloyd11}, as can clusters at all redshifts via the Sunyaev-Zeldovich effect \citep[e.g.,][]{bleem15,planck15}. However, these methods also require some member galaxy redshifts to determine the structure's redshift.  

Other group finders rely on additional sources of data, such as photometry and/or photometric redshifts, to locate groups using adaptive matched filters \citep[e.g.,][]{szabo11}, red sequence identification \citep[e.g.,][]{williams06,koester07}, or low surface brightness enhancement identification \citep[e.g.,][]{gonzalez01,white05}.  These methods can be useful, especially when the spectroscopic data are sparse compared with the amount of photometric information available, but are typically more sensitive to massive, evolved systems.  

Since even low mass groups can affect gravitational lensing models, we require a method that is sensitive to less massive systems as well as clusters.  One such approach is to compile large spectroscopic samples and then search for spatial and redshift clustering.  Two common methods are the friends-of-friends \citep{huchra82} and the Voronoi-Delaunay \citep{marinoni02} methods.  These techniques can be useful for uniformly complete redshift samples but cannot be used for samples with large spatial variations in spectroscopic completeness, which we have; in this situation, iterative methods, such as that of \citet{wilman05}, are more appropriate.

\subsection{Algorithm}
\label{ssec:method}

We use the iterative approach of \citet{ammons14} after modification to accommodate our spectroscopic sample and to find less massive structures.  We develop a two-phase process where we identify candidate density peaks based on galaxy overdensities in velocity and spatial projection and then iterate for the galaxy membership and group properties.  The main differences between our method and that of \citet{ammons14} are the method of generating candidate density peaks, the projected spatial criteria in the membership iteration, and the parameter values chosen.  We automate the peak finding process.  We do not use any magnitude or color information to identify candidate groups, because we want to be able to find lower mass groups without prominent red sequences as well as rich clusters.  The resulting group catalog is produced algorithmically, with parameter choices justified in this section and in Appendix \ref{appendix:choosingparams}; only one system, a complex supergroup, is altered after the fact (see Section \ref{sssec:supergroup}). Our algorithm is tuned to select groups that we visually identify but applies criteria uniformly so that our resulting group catalog is reproducible.  It also finds some groups with similar properties that were not visually identified.

There are nine parameters in our method of selecting candidate peaks and determining group membership, some of which are degenerate, leading to multiple ways to achieve roughly the same ends.  We want our catalog to include most of the likely structures we identify by eye as obvious overdensities of galaxies in velocity and projected spatial position (see Section \ref{sssec:visualid}).   We also want to minimize the number of structures with unphysically large velocity dispersions.  Although there is some variation in the exact group catalog, most of the groups are stable to small changes in the parameters.  We test the ``realness'' of our groups in Section \ref{sec:evaluation}.

The steps of our algorithm are as follows:

\begin{itemize}

\item

We identify overdensities of galaxies in velocity by binning the galaxies in 1200 km s$^{-1}$ wide bins (three times the velocity dispersion of that of typical groups we expect).  This method relies on the uncertainties in redshift, and consequently velocity, being small compared to the velocity bin width, as is the case for our sample, so galaxies can be robustly binned. We use the same bin width for all redshifts, although this velocity interval will correspond to a varying rest frame velocity dispersion.  We select velocity bins with at least five galaxies. We then shift the velocity bins five times by steps of 200 km s$^{-1}$, half the expected velocity dispersion for a typical group in our sample, and again select those velocity bins with at least five galaxies.  These multiple velocity bin positions reduce the chance of a group being missed because its members were split by a velocity bin boundary and increases the chance for groups to be well centered in a velocity bin.  We do not use a spatial selection in this step but do in the next step below.

\item

We look for spatial overdensities of the galaxies within these velocity bins.

We divide the observed field into grid squares 3.3$r_{vir}$ on a side, assuming a group-like velocity dispersion of 400 km s$^{-1}$ and using the formula for $r_{200}$ given by Carlberg et al. (1997) as an approximation of the virial radius:

\begin{equation}
r_{vir,\sigma} \approx r_{200} = \frac{\sqrt{3} \sigma_{grp}}{10 H(z)}, 
\end{equation}
where
\begin{equation}
H(z) = H_0 \sqrt{ \Omega_m (1 + z_{gal})^3 + \Omega_{\Lambda}}
\end{equation}

\noindent for a flat cosmology, and $z_{gal}$ corresponds to the lower boundary of the velocity bin.  We center the grid on the median RA and Dec of the galaxies in each bin.  We do not treat the lens galaxy as special in this or in any other step of the group finder.  We define each grid square with five or more galaxies as a ``candidate peak'' and calculate its median redshift, RA, and Dec.  We use these quantities as starting points for our iterative group member identification.  We repeat this candidate peak finder nine more times, centering the grid in different positions (shifting it a quarter of a grid square in RA, a quarter of a grid square in Dec, and a quarter of a grid square in both RA and Dec).   These multiple positions increase the likelihood of identifying multiple overdensities of galaxies in a single velocity bin as candidate peaks if one (or more) is split by a grid square boundary for one grid position and increases the likelihood of groups being well-centered in at least one grid position.  We use all the candidate peaks generated in all of these grid positions for all of these velocity bin phases, even though some are different parts of the same structures.  The combination of velocity bin width, grid square size, velocity dispersion, and number of galaxies per bin or grid square is chosen so the resulting group catalog best reproduces our initial visual identifications discussed in the next section.

\item

We optimize the group membership using these candidate peaks as starting points.  We select galaxies within a redshift interval $\Delta z$ of the candidate peak redshift and 3$r_{vir}$ from the peak's projected spatial centroid assuming a velocity dispersion of 400 km s$^{-1}$.  We choose a $\Delta z$ that corresponds to 6600 km s$^{-1}$ at the candidate peak's redshift, which corresponds to $\pm$ three times the velocity dispersion of a Coma-sized cluster.  We select a large $\Delta z$ compared with our initial assumed velocity dispersion so we do not exclude galaxies in the high velocity tails should the identified peak actually correspond to a higher velocity dispersion structure, as would become evident during the membership optimization process described below.  Using these galaxies, we calculate the bi-weight redshift ($z_{grp}$), mean position, and the cosmologically corrected bi-weight line-of-sight velocity dispersion ($\sigma_{grp}$). We then select the galaxies within 3$\sigma_{grp}$ of this new mean redshift after transforming the velocity dispersion to a redshift interval using
\begin{equation}
{\Delta}z = \frac{3\sigma_{grp}}{c} ,
\end{equation} 
and within 3$r_{vir}$, using $z_{grp}$, of the spatial centroid.

In the first iteration, we also reject members that are more than a velocity dispersion away from their nearest neighbor in velocity, as they are unlikely to be members.  This choice has an effect on roughly 20\% of the catalog, generally one or more galaxies being removed in the outskirts of the group as projected on the sky; the new group redshift and/or velocity dispersion differs by more than three times the error in $z_{grp}$ or $\sigma_{grp}$, respectively, for $\sim$ 1\% of the groups. There are an additional 19 groups with at least five members when clipping is not performed, but a structure with an unlikely large velocity dispersion is also identified.  Thus, we choose to perform this clipping to reduce the identification of galaxies as group members that are in the tails of both the group velocity and spatial distributions and minimize unrealistically large structures.

Using the galaxies that satisfy these cuts in velocity and position, we recalculate the mean redshift and spatial centroid and find the resulting membership.  We recalculate the group parameters and redetermine the galaxies within 3$\sigma_{grp}$ of the mean redshift and 3$r_{vir}$ of the spatial centroid for ten total iterations.  We consider a group to be converged if the membership does not change in the last three iterations.  Given our field boundaries, groups at low redshift and/or with large velocity dispersions may be sampled to less than 3$r_{vir}$.  We flag any group not sampled to at least 1$r_{vir}$ in the catalog tables (Tables \ref{table:grps5andup}, \ref{table:grps3and4}, and \ref{table:candgrps}).

\item

Large structures may have member galaxies spread over a projected area larger than 3$r_{vir}$ when a velocity dispersion of 400 km s$^{-1}$ is assumed, so our binning and gridding systems might generate multiple candidate peaks for a single structure. Because our iterative procedure can find the actual larger group, these multiple candidate peaks may converge on identical or overlapping groups.  Thus, we must remove duplicate groups and combine nearly identical groups (see Appendix \ref{appendix:choosingparams} for details).  If multiple candidate peaks converge on multiple identical group memberships, we select one of the identical groups without rerunning the algorithm.  If in more than 75\% of the trials a group converges with the same membership, we add that group to our catalog.  In other cases where a group shares at least 50\% of its members with another, we determine the mean of those groups' spatial centroids, redshifts, velocity dispersions, and virial radii, and reiterate on the membership to produce a single group for the catalog.  If between zero and 50\% of a group's members are also assigned to another group, we determine to which group the galaxies are more likely to be bound and remove them from the other (see Appendix \ref{appendix:duplicateremoval}).

\end{itemize}

For these groups, we calculate virial masses.  There are two ways to determine the masses of structures: via the projected harmonic mean radius ($R_{PV}$) and solely via the velocity dispersion.  \citet{biviano06} find that the $M_{vir,R_{PV}}$ and $r_{vir,R_{PV}}$ are overestimated for small sample sizes, whereas $M_{vir,\sigma}$ and $r_{vir,\sigma}$ are not biased.

As for the virial radius, to approximate the virial mass we use the formula for $M_{200}$ derived by \citet{munari13} from simulations, fit over the mass range of $10^{13} \lesssim M_{200} \lesssim 10^{15} M_{\odot}$:

\begin{equation}
M_{vir,\sigma} \approx M_{200} = \frac{10^{15}M_{\odot}}{h(z)}\left(\frac{\sigma_{grp}}{A_{1 D}}\right)^{(1/\alpha)},
\end{equation}
where $h(z) = H(z)/(100$ km s$^{-1}$ Mpc$^{-1})$.  We use $A_{1 D} = 1177$ km s$^{-1}$ and $\alpha = 0.364$ from their simulations using AGN feedback and galaxy tracers.  Compared to these, $R_{PV}$-based virial masses ($M_{vir,R_{PV}}$) using the equations of \citet{girardi98} and \citet{limber60} are systematically higher by $\sim$ 0.6 dex over most of the mass range of our final catalog.  This offset creates a high mass tail, consisting of 14 groups with $M_{vir} > 10^{15} M_{\odot}$ for 0.2 $< z_{grp} <$ 0.5, that is not in agreement with the high mass slope of the Millennium-XXL (MXXL) cosmological simulation \citep{angulo12} given any arbitrary volume normalization (see Section \ref{sssec:Millennium}). Furthermore, $r_{vir,R_{PV}}$ is about 1.5 times larger than both $r_{vir,\sigma}$ and that expected for NFW halos. 

Thus, our data support the conclusions of \citet{biviano06} and motivate our use of $r_{vir,\sigma}$ and $M_{vir,\sigma}$.  These choices differ from those made in \citet{ammons14}, who calculate $R_{PV}$ and $M_{vir,R_{PV}}$.

We only report masses $\ge 10^{13} M_{\odot}$.  The $M_{vir}$ relation we adopt has not been calibrated for smaller masses.  Smaller masses (i.e., those in the range normally expected for single massive galaxy halos rather than virialized groups) are likely to be underestimated due to incomplete group member identification.  

All uncertainties in group quantities are calculated using the bootstrap method with 1000 iterations.  We resample with replacement the identified group members and recalculate the group properties using the resampled members.  We then subtract the median value of each property over all the iterations and find the 16th and 84th percentile values, which are listed in the tables with the properties measured using all the algorithmically identified group members.

Our group catalog and individual group memberships are likely to be somewhat incomplete, although we later show that our catalog reproduces the distributions of group kinematic properties and the fraction of group members to field galaxies of other catalogs in the literature (Section \ref{sssec:statcomp}).  Here we perform a simple test of how incompleteness might affect the derived group centroids, velocity dispersions, and masses.  For the 32 groups with at least 20 members, we randomly select 25\%, 50\%, and 75\% of their members without replacement 1000 times and find the median group parameter values.  For all three incompleteness levels, the median values are within three standard deviations of the original values using the previously calculated uncertainties, although the effect can be systematic and does not account for groups whose incompleteness would have precluded their identification in the first place.  For 25\% completeness, the velocity dispersions are underestimated \citep[see][]{zabludoff98} by a median of 3\% and the masses by a median of 9\%.  We note that the groups with at least 20 members are well-sampled and/or rich compared with the other groups in our catalog.  The 20$+$ member groups have a median velocity dispersion of 630 km s$^{-1}$ compared with the median of 350 km s$^{-1}$ for the full sample of 5$+$ member groups.  However, there is considerable overlap in the velocity dispersion ranges of these two samples, 290 - 1180 km s$^{-1}$ compared with 60 - 1200 km s$^{-1}$ for the 20$+$ and 5$+$ member groups, respectively.  More importantly, the systematic offset between the measured velocity dispersion for the full 20$+$ member groups and for the 25\% completeness subsamples does not depend on velocity dispersion; the velocity dispersion generally is underestimated in the 25\% subsamples within one standard deviation over the full range of velocity dispersions.  Hence, we conclude that the impact of incompleteness, even on the lower sigma groups in our catalog, falls generally within the stated errors.

While we make choices to avoid fragmenting rich clusters with substructure, distinct but neighboring structures might also be artificially combined, resulting in more high velocity dispersion (higher mass) groups in our catalog than expected.  Thus, we examine the four clusters with $M_{vir} \ge 10^{15} M_{\odot}$ in our catalog to evaluate whether they likely are real or the artificial combination of multiple groups and/or large scale structure.  

To do so, we see if these groups are stable to changes in the velocity dispersion initial guess by comparing them in our catalog to that created using a initial velocity dispersion of 500 km s$^{-1}$.  Of the four massive clusters, three are identical in both catalogs, and the other has group parameters that agree within the $3\sigma$ uncertainties.  Additionally, three of these four have red sequences in their color magnitude diagrams that suggest evolved clusters.  The other might have a red sequence, but it is at low redshift and is thus poorly sampled.  One of these also is a supergroup (see Section \ref{sssec:supergroup}).

We also look at the 16 groups with the most bi-modal member galaxy velocity distributions to see if each velocity peak corresponds to a distinct substructure on the sky. There is only one group that displays a well defined distinction, which we identify as a supergroup (see Section \ref{sssec:supergroup}).

\subsection{Comparison to Visual Identification}
\label{sssec:visualid}

Early visual identification was used to construct a target group catalog for refining our group finding algorithm.  Two of us (AIZ and MLW) looked at the galaxy velocity histograms, as well as the individual galaxy velocities and projected spatial positions, in each field.  We identified likely groups in 1500 km s$^{-1}$ wide velocity bins where there were at least five (AIZ) or ten (MLW) galaxies clustered on the sky relative to the surveyed field.

We used cases where our algorithm did not find visually identified groups to guide our parameter tuning and to understand the intrinsic limitations of the algorithm for this sample (e.g., the difficulty of identifying groups at higher redshifts).  This check, of course, assumes that everything we identify visually is a real structure.

Our algorithm identifies visually identified groups in 69\% of peaks with at least five galaxies (91\% of peaks with at least ten galaxies) and with strong clustering in both RA and Dec.  There are two possible general types of failures: the algorithm could be missing real structures that are well represented in our data (algorithm failure) or the algorithm could be missing structures because they are inadequately sampled (data failure).  Usually the visually identified groups that are not identified by the algorithm are near the field edges, at higher redshift, or are in regions of their field that is more densely sampled, artificially making galaxies look clustered.  Thus, data failure is the dominant cause for our algorithm not identifying all our visually identified groups.  Of all the velocity peaks in which our algorithm identifies structures, 11\% were not visually identified.  These groups do not have different properties than others found both visually and by our algorithm except for those that have fewer members than we identified visually (the three and four member algorithmically identified groups).

\subsection{Additional Candidate Groups}
\label{ssec:candgrps}

Our magnitude-limited, multi-object spectroscopy is less successful at higher redshifts and in overdense cluster cores, reducing the efficacy of our group-finding algorithm.  To include additional structures for future spectroscopic followup or modeling in these fields, we create a supplemental candidate group catalog from our visual identifications. We focus on 58 single redshift bins with $\ge$ 10 galaxies clustered on the sky. For the additional candidate groups, we apply a simple 3$\sigma$ clipping procedure \citep{yahil77} to the galaxy redshift distribution within the bin to estimate group membership.  We use no projected spatial criteria in this case. Once the 3$\sigma$ clipping converges, we calculate the group virial radius.  We then discard any members projected more than 3$r_{vir}$ from the projected spatial centroid and recalculate all the group properties, including $r_{vir}$.

Four groups converge that include galaxies already assigned membership to algorithmically selected groups (two candidate group members in one case, three in two, and all 13 in one).  These galaxies are removed from the candidate groups, the group properties are recalculated, and the three that remain in the final catalog are flagged accordingly.  Of the 58 redshift bins, 26 are rejected because their memberships did not converge or converged to groups with fewer than three members, they had fewer than three members within 3$r_{vir}$, or they overlapped completely with a group in our algorithmic catalog.  

We check the remaining groups with the most distinctly bi-modal velocity distributions to see if they are separable on the sky; none have velocity substructure that corresponds to distinct spatial substructure, so we do not remove any groups.  Since the mean pairwise radial velocities at $r \sim$ 1 Mpc for dark and luminous matter at 0 $< z <$ 1 is around 200 km s$^{-1}$ in the simulations of \citet{weinberg04}, we discard an additional 9 candidate groups with $\sigma_{grp} <$ 200 km s$^{-1}$, as they could be merely cuts through filaments or sheets of galaxies rather than virialized groups.

We are left with 23 non-algorithmically selected candidate groups in our supplemental catalog.  While we do not use color information to identify these structures, five have red sequence-like galaxy overdensities in color-magnitude space.  Thus, while incomplete sampling might make their identification and characterization more difficult, their assigned candidate group members likely do trace evolved structures.

There are still some redshift bins at higher redshift with fewer than ten galaxies but that likely are parts of rich structures, since there are overdensities where our sensitivity is low.  These peaks are in field b1422 at $z \sim$ 0.78, b1600 at $z \sim$ 0.77, he1104 at $z \sim$ 0.73, and sbs1520 at $z \sim$ 0.72 and 0.82.  While we have enough galaxies in these redshift bins that there are noticeable peaks, we do not have enough to calculate meaningful candidate group quantities, so we make no further attempt to include them.

\section{Testing the Group Catalog}
\label{sec:evaluation}

We perform tests to evaluate the quality of the group catalog produced by the group finding algorithm.

\subsection{Comparison of Individual Groups to the Literature}
\label{sssec:comptolitgrpcat}

We compare our algorithmic catalog to group samples in the literature that overlap our fields.  As the methods for identifying structures differ, this comparison tests the robustness of our catalog.

We compare our groups to groups and clusters in NED\footnote[1]{The NASA/IPAC Extragalactic Database (NED) is operated by the Jet Propulsion Laboratory, California Institute of Technology, under contract with the National Aeronautics and Space Administration.} within 20$^{\prime}$ of our field centers with spectroscopic redshifts that are not from previous work by our group.  We consider a NED structure to be found if our group has a redshift within 3$\sigma_z$ of the NED spectroscopic value.  We do not compare other group properties, because ways of calculating them vary widely, they are not uniformly reported, and the survey footprints vary in size and position on the sky.

Three groups from NED that we do not recover are at either too high (two) or too low (one) redshift to be readily detected given our redshift survey's sensitivity, and four are in regions of their fields where we have little to no spectroscopic coverage.  Of the remaining 26 entries, we find 17.  For another five, we find either substructures of the previously reported group or we combine known groups into one.  We do not find four groups, which we explain below.

Below we discuss those of our fields with groups listed in NED.

\medskip

\textbf{b1600}: \citet{auger07} find four groups in this field in their 40 galaxy redshift sample.  We find a group at $z \sim$ 0.29 (0.2893 $^{+0.0005}_{-0.0004}$ which agrees with their 0.291 within their reported precision).  We also find a group similar to their $z =$ 0.415 group ($z_{grp}$ = 0.4146 $\pm$ 0.0002); five of our six group galaxies are added into our redshift sample from their work via NED.  We do not identify groups at $z =$ 0.540 or 0.623, as they do; none of the galaxy redshifts from their $z =$ 0.291, 0.540, or 0.623 groups are in \citet{momcheva15}.  

\textbf{h12531}:  We find a group at $z_{grp}$ = 0.0531 $^{+0.0007}_{-0.0008}$, a similar redshift to Abell 3528 of the Shapley Supercluster (which has three entries in NED at similar redshifts).  Our group is likely part of this larger cluster and has 13 of 22 member galaxies that have redshifts from NED.  Since the Abell cluster centroid is at the very edge of our field, we caution that the properties for our group might not be well determined.

\textbf{hst14113}: We recover a group at $z =$ 0.0808 $\pm$ 0.0008, near the $z =$ 0.0809 group reported by \citet{miller05}; two of our 26 members have redshifts added from NED.

We also recover part of the 3C 295 cluster originally identified by \citet{dressler92} at $z =$ 0.4599 with 112 galaxies and a velocity dispersion of 1300 km s$^{-1}$.  We recover a group with 55 galaxies at slightly higher redshift (but still within the uncertainties: $z =$ 0.4603 $\pm$ 0.0003) and two small groups at slightly lower redshift (seven and six members at $z =$ 0.4504$^{+0.0009}_{-0.0012}$ and 0.4592$^{+0.0004}_{-0.0003}$, respectively).  In our structures, only nine galaxies (four of seven members of the $z =$ 0.4504 group and five of 55 members of the $z =$ 0.4603 group) are added from NED.

\textbf{lbq1333}:  We find a group at $z =$ 0.0856 $\pm$ 0.0003, near the $z =$ 0.085 group reported by \citet{mcconnachie09}; three of the five members have redshifts from NED.

\textbf{mg0751}: NED lists three groups with $z \sim$ 0.0266.  Although we are relatively insensitive to structures at such low redshift, we do identify a four galaxy group at $z =$ 0.0265$^{+0.0002}_{-0.0001}$; one of these members has a redshift from NED.

\textbf{mg1549}: We identify a group at $z =$ 0.0709, as reported by \citet{lopes09}; three of our seven members are from NED.

\textbf{pg1115}: We find a structure at $z =$ 0.1600, like that reported by \citet{barkhouse06} and \citet{mason00}, including six (of 11) galaxies with redshifts from NED.  

We do not recover a group at $z =$ 0.477, as attributed to WARPS-II by \citet{mullis03}, but the more recent WARPS-II catalog of \citet{horner08} report a redshift of 0.4859, where we do find a group; none of the 11 members have redshifts from NED.

\citet{carrasco07} find a cluster at z = 0.485 made up of at least two substructures along the line of sight.  We recover the substructure redshifts at $z =$ 0.48218 and 0.49191 (our $z_{grp} =$ 0.4819$^{+0.0007}_{-0.0006}$ and 0.4922$^{+0.0007}_{-0.0005}$), but we do not find the two to be part of some larger cluster.  All but three of the galaxies in these two of our groups (of 25 total members) are added from NED.

\textbf{rxj1131}: \citet{tucker00} list four groups at $z =$ 0.0534, 0.0966, 0.1014, and 0.1032.  We find a group at 0.0531$^{+0.0007}_{-0.0003}$, consistent with their lowest redshift group; none of our group members have redshifts from NED.  We also recover a group at $z =$ 0.1021 $\pm$ 0.0004 that is a combination of their other three groups but has a redshift within 3$\sigma_z$ of their two higher redshift groups; seven of the 66 galaxies in our group are from NED.

\textbf{sbs1520}: \citet{auger08} find three groups ($z =$ 0.716, 0.758, and 0.818).  We only identify a group that is similar to the middle one; all six of our members are added from \citet{auger08}.  Our group at $z =$ 0.7590$^{+0.0004}_{-0.0006}$ is much smaller than theirs (six group galaxies to their 13).  One complication is that we do not include the lens galaxy in our sample, because its redshift is ambiguous, while they do and assign it to this group.  The most likely cause for the discrepancy, however, is their more permissive spatial selection.  We are capable of recovering groups similar to their $z =$ 0.716 and 0.818 ones if we accept galaxies up to 5$r_{vir}$ from the group centroid, which may not be bound to the group.

\medskip

In summary, our group finder finds similar, albeit not always identical, groups to those in the literature in most cases.

\subsection{Statistical Comparison of Group Properties to the Literature}
\label{sssec:statcomp}

While it is valuable to compare specific cases to ensure our algorithm can recover previously-reported structures, we also test whether the statistical properties of our group catalog are similar to those elsewhere.  Although we only discuss our algorithmic catalog here, including our non-algorithmic candidate groups does not significantly affect these comparisons.

\subsubsection{Distribution of Group Velocity Dispersions}
\label{sssec:zCOSMOS}

We compare our group catalog to that of zCOSMOS \citep{knobel12} for groups with reported velocity dispersions and $0.2 < z_{grp} < 0.5$.  We limit both samples to those with at least five spectroscopic members. 

\begin{figure}
\includegraphics[clip=true, width=9cm]{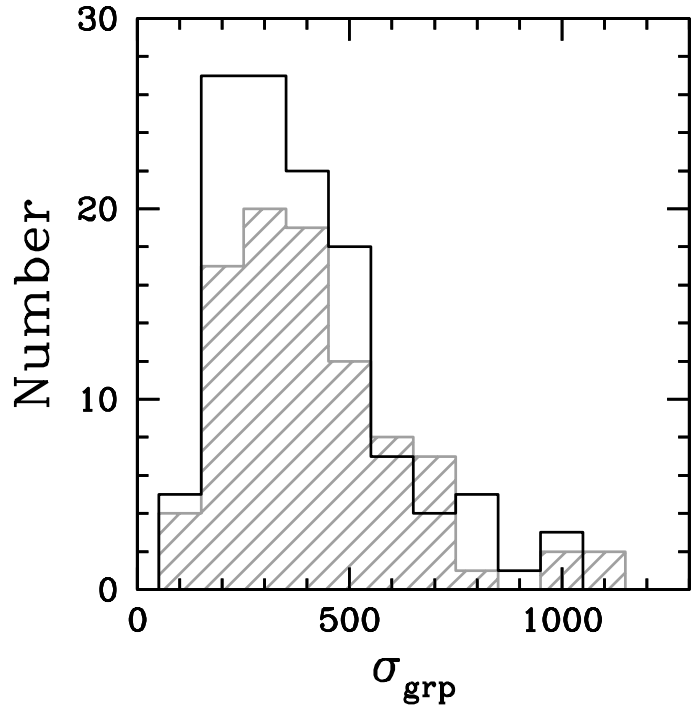}
\caption{Comparison of the distribution of velocity dispersions between our group catalog (black histogram) and that of zCOSMOS (shaded gray histogram).  For both catalogs, we select groups with $0.2 < z_{grp} < 0.5$ with at least five spectroscopically identified members and reported velocity dispersions.  These distributions cannot be distinguished at the 95\% confidence limit using a K-S test.}
\label{fig:sigmacosmoscompare}
\end{figure}

Using a Kolmogorov-Smirnov (K-S) test, we find the resulting distributions of velocity dispersions of the two catalogs are not distinguishable at the 95\% confidence level (see Figure \ref{fig:sigmacosmoscompare}).

\subsubsection{Frequency of Significant Substructure}
\label{sssec:DStest}

For those of our groups with at least ten members, we test for substructure using the Dressler-Shectman test \citep{dressler88}.  This test calculates the difference between the average velocity of a galaxy and its nearest neighbors as projected on the sky compared with the mean velocity of the entire structure, as well as the difference between the local and global velocity dispersion.  Spatial overdensities that correspond to velocity overdensities indicate substructure.  We use 5000 Monte Carlo shuffles when calculating the probability that the Dressler-Schectman statistic $\Delta$ is observed due to chance.  Of our main group sample with $0.2 < z_{grp} < 0.5$, 55 groups have at least ten members.  Of those, 21 have substructure detected (38\%) at the $P <$ 0.05 significance level.

Restricting our sample to only consider groups with at least 20 members (23 groups) and identifying $P <$ 0.01 as significant, as do \citet{hou12}, we find six have substructure (26\%), in good agreement with the 27\% (four of 15) found by \citet{hou12}.

\subsubsection{Distribution of Group Masses}
\label{sssec:Millennium}

\begin{figure}
\includegraphics[clip=true, width=9cm]{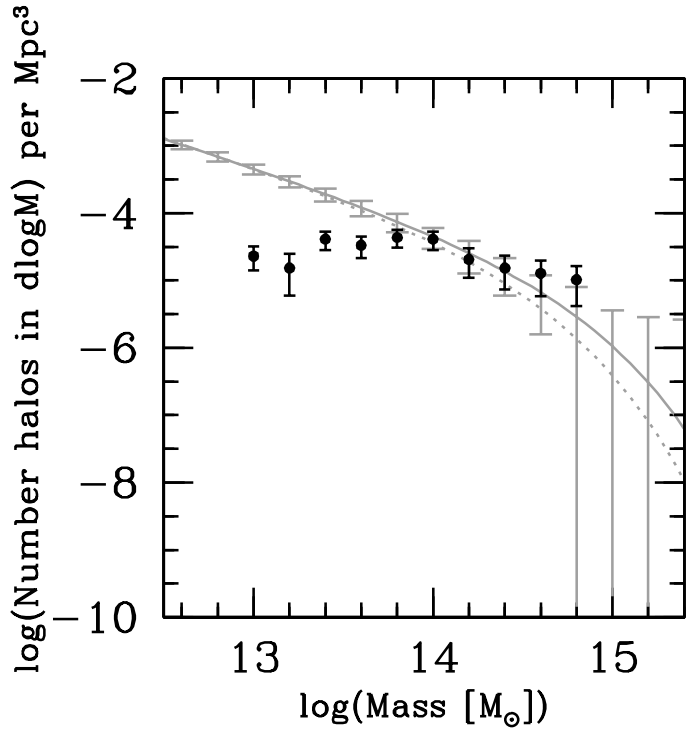}
\caption{Comparison of our mass function for groups with $N_{m} \ge 5$ and $0.2 < z < 0.5$ (black points) with the Millennium simulation mass functions (gray curves) at $z =$ 0.2 (solid) and 0.5 (dotted).  Bins are independent.  Black error bars are the 1$\sigma$ spread in the number of groups in a bin (not the error on the mean) using 1000 bootstrap realizations of our group masses and Poisson statistics.  Gray error bars are the expected uncertainty in the observed counts due to cosmic variance given the simulation's mass function at $z = 0.2$ using the Cosmic Variance Calculator \citep[][]{trenti08}.  The black points are normalized arbitrarily along the vertical axis so the high mass slope can be compared with the models (see text for details).  The two highest mass bins do not have any observed groups which is consistent with the model given the shot noise.  The high mass slope of our mass function agrees well, within the uncertainties, with that of the simulation.  So, provided the normalization is reasonable, we are reliably finding high mass structures.  Our catalog begins to suffer from incompleteness at masses lower than $\sim 10^{14}{ M_\odot}$.  }
\label{fig:millennium}
\end{figure}

We compare our group mass function to that of the Millennium-XXL (MXXL) Simulation (see Figure \ref{fig:millennium}).  MXXL is a high resolution, large volume, cosmological N-body simulation aimed at studying simulated cluster properties \citep{angulo12}.  We calculate the mass function of our algorithmic catalog for those groups with $0.2 < z < 0.5$, $N_{m} \ge 5$, and $M_{vir} \ge 10^{13} M_{\odot}$.

We wish to test 1) the shape of our mass function at high masses, where our group sample should be mostly complete, and 2) at what lower masses our sample is incomplete.  Our mass function should be normalized by the volume probed by our sample to compare it to the simulations.  However, because our redshift sample is not volume-limited, and there are large variations in spectroscopic completeness over each individual field and among fields, this normalization is not straightforward. Therefore, we ask whether there is a normalization that allows the mass distribution's shape to match the simulations in the highest mass bins and that represents a reasonable effective sample volume.  We select a normalization using a $\chi^2$ fit of the data at $M \gtrsim 10^{14} M_{\odot}$ assuming a two-sided Gaussian for the uncertainties in the data.  The resulting fit has a $\chi^2 \sim$ 1 and a $\sim 5^{\prime}$ radius circular footprint per field.  This footprint size corresponds to the $5'$ radius regions around the lenses that have the highest spectroscopic completenesses.  Figure \ref{fig:millennium} shows this normalization of our data that both matches the simulations at high masses and predicts a sensible field size.  We conclude that the matched normalization is reasonable.

We calculate the simulation's mass function using the Millennium simulation cosmology and the fitting function of \citet{angulo12} using HMFcalc \citep{murray13}.  We calculate their mass functions for $z =$ 0.2 and $z =$ 0.5 to bracket the redshift range for which we calculate our mass function.  Our high mass tail's slope agrees with MXXL's to within the uncertainties.  Thus, we conclude that the number of high mass objects we find is in agreement with theoretical expectations, provided our normalization of our mass function is reasonable, above roughly $10^{14}{ M_\odot}$.  Our mass function turns over at masses $\sim 10^{14}{ M_\odot}$ and lower, indicating incompleteness in our group catalog at these masses.

\subsubsection{Fraction of Galaxies in Groups}

For our algorithmic group catalog, we calculate the fraction of all galaxies that are group members and compare it to the literature.

\citet{carlberg01}, over similar redshift (0.1-0.6) and velocity dispersion (40-700 km s$^{-1}$) ranges, find that 691 of 3290 (21\% $\pm$ 1\%, assuming Poisson errors) of their galaxies lie in groups of at least three members, where membership is defined within 1.5$r_{vir}$ of the group centroid.  To make a fair comparison, we consider our algorithmic groups with three or more members and similarly limit their membership to within 1.5$r_{vir}$.  We find that 1837 of our 9432 galaxies (19.5\% $\pm$ 0.5\%) are group members, consistent with \citet{carlberg01}.

\section{Results}
\label{sec:results}

\subsection{Group Catalog Properties}
\label{sec:catalogsummary}

\begin{figure*}
\includegraphics[clip=true, width=18cm]{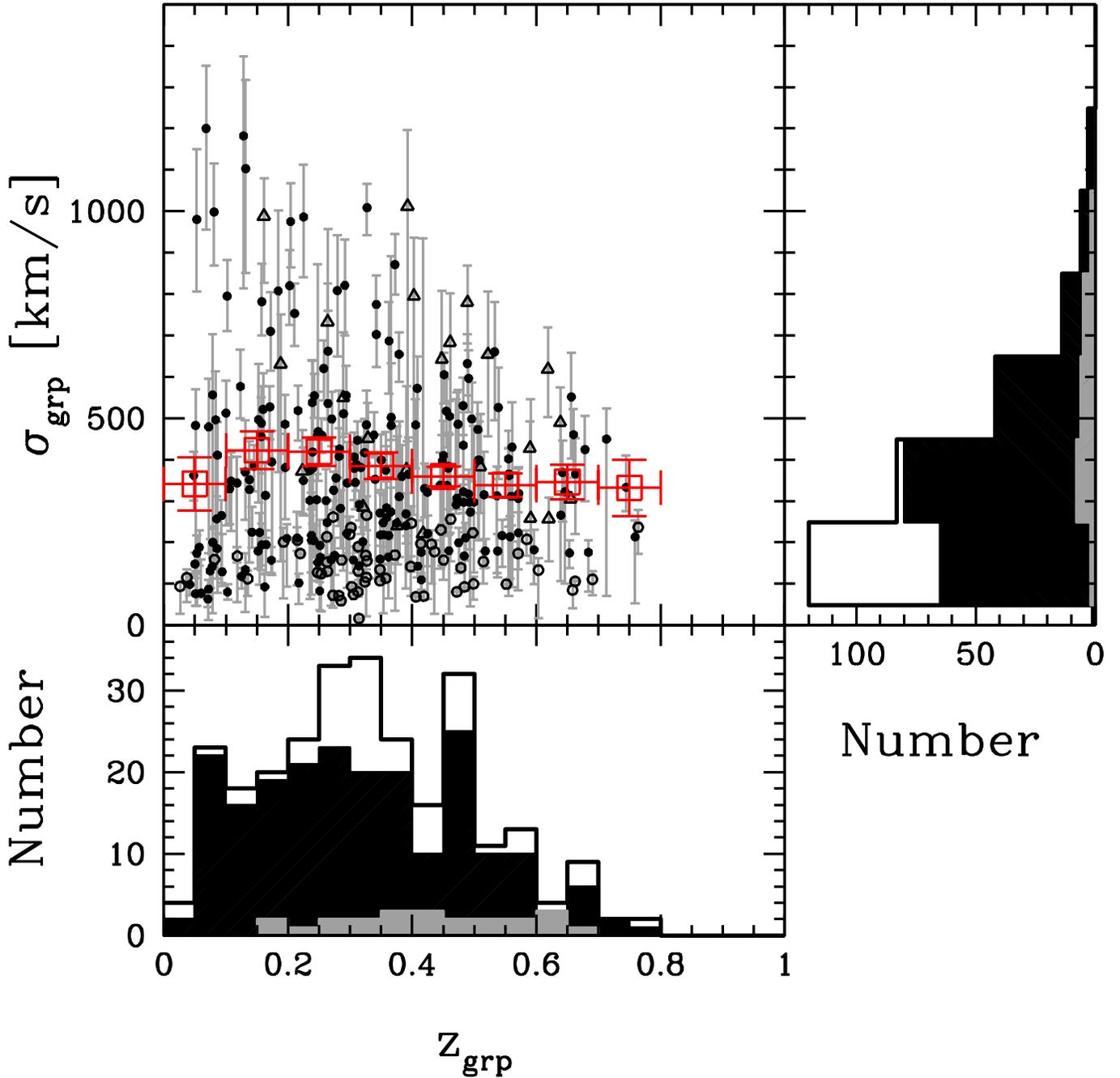}
\caption{Velocity dispersion and redshift distributions for our group sample. \textit{Upper left:} We show groups with at least five members (filled circles), those with three and four members (open circles), and the additional group candidates that are not found by our algorithm, but visually identified (open triangles; see Sections \ref{ssec:candgrps} and \ref{sec:catalogsummary}). Open red squares denote the mean $\sigma_{grp}$'s derived from the filled circles; their red horizontal error bars show the 0.1 redshift bin width and vertical error bars the standard error in the mean.  \textit{Lower left:} Redshift distributions for the 3$+$ member group sample (open histogram), the 5$+$ member group sample (solid black), and the additional group candidate sample (solid gray). The dip at $z_{grp} \sim$ 0.4 corresponds to a decrease in the number of galaxy redshifts measured \citep[see Figure 6 of][]{momcheva15}, likely due to the Ca H and K absorption lines coinciding with the [OI] 5577\AA{ }sky line. \textit{Upper right:} Velocity dispersion distributions for the same three samples. Our group catalog spans a range of velocity dispersions and redshifts, although most are in the 100 km s$^{-1} < \sigma_{grp} <$ 600 km s$^{-1}$ and $0.05 < z_{grp} < 0.55$ ranges.  The dearth of higher redshift groups arises in part from the difficulty of spectroscopically sampling faint, overdense group cores.  The differences between the algorithmically-selected and supplemental, visually-identified samples are discussed in Sections \ref{ssec:candgrps} and \ref{sec:catalogsummary}.}
\label{fig:zsig}
\end{figure*}

We present the group catalog in Tables \ref{table:grps5andup} and \ref{table:grps3and4}, present group member galaxies in Appendix \ref{appendix:membertables} (Table \ref{table:alggrpgxys}), plot the redshift and velocity dispersion distributions in Figure \ref{fig:zsig}, and plot redshift histograms for each field in Figure \ref{fig:zhists}.  We also include sky plots and velocity histograms for each group in Appendix \ref{appendix:grpplots} (Figure \ref{fig:grpskyplotsandvelhists}).

In our algorithmic group catalog, we find 210 groups with at least five member galaxies, 186 of which are at $z \ge$ 0.1.   At redshifts below 0.1, our fields are not large enough to sample out to $r_{vir}$ for any but the smallest groups.  So, we present groups at $z <$ 0.1 with the caveat that the group properties, especially the group projected spatial centroid, might be poorly determined.  Of the 210 at all redshifts, 187 have not previously been identified in NED either by other authors or in work using earlier versions of our redshift catalog \citep{williams06}.  We identify groups at  0.04 $\le z \le$ 0.76 with our range of greatest sensitivity at $0.2 < z < 0.5$ and a median at $z_{grp} = 0.31$.  Our groups have 60 km s$^{-1} \le \sigma_{grp} \le 1200$ km s$^{-1}$, with most (84\%) with 100 km s$^{-1} < \sigma_{grp} < 600$ km s$^{-1}$.  The median value is $\sigma_{grp} =$ 350 km s$^{-1}$.

From our experiments with our group algorithm (see Appendix \ref{appendix:canpeakselection}), we are not confident in the robustness of groups of 3-4 galaxies.  Using geometric simulations of large-scale structure, \citet{ramella97} estimate that 50-75\% of their groups with three members are not gravitationally bound systems, while only 10-30\% of five or more member groups are spurious.  As our algorithm only looks for groups in velocity bins and grid squares with at least five galaxies, three and four member groups are found only when there is another group at a similar redshift and position on the sky or there are field galaxies nearby.  However, we tabulate three and four member groups for completeness and to inform future observational followup and gravitational lens modeling.  We find 59 groups with three or four members at $0.03 \le z \le 0.76$ and a median redshift of 0.33.  These groups have 20 km s$^{-1} < \sigma_{grp} < 270$ km s$^{-1}$ and a median of 140 km s$^{-1}$.

We find one supergroup that we discuss in Section \ref{sssec:supergroup} and present in Table \ref{table:supergroup}, Appendix \ref{appendix:membertables} (Table \ref{table:supergrpgxys}), and Figure \ref{fig:supergroup}.

We present 23 additional candidate groups in Table \ref{table:candgrps}, Appendix \ref{appendix:membertables} (Table \ref{table:candgrpgxys}), and Appendix \ref{appendix:grpplots} (Figure \ref{fig:grpskyplotsandvelhistscandgrps}) that were not identified by the algorithm.  They span a range of $0.16 \le z \le 0.66$ with a median of 0.42, 220 km s$^{-1} < \sigma_{grp} <$ 1010 km s$^{-1}$ with a median of 490 km s$^{-1}$, and their median number of members is eight.  The additional groups have, on average, somewhat higher velocity dispersions and mean redshifts than the $N_m \ge$ 5 groups selected by our algorithm.  Even at a given redshift, the non-algorithmic groups tend to have higher velocity dispersions.  These differences arise from the sorts of groups we are selecting with each method.  We require at least 10 galaxies to be at a similar redshift to identify a possible non-algorithmic group where there is no algorithmic group.  This choice selects for the richer groups missed by the algorithm.  Non-algorithmic groups tend to have higher average redshifts, mainly because we only report those not found by our algorithm and the algorithm identifies most of the groups at low redshift.

We compare our group catalog to several group and cluster catalogs from the literature: CNOC2 \citep{carlberg01}, EDisCS \citep{white05}, GAMA \citep{robotham11}, zCOSMOS \citep{knobel12}, and a flux-limited group catalog for SDSS galaxies \citep{tempel14} (see Figure \ref{fig:catcomp}).  While our catalog is similar to others in size, limiting magnitude, group velocity dispersions, and group redshifts, it is unique in that it is in fields of galaxy-scale strong gravitational lenses.

\subsection{Supergroup}
\label{sssec:supergroup}

In field b0712 our algorithm identifies a structure at $z_{grp} =$ 0.2941 with the largest membership found in our sample (230 galaxies).  However, it exhibits three clumps in projected spatial position and two distinct velocity peaks.  While having clumps in space and velocity in clusters is reasonable if groups have fallen in but have not completely virialized within the cluster, these clumps have projected spatial centroids that are several Mpc apart, on the order of or larger than the cluster's virial radius.  Thus, we conclude that this structure is a supergroup, remove it from our final group catalog, and present its properties and those of its three main substructures in Table \ref{table:supergroup}, Table \ref{table:supergrpgxys}, and Figure \ref{fig:supergroup}.  

Versions of this structure have been identified before.  \citet{fassnacht02} find a much smaller group (ten members with $\sigma_{grp}$ = 306 km s$^{-1}$) at $z_{grp} =$ 0.2909 with their smaller sample ($\sim$ 50 redshifts within a couple arcminutes of the lens). \citet{momcheva09}, using an earlier version of the redshift catalog we use, finds a large cluster (231 galaxies) but suggests it might alternatively be a sheet of galaxies or several groups that are merging, similar to our interpretation. 

In a different field, \citet{smit15} find a supergroup with similar properties to ours.  Their super galaxy group at $z =$ 0.37 has four component groups.  The difference in group average velocity between the lowest and highest redshift components is about twice as large as the difference between ours, although when their outlying component is excluded, the velocity difference between the remaining lowest  and highest redshift components is roughly five times smaller than ours.  The projected spatial separations of their components are similar to that of ours (a few Mpc).  The components' velocity dispersions are similar: ours range from 330 to 420 km s$^{-1}$ and theirs from 303 to 580 km s$^{-1}$.  However, our components have more identified members (39 to 54 galaxies compared to their 13 to 29).  Our spectroscopic completeness over most of our field is comparable to theirs \citep[as suggested by][]{gonzalez05}, although they have a much deeper limiting magnitude ($R \le$ 22.5), so our components are likely to be richer structures than theirs.

\begin{figure*}[!h]
\includegraphics[clip=true, width=18cm]{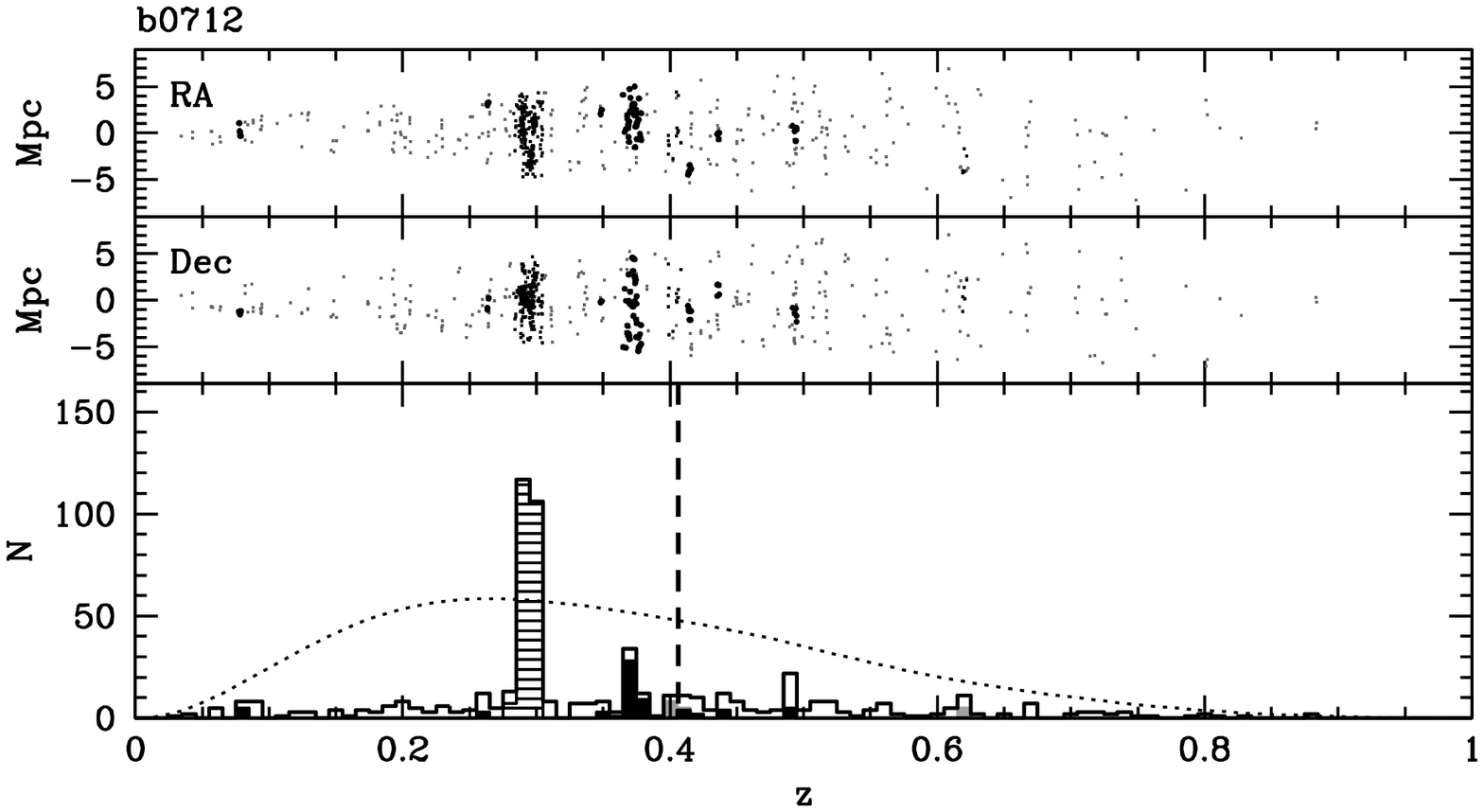}
\includegraphics[clip=true, width=18cm]{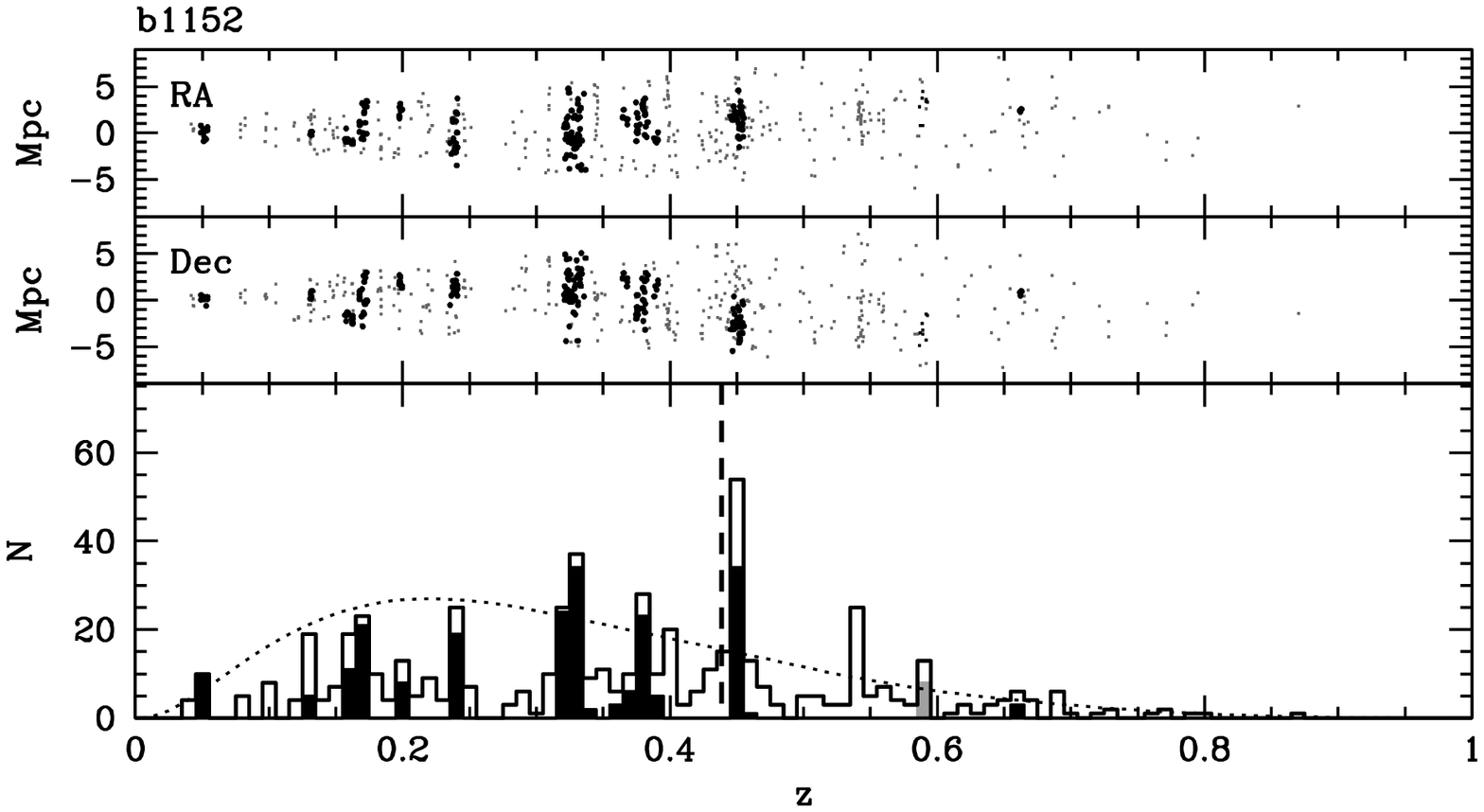}
\caption{Redshift histograms for the 26 fields we include in our group catalog.  The top two panels are the RA (top) and Dec (middle) offsets from the center of the field in Mpc at the redshift of the galaxy; the opening angles of the beams are exaggerated to show structure.  The bottom panel shows the redshift distribution of the sample in redshift bins 0.01 wide, which corresponds to $\sim$ 2000 km s$^{-1}$ at z = 0.3. The gray points and open histograms denote all the galaxies in our redshift catalog, the black circles and solid black histograms show the galaxies in our algorithmic group catalog (those listed in Tables \ref{table:grps5andup} and \ref{table:grps3and4}), and the small black points and the solid gray histograms are our non-algorithmic candidate groups (Table \ref{table:candgrps}).  The large peak with the horizontally shaded histogram in field b0712 at $z \sim$ 0.29 is the supergroup discussed in Section \ref{sssec:supergroup}, Table \ref{table:supergroup}, and Figure \ref{fig:supergroup}.  The black dashed and dotted vertical lines indicate spectroscopic and photometric lens redshifts, respectively.  The dotted curves are redshift selection functions arbitrarily normalized based on each field's most populated redshift bin.  These functions are calculated using the spectroscopic field size and how the spectroscopic completeness varies over our I-band magnitude range for a homogeneous universe using the luminosity functions of \citet{faber07} for 0 $< z <$ 1, as in \citet{ammons14}.  These functions suggest where in redshift the spectroscopic samples for each field are most sensitive.  Even when a group is not found at $z_{lens}$, there usually is more than one galaxy present.}
\label{fig:zhists}
\end{figure*}
\begin{figure*}[!h]
\ContinuedFloat
\includegraphics[clip=true, width=18cm]{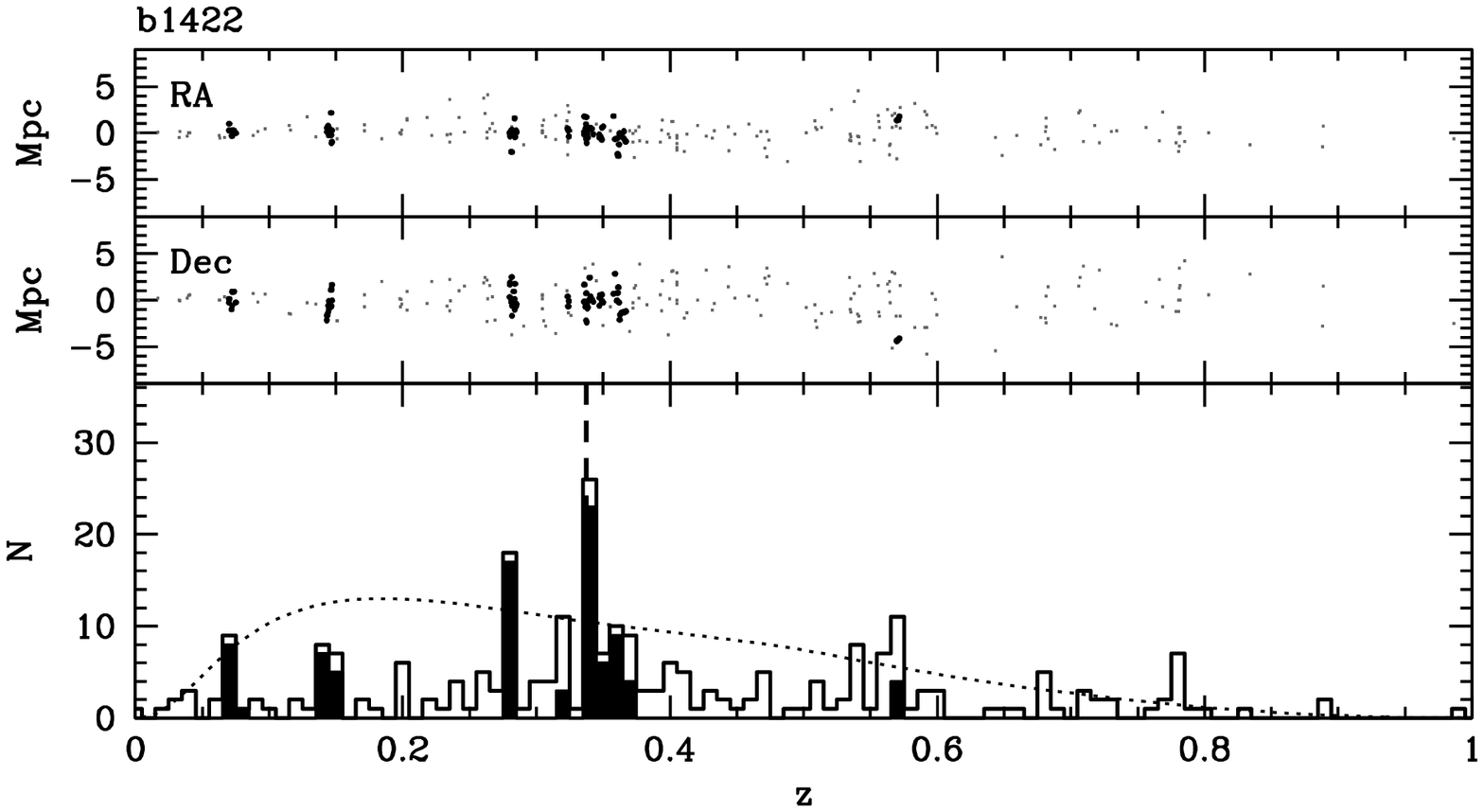}
\includegraphics[clip=true, width=18cm]{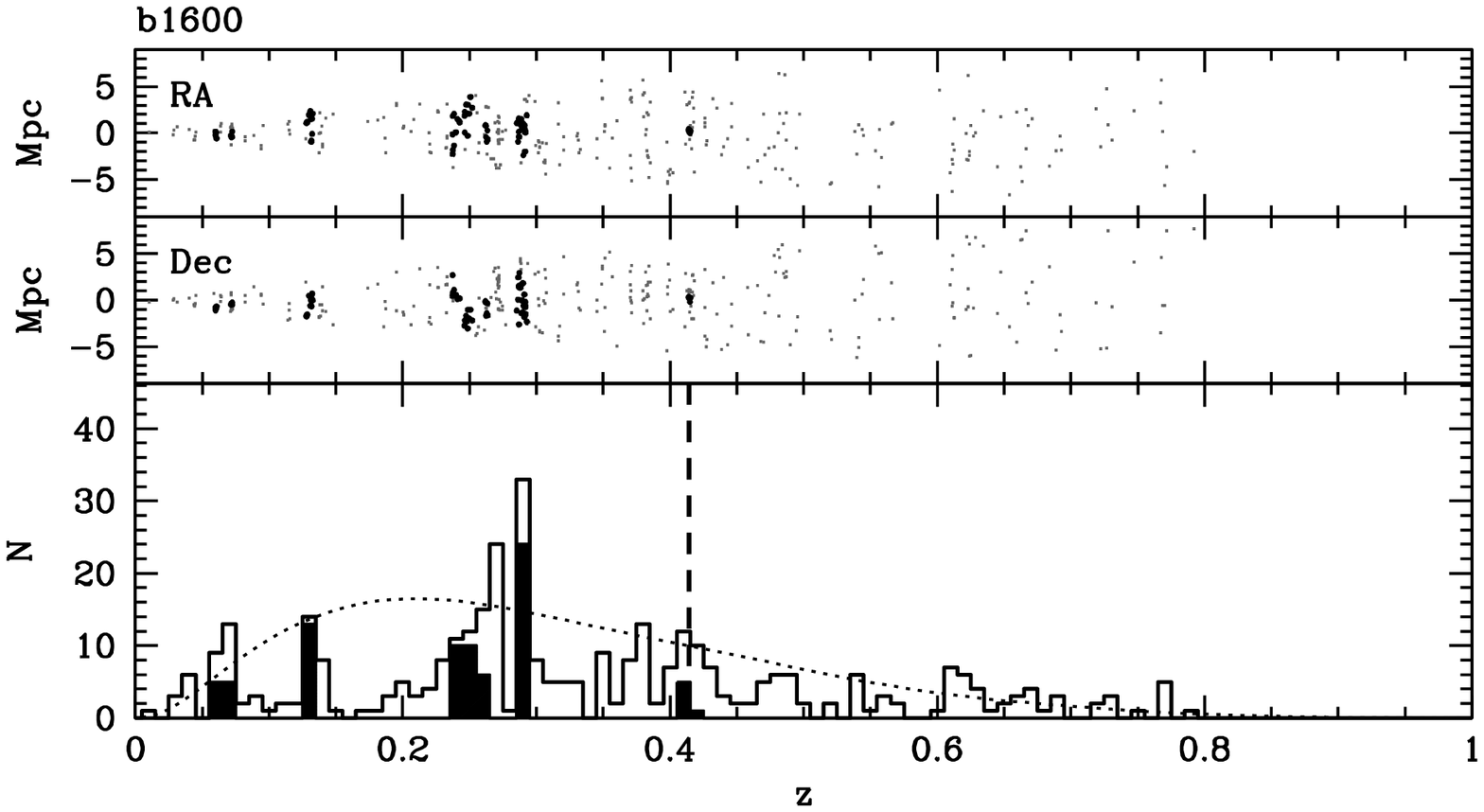}
\caption{Continued.}
\end{figure*}
\begin{figure*}[!h]
\ContinuedFloat
\includegraphics[clip=true, width=18cm]{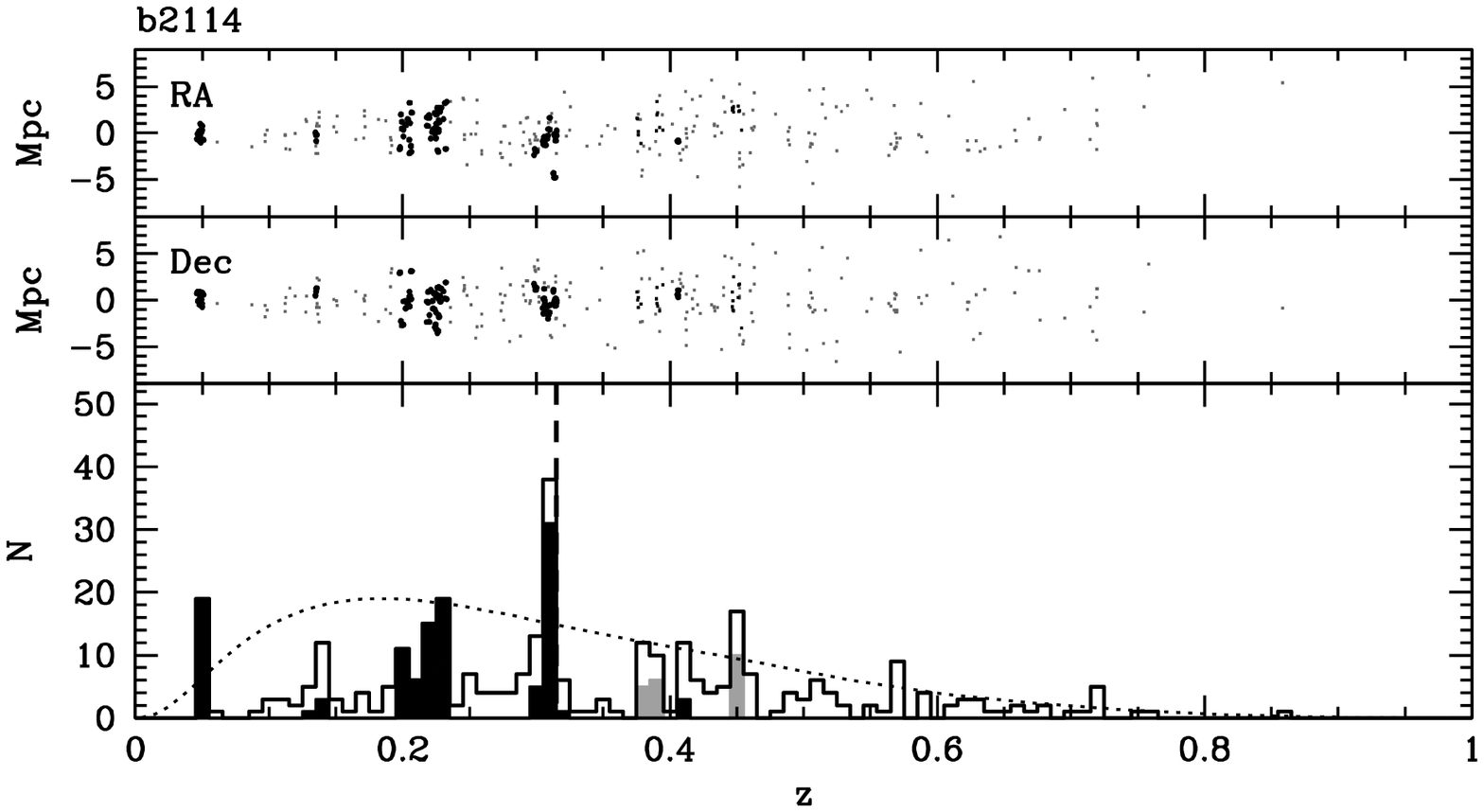}
\includegraphics[clip=true, width=18cm]{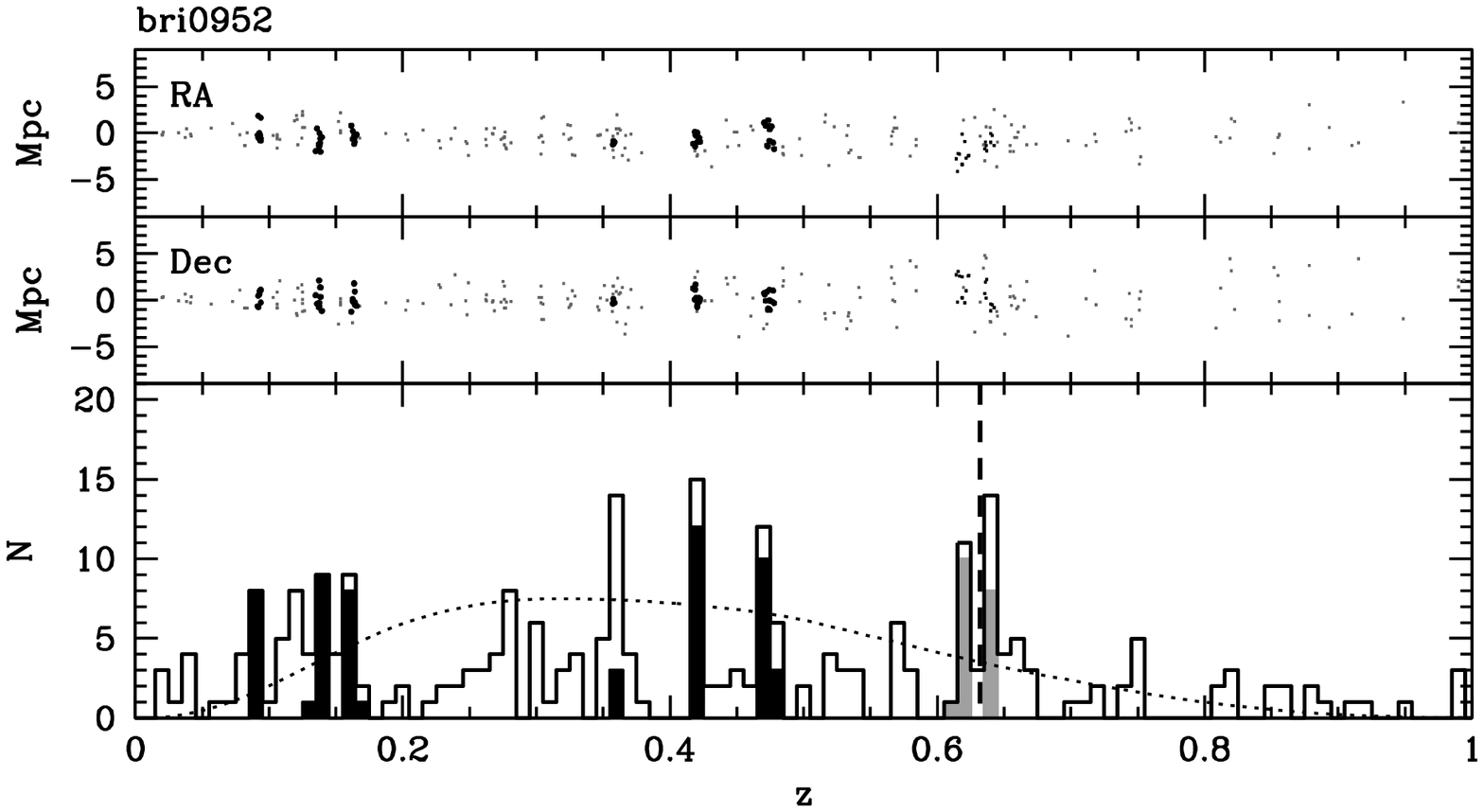}
\caption{Continued.}
\end{figure*}
\begin{figure*}[!h]
\ContinuedFloat
\includegraphics[clip=true, width=18cm]{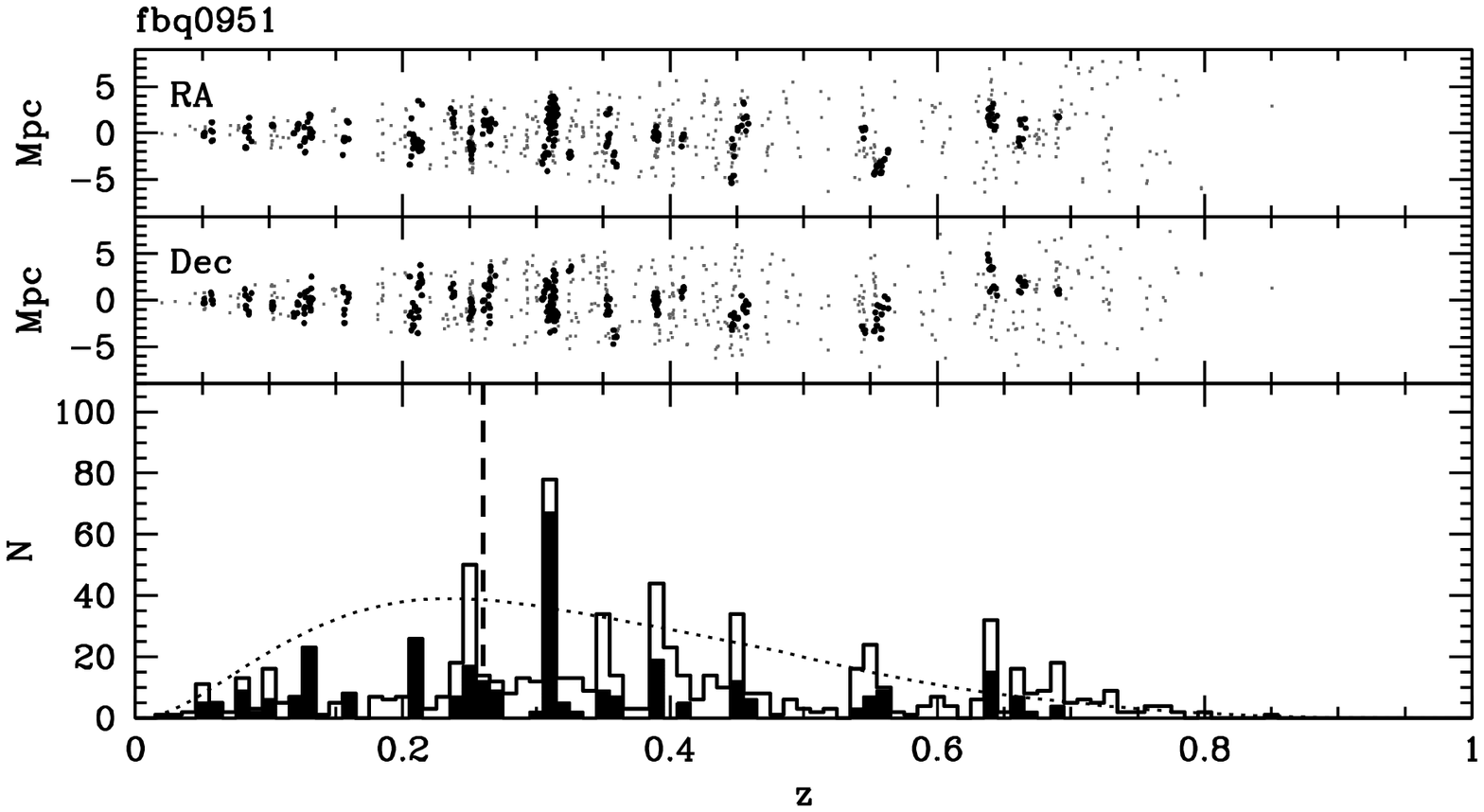}
\includegraphics[clip=true, width=18cm]{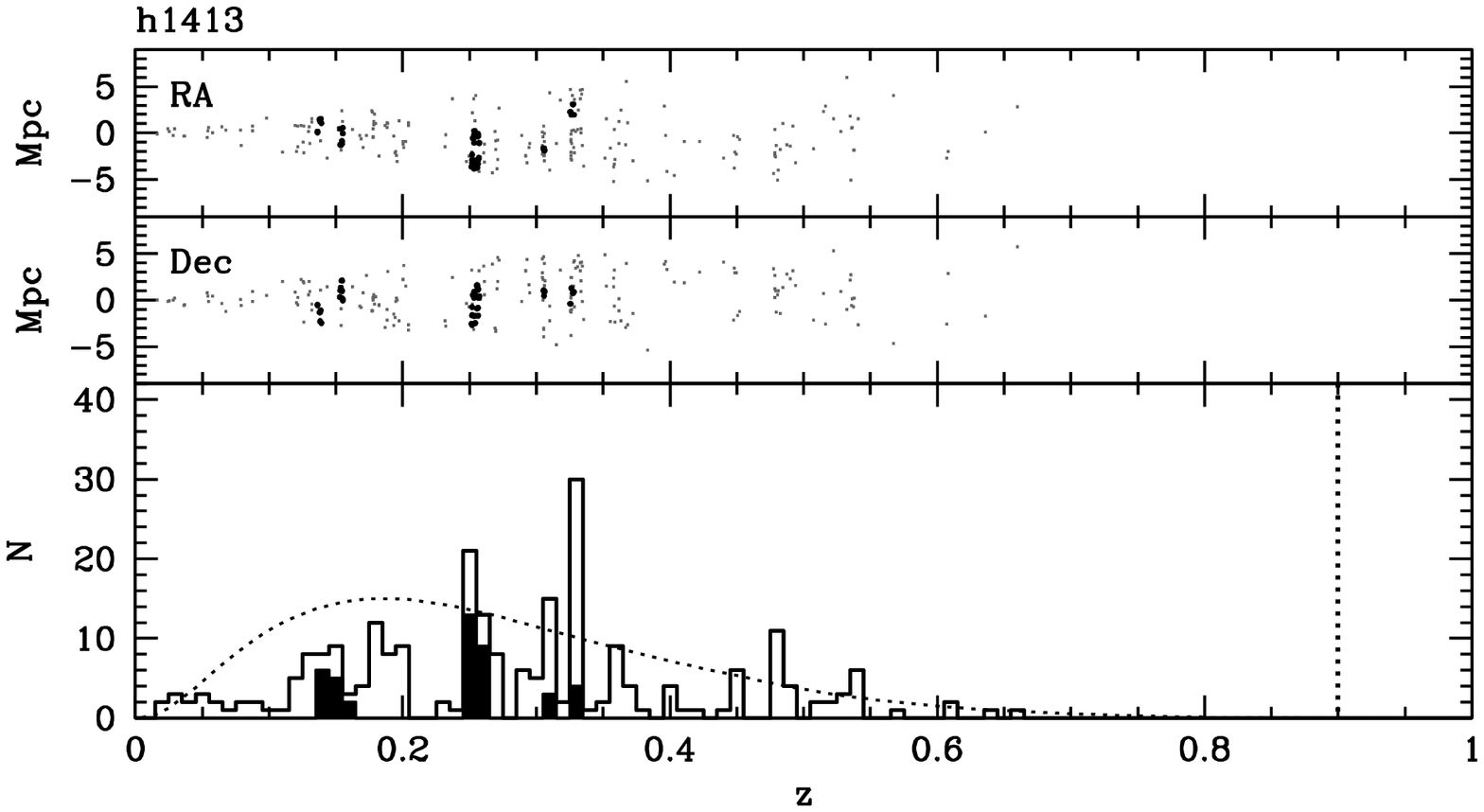}
\caption{Continued.}
\end{figure*}
\begin{figure*}[!h]
\ContinuedFloat
\includegraphics[clip=true, width=18cm]{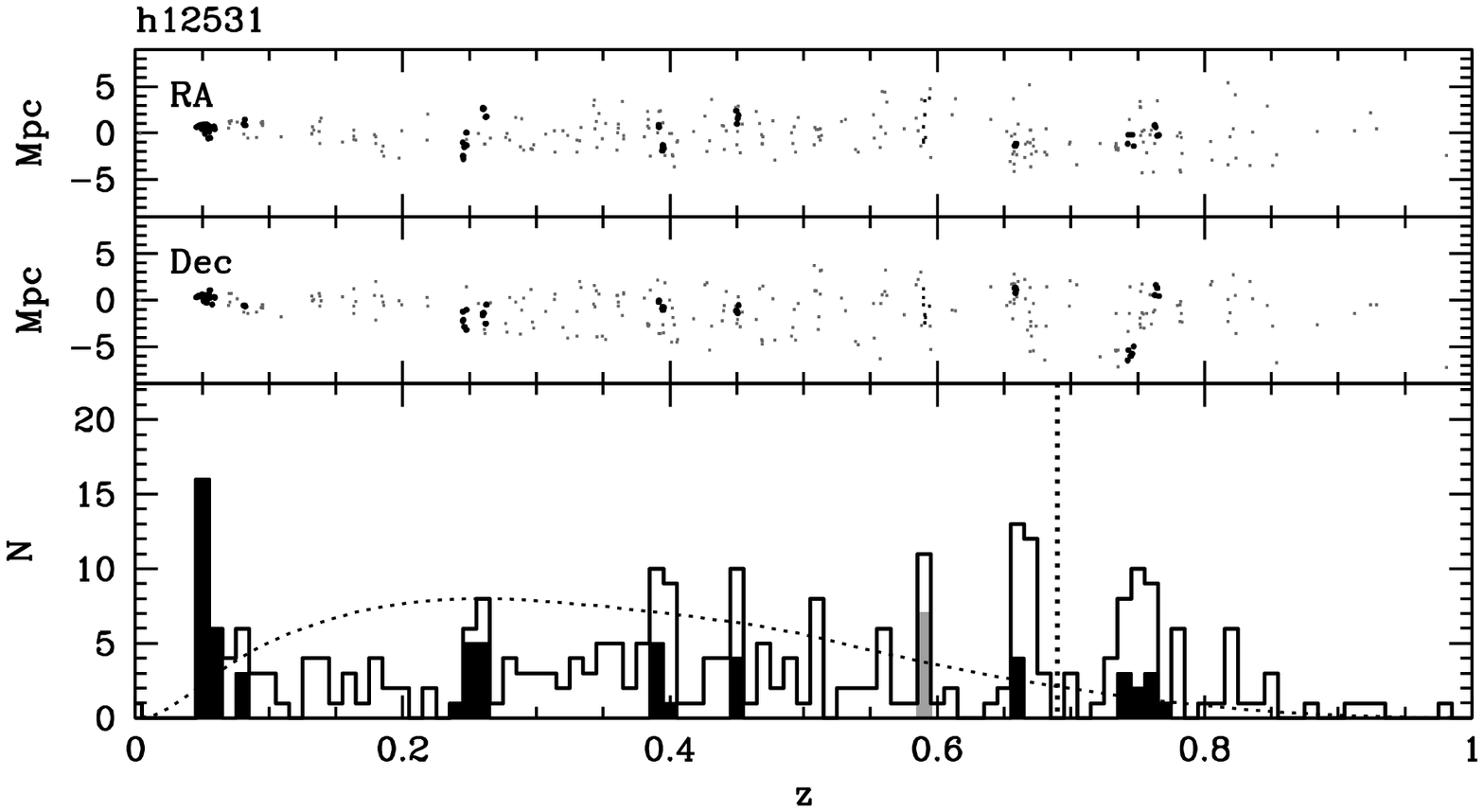}
\includegraphics[clip=true, width=18cm]{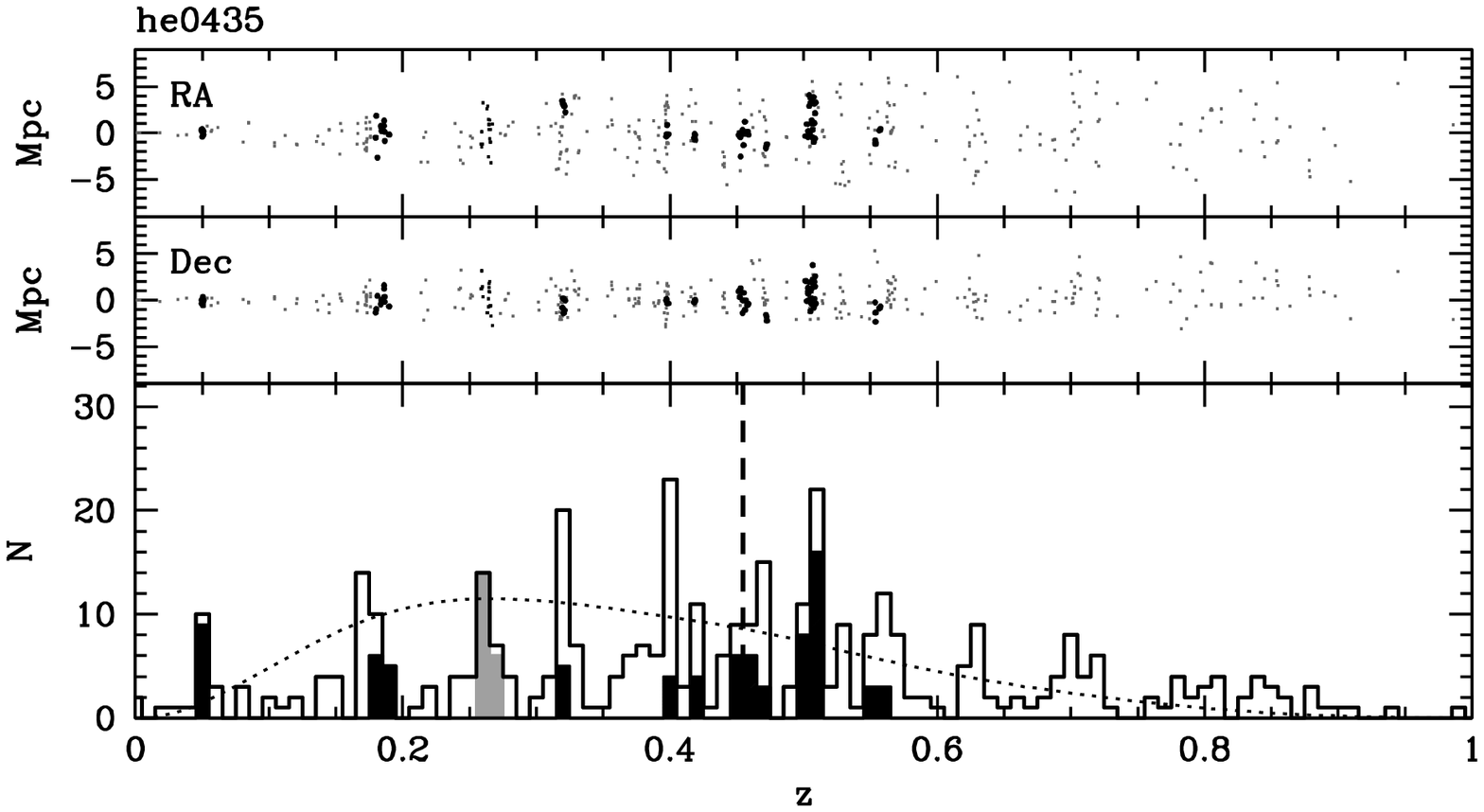}
\caption{Continued.}
\end{figure*}
\begin{figure*}[!h]
\ContinuedFloat
\includegraphics[clip=true, width=18cm]{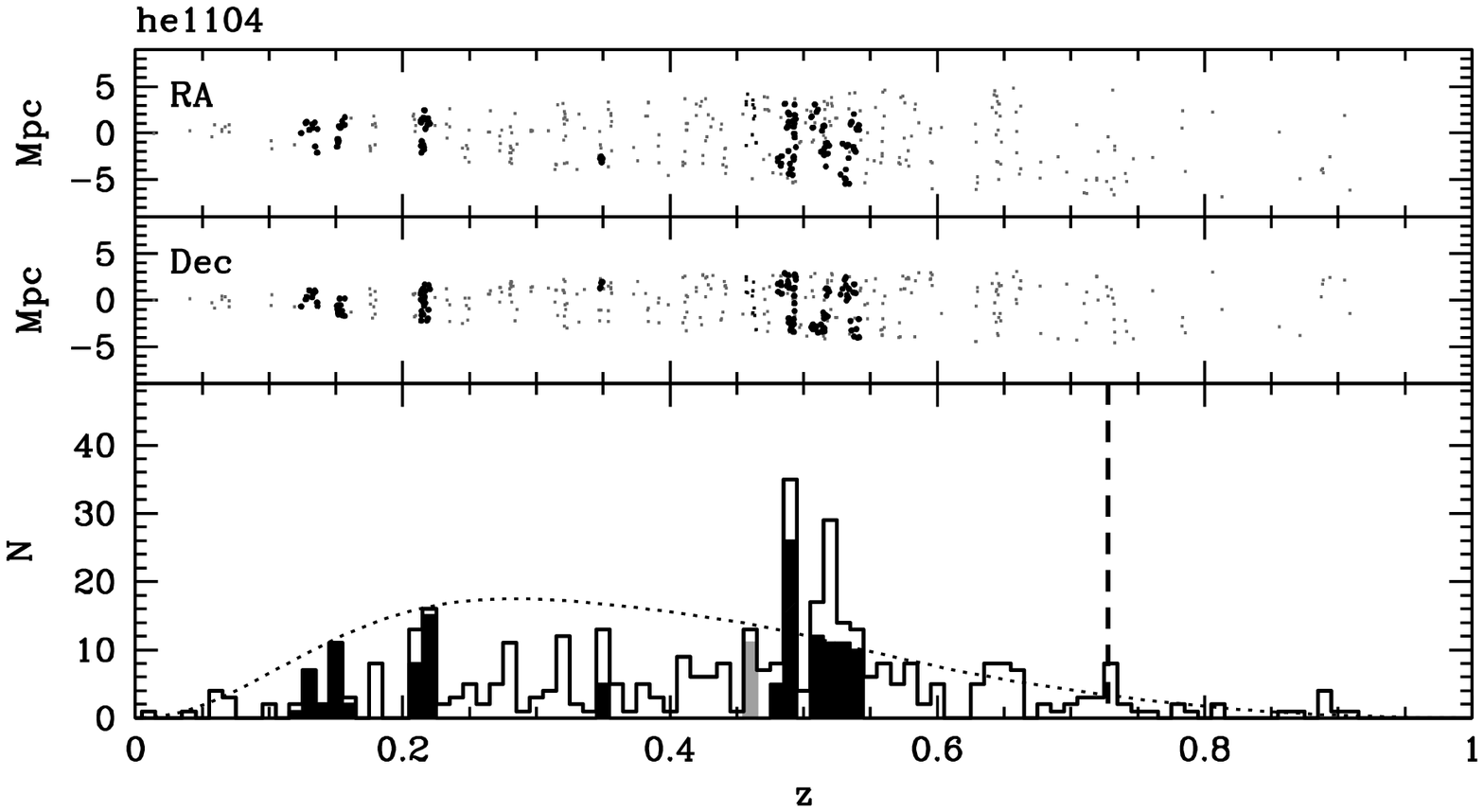}
\includegraphics[clip=true, width=18cm]{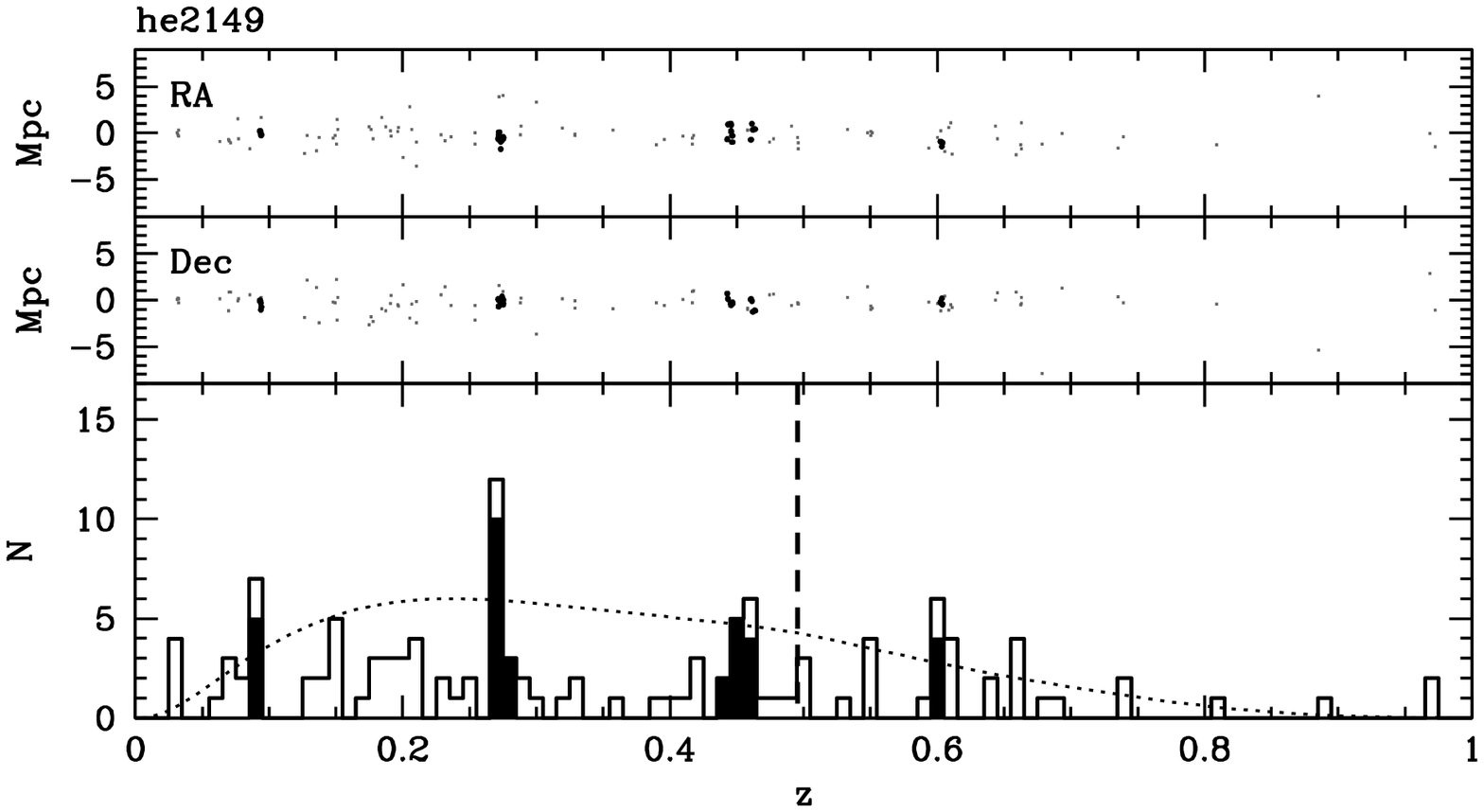}
\caption{Continued.}
\end{figure*}
\begin{figure*}[!h]
\ContinuedFloat
\includegraphics[clip=true, width=18cm]{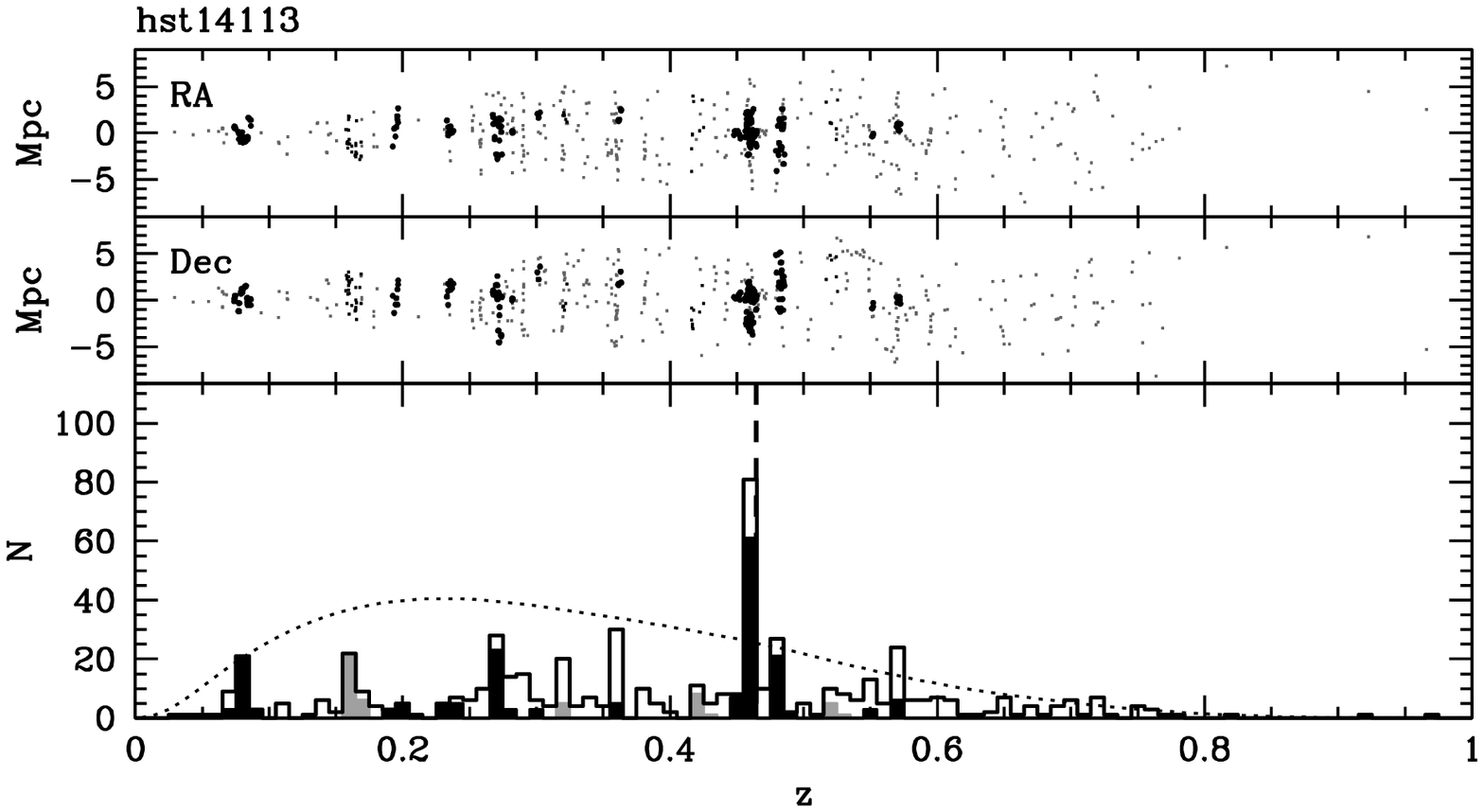}
\includegraphics[clip=true, width=18cm]{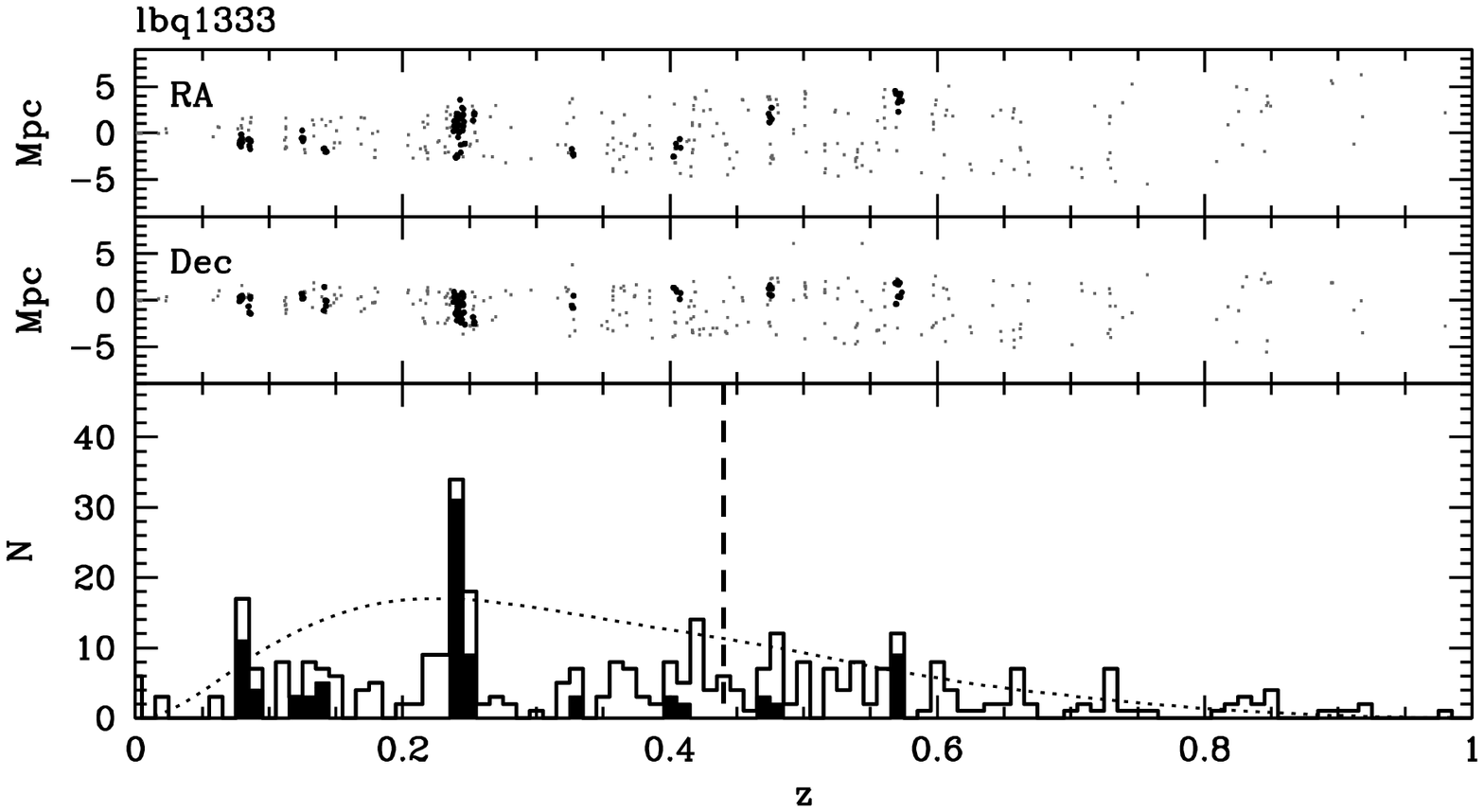}
\caption{Continued.}
\end{figure*}
\begin{figure*}[!h]
\ContinuedFloat
\includegraphics[clip=true, width=18cm]{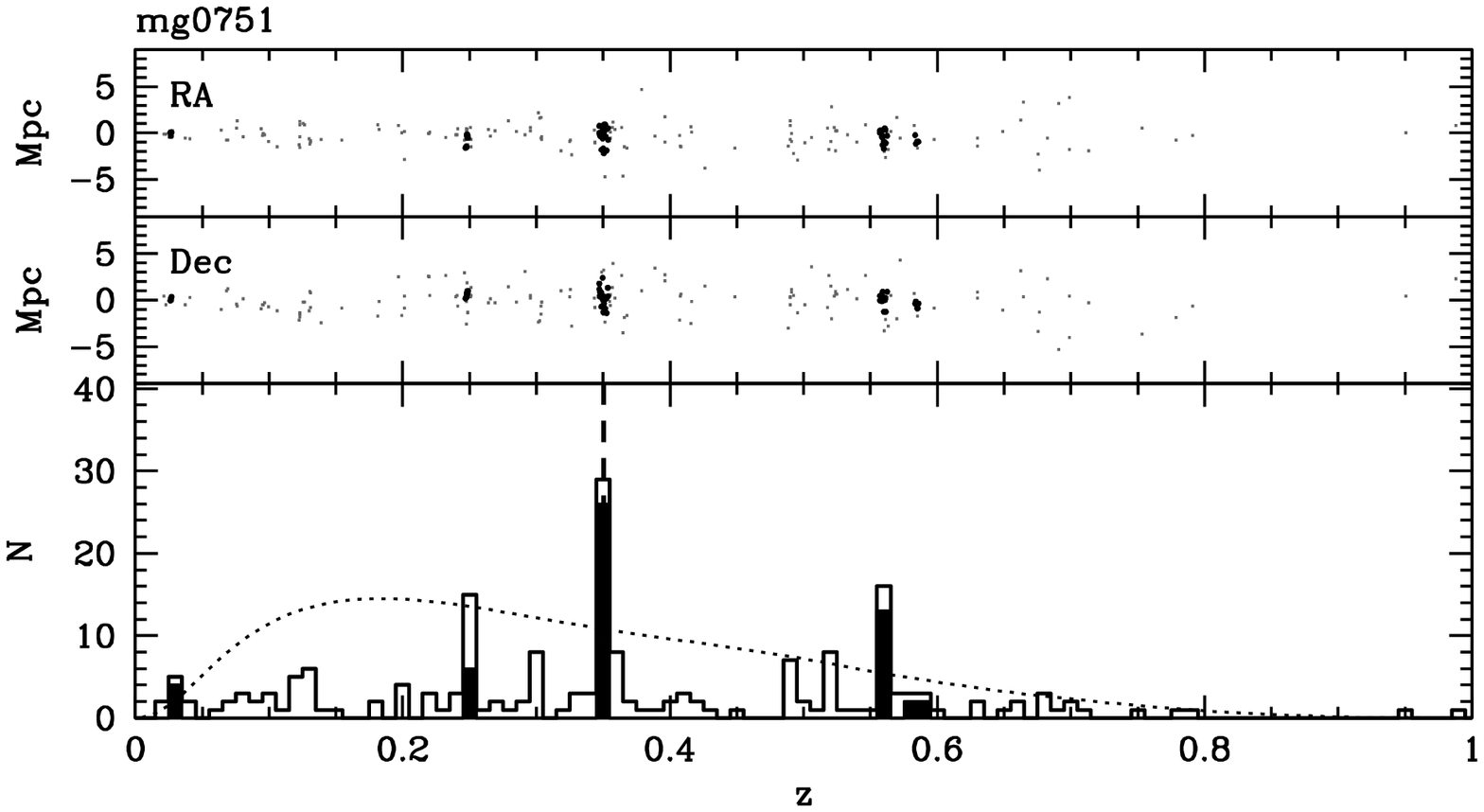}
\includegraphics[clip=true, width=18cm]{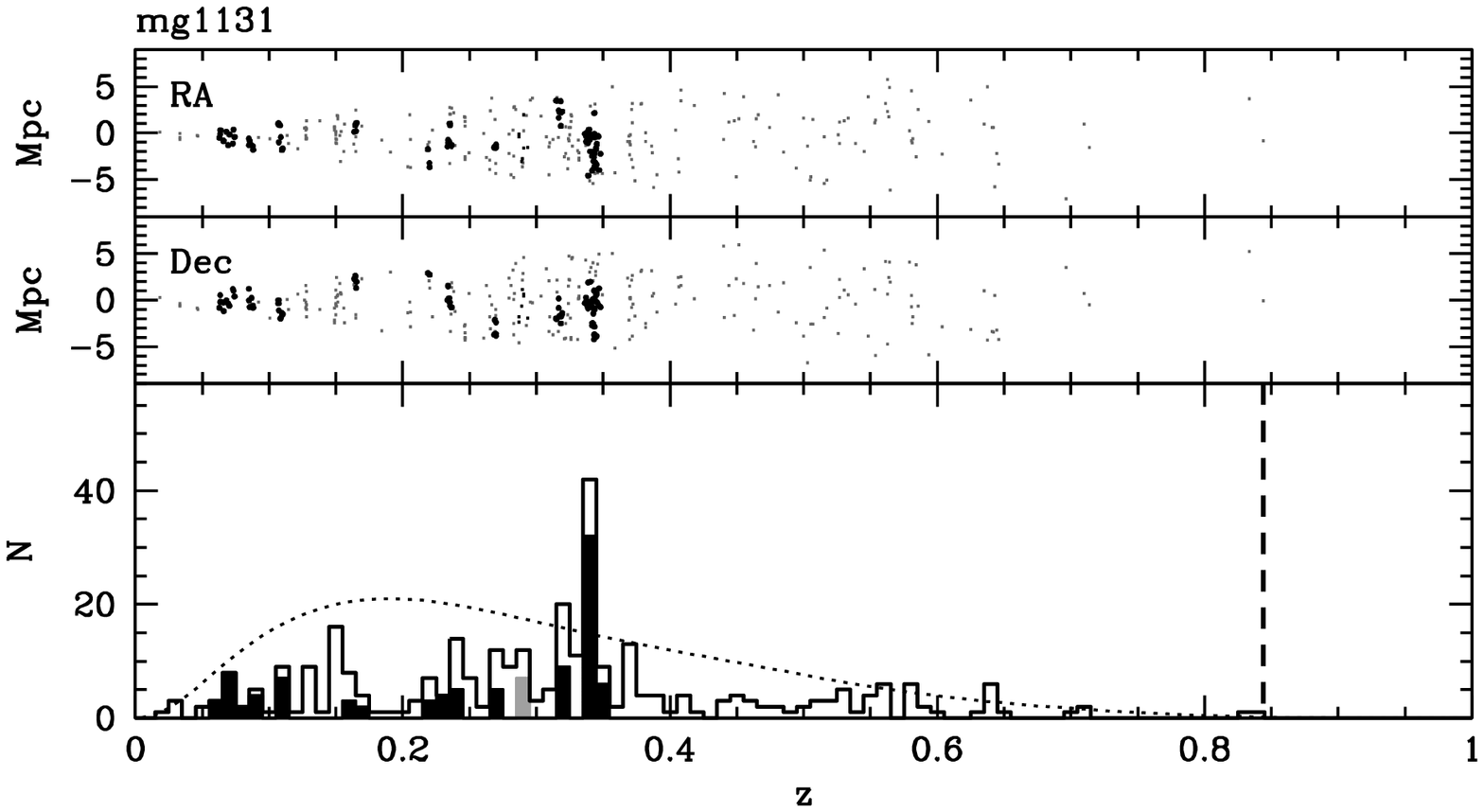}
\caption{Continued.}
\end{figure*}
\begin{figure*}[!h]
\ContinuedFloat
\includegraphics[clip=true, width=18cm]{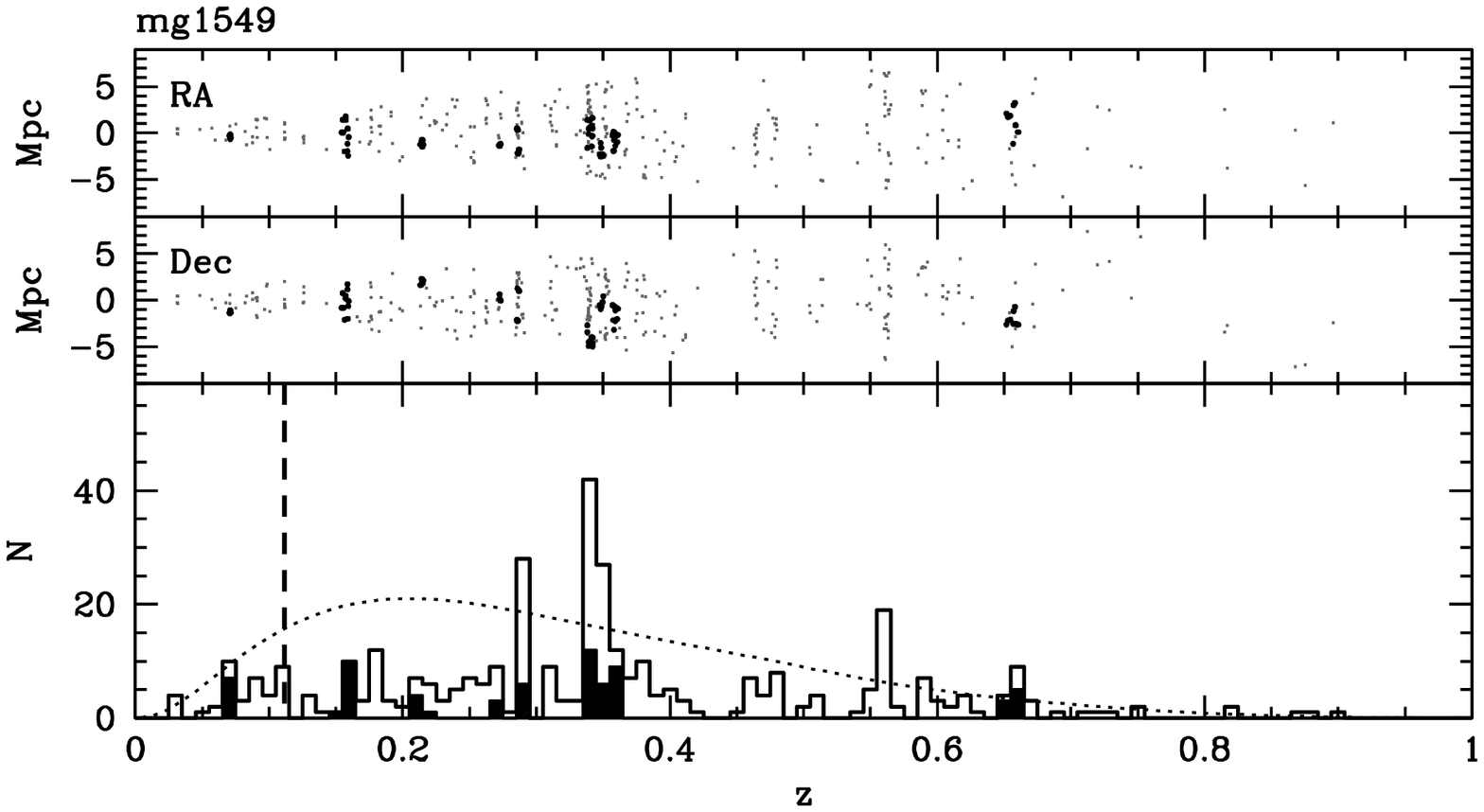}
\includegraphics[clip=true, width=18cm]{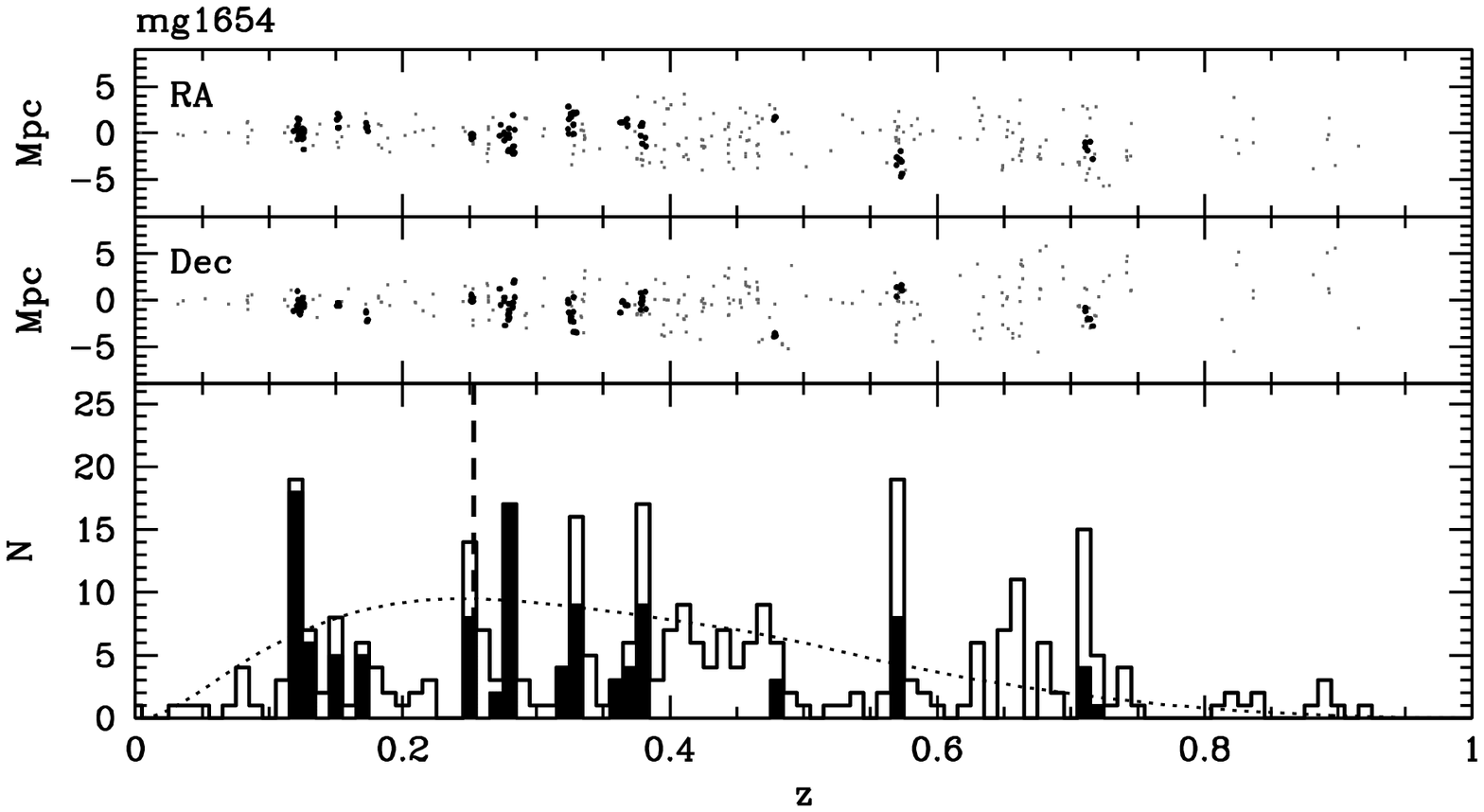}
\caption{Continued.}
\end{figure*}
\begin{figure*}[!h]
\ContinuedFloat
\includegraphics[clip=true, width=18cm]{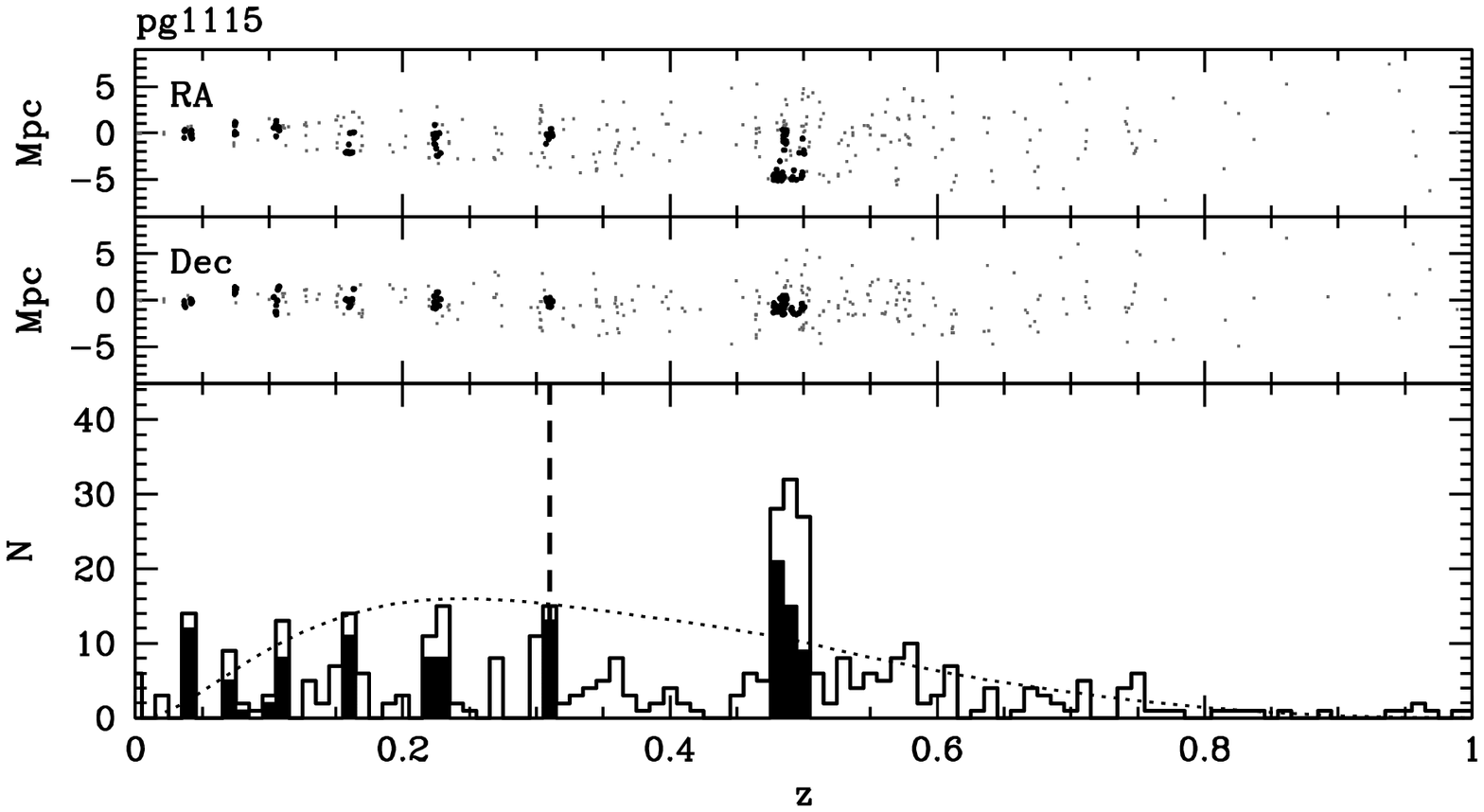}
\includegraphics[clip=true, width=18cm]{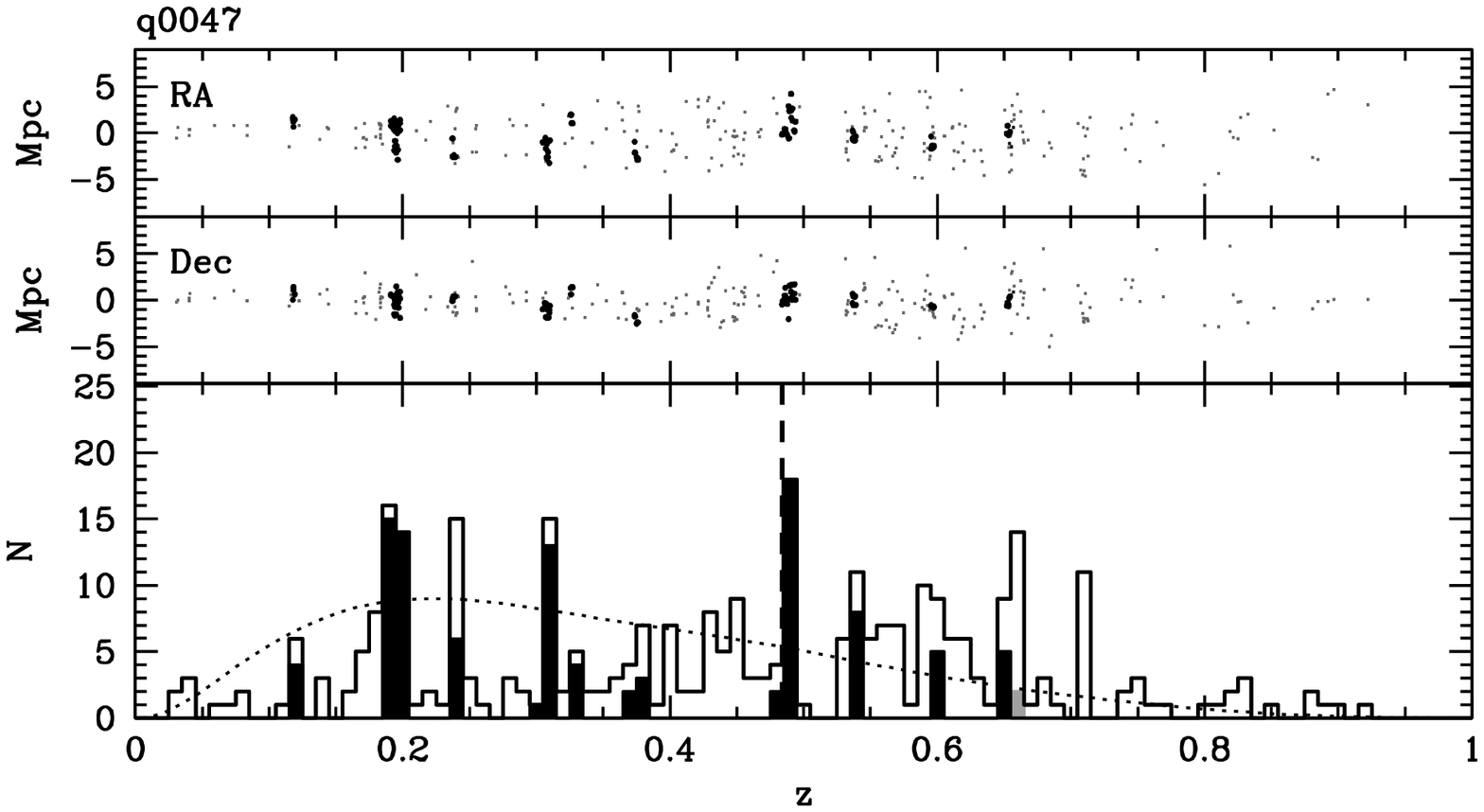}
\caption{Continued.}
\end{figure*}
\begin{figure*}[!h]
\ContinuedFloat
\includegraphics[clip=true, width=18cm]{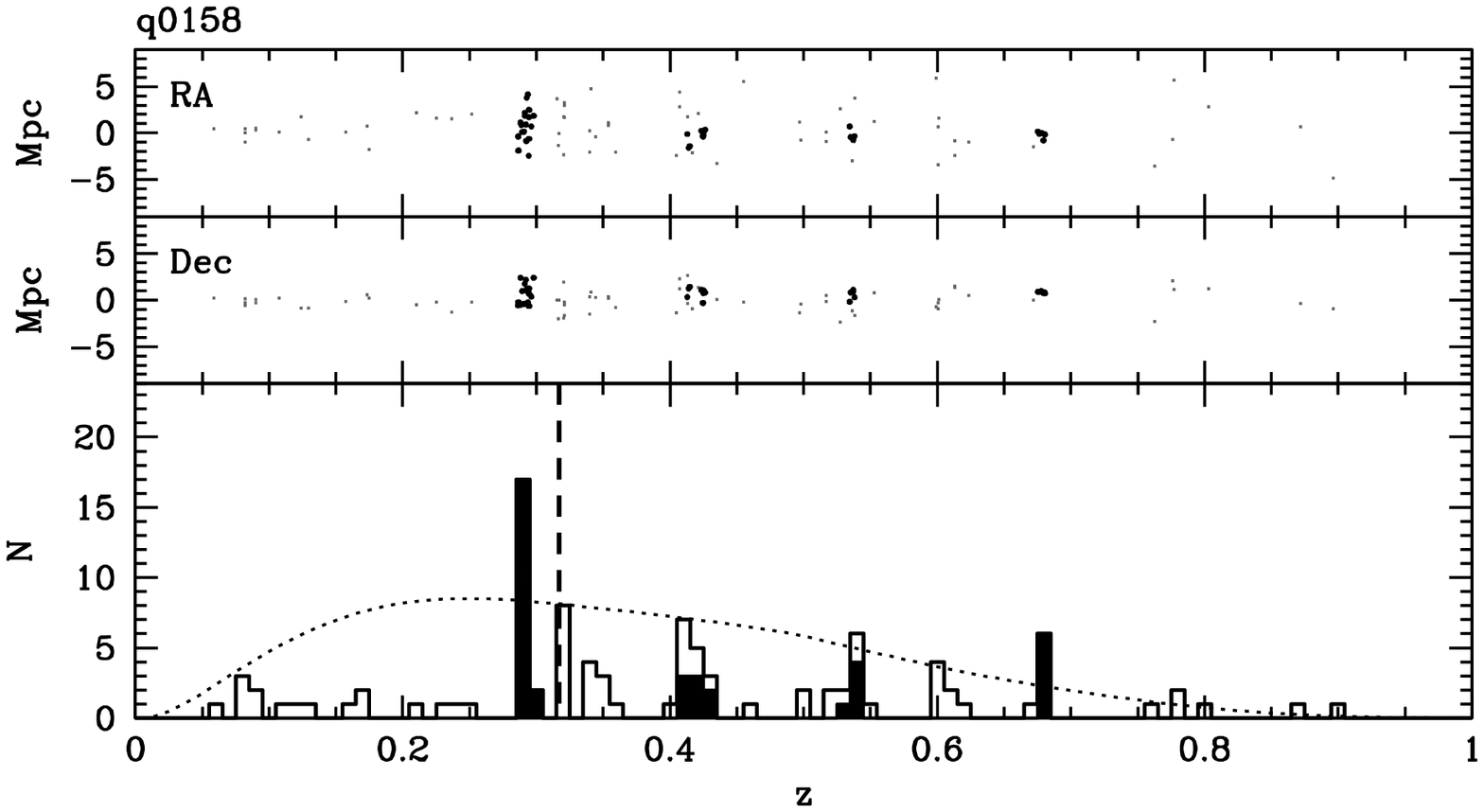}
\includegraphics[clip=true, width=18cm]{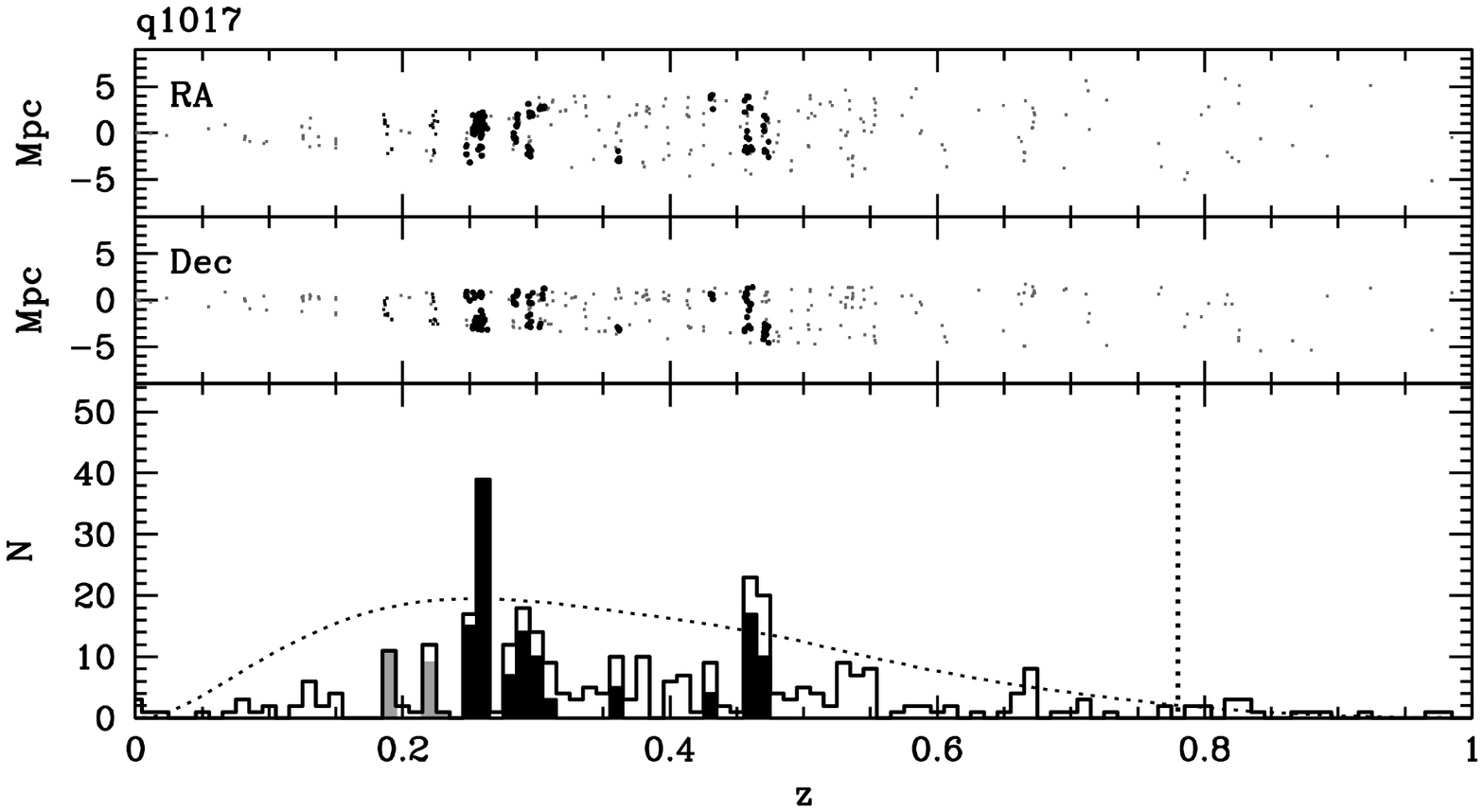}
\caption{Continued.}
\end{figure*}
\begin{figure*}[!h]
\ContinuedFloat
\includegraphics[clip=true, width=18cm]{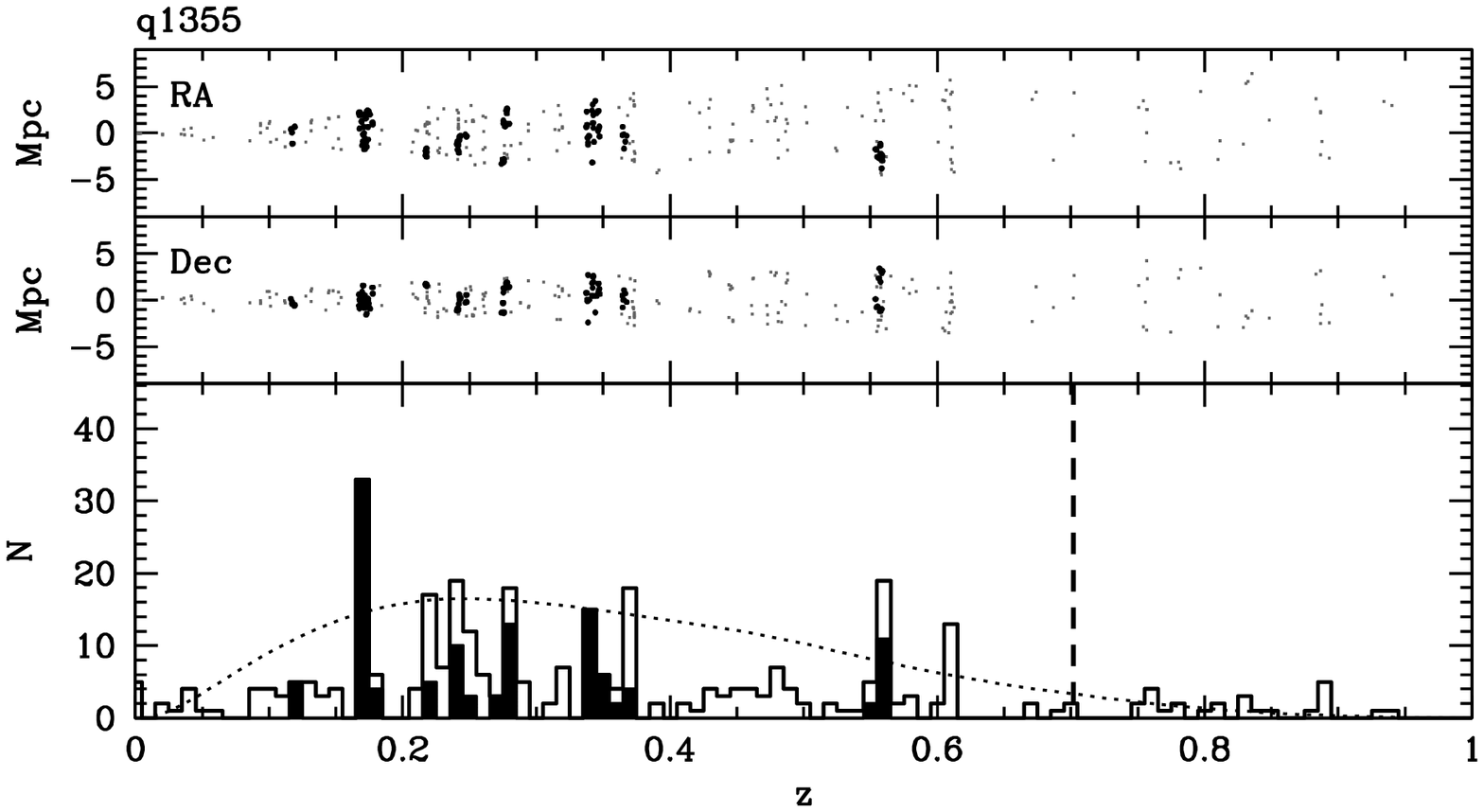}
\includegraphics[clip=true, width=18cm]{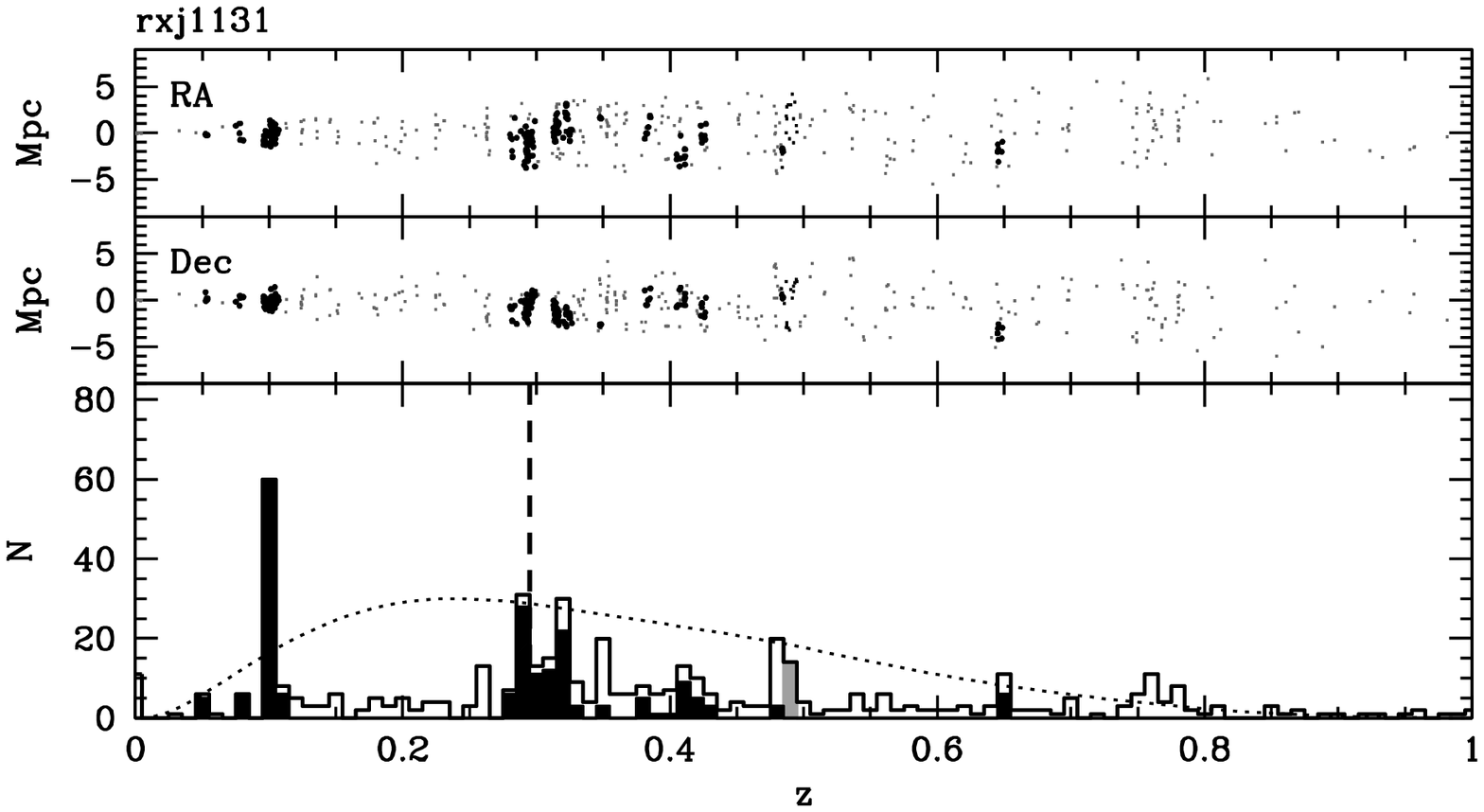}
\caption{Continued.}
\end{figure*}
\begin{figure*}[!h]
\ContinuedFloat
\includegraphics[clip=true, width=18cm]{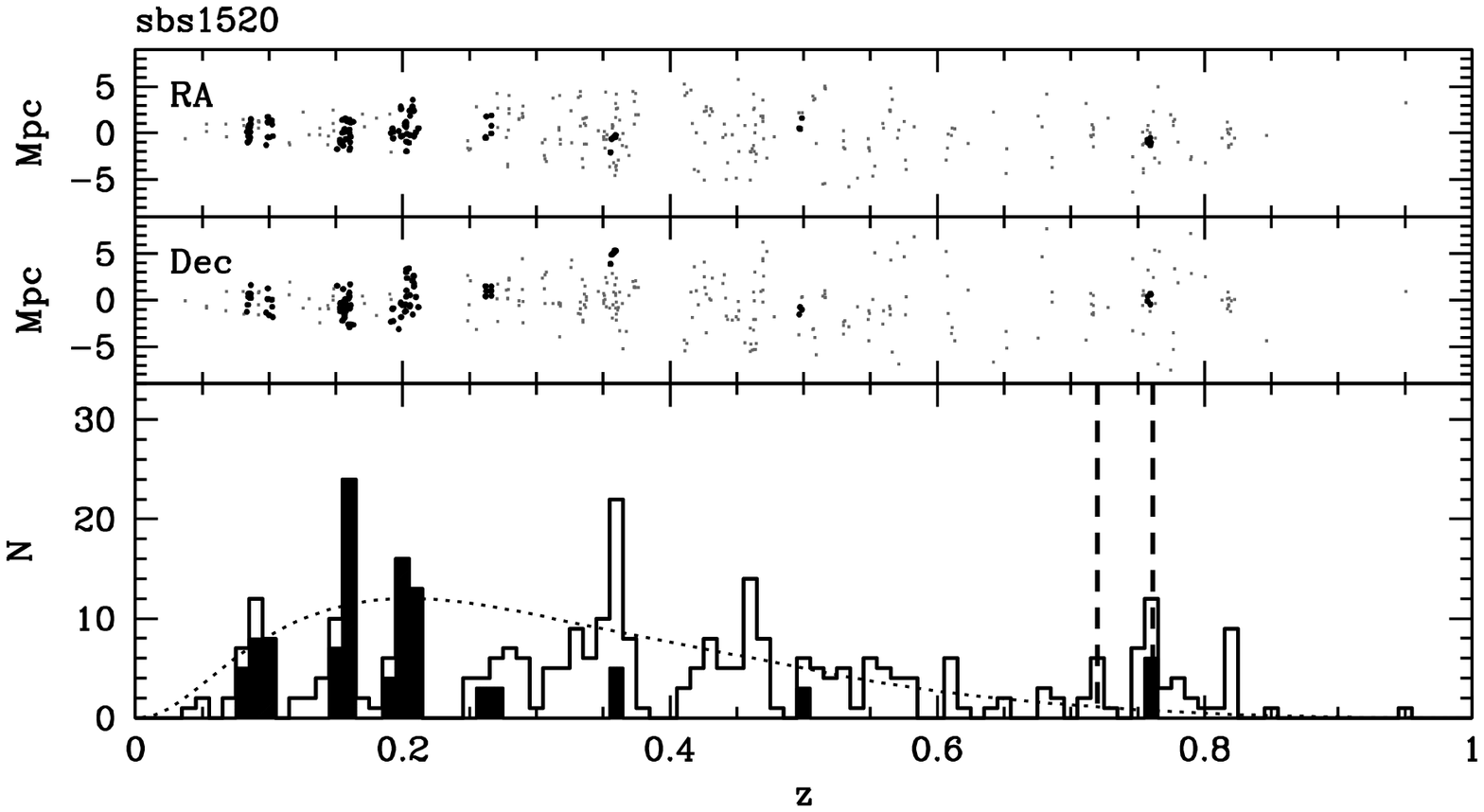}
\includegraphics[clip=true, width=18cm]{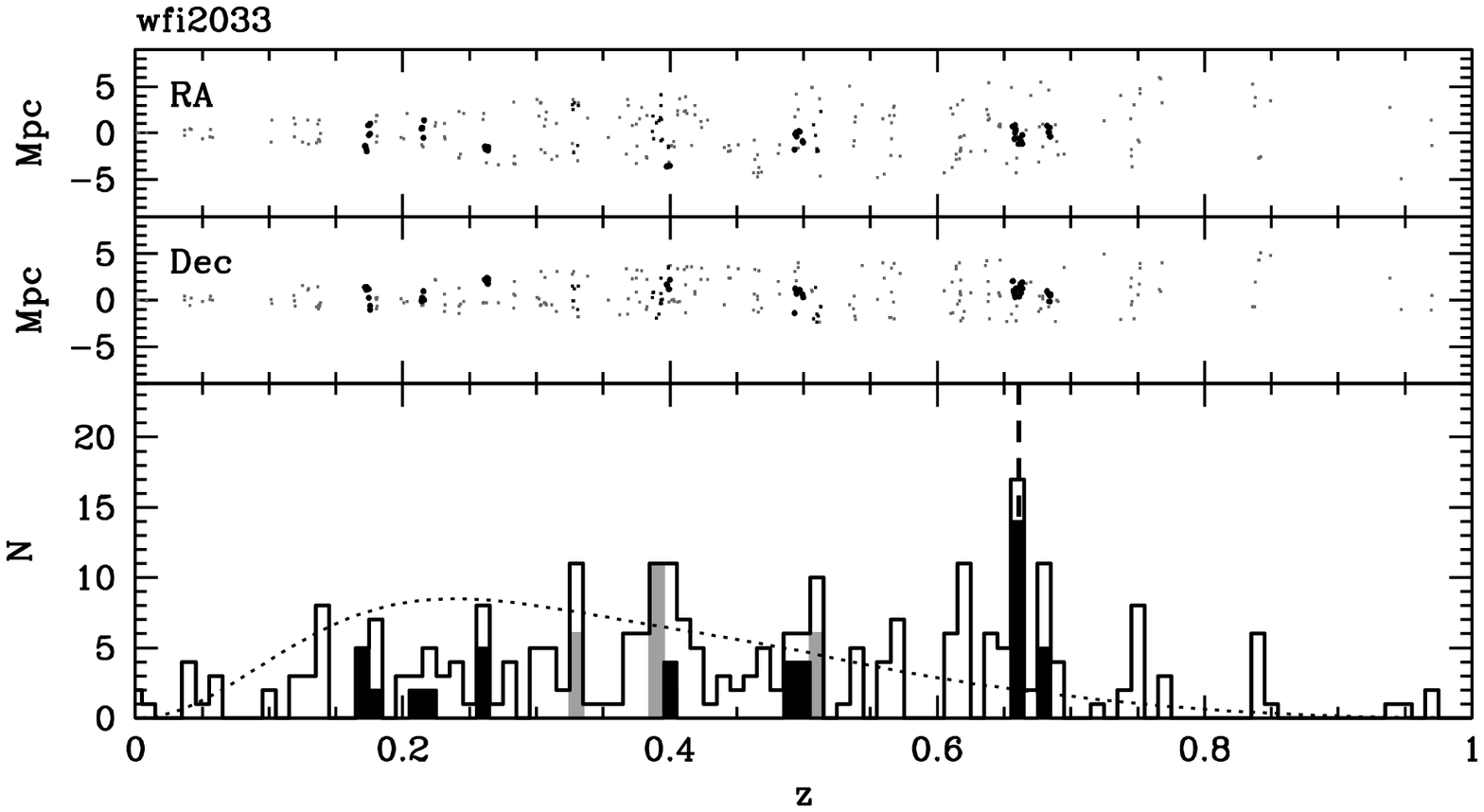}
\caption{Continued.  The two dashed lines in the sbs1520 field denote the two spectroscopic redshifts available for the lens, neither of which appear in our redshift sample.}
\end{figure*}
\clearpage

\begin{figure*}
\includegraphics[clip=true, width=18cm]{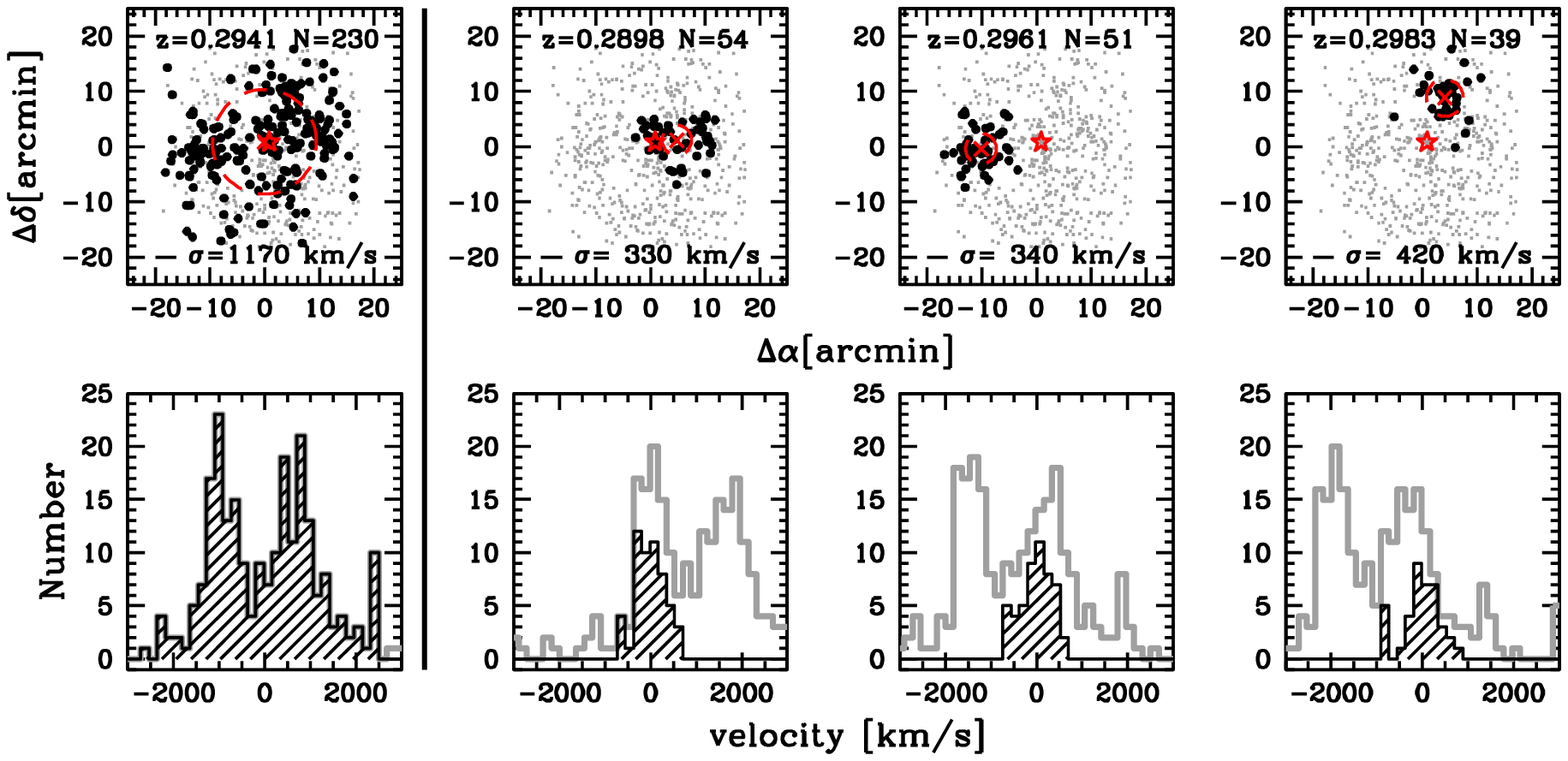}
\caption{\emph{Top row:} Sky plots for the entire supergroup (leftmost column) and its three main substructures (other columns) in field b0712.  Black points are group members, and gray points are the rest of the galaxies in the field in our redshift sample.  Group redshifts, group velocity dispersions, and the number of group galaxies are marked.  We also show group projected spatial centroid (red cross), $r_{vir}$ (red dotted circle), the lens (red star), and the angular size of 1 Mpc at the redshift of the group (black bar).  \emph{Bottom row:} Velocity histograms for these structures shifted so the group mean velocity is at 0 km s$^{-1}$.  Bin widths are 180 km s$^{-1}$, which corresponds to about three times the mean redshift error ($\sim$ 0.0002, or $\sim$ 60 km s$^{-1}$) at $z =$ 0.3.  Gray histograms are the full sample for the field, and black histograms are group galaxies.  The group finder finds substructures that correspond to distinct regions both on the sky and in velocity.}
\label{fig:supergroup}
\end{figure*}

\begin{figure*}
\includegraphics[clip=true, width=18cm]{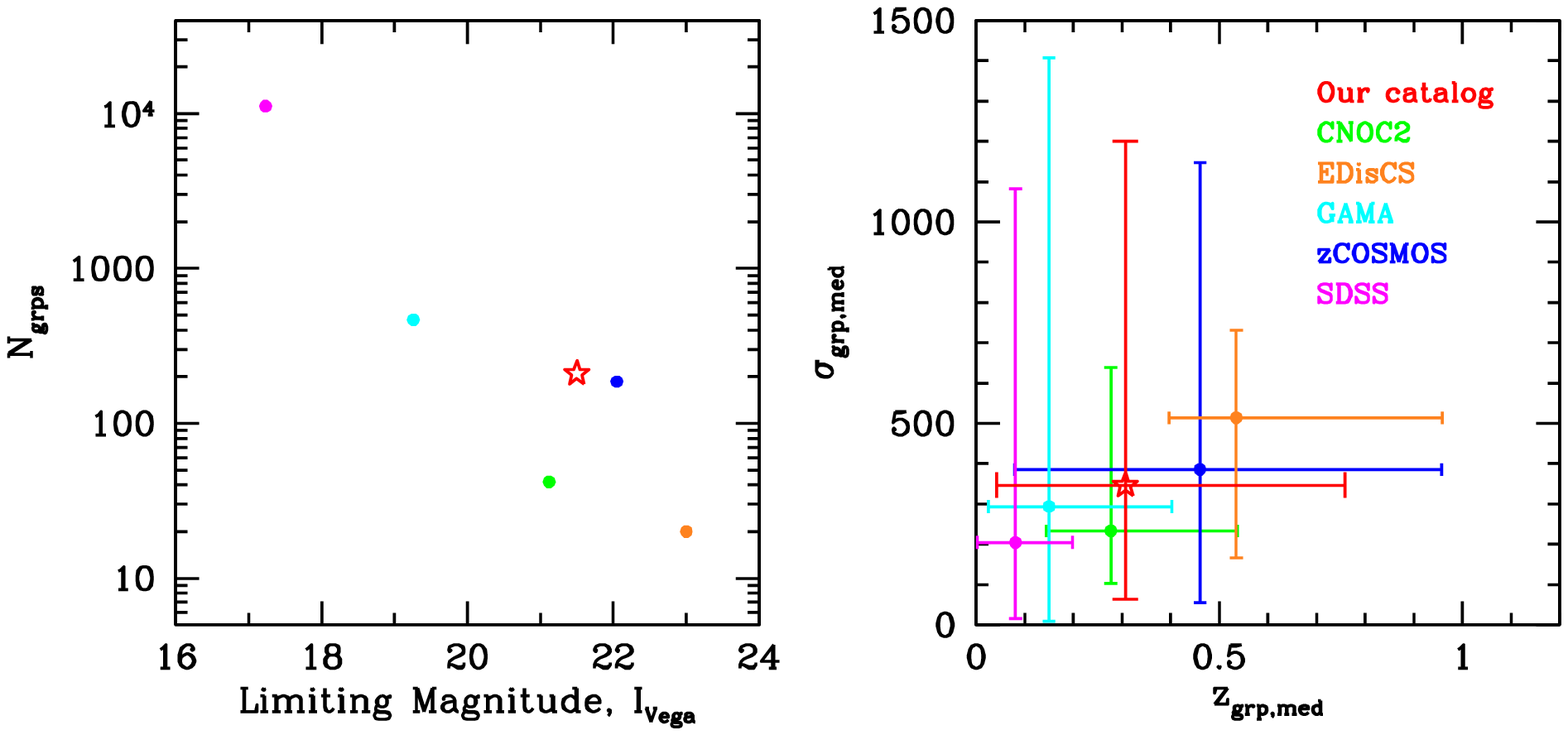}
\caption{Comparison of our algorithmic group catalog (red stars) to others, including CNOC2 (green circles), EDisCS (orange), GAMA (cyan), zCOSMOS (blue), and an SDSS flux-limited catalog (magenta).  Here we include only groups with $\ge$ five spectroscopic members and a reported velocity dispersion.  {\it Left:} Approximate limiting magnitudes for the redshift surveys used for each catalog transformed to $I_{Vega}$ and number of groups found.  The red star represents our fainter limiting magnitude, which was used for the majority of the survey \citep{momcheva15}.  {\it Right:} Median velocity dispersions versus redshifts for these group catalogs.  Bars represent the total range in group properties.  Our catalog has a comparable number of groups for its limiting magnitude, redshift range, and velocity dispersion range to the zCOSMOS group catalog.  The uniqueness of our catalog is that it targets the fields of strong gravitational lensing galaxies.}
\label{fig:catcomp}
\end{figure*}

\subsection{Lens Galaxy Environments}
\label{ssec:lensenvs}

We find 12 of 26 (46\%) of the lens galaxies in our sample to be members of groups with at least five galaxies.  These lenses are b1422, b1600, b2114, fbq0951, he0435, hst14113, mg0751, mg1654, pg1115, q0047, rxj1131, and wfi2033 (see Table \ref{table:grps5andup} for group properties).  One additional lens, b0712, is identified as a member of a non-algorithmic candidate group (see Table \ref{table:candgrps}).  While there are two non-algorithmic candidate groups near the lens redshift in field bri0952, the lens galaxy is not a member.  Our percentage of lenses in groups with at least five galaxies agrees well with the 42\% found by \cite{momcheva09}, which used redshifts drawn from an earlier version of \citet{momcheva15}.

The remaining 13 lenses are not identified as group members but likely are not isolated.  Of those nine lenses with firm spectroscopic redshifts, five have at least five other galaxies in the same redshift bin in Figure \ref{fig:zhists}.  Four (bri0952, he2149, mg1131, and q1355) are not in overdensities of galaxies, although three are at $z > 0.5$ where our redshift sample is much less sensitive.  Because of the ambiguity of the lens redshift in field sbs1520, we cannot say if it is located in a group, but we do find galaxies near both possible redshifts.  The remaining three lenses (h1413, h12531, and q1017) have only photometric redshifts, and they are not located in peaks of galaxies in the redshift number distribution.

\subsection{Significant Line-of-Sight Structures in Lens Fields}
\label{ssec:lenslos}

The importance of a line-of-sight structure to the lensing potential is affected by the structure's mass and its proximity to the lens both in redshift and projected on the sky \citep[][]{wong11,wong12,mccully16}.  We select groups and candidate groups within 2$^{\prime}$ of the lens, as \citet{wong11} found most line-of-sight shear was caused by objects within this distance, and a high mass threshold of $\sigma_{grp} \geq$ 500 km s$^{-1}$ to estimate how often line-of-sight structure might be important when determining the lensing potential.  We do not include field b0712 because of the ambiguity the supergroup provides; if this structure is one monolithic structure, it likely is significant for the lensing, but if the three main substructures are better tracers of the mass, they are not as significant.  Therefore, for this analysis we use a sample of 25 fields.

All but two of the systems for which a source redshift estimate is available have $z_{source} >$ 1.  So, our group catalog is not sensitive to all mass between the observer and the source for most of these fields.  \cite{french14} find that their beams that most effectively lens $z \sim$ 10 sources generally have significant halos with 0.1 $< z <$ 0.6 (single halo beams) or 0.3 $< z <$ 1.0 (multi-halo beams).  Also, the mass between the main lens and the observer typically affects the lensing potential more than those between the main lens and the source \citep{mccully14}.  Since all our lens galaxies lie between $z =$ 0.1 and 0.9, we likely are sensitive to the redshift range over which single groups might significantly affect the lensing potential as well as most of the redshift range for which multiple line-of-sight masses could be important.

In at least six of our 25 (24\%) fields, there is at least one group or candidate group, of which the lens is not a member, that is likely to significantly affect the lensing potential.  In future work, we will further quantify the importance of lens environments and and line-of-sight structures to the lensing models of these fields.

\section{Using Our Group Catalog}
\label{ssec:usingthecatalog}

While we make choices in our group finding algorithm to balance completeness and contamination, we provide several tables and quality flags so a user can tailor the group catalog for a particular science case.  Here we give some suggestions of how to select subsamples of our catalog for various purposes.

Some of the groups in Table \ref{table:grps5andup} have member galaxies quite close to the edge of their fields and thus might have group properties (primarily the projected spatial centroid) that are not well determined.  Some also are at redshifts such that they are not sampled out to at least $r_{vir}$.  Therefore, if one desires the highest confidence group sample with the best determined group properties, we suggest using only groups in Table \ref{table:grps5andup} with no flag (column 11).  Such a subset would be well-suited to studies of environmental effects on galaxy properties where one would like to minimize contamination.  Additionally, if one does not mind excluding the poorest groups in order to remove those most likely to be spurious, one could also remove groups with $\sigma_{grp} < 200$ km s$^{-1}$.

For lensing analyses of these fields, we suggest using everything in Tables \ref{table:grps5andup} and \ref{table:candgrps}.  Note that this choice will include groups at low redshift that we might not sample out to a virial radius or groups near the edge of our field with likely poorly-determined spatial centroids.

To prioritize additional spectroscopic followup in these fields and thus to select galaxies likely to be at redshifts near our candidate groups, use Tables \ref{table:grps3and4} and \ref{table:candgrps}.  If one intends to expand the footprint on the sky and thus would be interested in groups that are detected on the edge of our observed fields or for which we do not sample out to at least $r_{vir}$, one could also include those from Table \ref{table:grps5andup} with flags of 1, 2, or 3.

Users who wish to know of anything in these fields that could even possibly be a structure should use everything in Tables \ref{table:grps5andup}, \ref{table:grps3and4}, and \ref{table:candgrps} as well as either the first line (if one prefers the overall supergroup of b0712) or the second through fourth lines (if one desires the supergroup's main substructures) in Table \ref{table:supergroup}.  Note, however, that the resulting composite will include groups identified differently.

\section{Conclusions}

We create a group catalog for 26 of the 28 gravitational lens fields in \citet{momcheva15}. This catalog improves the characterization of the environments of the gravitational lenses and of line-of-sight groups and clusters in these fields, which will constrain how much such structures affect the lensing potentials.  

We develop an iterative group finder for redshift samples with variable spectroscopic completeness levels and sampling footprints. We find 210 groups with at least five member galaxies in these fields.  We newly identify 187.  These groups span 0.04 $\le z \le$ 0.76 with our range of greatest sensitivity at $0.2 < z < 0.5$ and a median of 0.31.  Our groups have a velocity dispersion range of 60 to 1200 km s$^{-1}$ with most between 100 and 600 km s$^{-1}$ and a median of 350 km s$^{-1}$.  Our groups have from five to 66 members with a median of eight.  

We test our group finder by comparing our group catalog to those in the same fields in the literature.  We find groups similar to most ($\sim$ 85\%) that were previously reported within our redshift range of sensitivity.

We also compare our groups' distribution of $\sigma_{grp}$, distribution of $M_{vir}$, incidence of substructure, and fraction of all sample galaxies that are group members to the literature.  Our catalog agrees well with the empirical and theoretical expectations.

Our main results are as follows:

\begin{itemize}

\item
We identify a supergroup in field b0712 with $z = $ 0.29.  This structure has three main components with 39 to 54 members and $\sigma_{grp}$ from 330 to 420 km s$^{-1}$ that occupy different regions projected on the sky and in radial velocity.

\item
Of the 26 lenses in our sample for which we produce group catalogs, 12 (46\%) are members of $\ge$ 5 member groups (b1422, b1600, b2114, fbq0951, he0435, hst14113, mg0751, mg1654, pg1115, q0047, rxj1131, and wfi2033), and one (b0712) is identified as a member of an additional, visually-identified group candidate.  These structures, which have $\sigma_{grp}$ from 110 to 800 km s$^{-1}$, are likely to affect lensing models. 

\item
In six of 25 (24\%) of these fields that do not have the supergroup, there is at least one group or candidate group that does not include the lens galaxy with $\sigma_{grp} \ge$ 500 km s$^{-1}$ projected within 2$^{\prime}$ of the lens, which likely affects the lensing potential.

\end{itemize}

Further analysis of the environments and line-of-sight structures of these lens fields will be discussed in a forthcoming paper.

\acknowledgments

We thank the anonymous referee for helpful comments.  We also thank Dan Marrone, George Rieke, Dennis Zaritsky, and K. Decker French for helpful discussions.  MLW and AIZ acknowledge support from NASA grants ADP-NNX10AD476 and ADP-NNX10AE88G, as well as NSF grant AST-0908280.  MLW also thanks the Technology and Research Initiative Fund (TRIF) Imaging Fellowship program for its support.  SMA thanks NASA through Hubble Fellowship grant HST-HF-51250.01-A from the Space Telescope Science Institute, which is operated by the Association of Universities for Research in Astronomy, Incorporated under NASA contract NAS5-26555.  Portions of this work were performed under the auspices of the U.S. Department of Energy by Lawrence Livermore National Laboratory under Contract DE-AC52-07NA27344.  CRK acknowledges support from NSF grant AST-1211385.

{\it Facilities:} \facility{Mayall (Mosaic-I)}, \facility{Blanco (Mosaic-II)}, \facility{Magellan-1 (LDSS-2, LDSS-3)}, \facility{Magellan-2 (IMACS)}, \facility{MMT (Hectospec)}.

\clearpage

\LongTables
\begin{deluxetable*}{llllrrrlllll}
\tablecaption{Algorithmically Identified Groups With $\ge$ 5 Members \label{table:grps5andup}}
\tablehead{
\colhead{Field} & \colhead{Group} & \colhead{$N_{m}$} & \colhead{$z_{grp}$} & \colhead{$\overline{RA}$} & \colhead{$\overline{Dec}$} & \colhead{R\tablenotemark{a} } & \colhead{$\sigma_{grp}$} & \colhead{$M_{vir}$\tablenotemark{b}} & \colhead{$r_{vir}$} & \colhead{Flag\tablenotemark{c}} & \colhead{Match\tablenotemark{d}} \\
\colhead{ } & \colhead{ } & \colhead{ } & \colhead{ } & \colhead{[deg]} & \colhead{[deg]} & \colhead{[$^{\prime}$]} & \colhead{[km s$^{-1}$]} & \colhead{[10$^{13}$M$_{\odot}$]} & \colhead{[Mpc]} & \colhead{ } & \colhead{} 
}
\startdata

   b0712 &   0 &    5 &   0.0784 $^{+  0.0002}_{-  0.0003}$ &   109.0365 $^{+ 0.0541}_{- 0.0524}$ &    46.8814 $^{+ 0.0112}_{- 0.0112}$ &  15.99 &    140 $^{+    30}_{-   110}$ &      -                         &   0.34 &   1,3 & \\ 
         &   3 &   34 &   0.3723 $^{+  0.0007}_{-  0.0007}$ &   109.1105 $^{+ 0.0226}_{- 0.0225}$ &    47.0370 $^{+ 0.0263}_{- 0.0226}$ &   7.68 &    870 $^{+    70}_{-    70}$ &     51.5 $^{+    12.5}_{-    10.6}$ &   1.78 &     4 & \\ 
         &   5 &    6 &   0.4146 $^{+  0.0003}_{-  0.0003}$ &   108.6920 $^{+ 0.0097}_{- 0.0102}$ &    47.0648 $^{+ 0.0126}_{- 0.0108}$ &  14.10 &    180 $^{+    40}_{-    60}$ &      -                         &   0.35 &       & \\ 
         &   7 &    5 &   0.4938 $^{+  0.0007}_{-  0.0006}$ &   109.0083 $^{+ 0.0171}_{- 0.0172}$ &    47.0681 $^{+ 0.0086}_{- 0.0086}$ &   4.77 &    270 $^{+    50}_{-   230}$ &      2.0 $^{+     0.9}_{-     1.3}$ &   0.52 &       & \\ 
   b1152 &   0 &   10 &   0.0514 $^{+  0.0006}_{-  0.0003}$ &   178.8423 $^{+ 0.0505}_{- 0.0486}$ &    19.7020 $^{+ 0.0247}_{- 0.0269}$ &   2.58 &    480 $^{+    90}_{-   350}$ &     11.9 $^{+     6.2}_{-     9.9}$ &   1.15 &   1,2 & \\ 
         &   1 &    5 &   0.1314 $^{+  0.0002}_{-  0.0002}$ &   178.8226 $^{+ 0.0079}_{- 0.0079}$ &    19.7307 $^{+ 0.0145}_{- 0.0145}$ &   4.14 &    130 $^{+    40}_{-   100}$ &      -                         &   0.31 &       & \\ 
         &   2 &    5 &   0.1579 $^{+  0.0002}_{-  0.0003}$ &   178.7729 $^{+ 0.0250}_{- 0.0249}$ &    19.4936 $^{+ 0.0140}_{- 0.0140}$ &  10.53 &    190 $^{+    50}_{-    50}$ &      -                         &   0.44 &       & \\ 
         &   3 &    6 &   0.1624 $^{+  0.0000}_{-  0.0003}$ &   178.7169 $^{+ 0.0050}_{- 0.0050}$ &    19.4548 $^{+ 0.0132}_{- 0.0135}$ &  13.88 &     90 $^{+    30}_{-    70}$ &      -                         &   0.21 &       & \\
\enddata
\tablenotetext{*}{A lens group.}
\tablenotetext{a}{Projected distance between the lens and the group centroid.}
\tablenotetext{b}{Masses are reported when $\ge 10^{13} M_{\odot}$.}
\tablenotetext{c}{1: $z_{grp} <$ 0.1 (groups at lower redshift likely are not sampled out to $r_{vir}$ due to the spectroscopic field size); 2: $r_{vir}$ intersects the edge of the well-sampled region of the field because the group is large and/or at low $z$; 3: $r_{vir}$ extends beyond the well-sampled region of the field because the group centroid or several of its members are near the edge of the field; 4: Group membership was adjusted manually to remedy overlapping membership.}
\tablenotetext{d}{Groups in the literature that have a redshift within 3$\sigma_{z}$ of a group in our catalog.  The reference abbreviations are as follows: AFALS07 = \citet{auger07}; AFW08 = \citet{auger08}; CCN07 = \citet{carrasco07}; DG92 = \citet{dressler92}; KSG = \citet{katgert96}, \citet{struble99}, and \citet{gastaldello03}; LCKJ09 = \citet{lopes09}; MB = \citet{mason00} and \citet{barkhouse06}; MPES09 = \citet{mcconnachie09}; MNR05 = \citet{miller05}; TOH00 = \citet{tucker00}; WMKZL06 = \cite{williams06}; and WH = \citet{williams06} and \citet{horner08}.
}

\end{deluxetable*}

\begin{deluxetable*}{llllrrrlllll}
\tablecaption{Group Catalog Supplement: Algorithmically Identified Groups With 3-4 Members \label{table:grps3and4}}
\tablehead{
\colhead{Field} & \colhead{Group} & \colhead{$N_{m}$} & \colhead{$z_{grp}$\tablenotemark{a}} & \colhead{$\overline{RA}$\tablenotemark{a}} & \colhead{$\overline{Dec}$\tablenotemark{a}} & \colhead{R\tablenotemark{b} } & \colhead{$\sigma_{grp}$\tablenotemark{a}} & \colhead{$M_{vir}$\tablenotemark{a,c}} & \colhead{$r_{vir}$} & \colhead{Flag\tablenotemark{d}} & \colhead{Match\tablenotemark{e}} \\
\colhead{ } & \colhead{ } & \colhead{ } & \colhead{ } & \colhead{[deg]} & \colhead{[deg]} & \colhead{[$^{\prime}$]} & \colhead{[km s$^{-1}$]} & \colhead{[10$^{13}$M$_{\odot}$]} & \colhead{[Mpc]} &   \colhead{ } & \colhead{} 
}
\startdata

   b0712 &   1 &    3 &   0.2635                       &   109.3152 ~~~~~~~~~\,          &    47.0962 ~~~~~~~~~\,         &  12.63 &    130                    &      -                         &   0.28 &       & \\ 
         &   2 &    3 &   0.3483                       &   109.1790 ~~~~~~~~~\,          &    47.1236 ~~~~~~~~~\,         &   6.84 &    110                    &      -                         &   0.22 &       & \\ 
         &   4 &    4 &   0.3724 $^{+  0.0003}_{-  0.0002}$ &   109.2126 $^{+ 0.0157}_{- 0.0158}$ &    47.3602 $^{+ 0.0166}_{- 0.0166}$ &  15.09 &    170 $^{+    60}_{-   130}$ &      -                         &   0.34 &       & \\ 
         &   6 &    4 &   0.4360 $^{+  0.0004}_{-  0.0006}$ &   108.9779 $^{+ 0.0107}_{- 0.0107}$ &    47.1869 $^{+ 0.0127}_{- 0.0127}$ &   2.82 &    180 $^{+    30}_{-   150}$ &      -                         &   0.35 &       & \\ 
   b1152 &  13 &    3 &   0.6623                       &   178.9290 ~~~~~~~~~\,          &    19.6922 ~~~~~~~~~\,         &   6.08 &    110                    &      -                         &   0.18 &       & \\ 
   b1422 &   3 &    3 &   0.3239                       &   216.1767 ~~~~~~~~~\,          &    22.9344 ~~~~~~~~~\,         &   0.99 &    100                    &      -                         &   0.22 &       & \\ 
         &   7 &    4 &   0.5706 $^{+  0.0003}_{-  0.0006}$ &   216.2366 $^{+ 0.0044}_{- 0.0044}$ &    22.7597 $^{+ 0.0021}_{- 0.0021}$ &  11.29 &    170 $^{+    30}_{-    90}$ &      -                         &   0.32 &     3 & \\ 
   b1600 &   5 &    3 &   0.2624                       &   240.4598 ~~~~~~~~~\,          &    43.2505 ~~~~~~~~~\,         &   2.51 &    210                    &      1.1                       &   0.46 &       & \\
\enddata
\tablenotetext{a}{Uncertainties are not calculated for $N_m <$ 4}
\tablenotetext{b}{Projected distance between the lens and the group centroid.}
\tablenotetext{c}{Masses are reported when $\ge 10^{13} M_{\odot}$.}
\tablenotetext{d}{1: $z_{grp} <$ 0.1 (groups at lower redshift likely are not sampled out to $r_{vir}$ due to the spectroscopic field size); 2: $r_{vir}$ intersects the edge of the well-sampled region of the field because the group is large and/or at low $z$; 3: $r_{vir}$ extends beyond the well-sampled region of the field because the group centroid or several of its members are near the edge of the field; 4: Group membership was adjusted manually to remedy overlapping membership.}
\tablenotetext{e}{Groups in the literature in these fields that have a redshift within 3$\sigma_{z}$ of a group in our catalog.  The reference abbreviations are as follows: BHS = \citet{berlind06}, \citet{haynes11}, and \citet{smith12}; and WMKZL06 = \cite{williams06}.}
\end{deluxetable*}

\begin{deluxetable*}{lllllrlll}
\tablecaption{Group Catalog Supplement: Supergroup and its Substructures in Field b0712 \label{table:supergroup}}
\tablehead{
\colhead{Group} & \colhead{$N_{m}$} & \colhead{$z_{grp}$} & \colhead{$\overline{RA}$} & \colhead{$\overline{Dec}$} & \colhead{R\tablenotemark{a} } & \colhead{$\sigma_{grp}$} & \colhead{$M_{vir}$\tablenotemark{b}} & \colhead{$r_{vir}$} \\
\colhead{ } &\colhead{ } & \colhead{ } & \colhead{[deg]} & \colhead{[deg]} & \colhead{[$^{\prime}$]} & \colhead{[$km _{   } s^{-1}$]} & \colhead{[10$^{13}$M$_{\odot}$]} & \colhead{[Mpc]} 
}
\startdata
 0 &  230 &  0.2941 $^{+  0.0004}_{ -0.0003}$ &   108.9931 $^{+ 0.0152}_{-0.0137}$ &    47.1480 $^{+ 0.0079}_{-0.0075}$ &   0.90 &   1170 $^{+    40}_{   -50}$ &    119.7 $^{+    12.0}_{   -13.6}$ &   2.48 \\ 
\hline
 1 &   54 &  0.2898 $^{+  0.0002}_{ -0.0002}$ &   109.1119 $^{+ 0.0144}_{-0.0133}$ &    47.1527 $^{+ 0.0070}_{-0.0071}$ &   3.96 &    330 $^{+    30}_{   -30}$ &      3.7 $^{+     1.1}_{    -0.9}$ &   0.70 \\ 
 2 &   51 &  0.2961 $^{+  0.0002}_{ -0.0002}$ &   108.7460 $^{+ 0.0106}_{-0.0098}$ &    47.1269 $^{+ 0.0070}_{-0.0078}$ &  11.06 &    340 $^{+    30}_{   -30}$ &      4.2 $^{+     1.2}_{    -1.0}$ &   0.73 \\ 
 3 &   39 &  0.2983 $^{+  0.0003}_{ -0.0003}$ &   109.0947 $^{+ 0.0111}_{-0.0110}$ &    47.2807 $^{+ 0.0083}_{-0.0087}$ &   8.63 &    420 $^{+    60}_{   -60}$ &      7.1 $^{+     3.0}_{    -2.4}$ &   0.88 \\ 

\enddata
\tablenotetext{a}{Projected distance between the lens and the group centroid.}
\tablenotetext{b}{Mass assuming virialization.}
\end{deluxetable*}

\begin{deluxetable*}{llllrrrlllll}
\tablecaption{Group Catalog Supplement: Visually Identified Candidate Groups \label{table:candgrps}}
\tablehead{
\colhead{Field} & \colhead{Group\tablenotemark{a}} & \colhead{$N_{m}$} & \colhead{$z_{grp}$\tablenotemark{b}} & \colhead{$\overline{RA}$} & \colhead{$\overline{Dec}$} & \colhead{R\tablenotemark{c} } & \colhead{$\sigma_{grp}$} & \colhead{$M_{vir}$} & \colhead{$r_{vir}$}  & \colhead{Flag\tablenotemark{d}}  \\
\colhead{ } & \colhead{ } & \colhead{ } & \colhead{ } & \colhead{[deg]} & \colhead{[deg]} & \colhead{[$^{\prime}$]} & \colhead{[km s$^{-1}$]} & \colhead{[10$^{13}$M$_{\odot}$]} & \colhead{[Mpc]} & \colhead{ } 
}
\startdata

   b0712 &   i*&   13 &   0.4030 $^{+  0.0009}_{-  0.0010}$ &   108.9890 $^{+ 0.0362}_{- 0.0306}$ &    47.1460 $^{+ 0.0250}_{- 0.0262}$ &   1.07 &    800 $^{+   140}_{-   530}$ &     39.5 $^{+    18.6}_{-    28.2}$ &   1.59 & \\ 
         &  ii &    5 &   0.6200 $^{+  0.0005}_{-  0.0005}$ &   108.7936 $^{+ 0.0442}_{- 0.0441}$ &    47.1569 $^{+ 0.0223}_{- 0.0223}$ &   9.06 &    260 $^{+    50}_{-   100}$ &      1.6 $^{+     0.8}_{-     0.9}$ &   0.46 & \\ 
   b1152 &   i &    8 &   0.5897 $^{+  0.0007}_{-  0.0007}$ &   178.9546 $^{+ 0.0350}_{- 0.0391}$ &    19.5294 $^{+ 0.0220}_{- 0.0218}$ &  10.75 &    430 $^{+    60}_{-   180}$ &      6.4 $^{+     2.7}_{-     4.2}$ &   0.77 & \\ 
   b2114 &   i &    5 &   0.3762 $^{+  0.0004}_{-  0.0005}$ &   319.2515 $^{+ 0.0294}_{- 0.0294}$ &     2.4426 $^{+ 0.0338}_{- 0.0363}$ &   2.51 &    240 $^{+   220}_{-    90}$ &      1.5 $^{+     3.0}_{-     0.1}$ &   0.49 & \\ 
         &  ii &    6 &   0.3911 $^{+  0.0007}_{-  0.0006}$ &   319.3062 $^{+ 0.0354}_{- 0.0289}$ &     2.4526 $^{+ 0.0438}_{- 0.0431}$ &   5.83 &    380 $^{+   200}_{-   160}$ &      5.0 $^{+     6.7}_{-     1.5}$ &   0.76 & \\ 
         & iii &   10 &   0.4479 $^{+  0.0009}_{-  0.0008}$ &   319.3432 $^{+ 0.0232}_{- 0.0215}$ &     2.4487 $^{+ 0.0451}_{- 0.0437}$ &   7.97 &    640 $^{+   160}_{-   400}$ &     21.4 $^{+    11.5}_{-     8.8}$ &   1.25 & \\ 
 bri0952 &   i &   11 &   0.6187 $^{+  0.0009}_{-  0.0010}$ &   148.6635 $^{+ 0.0077}_{- 0.0085}$ &    -1.4307 $^{+ 0.0149}_{- 0.0152}$ &   6.73 &    620 $^{+   100}_{-   120}$ &     17.5 $^{+     8.1}_{-     6.6}$ &   1.10 & \\ 
         &  ii &    8 &   0.6385 $^{+  0.0009}_{-  0.0008}$ &   148.7207 $^{+ 0.0070}_{- 0.0079}$ &    -1.4832 $^{+ 0.0173}_{- 0.0182}$ &   2.10 &    490 $^{+    80}_{-   120}$ &      9.1 $^{+     4.5}_{-     4.4}$ &   0.86 & \\
\enddata
\tablenotetext{*}{A lens group.}
\tablenotetext{a}{Numbering is not continued from Tables \ref{table:grps5andup} and \ref{table:grps3and4}.}
\tablenotetext{b}{Uncertainties are not calculated for $N_m <$ 4}
\tablenotetext{c}{Projected distance between the lens and the group centroid.}
\tablenotetext{d}{1: $z_{grp} <$ 0.1 (groups at lower redshift likely are not sampled out to $r_{vir}$ due to the spectroscopic field size); 2: $r_{vir}$ intersects the edge of the well-sampled region of the field because the group is large and/or at low $z$; 3: $r_{vir}$ extends beyond the well-sampled region of the field because the group centroid or several of its members are near the edge of the field; 4: Group membership was adjusted manually to remedy overlapping membership.}
\end{deluxetable*}

\clearpage

\appendix
\label{appendix}

\section{Choosing Parameters for Our Algorithm}
\label{appendix:choosingparams}

Here we describe the main parameters of our group finding algorithm and how we choose them.

\subsection{Candidate Peak Selection}
\label{appendix:canpeakselection}

There are several degenerate parameters that affect what kind of groups can be identified as candidate peaks. The width of the bins in the galaxy velocity distribution affects how easily groups projected along the same line of sight can be distinguished. The grid square width in the galaxy sky position likewise affects how easily groups with similar radial velocities but different locations projected on the sky can be identified as separate candidate peaks.  If velocity bins or grid squares are too small, large and/or sparsely sampled structures (i.e., clusters and higher redshift structures) will be artificially fragmented and might not be identified as candidate peaks. If the bins or grid squares are too large, multiple structures might be identified as a single candidate peak.  This conflation also can result in no group being identified if the average position and velocity of the galaxies are in sparsely populated regions. For both the velocity bins and sky grid, we need to decide how many galaxies must be in a bin or grid square to further consider that it might contain a group.  If the threshold is set too high, we will only ever be able to find the richest structures.  If it is set too low, we have a larger chance of including spurious groups in our sample. 

Because we define the grid square sizes in terms of $r_{vir}$, where we use the relation for $r_{200}$ given by \citet{carlberg97}:

\begin{equation}
r_{vir,\sigma} \approx r_{200} = \frac{\sqrt{3} \sigma_{grp}}{10 H(z)}, 
\end{equation}
the initial velocity dispersion guess affects the properties above.

How the grid is centered is another choice.  We place the center of a grid square at the median RA and Dec of the galaxies in each velocity bin to increase the likelihood that the best populated structure in each velocity bin is centered in a grid square.  However, the projected spatial distributions of galaxies in a given velocity bin are often complex, with multiple overdensities and/or a concentration plus a large diffuse component.  Thus, we shift the grid to nine other positions (in units of a quarter of the width of a grid square in RA, Dec, and both) to increase the likelihood that multiple overdensities, where present, are centered in a grid square in at least one of the grid positions.  Mismatches between galaxy overdensities and grid squares would primarily impact groups with few members in our redshift catalog and/or with members spread out on the sky; groups with up to eight members could be missed if they have their members split equally between two grid squares for all grid positions.  If an overdensity is identified as a candidate peak but is not well centered in a grid square, our subsequent iterative membership procedure can remedy this, as the projected spatial centroid is allowed to vary.

We test our methodology using velocity bins of 1000, 1200, 1500, 1600, and 2000 km s$^{-1}$ and several grid square widths from 1.5$r_{vir}$ to 5$r_{vir}$.  We set the galaxy number threshold at three and five galaxies for the velocity bins and three, four, and five galaxies for the spatial grid squares.  We test using velocity dispersion initial guesses of 300, 400, and 500 km s$^{-1}$.  Using 1200 km s$^{-1}$ wide velocity bins, 3.3$r_{vir}$ wide grid squares, galaxy thresholds of five galaxies for both velocity bins and grid squares, a velocity dispersion initial guess of 400 km s$^{-1}$, and multiple grid positions does a good job of distinguishing multiple visually identified groups and of not creating a large number of small, likely spurious groups or unphysically massive structures.  While the majority of the catalog is robust for a grid square size within the range of 3 to 3.6 $r_{vir}$ (tested in steps of 0.1 $r_{vir}$), we select the value that maximizes the number of recovered visually identified groups while minimizing the number of unphysically large structures.

Most of our groups have low velocity dispersions, with a peak in the number distribution at $\sim$ 300 km s$^{-1}$ (median value of 350 km s$^{-1}$ for groups with at least five members) and a tail to larger values (see Figure \ref{fig:zsig}).  To test how our assumption of 400 km s$^{-1}$ for the initial velocity dispersion affects the final $\sigma_{grp}$ distribution, we compare the results from our three velocity dispersion initial guesses.  Overall, the catalogs are similar; the majority of groups have similar or identical calculated group properties.  However, the 300 km s$^{-1}$ catalog finds fewer groups, including 24 fewer visually identified groups.  The 500 km s$^{-1}$ catalog finds double the number of groups with $M_{vir} \ge 10^{15} M_{\odot}$, including one with an unphysical $M_{vir} = 5.6 \times 10^{15} M_{\odot}$ ($\sigma_{grp} =$ 1990 km s$^{-1}$).  The 400 km s$^{-1}$ catalog agrees best with the velocity dispersion distribution from zCOSMOS, although none of these catalogs have distributions distinguishable from it using a K-S test.  The 300 km s$^{-1}$ catalog suffers from incompleteness when compared to the MXXL mass function at slightly larger masses than the 400 km s$^{-1}$ catalog. The high mass slope of the 500 km s$^{-1}$ catalog does not agree as well with that of MXXL due to its larger number of unphysically massive structures.  Thus, we choose to use an initial velocity dispersion of 400 km s$^{-1}$.

\subsection{Group Membership Iteration}

We must select a redshift range $\Delta z$ within which to look for possible group members. If $\Delta z$ is too small, the velocity tails of massive structures get clipped, but if it is too big, large scale structure might be misidentified as a group or cluster. We select $\Delta z$ equivalent to 6600 km s$^{-1}$ at the candidate peak's redshift (using the special relativistic velocities) to include galaxies up to 3$\sigma$ from the mean velocity for a cluster with a velocity dispersion of 1100 km s$^{-1}$, that of a Coma-like cluster. Very rare structures with even larger velocity dispersions can still be found by our algorithm, but their high velocity tails will be somewhat clipped.  Some small groups near larger structures in redshift also may be clipped away.  Even though most of our groups can be found using a smaller $\Delta z$, choosing 6600 km s$^{-1}$ ensures that we properly sample the few massive clusters in our fields, so their velocity dispersions and contributions to the lensing potentials are not underestimated.

In the first step of the group membership iteration, we must assume a velocity interval and a radius within which to consider possible group galaxies.  We assume a velocity dispersion of 400 km s$^{-1}$ and the virial radius relation given above for the same reasons given in the previous section.  

For the final group catalog, we include galaxies projected up to 3$r_{vir}$ from the group's center on the sky.  Including group members out to 3$r_{vir}$ will likely increase the contamination by nonmembers, but it improves both our group and group member completeness.  Both \citet{carlberg01} and \citet{wilman05} accept members out to 1.5$r_{vir}$.  The cosmological hydrodynamical simulations of \citet{bahe13} show that while beyond 3$r_{vir}$ almost all galaxies are infalling into the host halos for the first time, some galaxies now at $\sim$ 3$r_{vir}$ passed within $\sim$ 1$r_{vir}$ earlier and thus have been influenced by the host halo.

We require the velocity dispersion to be at least 50 km s$^{-1}$ for the structure to be considered a halo.

\subsection{Duplicate Removal}
\label{appendix:duplicateremoval}

We determine which groups include galaxies assigned to more than one group (overlapping members) in multiple trials. If at least 75\% of the trials converge on an identical group, we add that group to our catalog.  If a group is only found once among all the trials, it is discarded, because it is unlikely to be real.  Any group that shares more than 50\% of its membership with a group with at least twice as many members is discarded, which prevents the artificial fragmentation of rich clusters.  In other cases where a given group shares at least 50\% of its members with another, we determine the mean group spatial centroid, the mean group redshift, the maximum group velocity dispersion, and the maximum group virial radius and reiterate on the group membership to produce the final group for the catalog.

The main parameters here are the percentage of group membership overlap and the fraction of groups found for a given candidate peak that must be identical to accept that group as is.

If the percentage of group membership overlap is set too high, then many galaxies are assigned to multiple groups.  If it is set too low, we will combine groups that have most of their galaxies well separated but have a couple outliers that overlap, resulting in only one or neither group remaining in our catalog.

As most of the groups found among multiple trials either have identical membership or do not overlap at all, our catalog is fairly insensitive to our choice of group membership overlap fraction.  Out of 53652 group membership overlap comparisons, only 238 (0.4\%) have overlap fractions between 5\% and 75\%.

We rerun our membership iteration on those groups that have at least 50\% overlapping membership with another group.  Of those, 47 of 50 converge to 47 new groups.  We remove the remaining three.
 
After the second run of the membership iterator, there are seven pairs of groups that share a few member assignments (one galaxy in two cases, two galaxies in three cases, and three galaxies in two cases).  To determine to which group these remaining galaxies belong, we calculate the offset in velocity and projected spatial position of each shared galaxy from its two possible groups using

\begin{equation}
R^2=\left(v_{gxy}-v_{grp}\right)^2+\left(r_{gxy}-r_{grp}\right)^2,
\end{equation}

\noindent where $r_{gxy}$ is the position of the shared galaxy and $r_{grp}$ is the group projected spatial centroid.  The galaxy is assigned to the group for which $R^2$ is smaller, then the other group's properties are recalculated without that galaxy.  We perform no further cuts on minimum velocity distribution or iterate further on membership. The groups with adjusted membership are flagged in Tables \ref{table:grps5andup} and \ref{table:grps3and4}.

\clearpage

\section{Sky Maps and Velocity Distributions of Groups and Candidate Groups}
\label{appendix:grpplots}

\begin{figure*}[h!]
\includegraphics[clip=true, width=18cm]{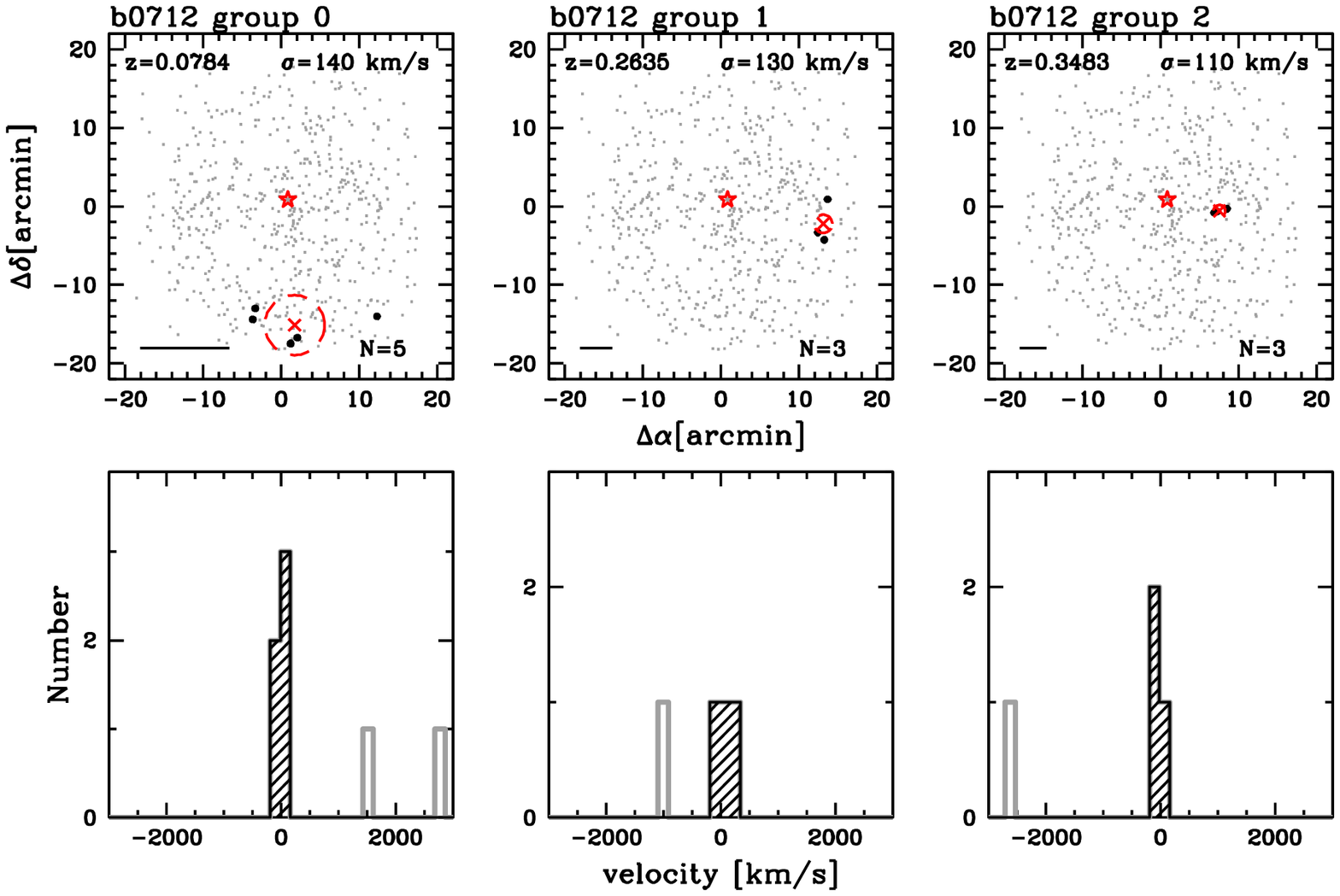}
\caption{\emph{First (and third, on subsequent pages) rows:} Sky plots for all groups in our algorithmic group catalog.  Black points are group members, and gray points are the rest of the galaxies in the field in our redshift catalog (including galaxies in other groups).  Group redshifts, group velocity dispersions, and the number of group galaxies are marked.  We also show group projected spatial centroids (red cross), $r_{vir}$ (red dashed circle), the lens (red star), and the angular size of 1 Mpc at the redshift of the group (black bar).  \emph{Second (and fourth) rows:} Velocity histograms for our groups shifted so the group mean velocity is at 0 km s$^{-1}$.  Bin widths are 180 km s$^{-1}$, which corresponds to about three times the mean redshift error ($\sim$ 0.0002, or $\sim$ 60 km s$^{-1}$) at $z =$ 0.3.  Gray histograms are all galaxies in the redshift sample within 3$r_{vir}$ of the group centroid, and black histograms are group galaxies.  The dashed (dotted) vertical lines mark the velocities of the spectroscopic (photometric) redshifts of the lenses, in the cases where the lens velocity is within 3000 km s$^{-1}$ of the group velocity centroid, whether or not we identify the lens as a group member.}
\label{fig:grpskyplotsandvelhists}
\end{figure*}
\clearpage
\begin{figure*}
\ContinuedFloat
\includegraphics[clip=true, width=18cm]{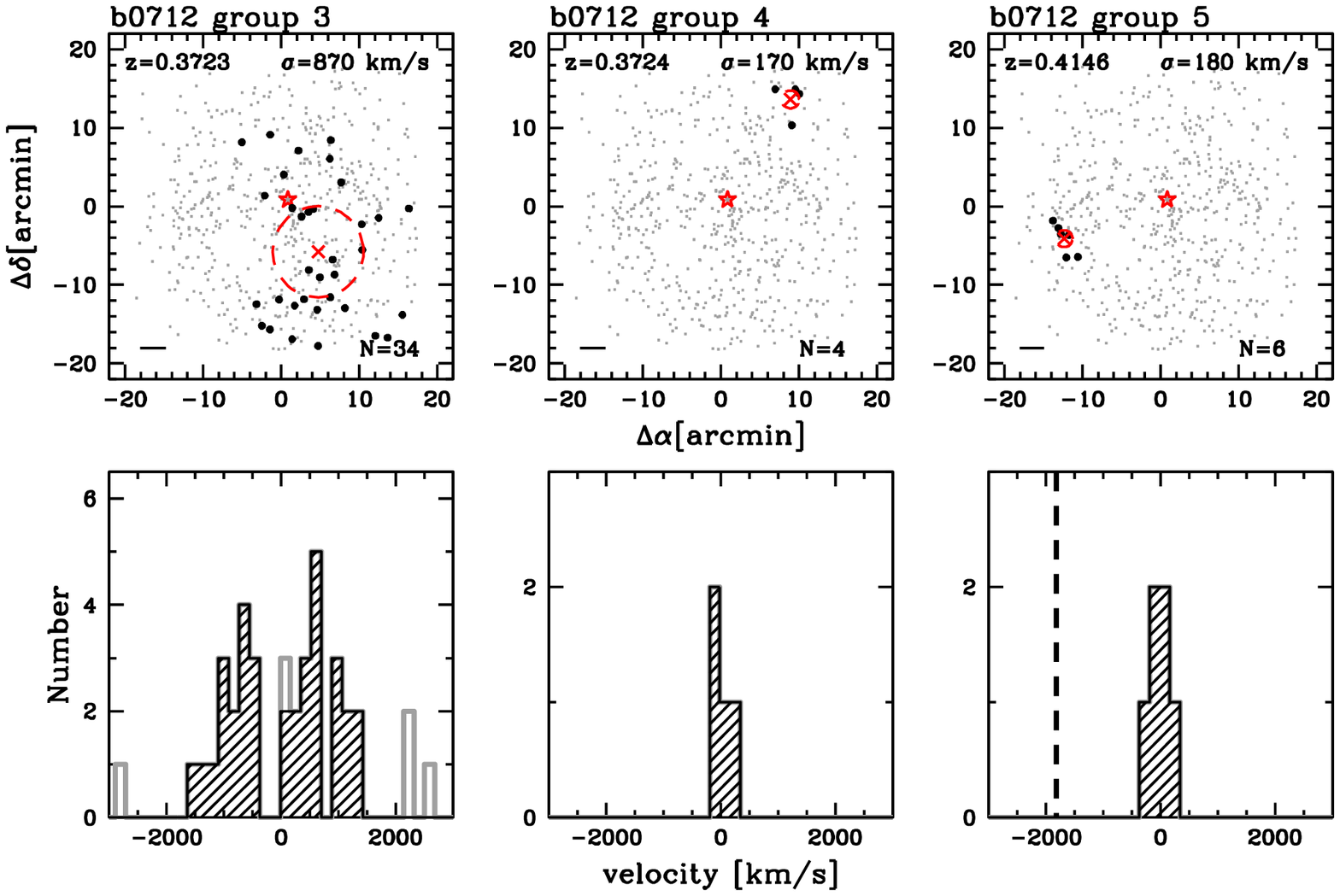}
\includegraphics[clip=true, width=18cm]{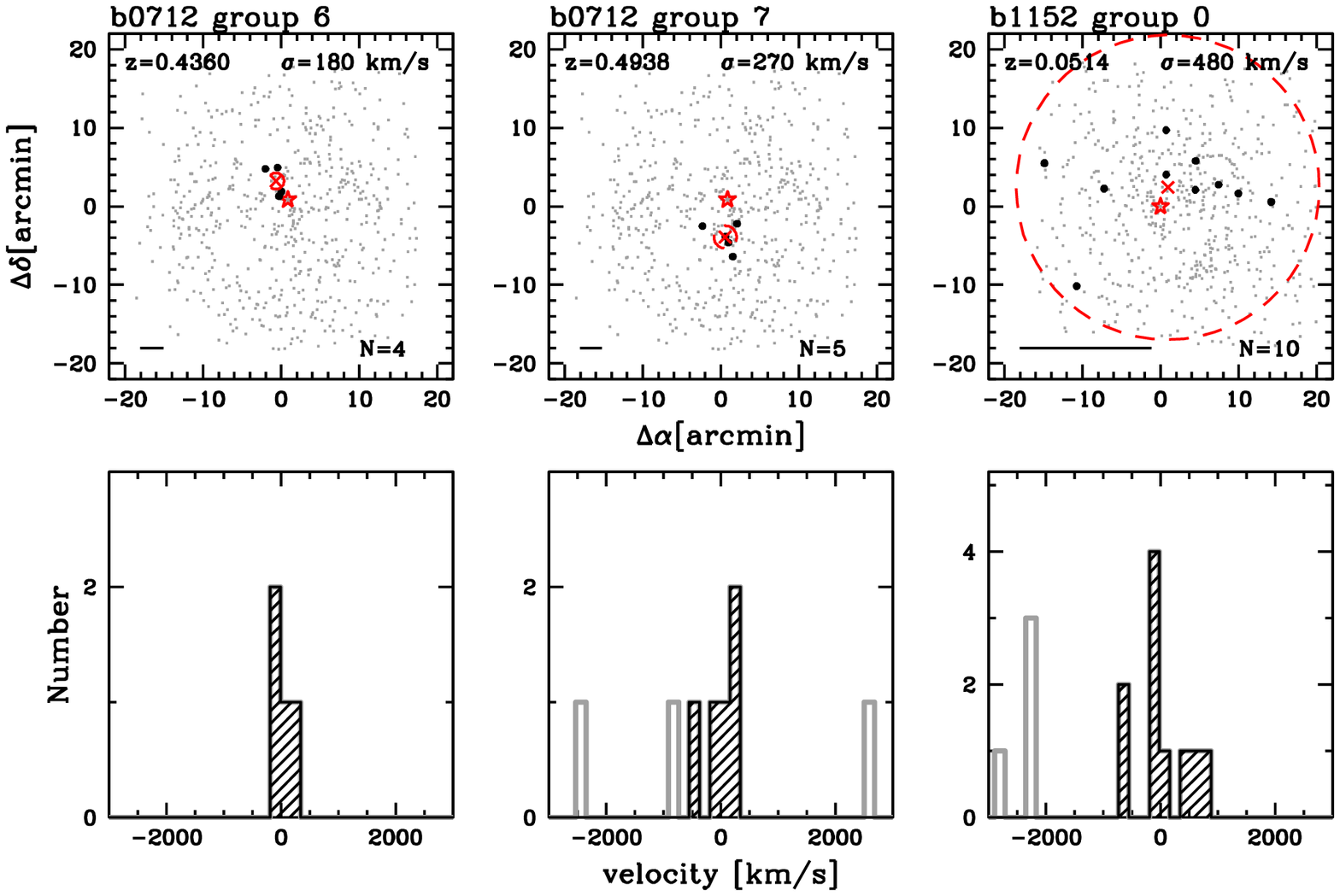}
\caption{Continued.}
\end{figure*}
\clearpage
\begin{figure*}
\ContinuedFloat
\includegraphics[clip=true, width=18cm]{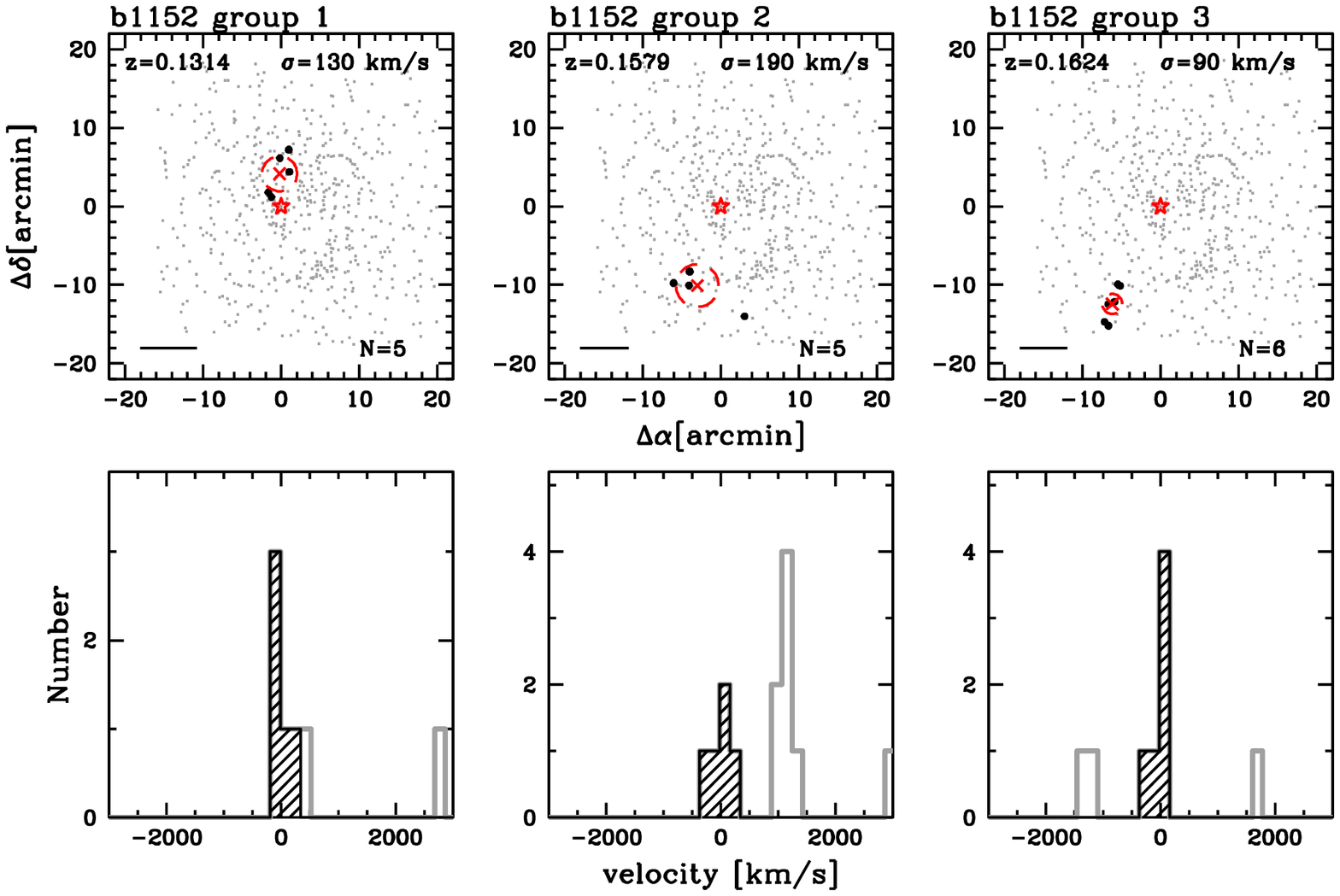}
\includegraphics[clip=true, width=18cm]{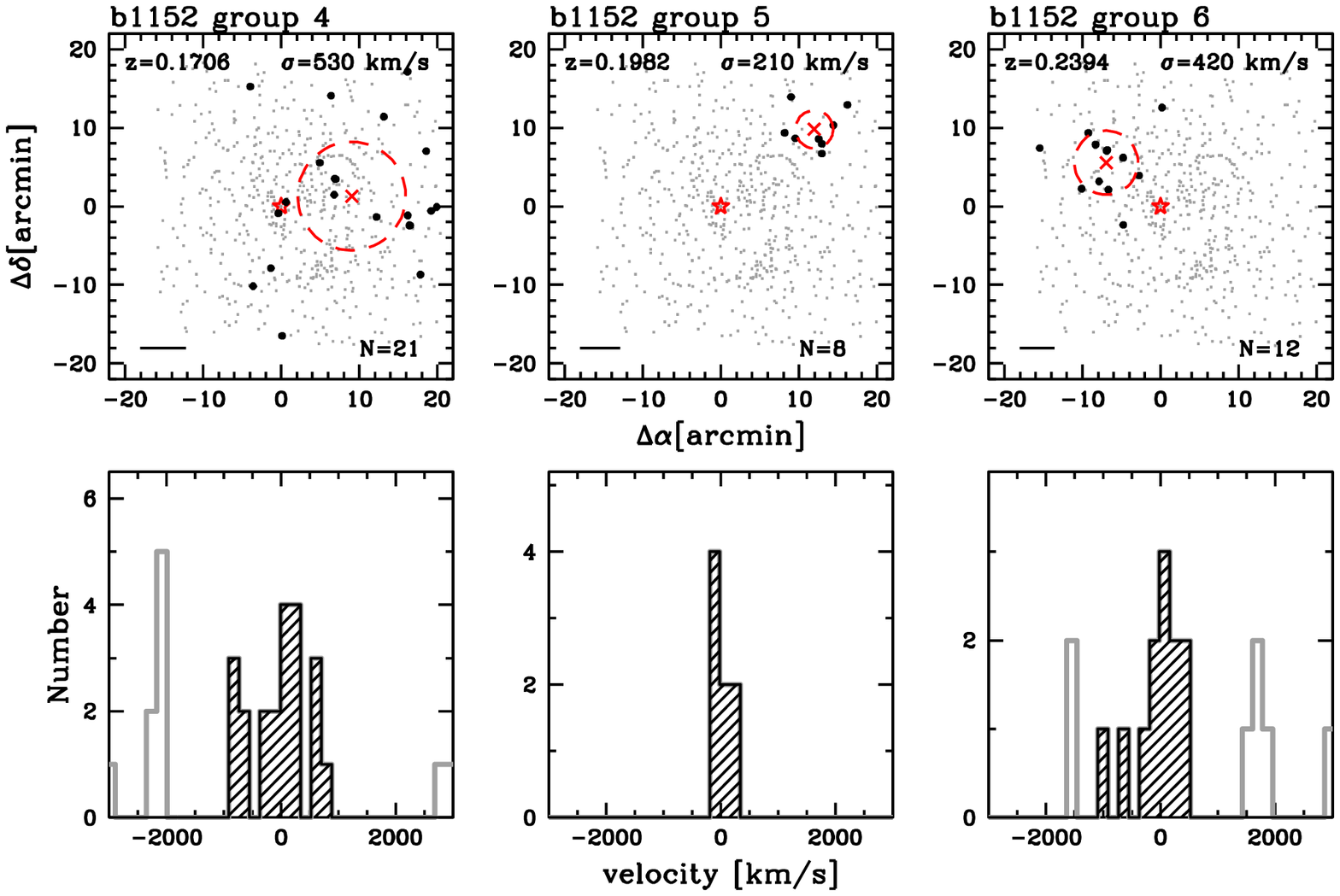}
\caption{Continued.}
\end{figure*}
\clearpage
\begin{figure*}
\ContinuedFloat
\includegraphics[clip=true, width=18cm]{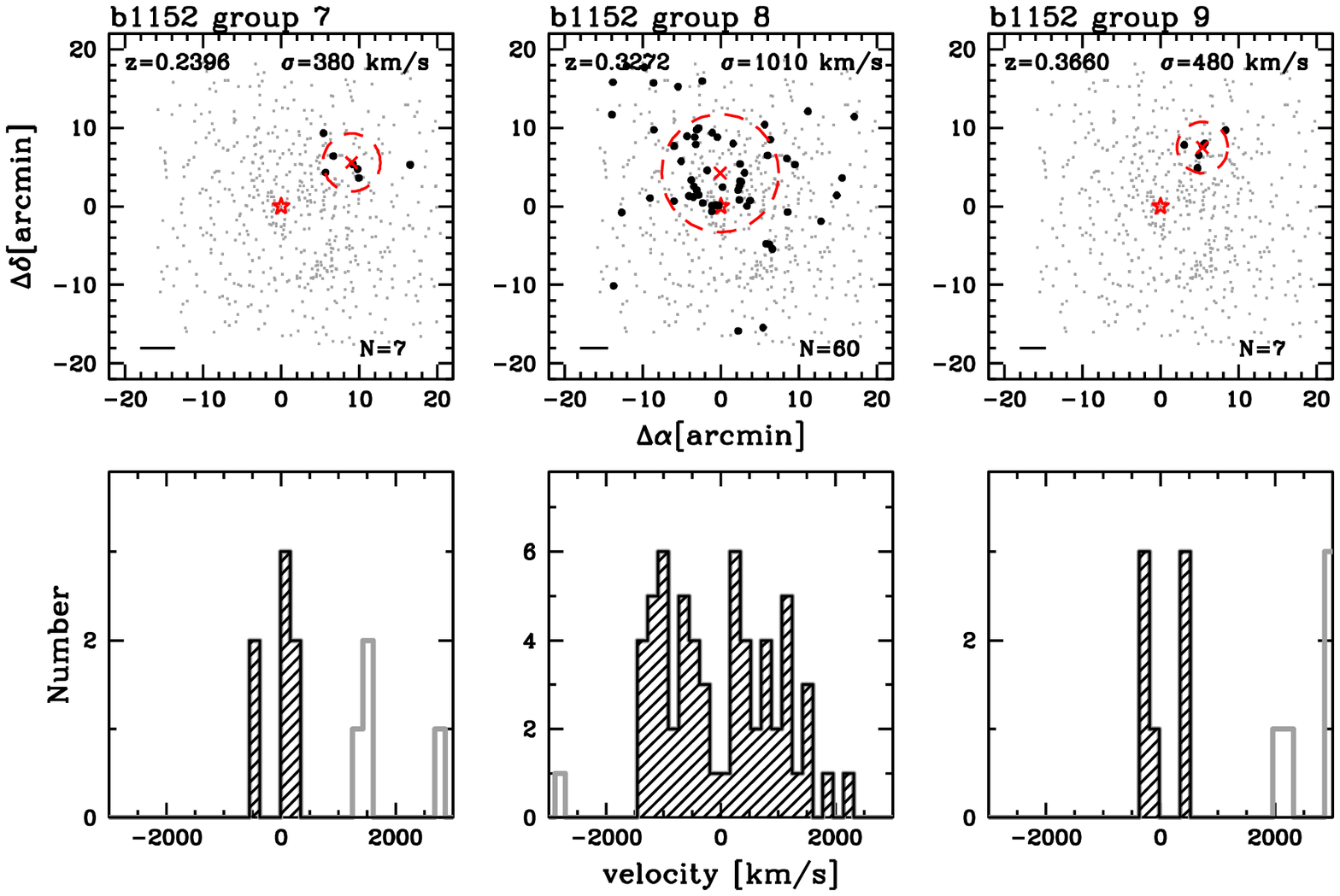}
\includegraphics[clip=true, width=18cm]{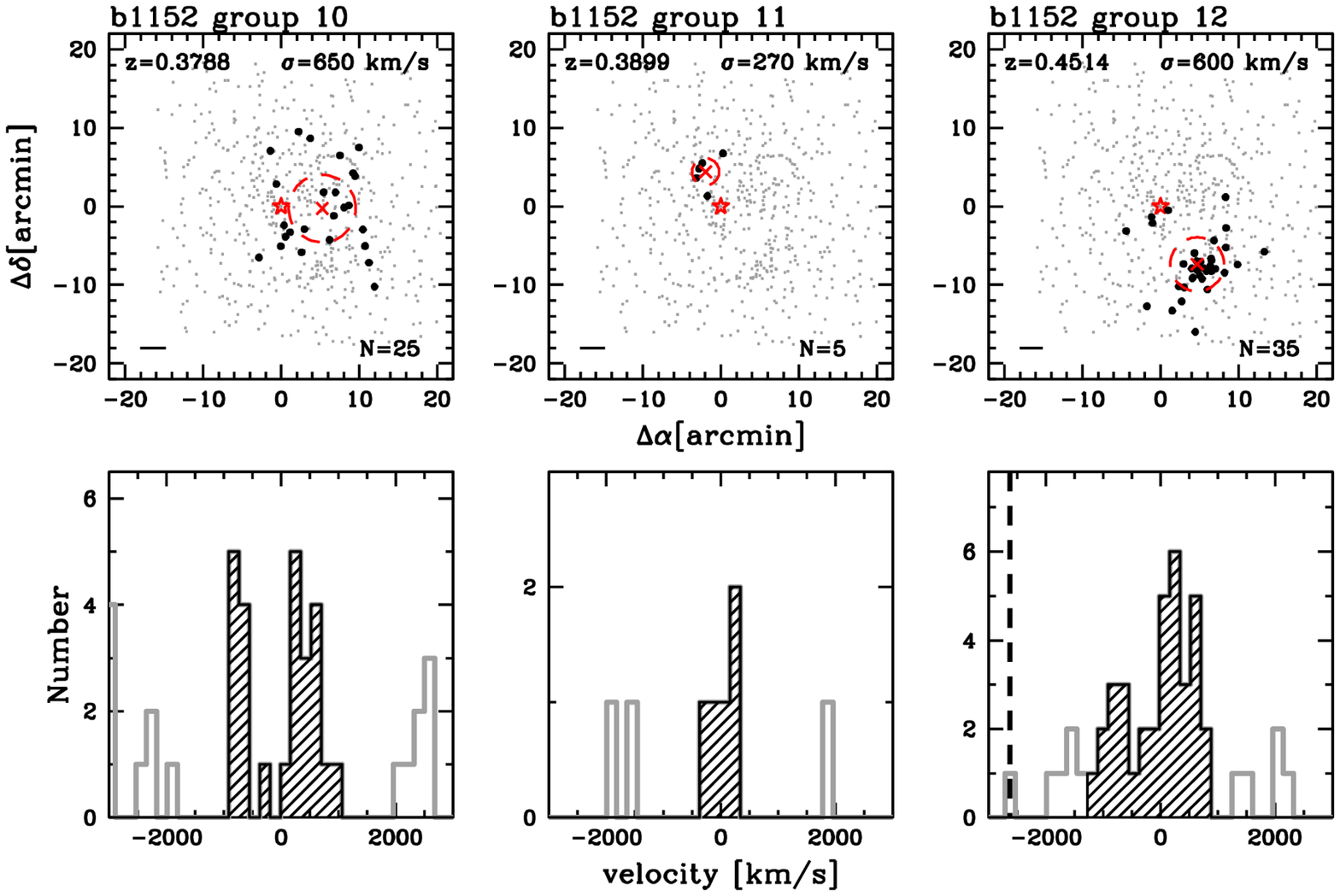}
\caption{Continued.}
\end{figure*}
\clearpage
\begin{figure*}
\ContinuedFloat
\includegraphics[clip=true, width=18cm]{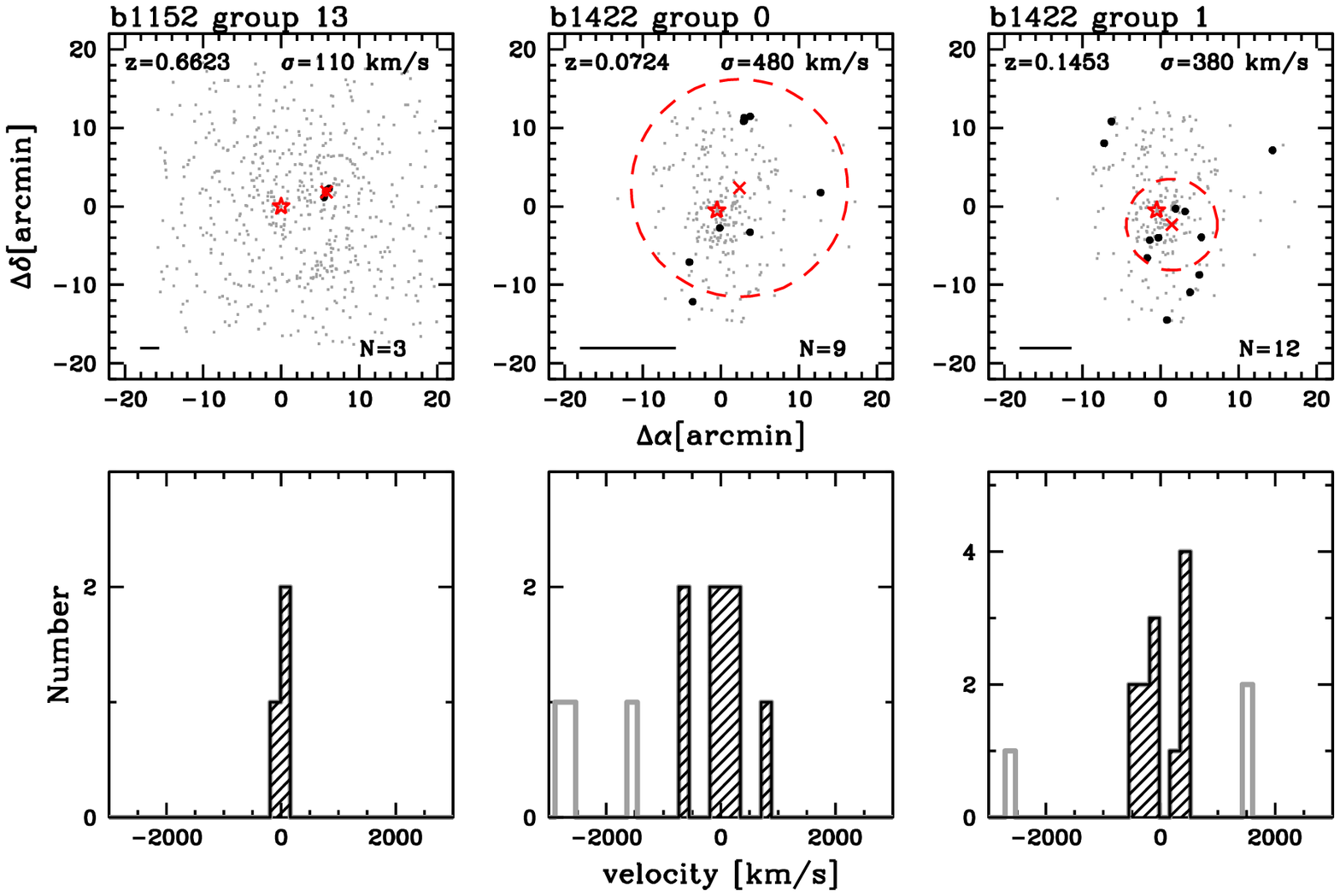}
\includegraphics[clip=true, width=18cm]{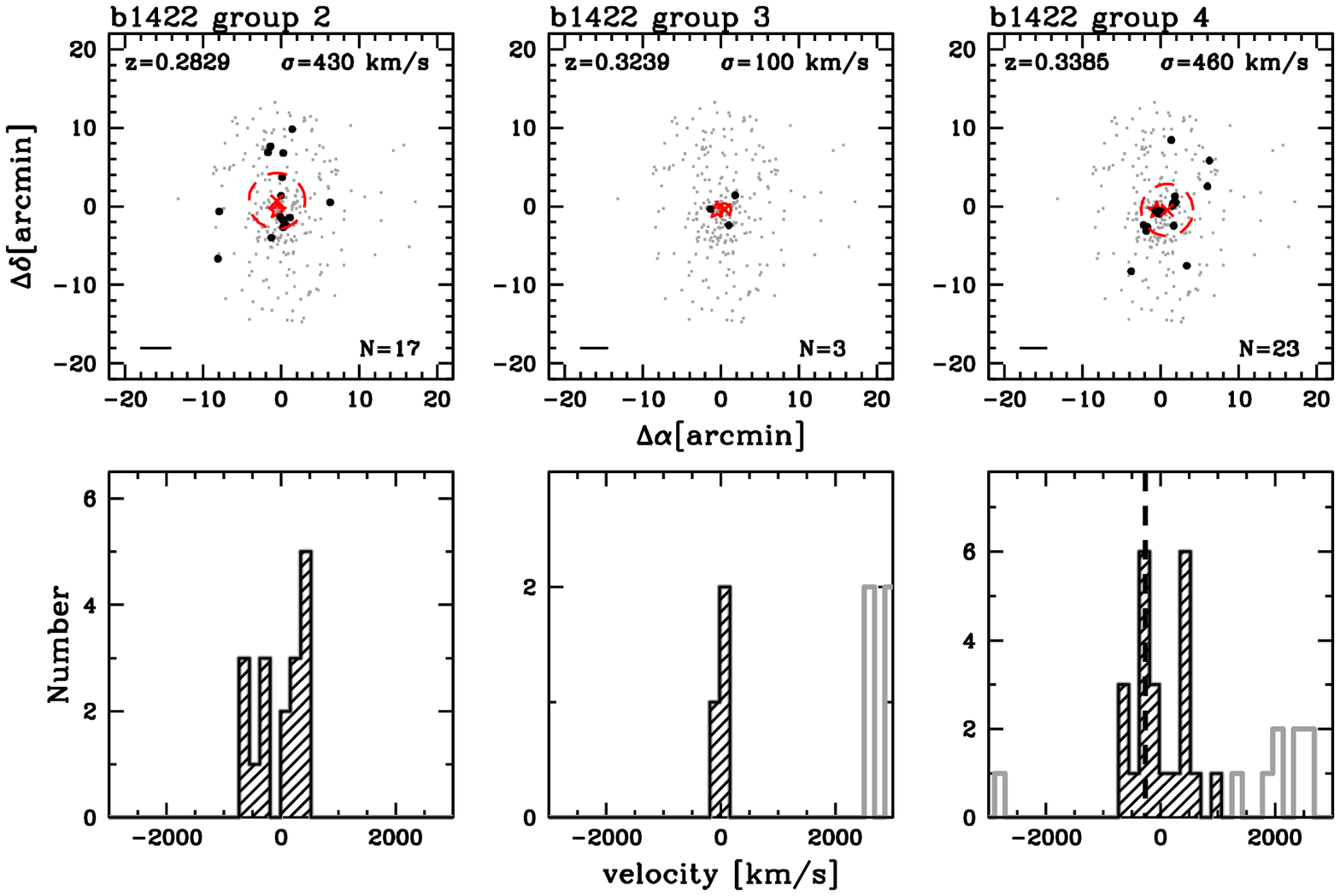}
\caption{Continued.}
\end{figure*}
\clearpage
\begin{figure*}
\ContinuedFloat
\includegraphics[clip=true, width=18cm]{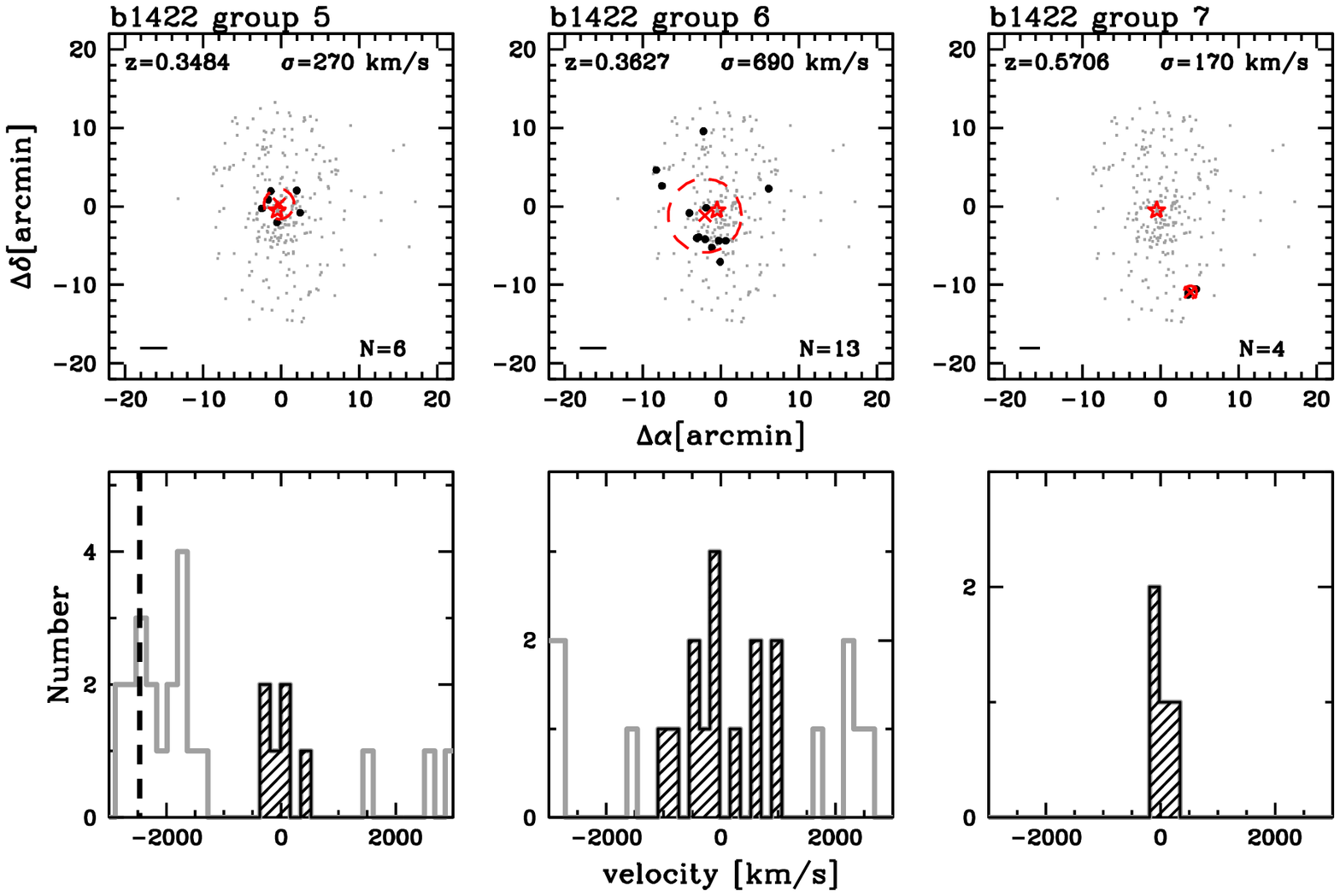}
\includegraphics[clip=true, width=18cm]{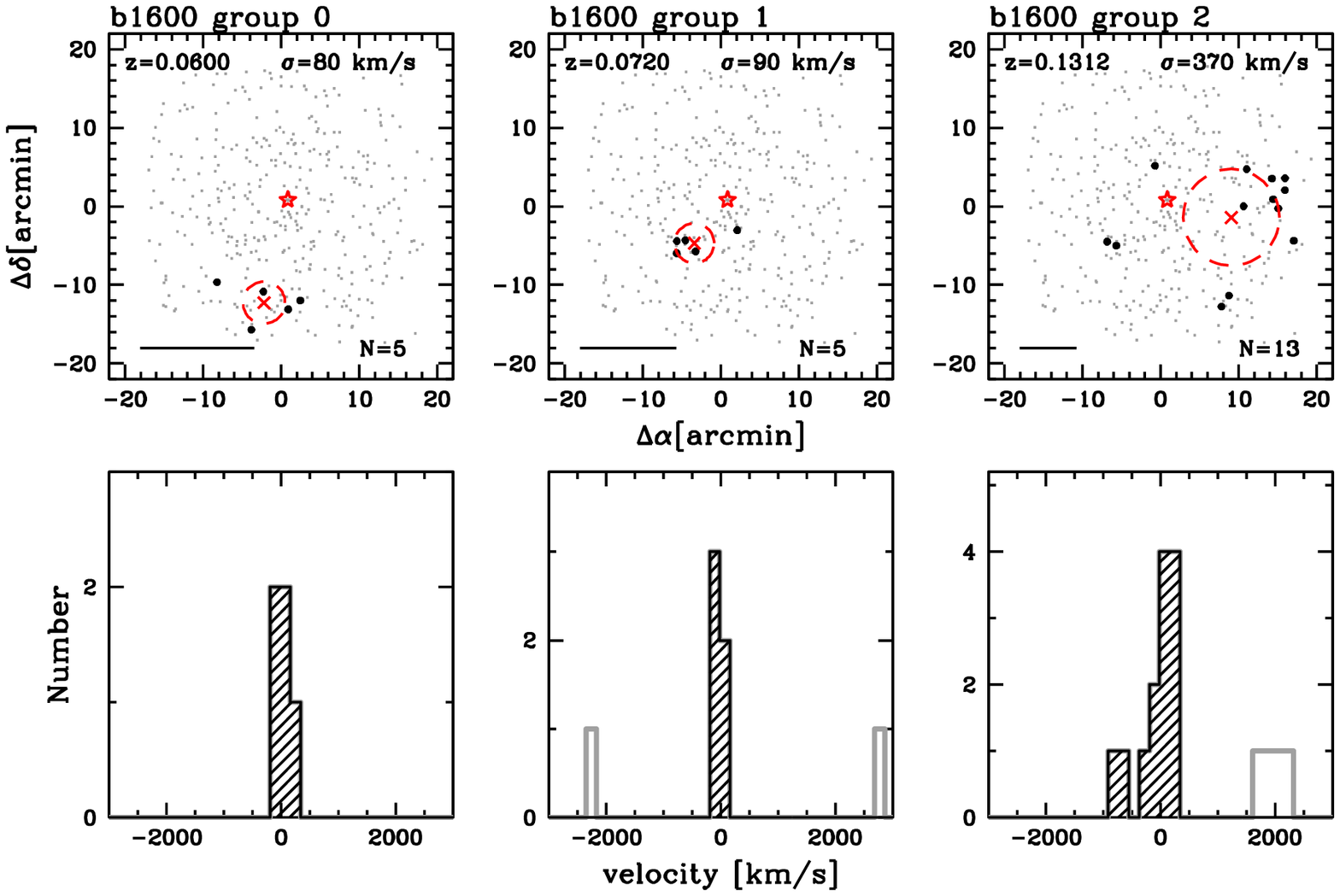}
\caption{Continued.}
\end{figure*}
\clearpage
\begin{figure*}
\ContinuedFloat
\includegraphics[clip=true, width=18cm]{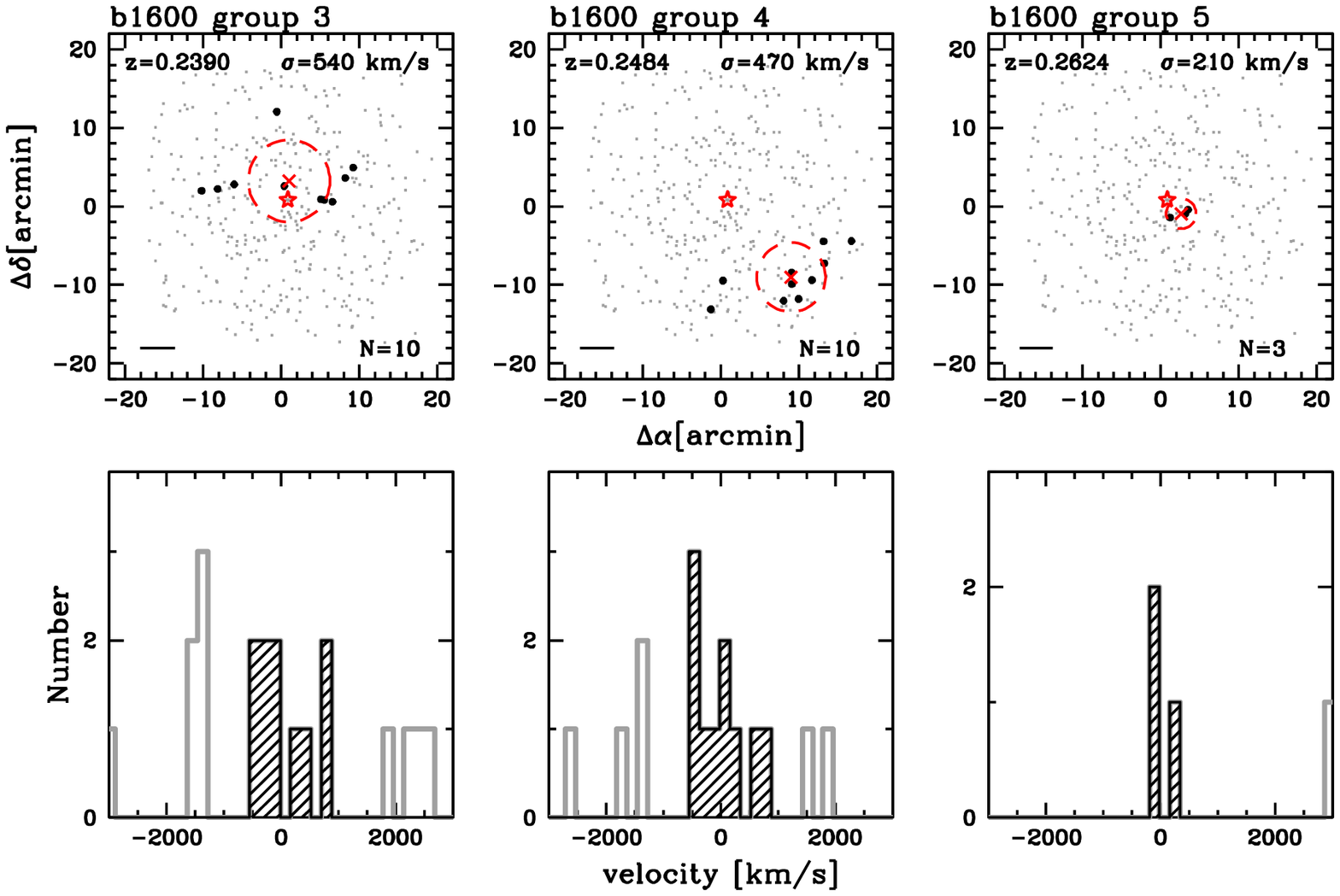}
\includegraphics[clip=true, width=18cm]{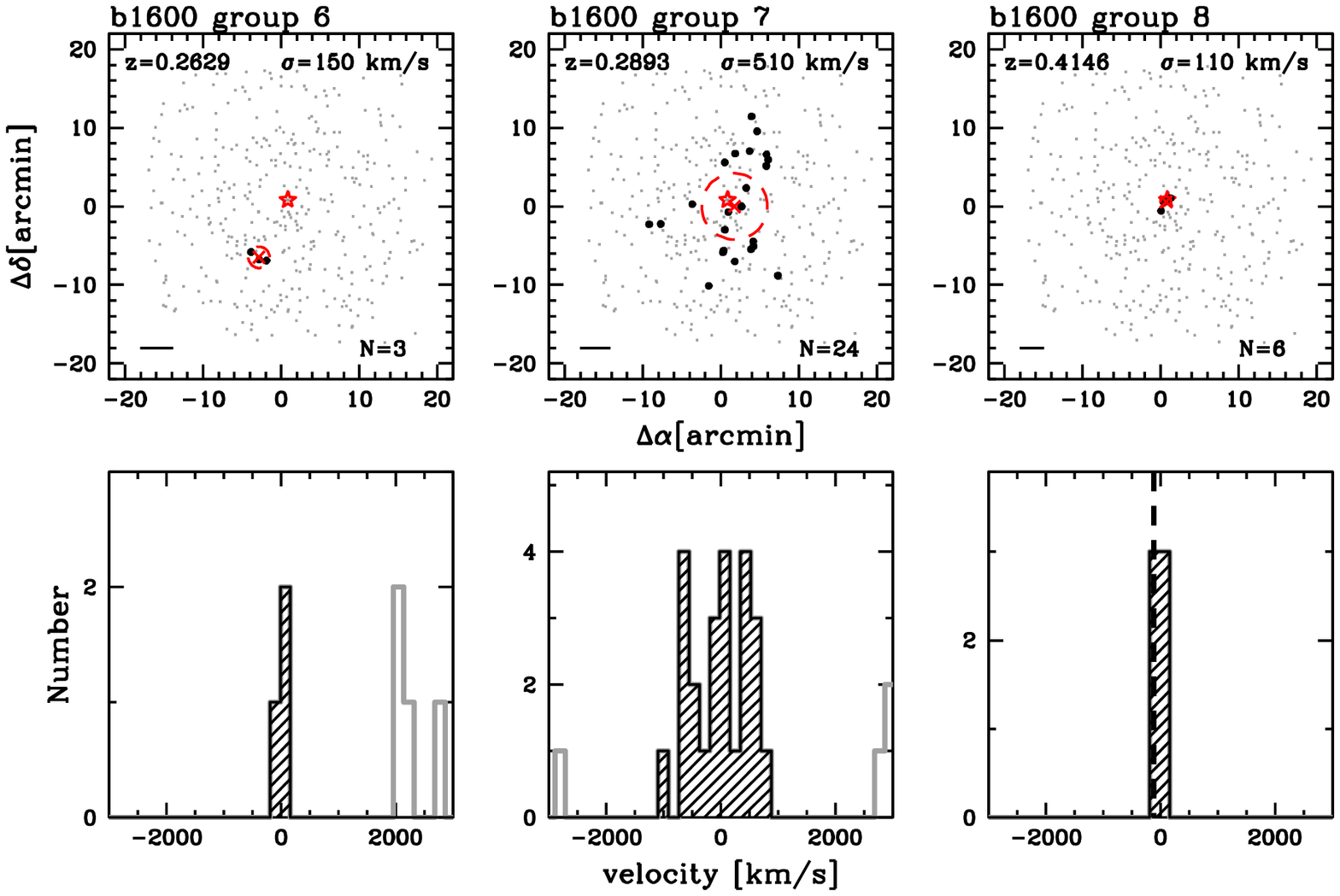}
\caption{Continued.}
\end{figure*}
\clearpage
\begin{figure*}
\ContinuedFloat
\includegraphics[clip=true, width=18cm]{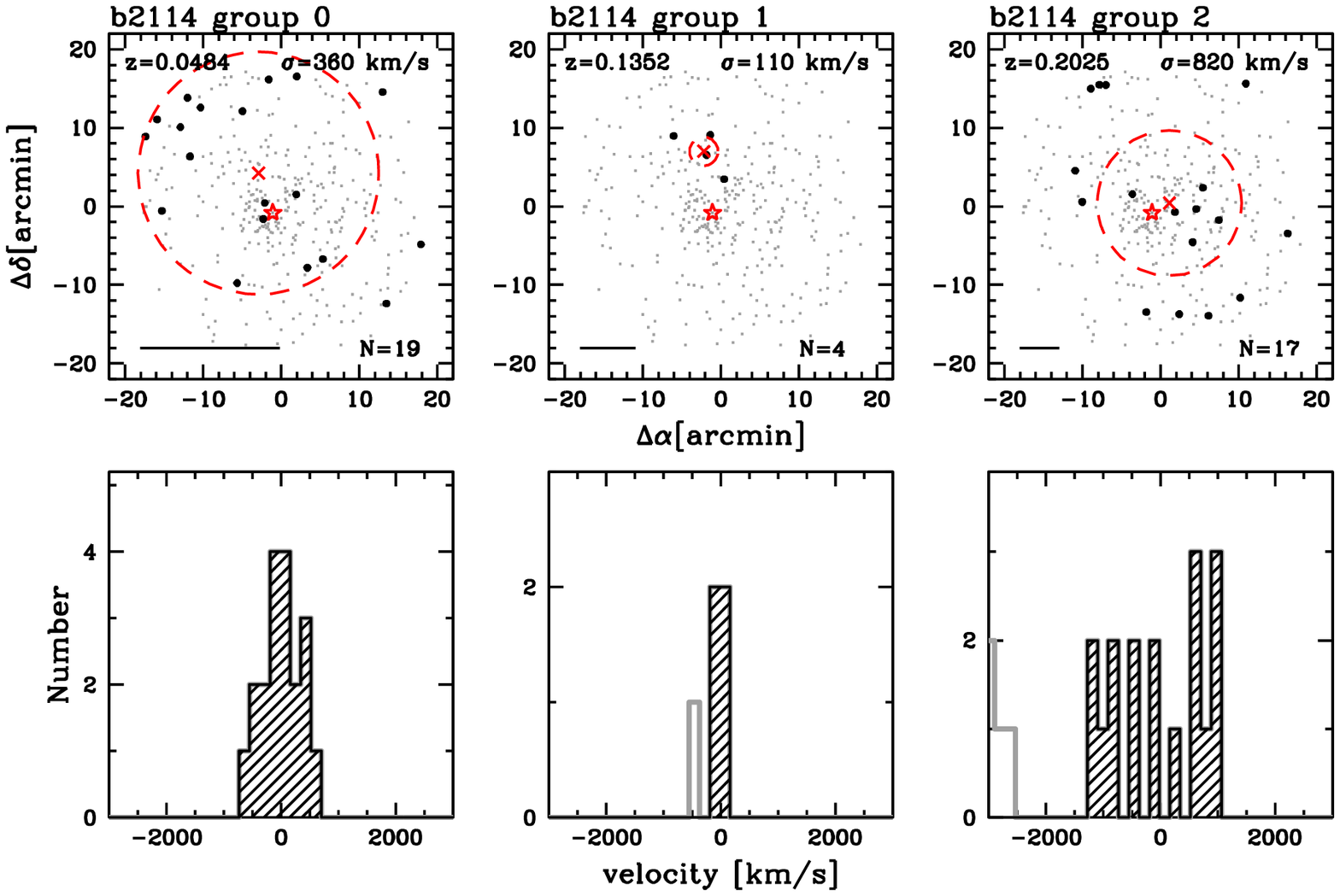}
\includegraphics[clip=true, width=18cm]{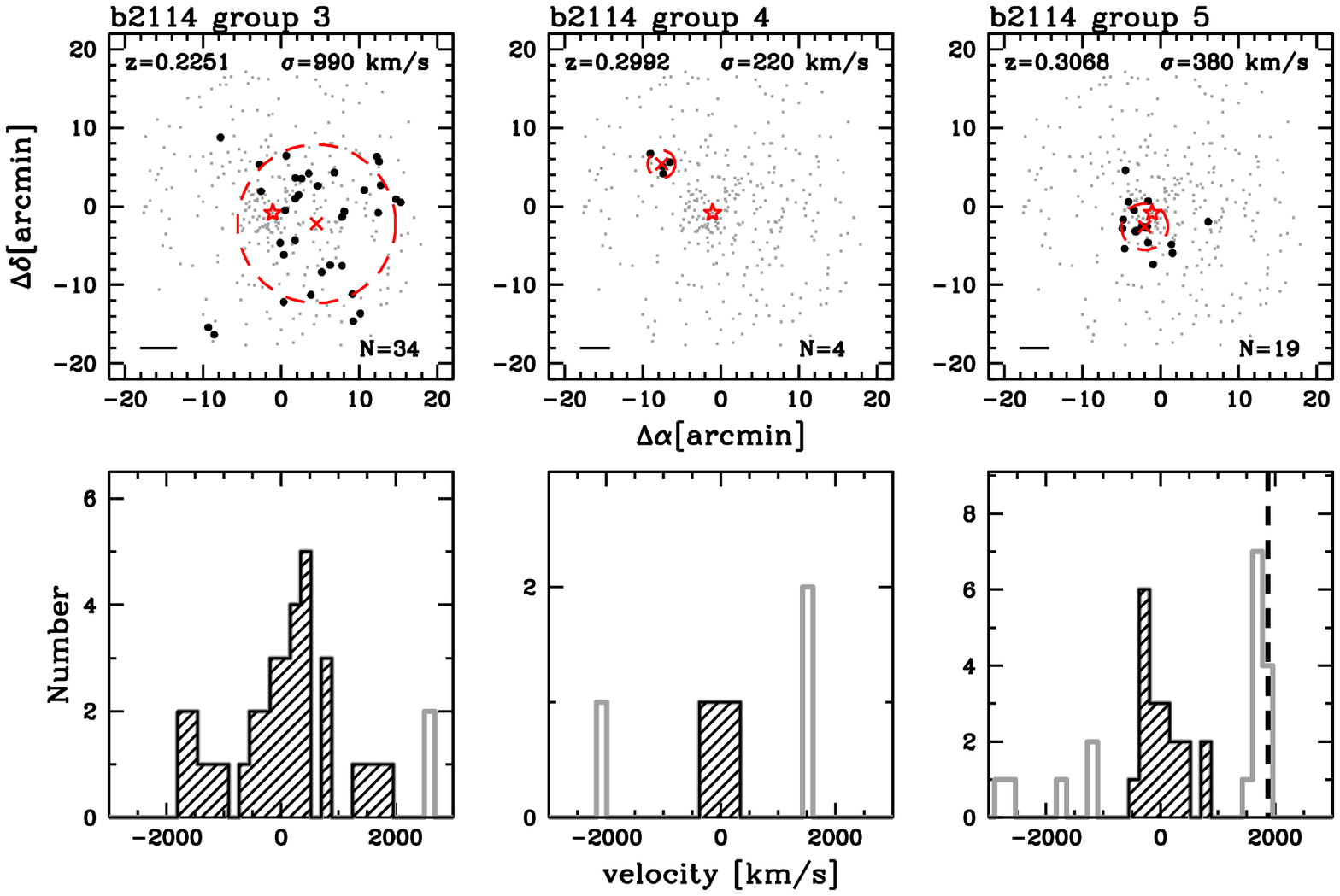}
\caption{Continued.}
\end{figure*}
\clearpage
\begin{figure*}
\ContinuedFloat
\includegraphics[clip=true, width=18cm]{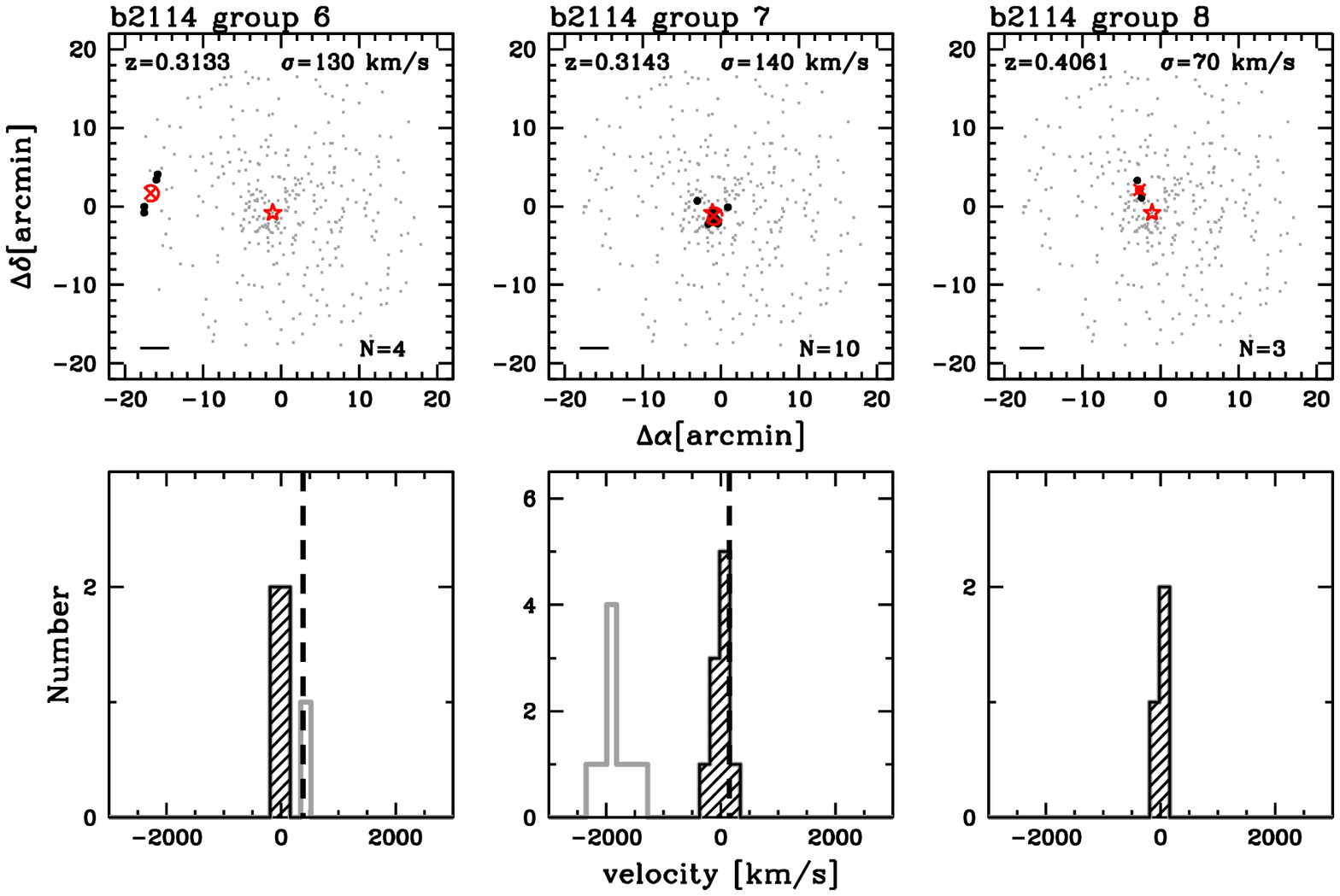}
\includegraphics[clip=true, width=18cm]{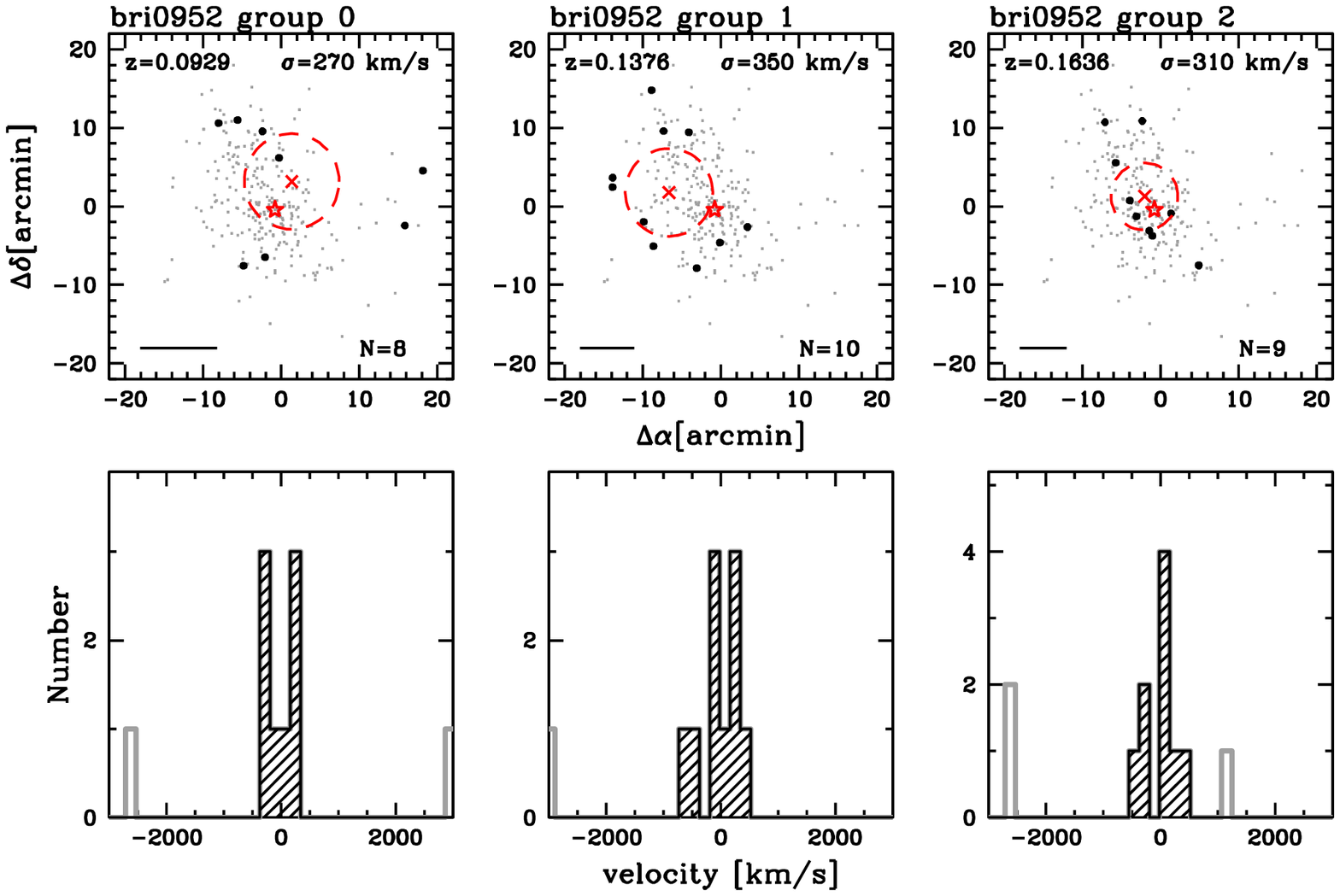}
\caption{Continued.}
\end{figure*}
\clearpage
\begin{figure*}
\ContinuedFloat
\includegraphics[clip=true, width=18cm]{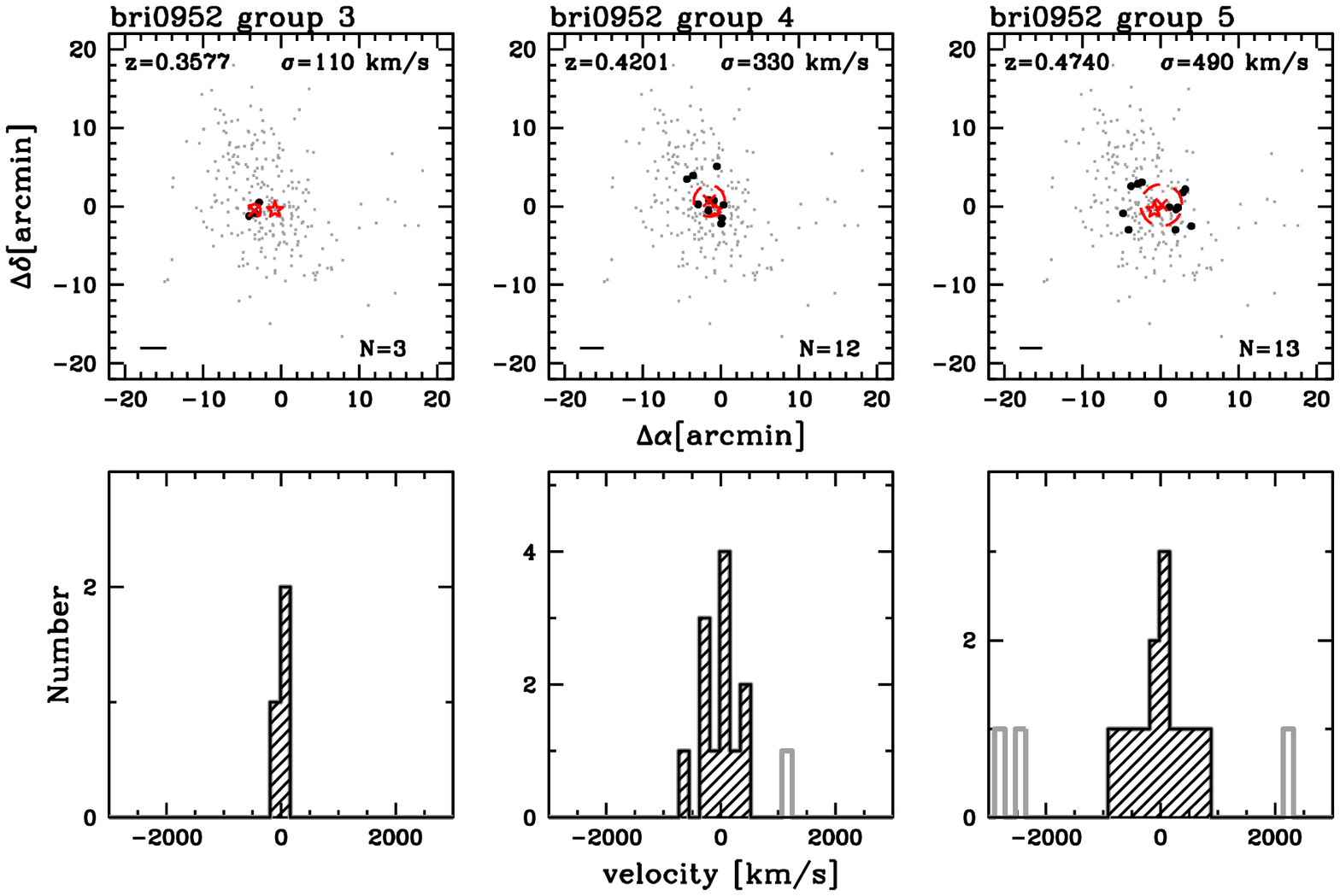}
\includegraphics[clip=true, width=18cm]{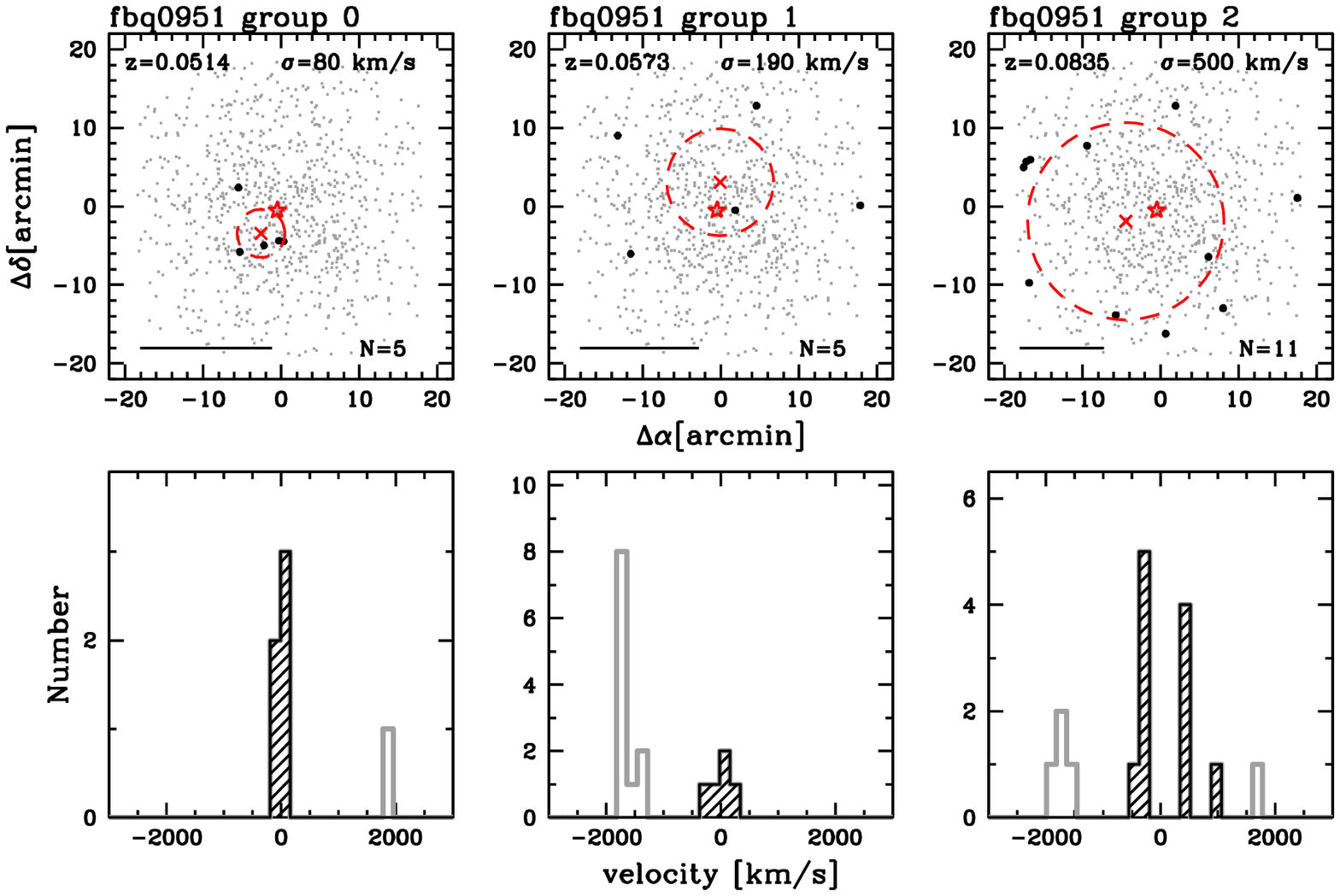}
\caption{Continued.}
\end{figure*}
\clearpage
\begin{figure*}
\ContinuedFloat
\includegraphics[clip=true, width=18cm]{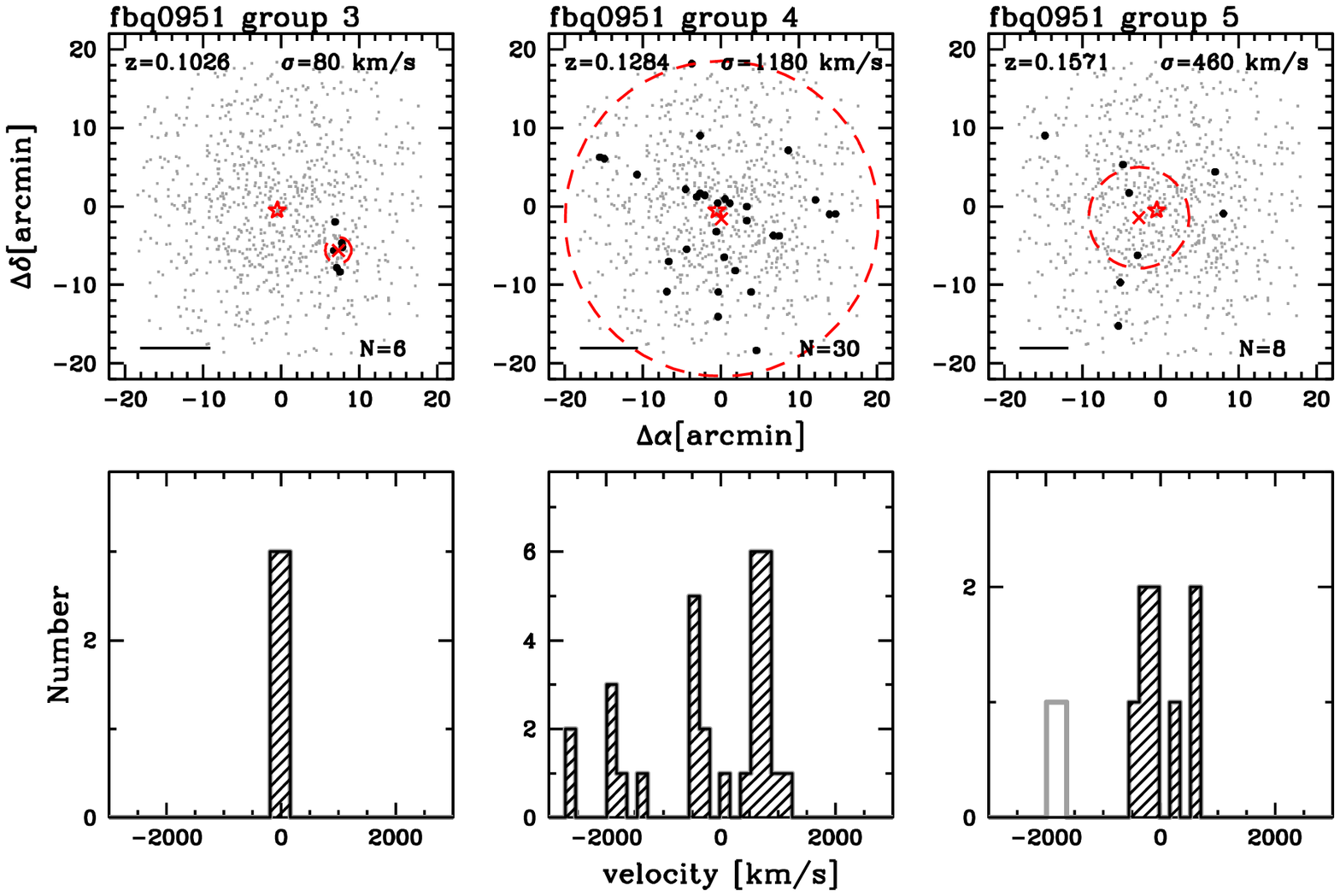}
\includegraphics[clip=true, width=18cm]{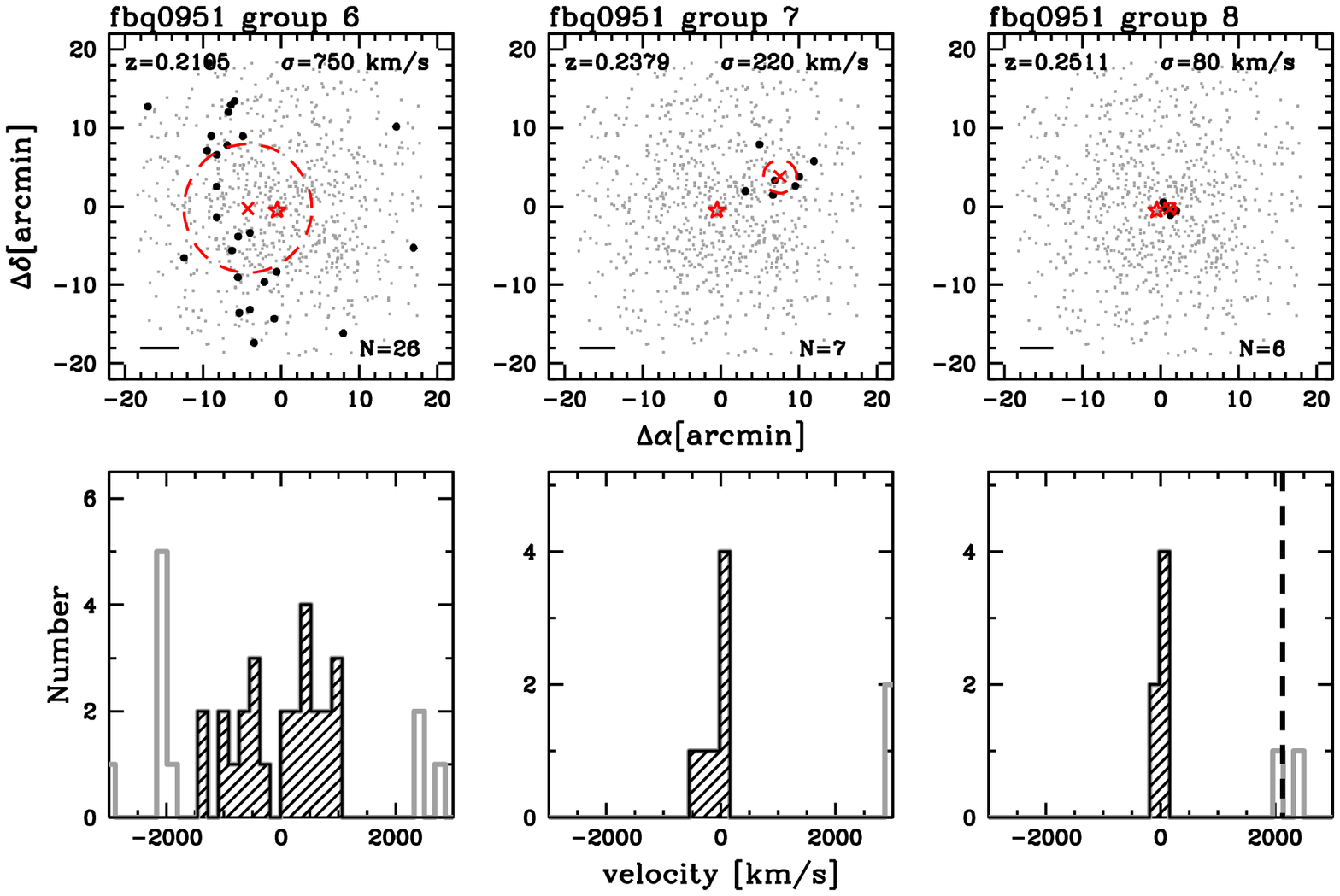}
\caption{Continued.}
\end{figure*}
\clearpage
\begin{figure*}
\ContinuedFloat
\includegraphics[clip=true, width=18cm]{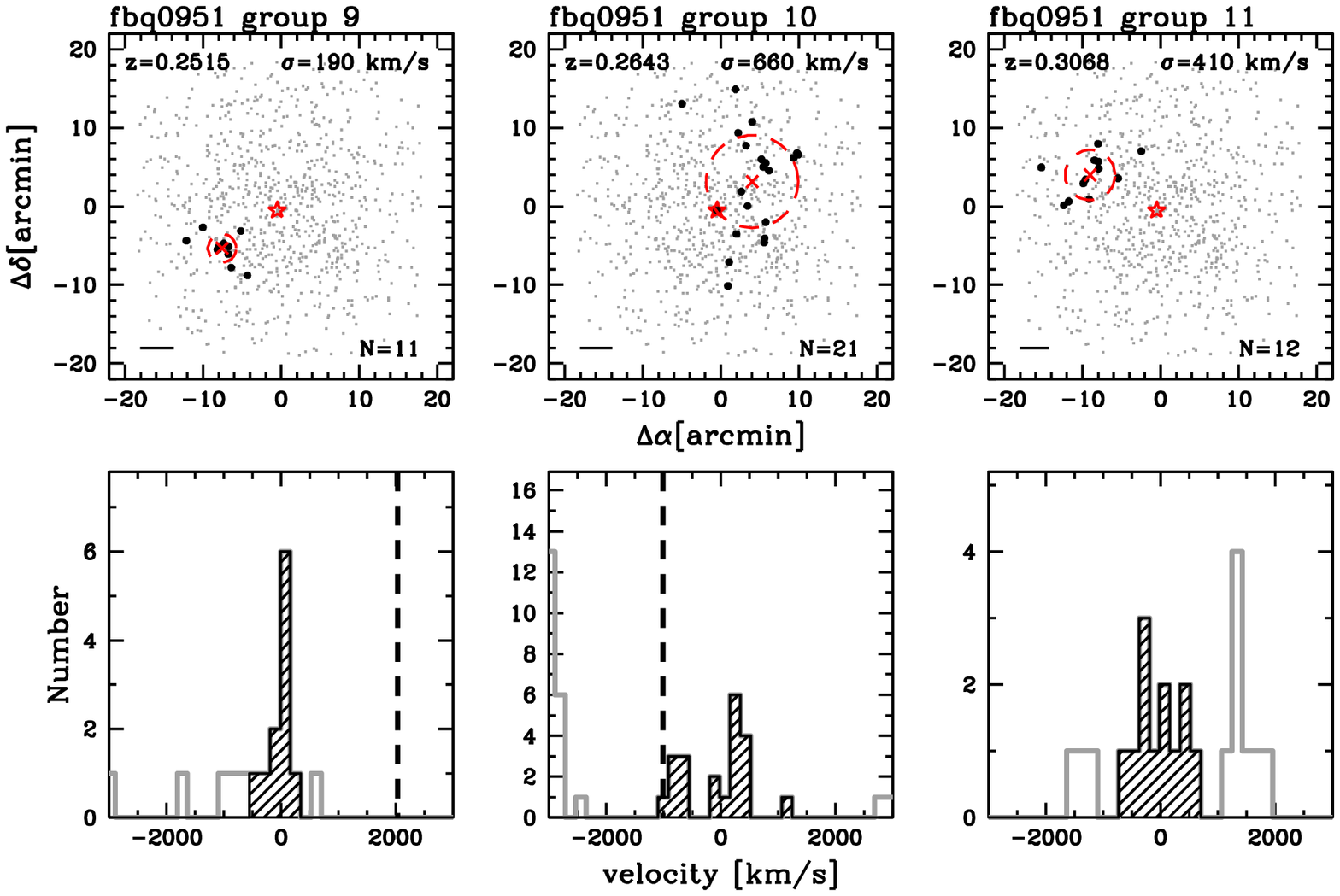}
\includegraphics[clip=true, width=18cm]{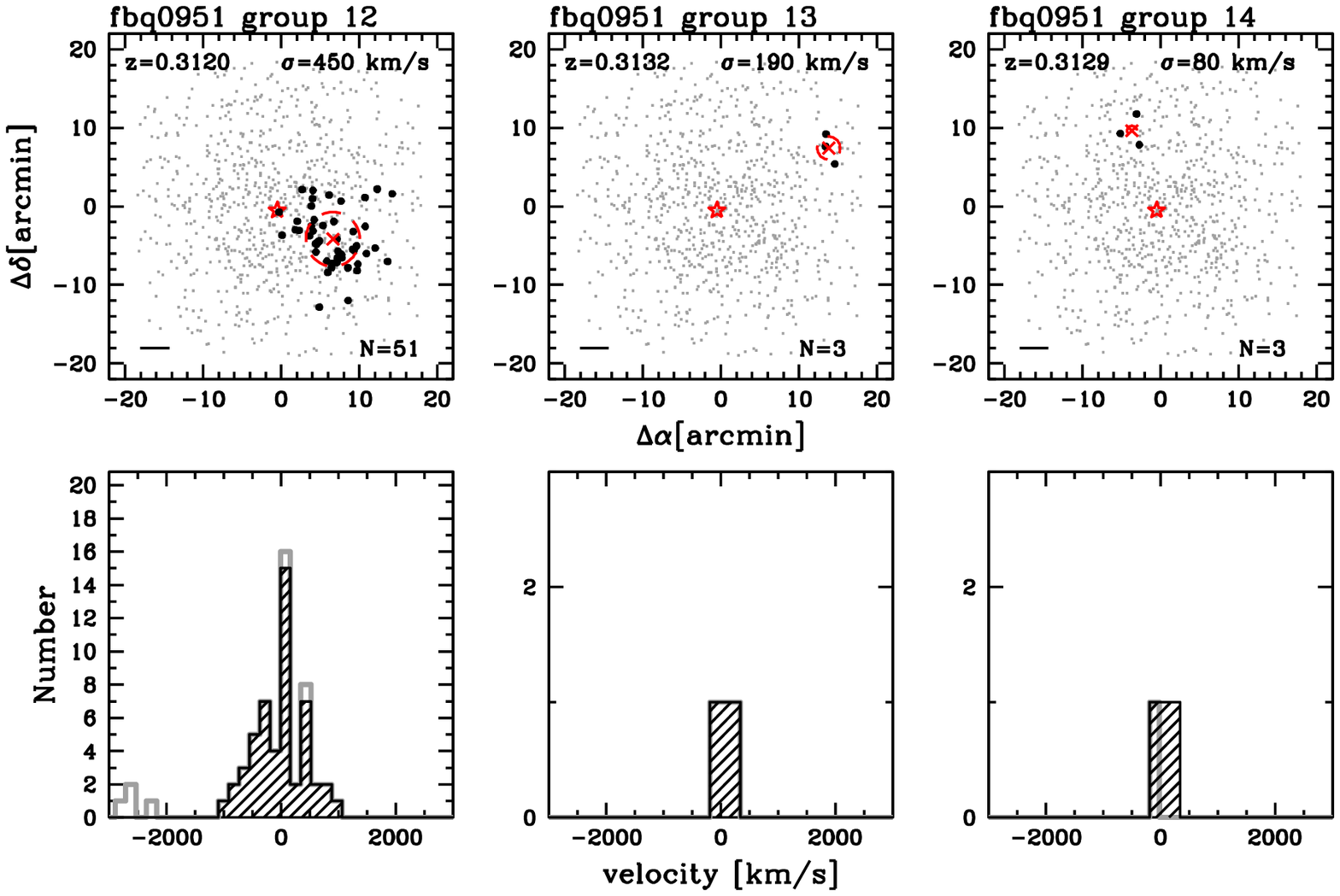}
\caption{Continued.}
\end{figure*}
\clearpage
\begin{figure*}
\ContinuedFloat
\includegraphics[clip=true, width=18cm]{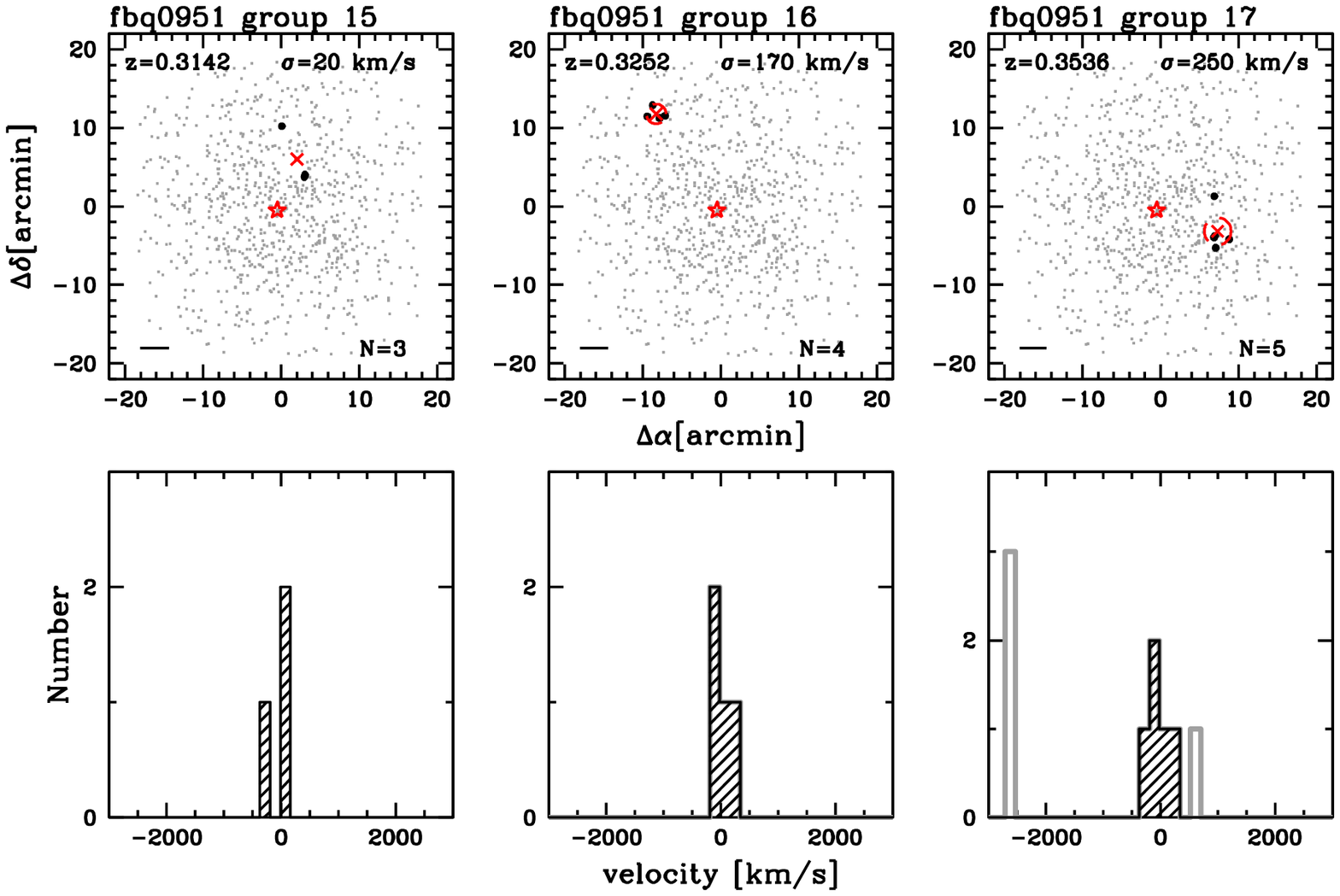}
\includegraphics[clip=true, width=18cm]{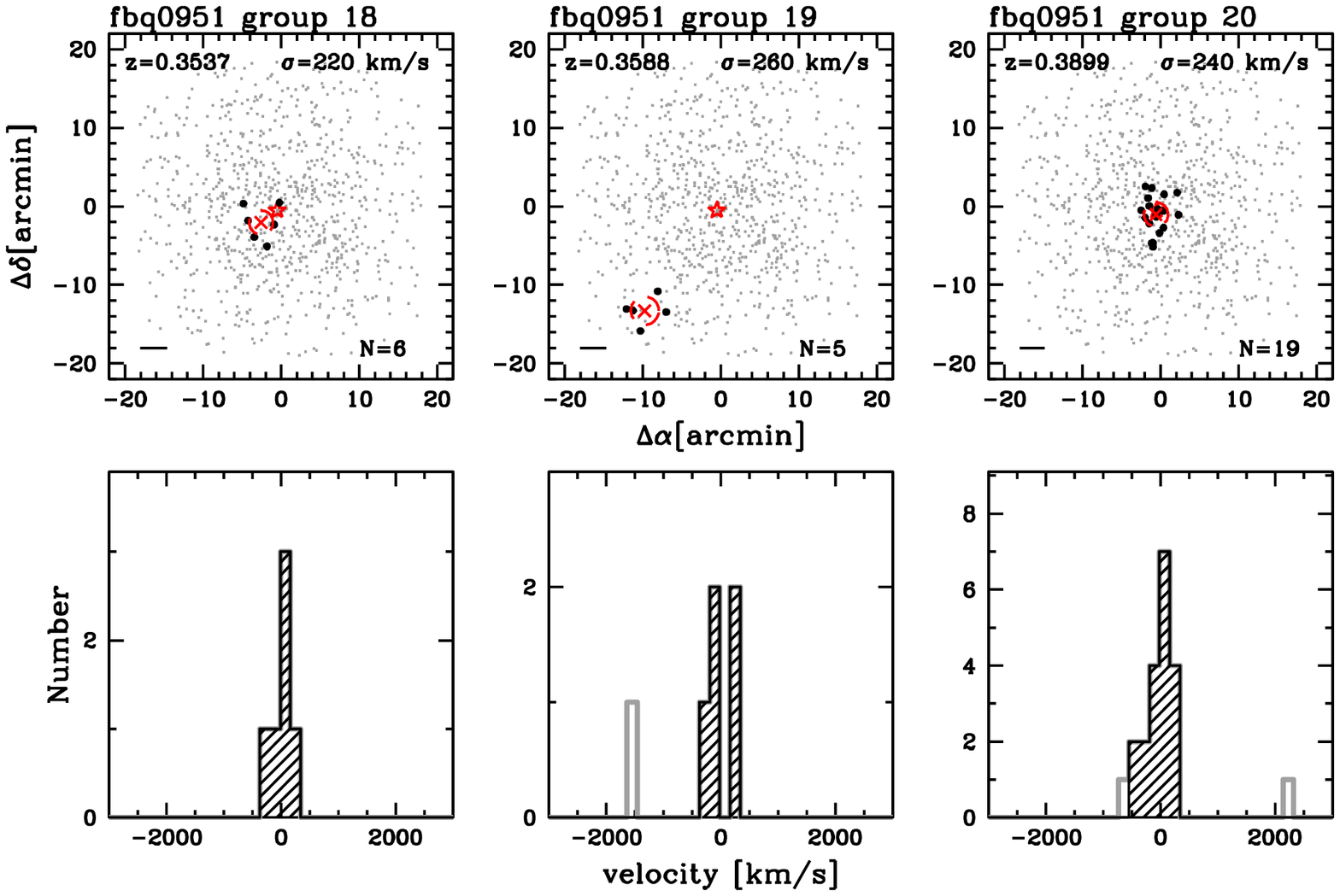}
\caption{Continued.}
\end{figure*}
\clearpage
\begin{figure*}
\ContinuedFloat
\includegraphics[clip=true, width=18cm]{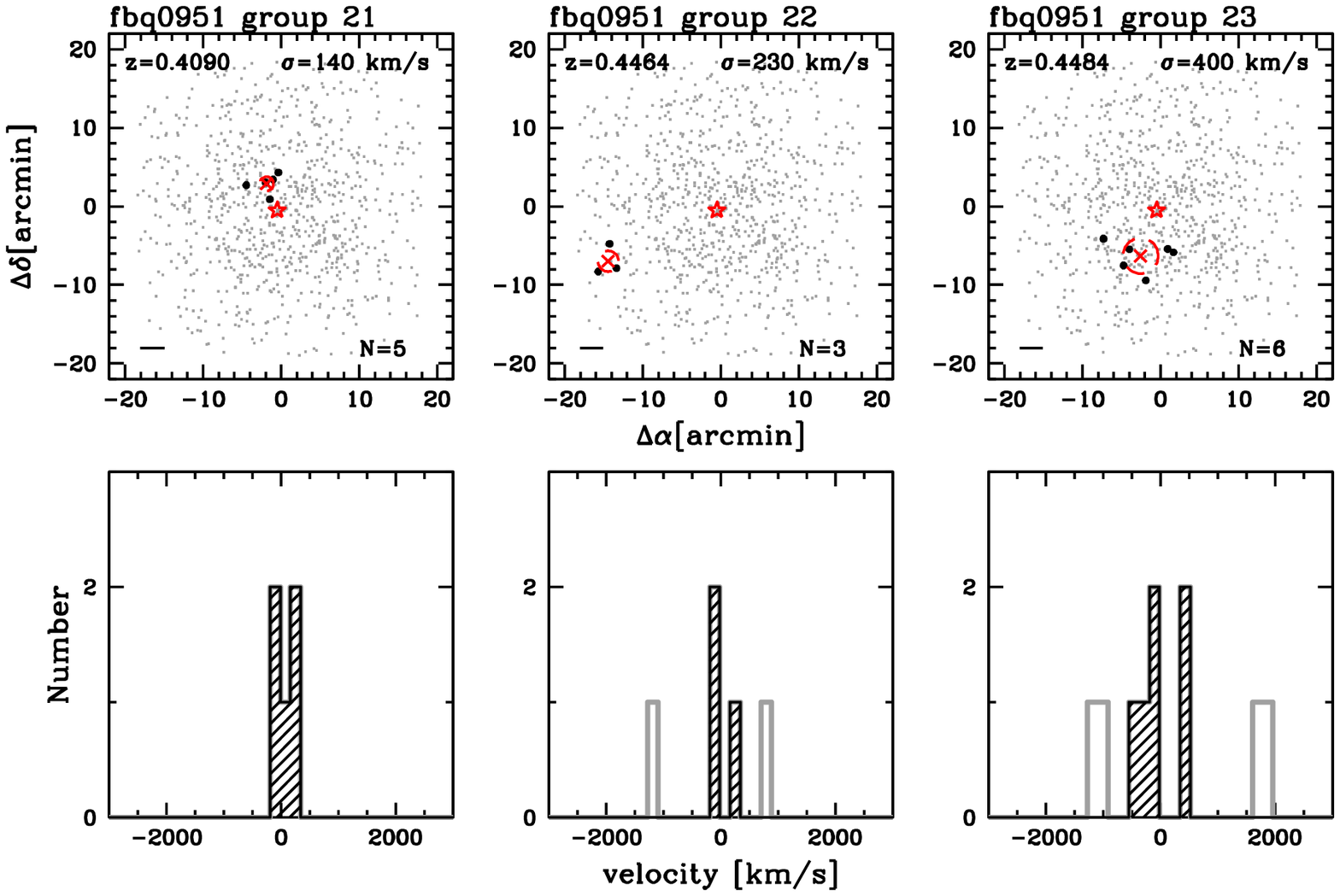}
\includegraphics[clip=true, width=18cm]{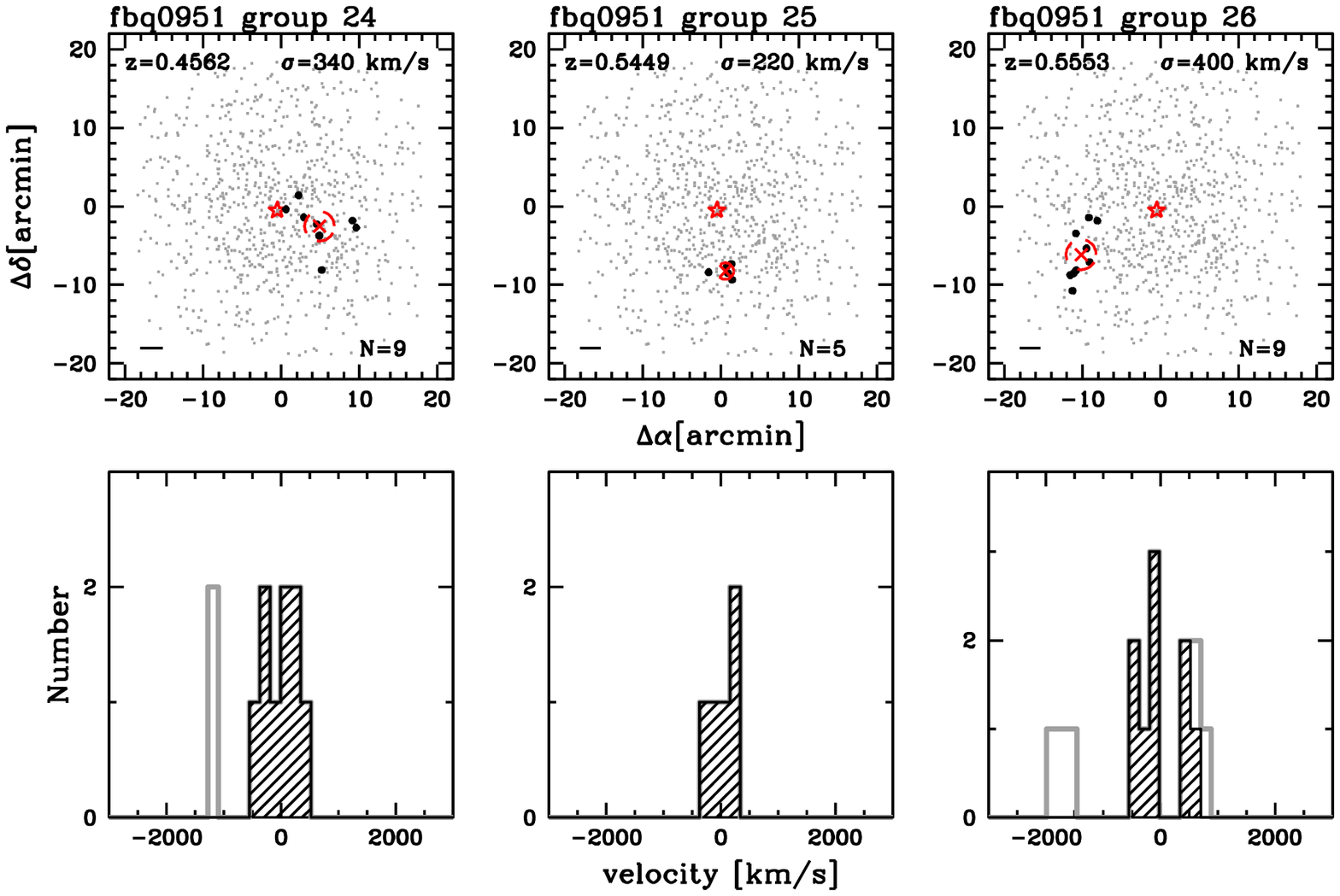}
\caption{Continued.}
\end{figure*}
\clearpage
\begin{figure*}
\ContinuedFloat
\includegraphics[clip=true, width=18cm]{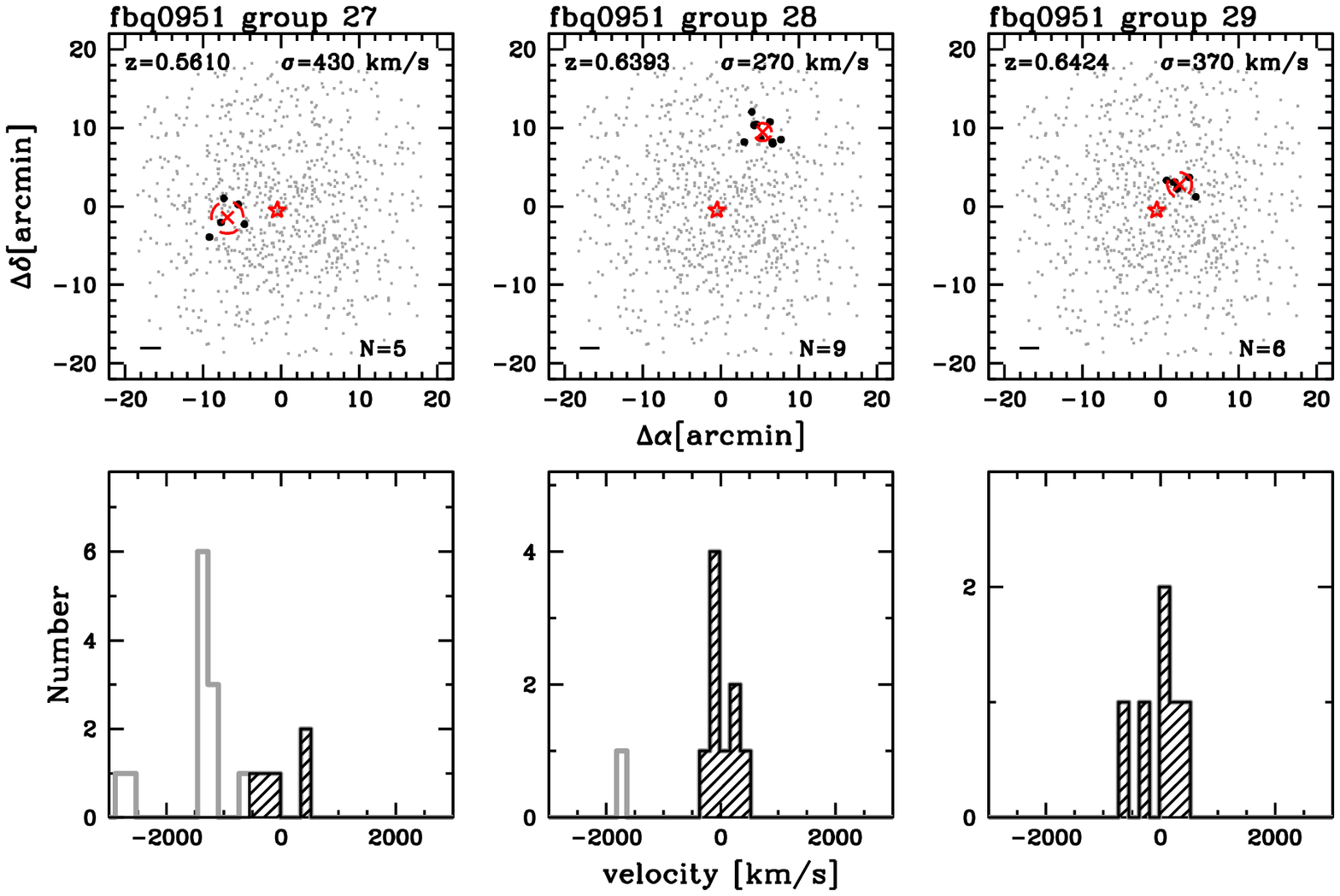}
\includegraphics[clip=true, width=18cm]{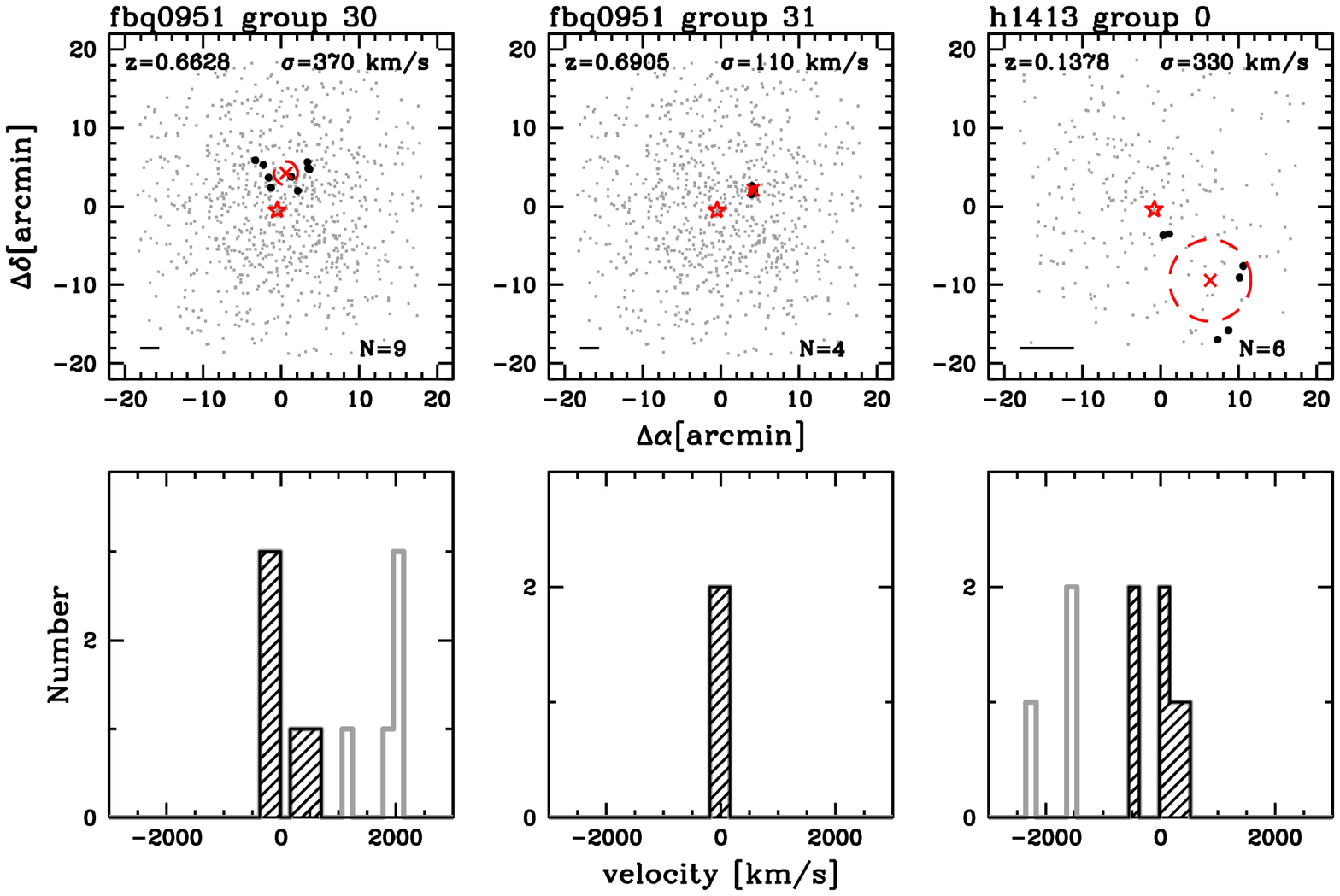}
\caption{Continued.}
\end{figure*}
\clearpage
\begin{figure*}
\ContinuedFloat
\includegraphics[clip=true, width=18cm]{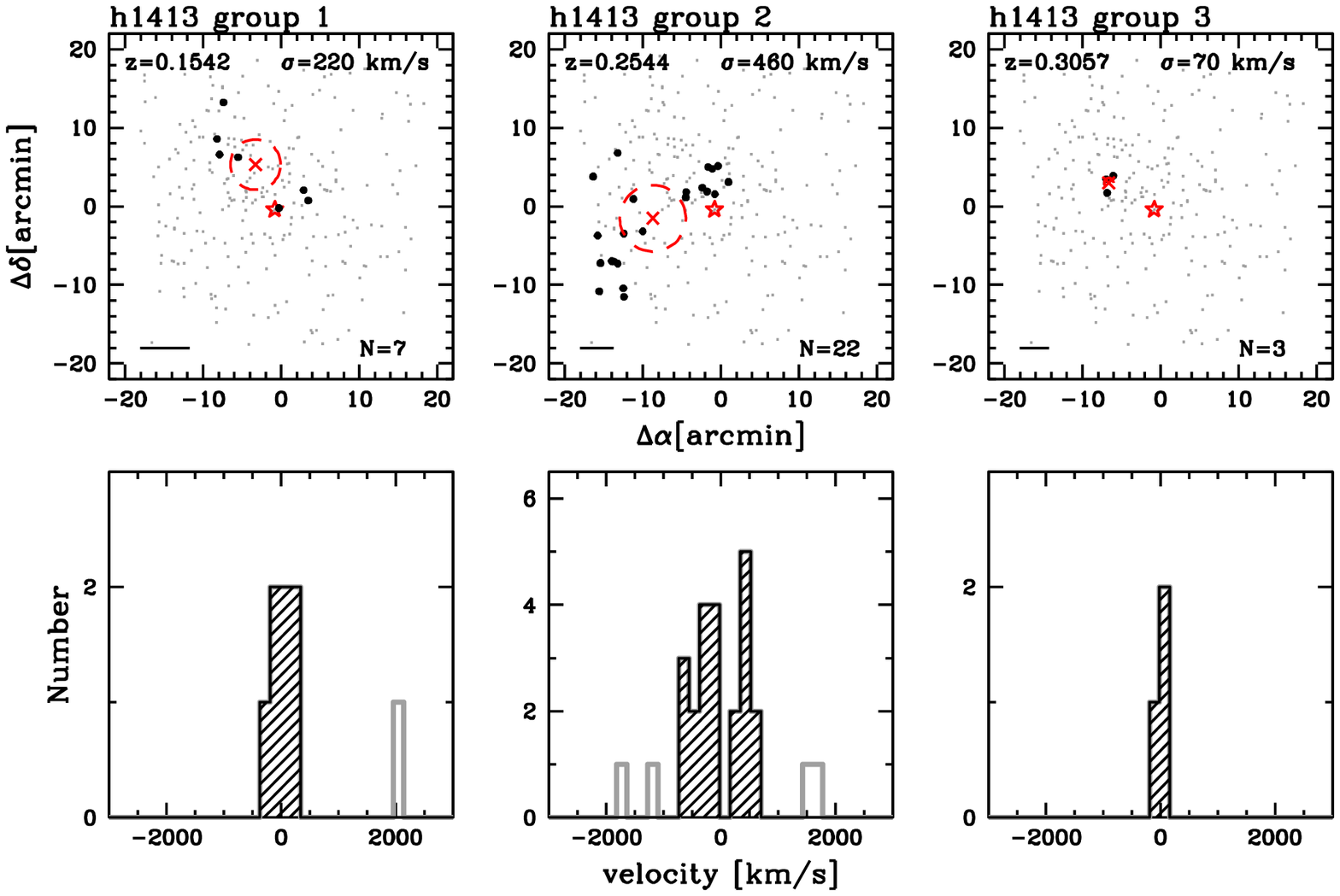}
\includegraphics[clip=true, width=18cm]{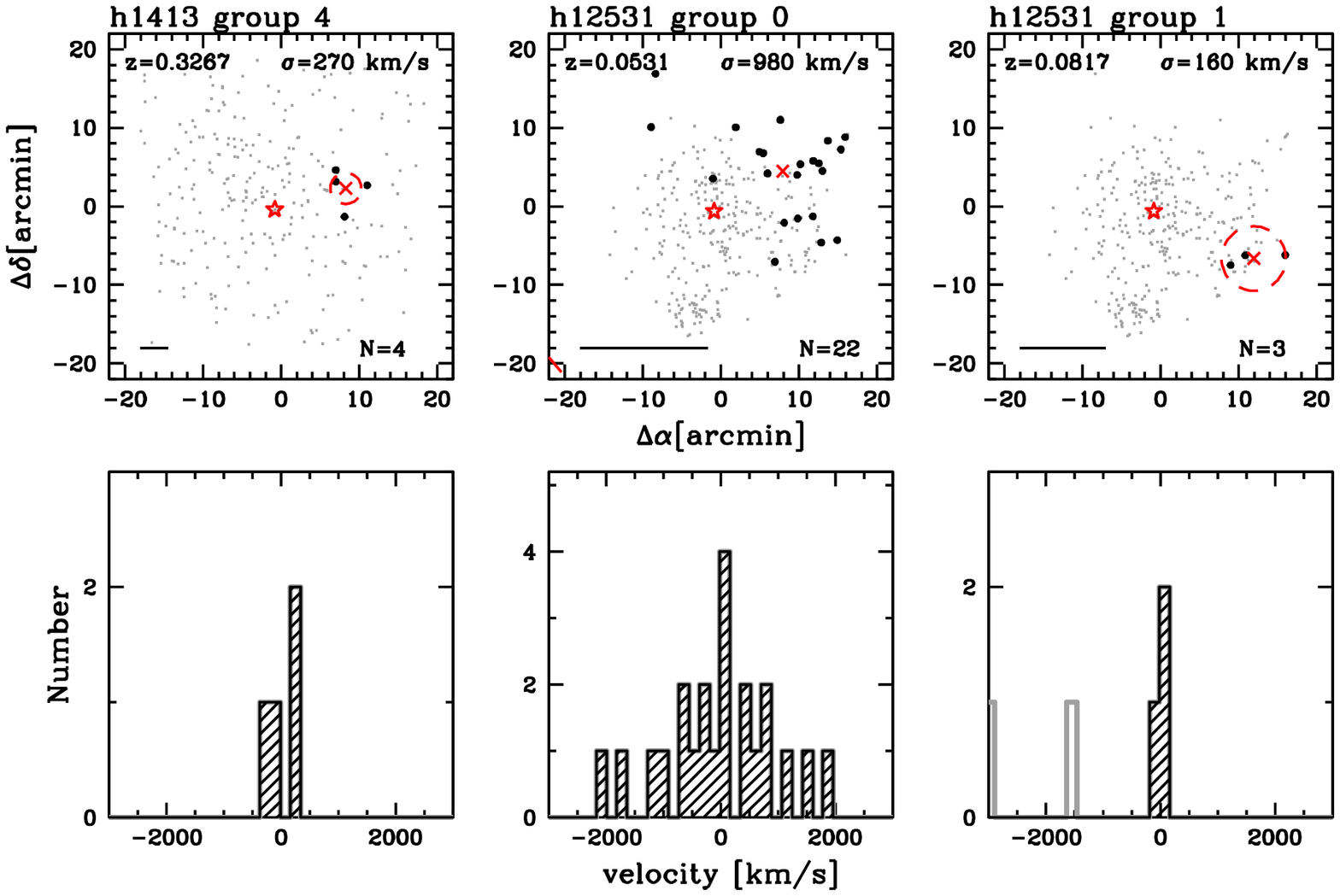}
\caption{Continued.}
\end{figure*}
\clearpage
\begin{figure*}
\ContinuedFloat
\includegraphics[clip=true, width=18cm]{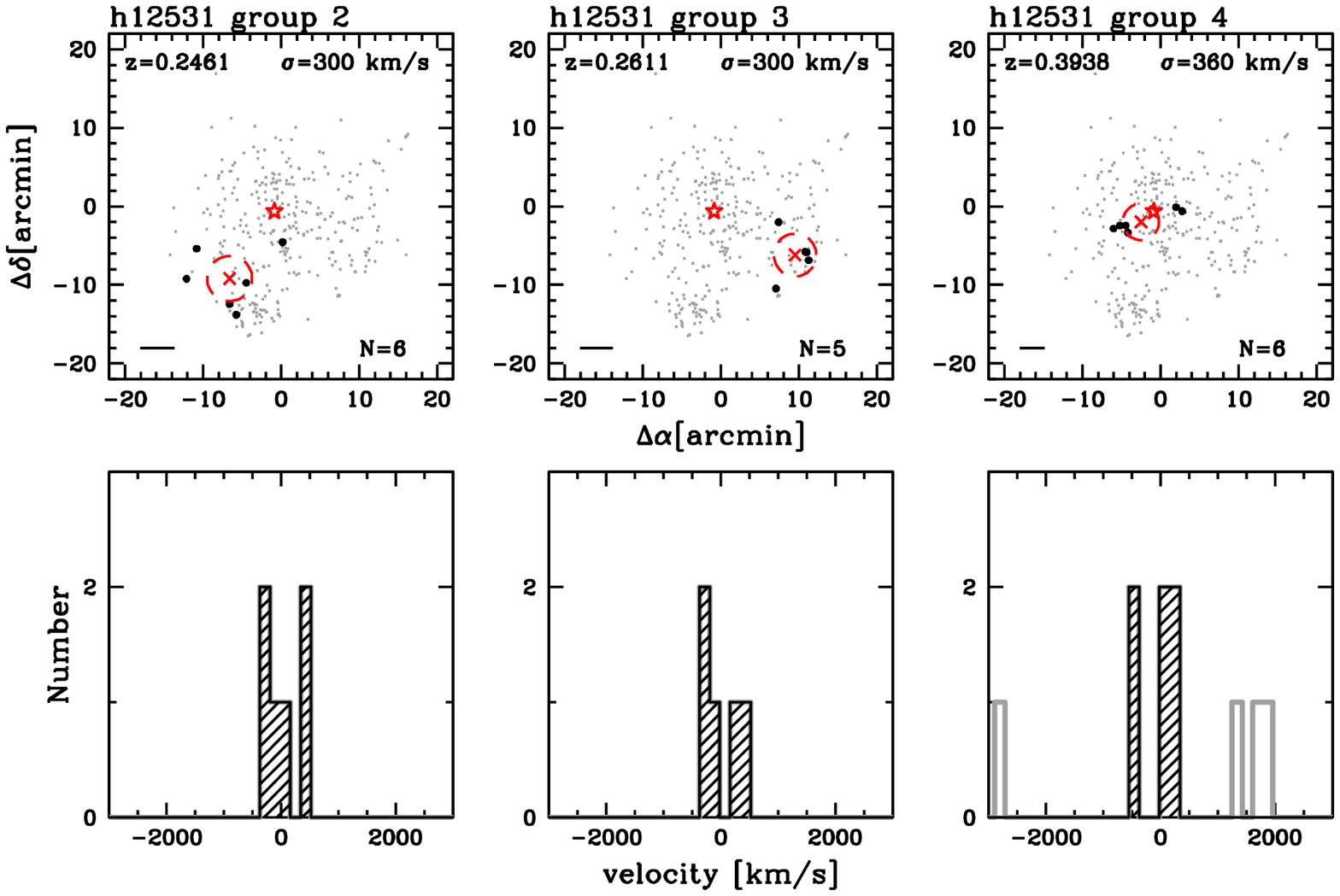}
\includegraphics[clip=true, width=18cm]{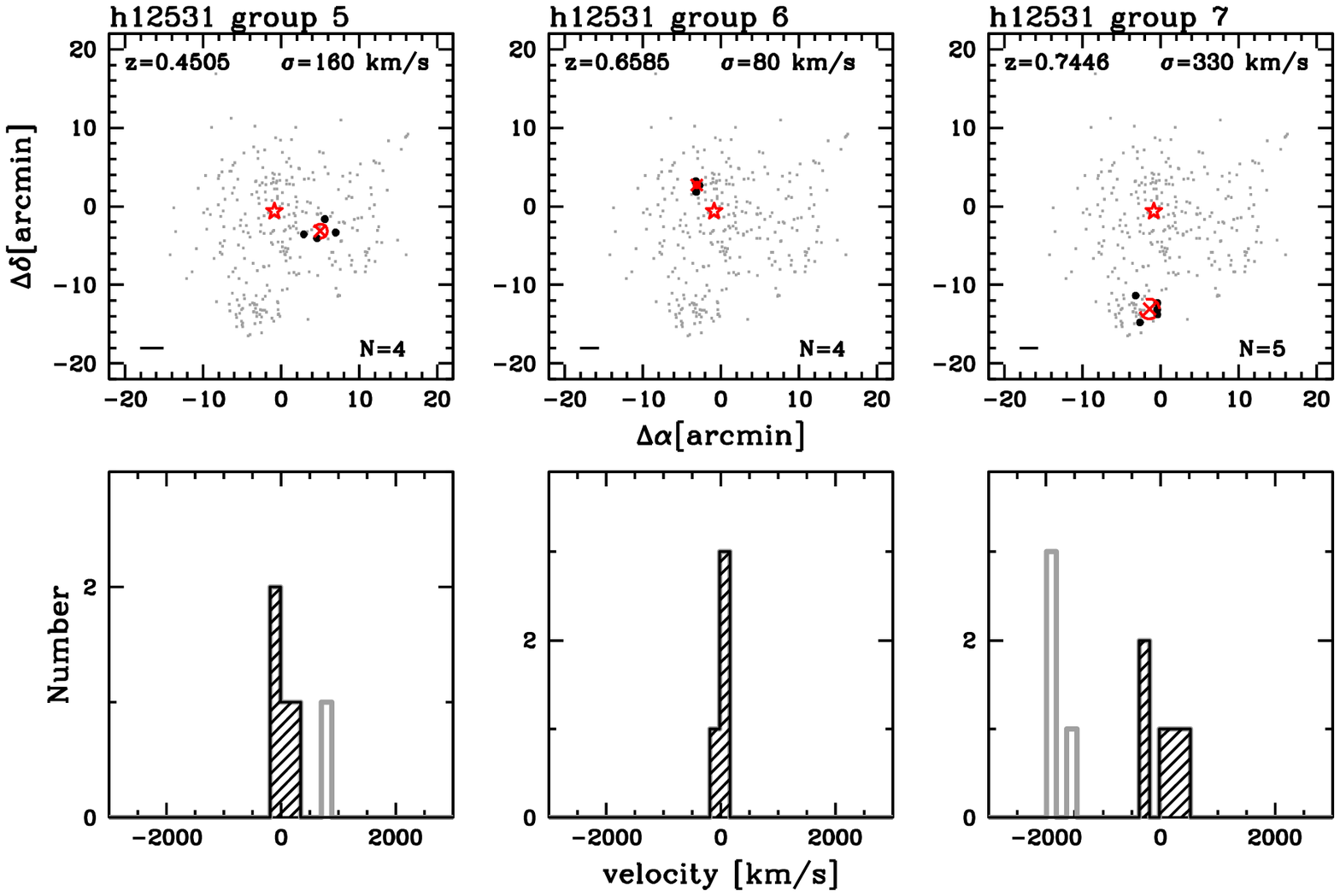}
\caption{Continued.}
\end{figure*}
\clearpage
\begin{figure*}
\ContinuedFloat
\includegraphics[clip=true, width=18cm]{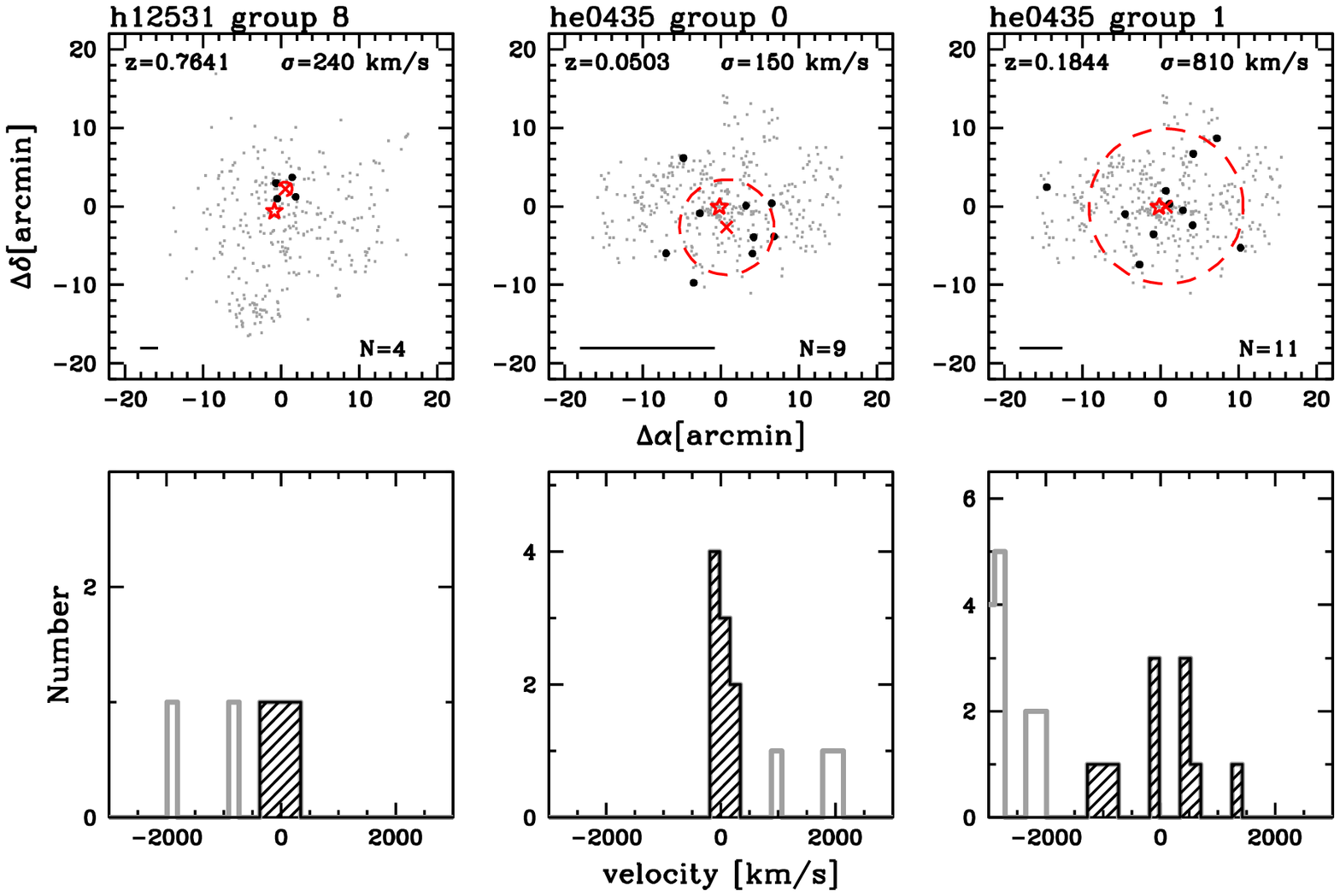}
\includegraphics[clip=true, width=18cm]{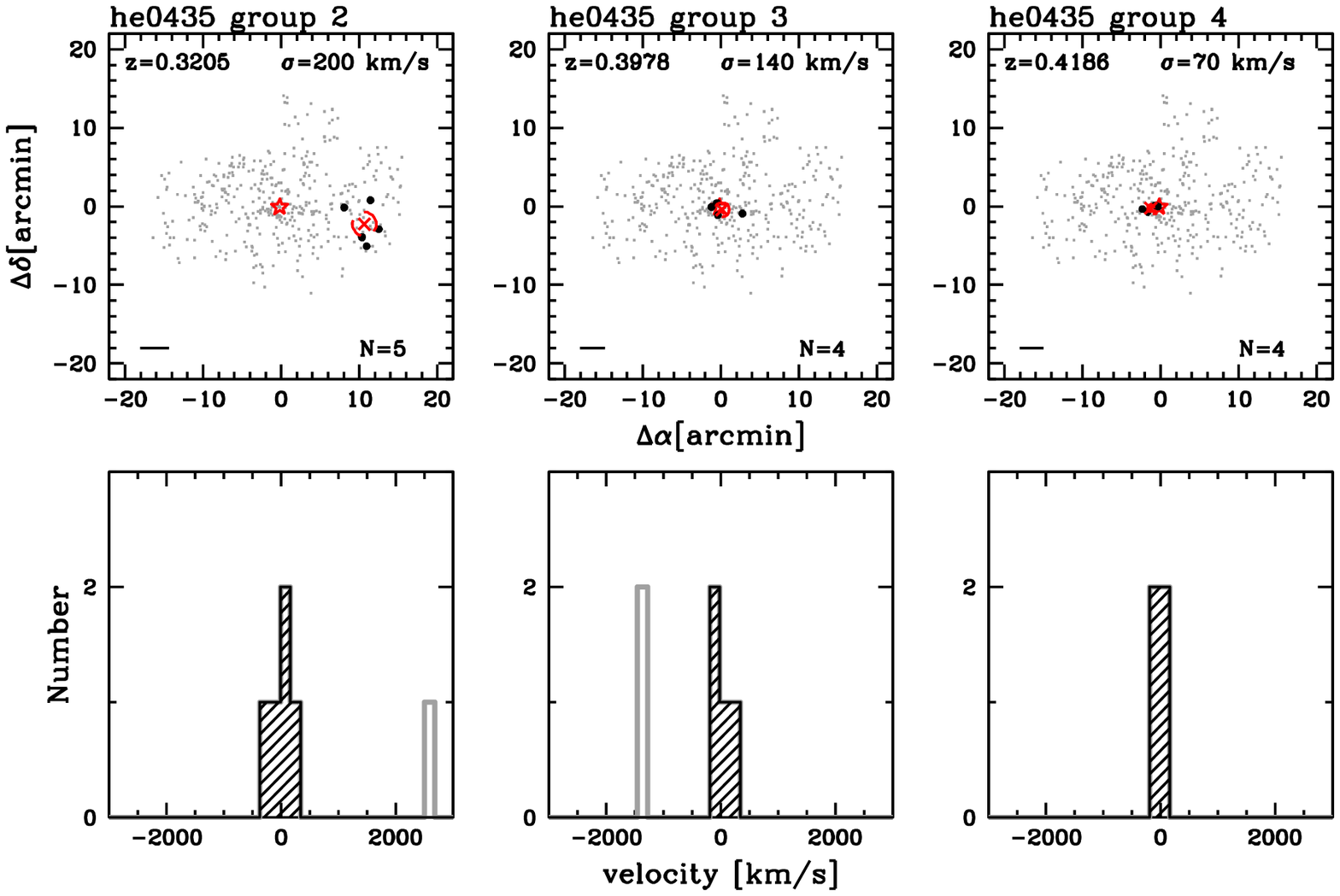}
\caption{Continued.}
\end{figure*}
\clearpage
\begin{figure*}
\ContinuedFloat
\includegraphics[clip=true, width=18cm]{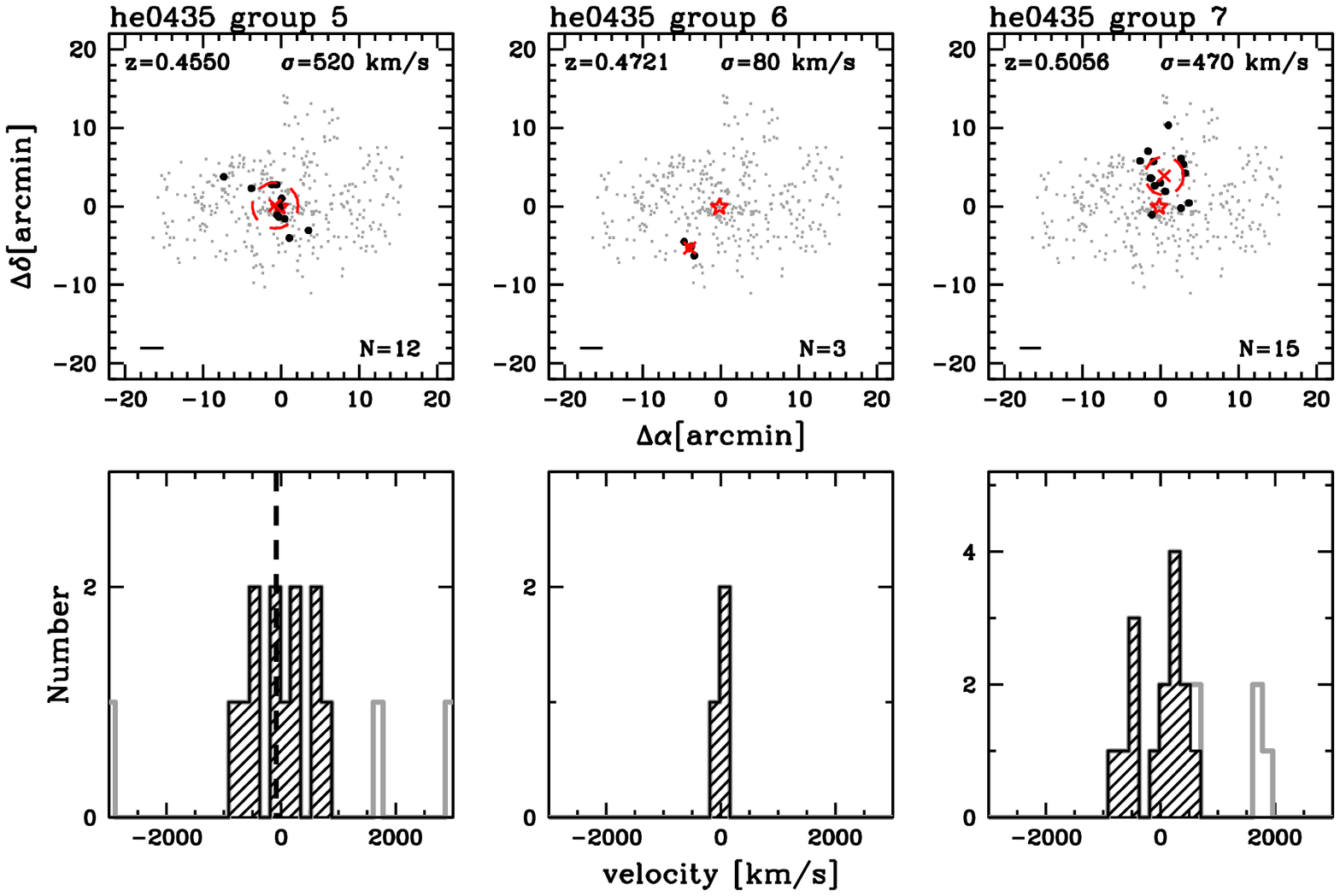}
\includegraphics[clip=true, width=18cm]{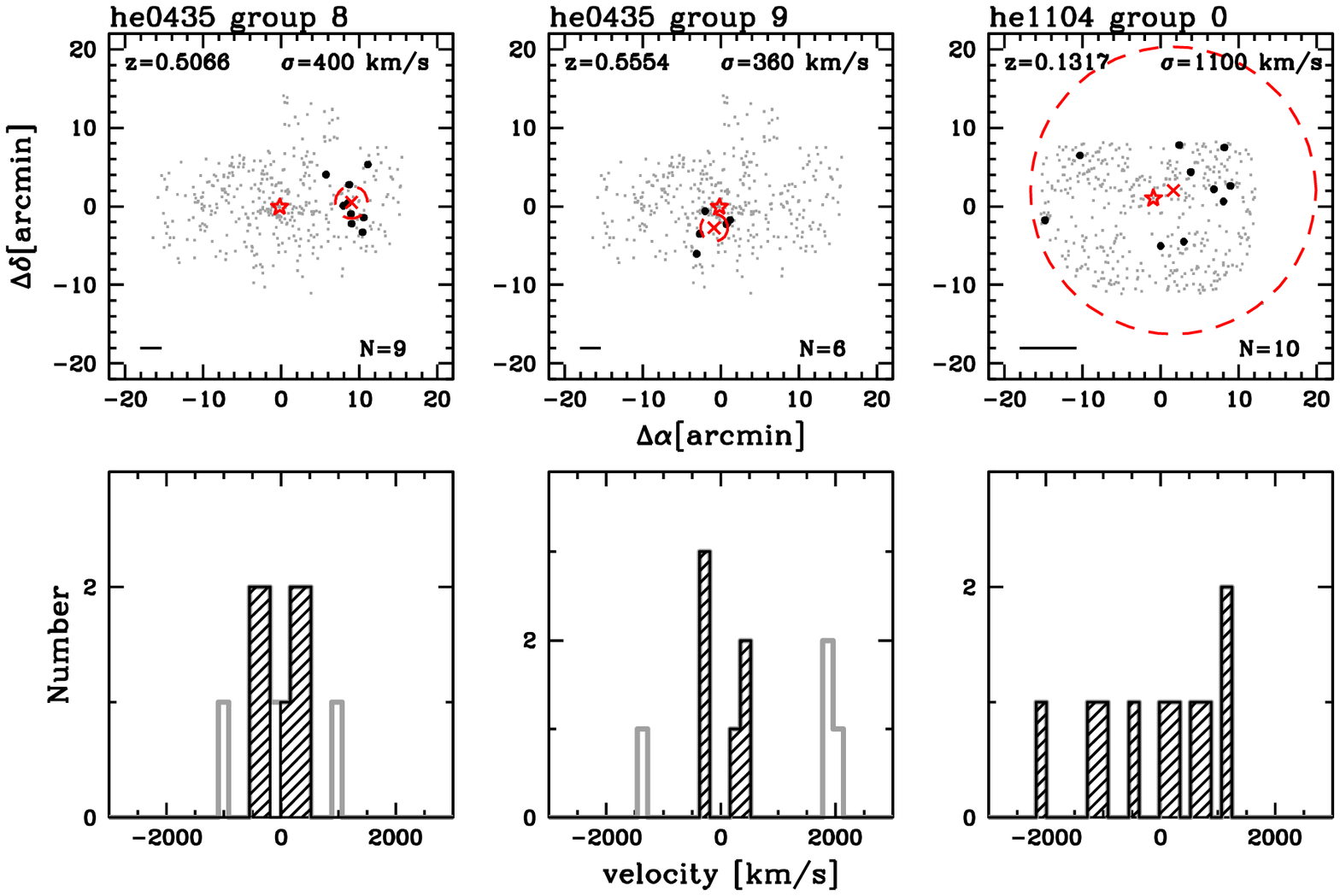}
\caption{Continued.}
\end{figure*}
\clearpage
\begin{figure*}
\ContinuedFloat
\includegraphics[clip=true, width=18cm]{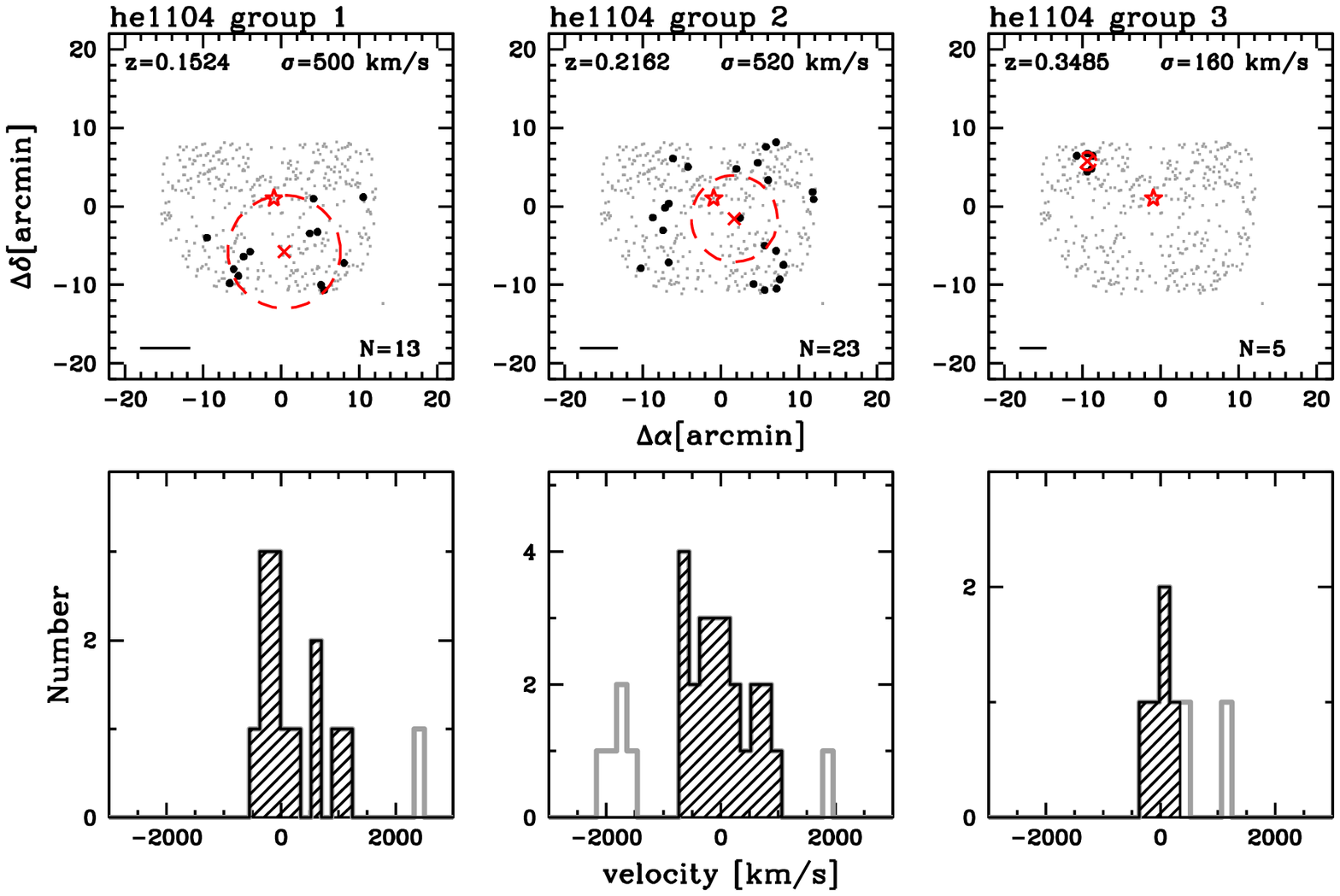}
\includegraphics[clip=true, width=18cm]{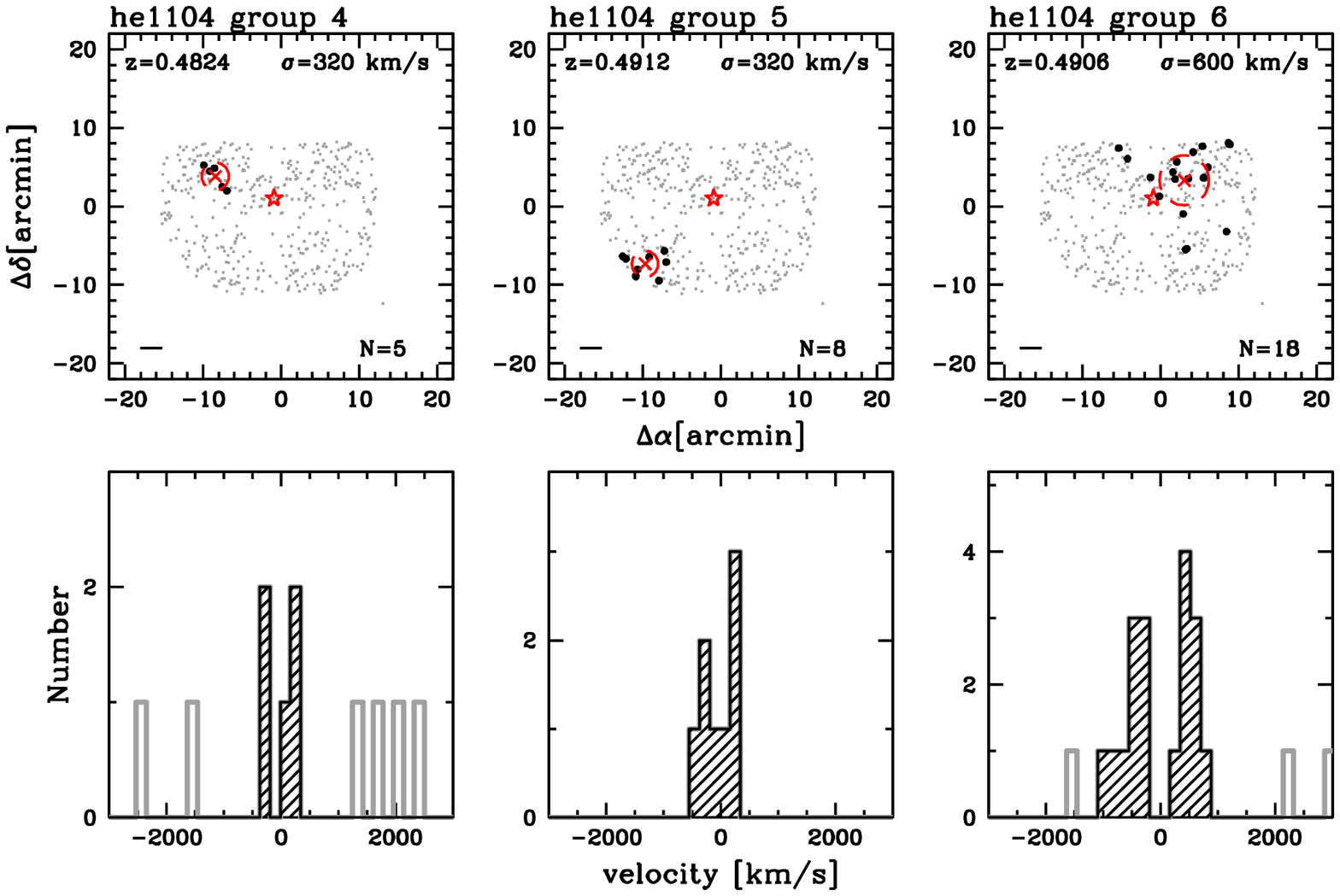}
\caption{Continued.}
\end{figure*}
\clearpage
\begin{figure*}
\ContinuedFloat
\includegraphics[clip=true, width=18cm]{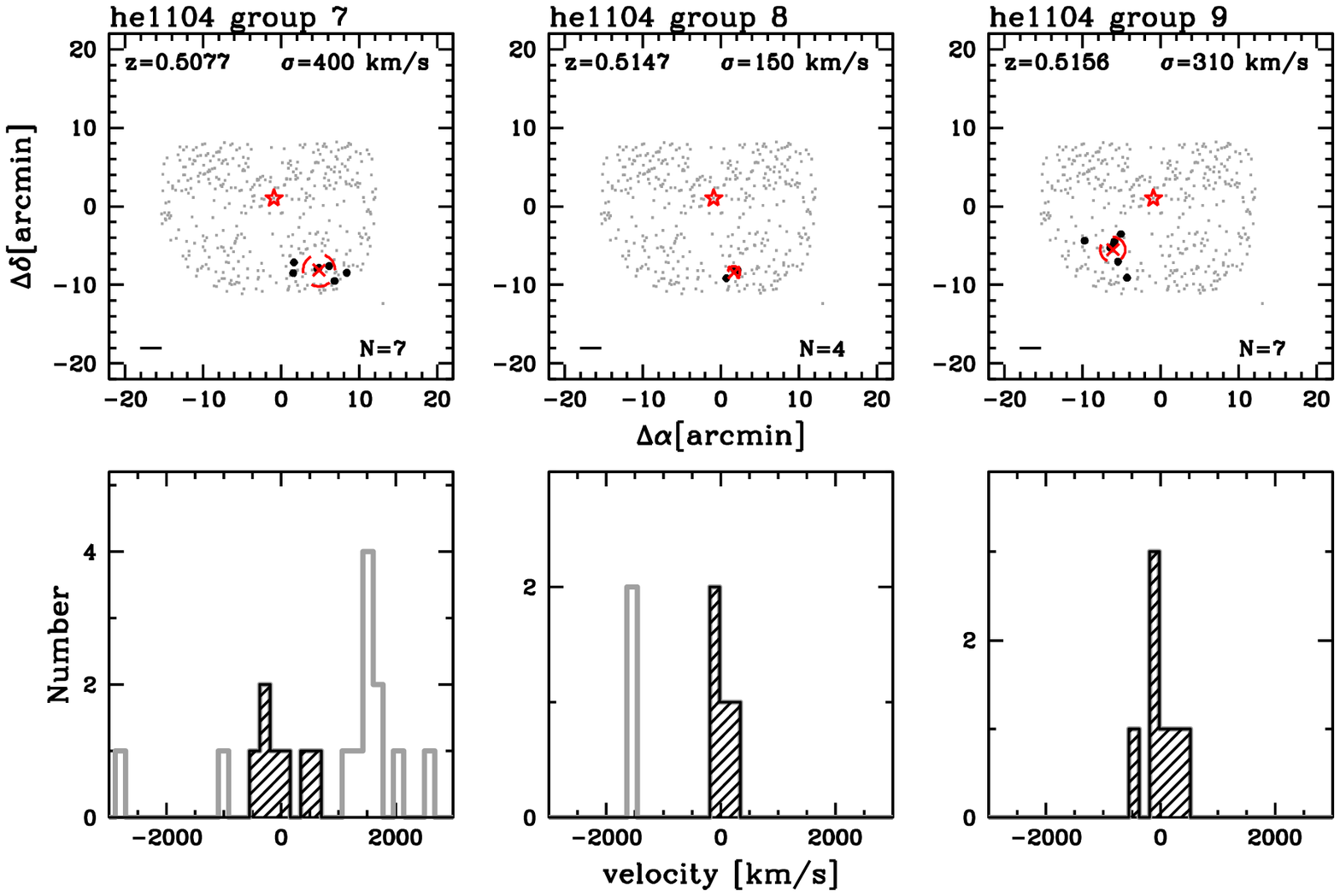}
\includegraphics[clip=true, width=18cm]{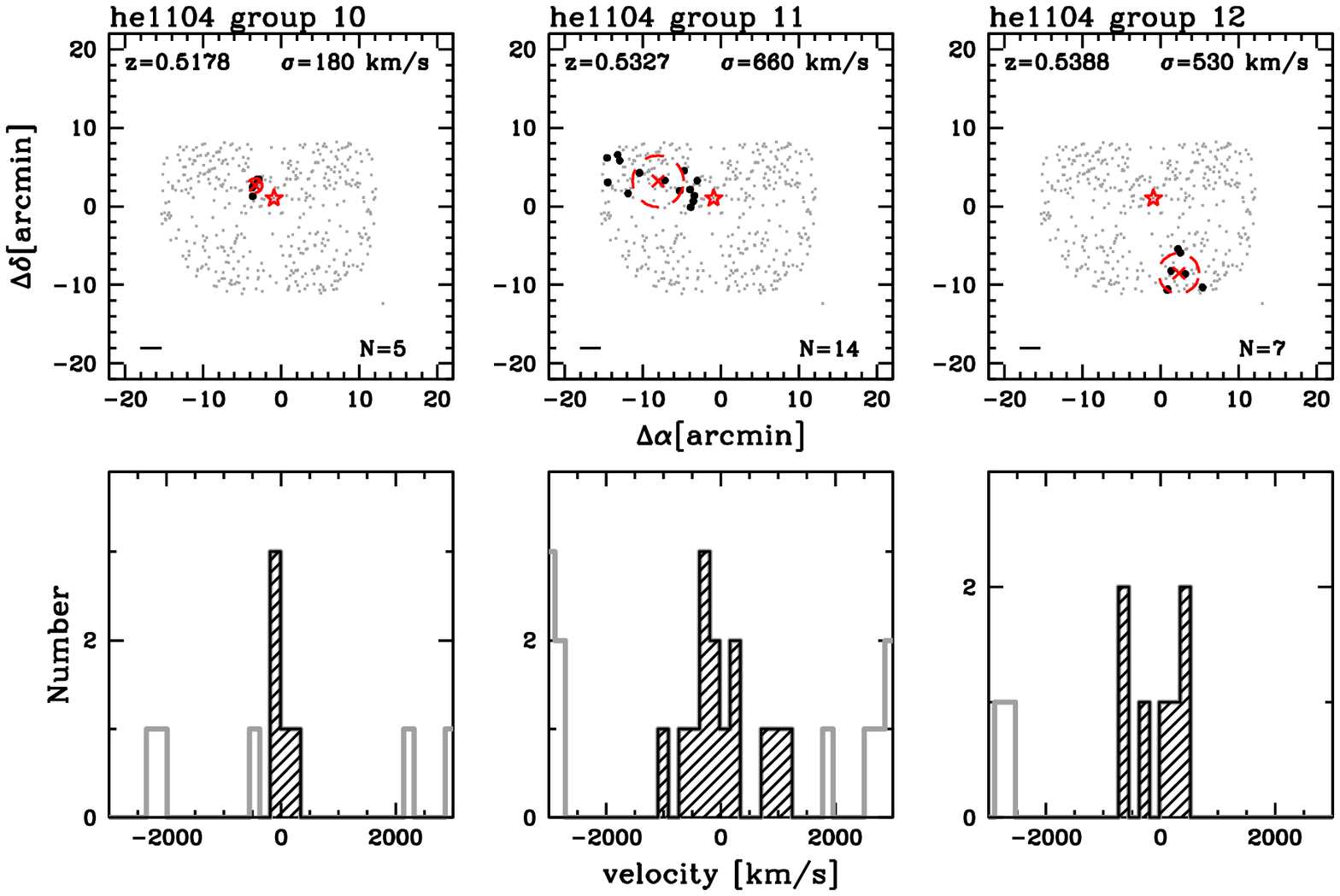}
\caption{Continued.}
\end{figure*}
\clearpage
\begin{figure*}
\ContinuedFloat
\includegraphics[clip=true, width=18cm]{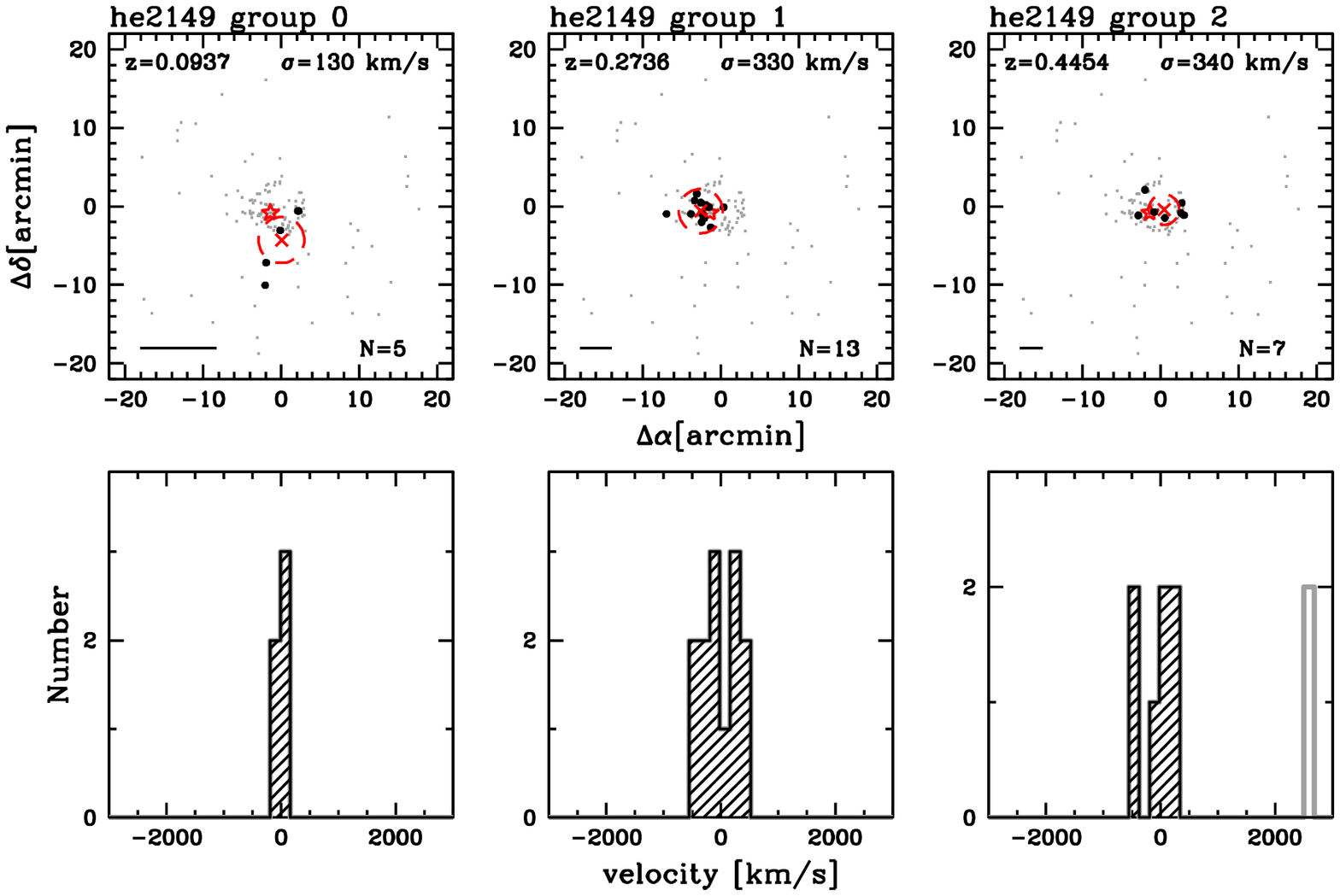}
\includegraphics[clip=true, width=18cm]{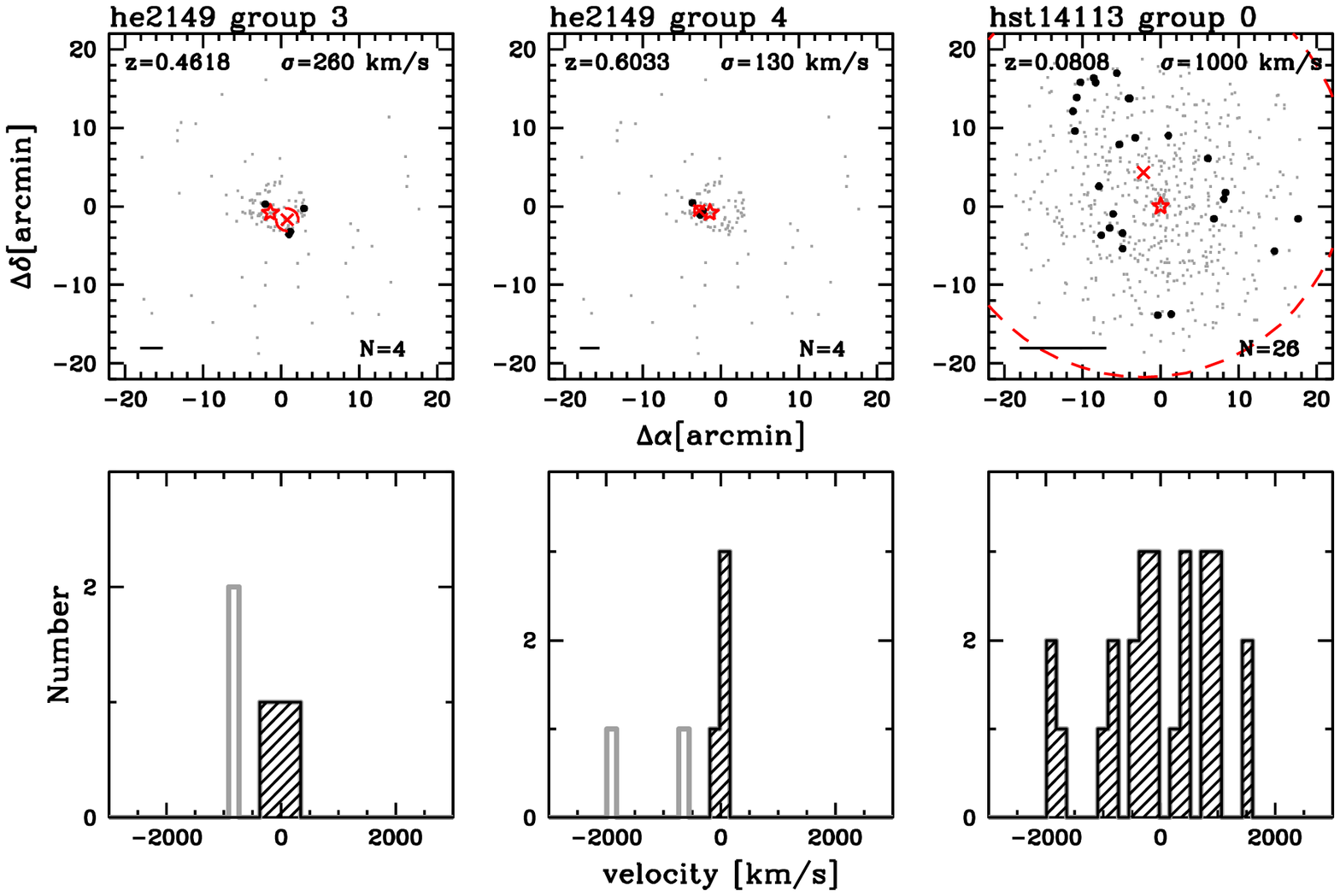}
\caption{Continued.}
\end{figure*}
\clearpage
\begin{figure*}
\ContinuedFloat
\includegraphics[clip=true, width=18cm]{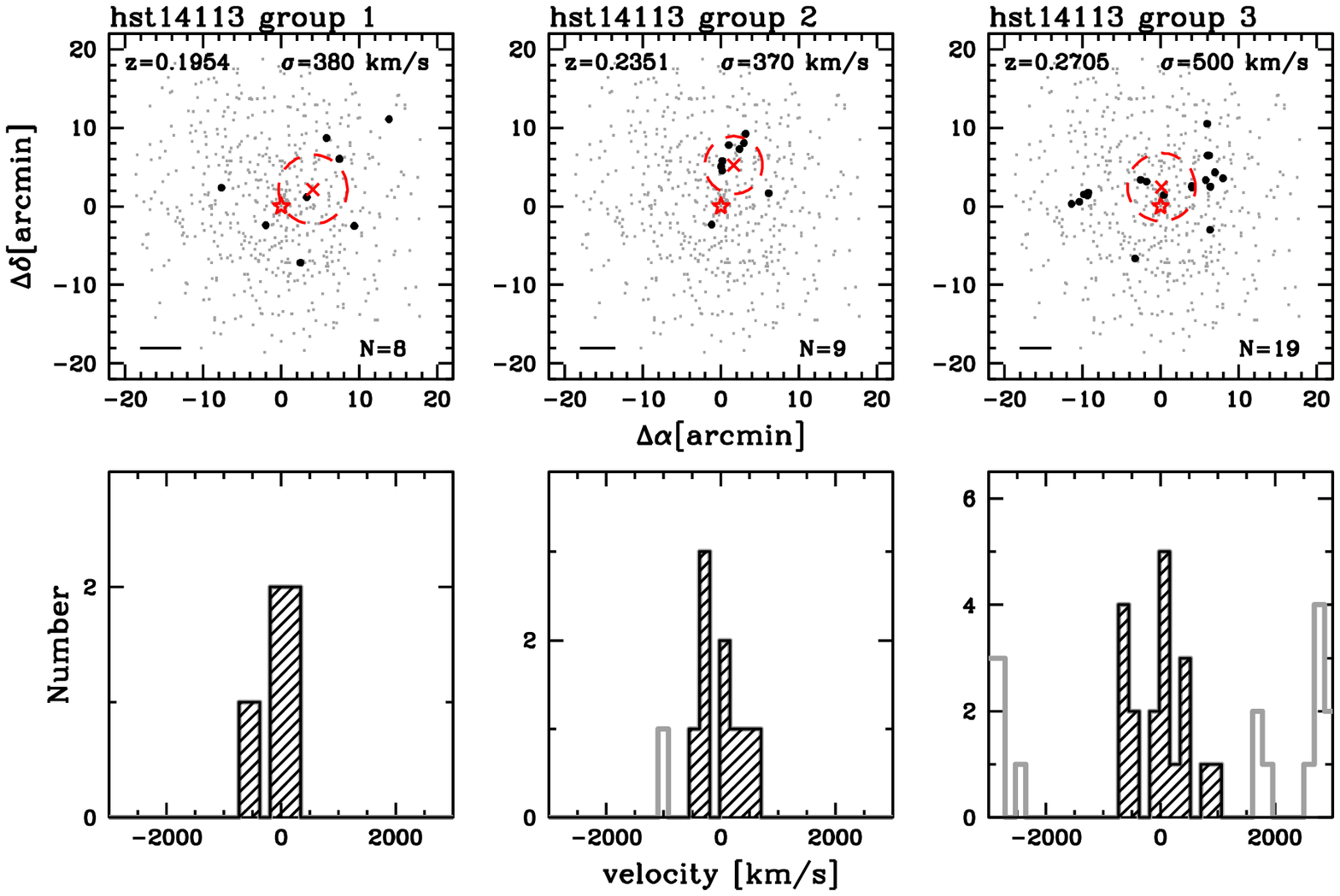}
\includegraphics[clip=true, width=18cm]{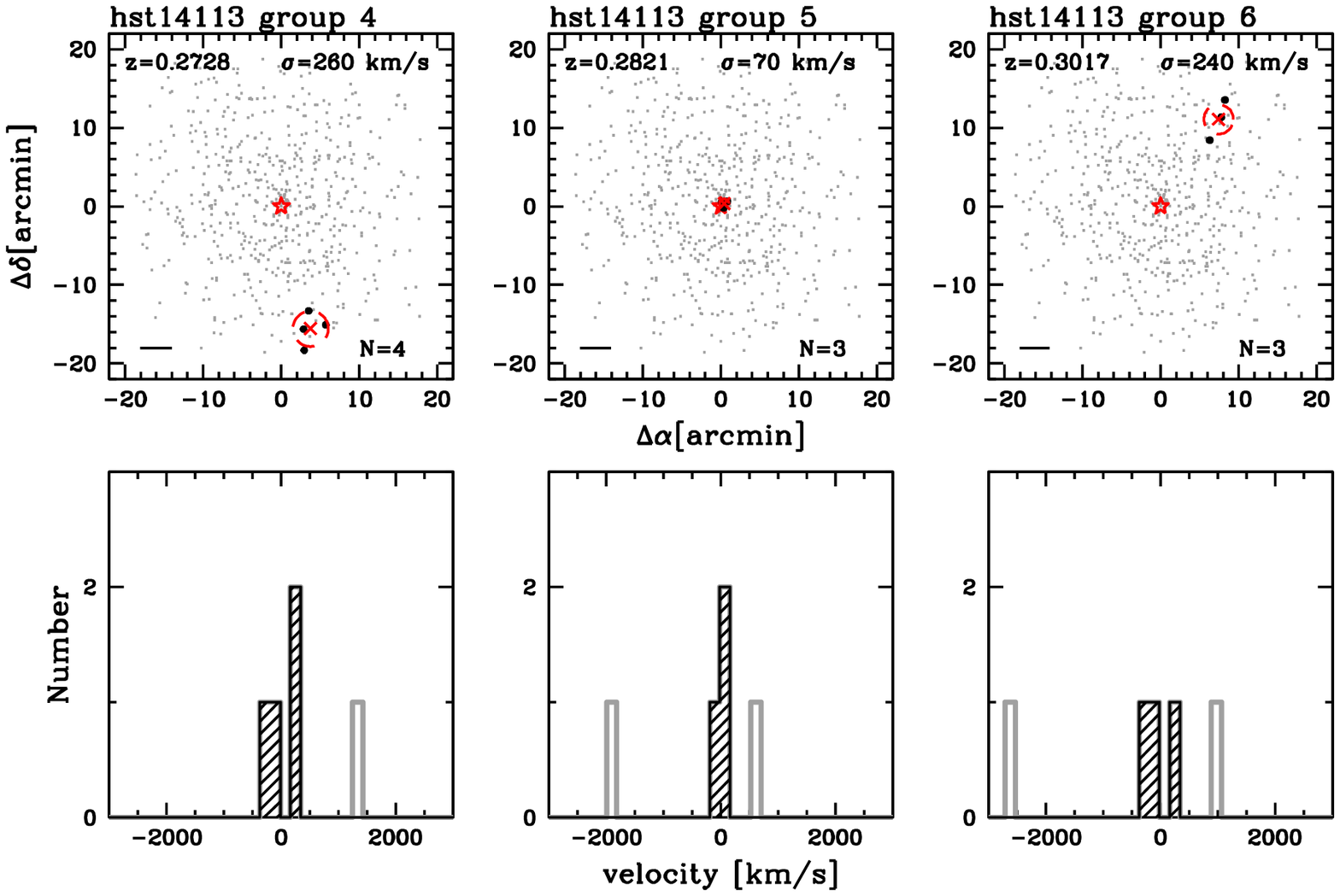}
\caption{Continued.}
\end{figure*}
\clearpage
\begin{figure*}
\ContinuedFloat
\includegraphics[clip=true, width=18cm]{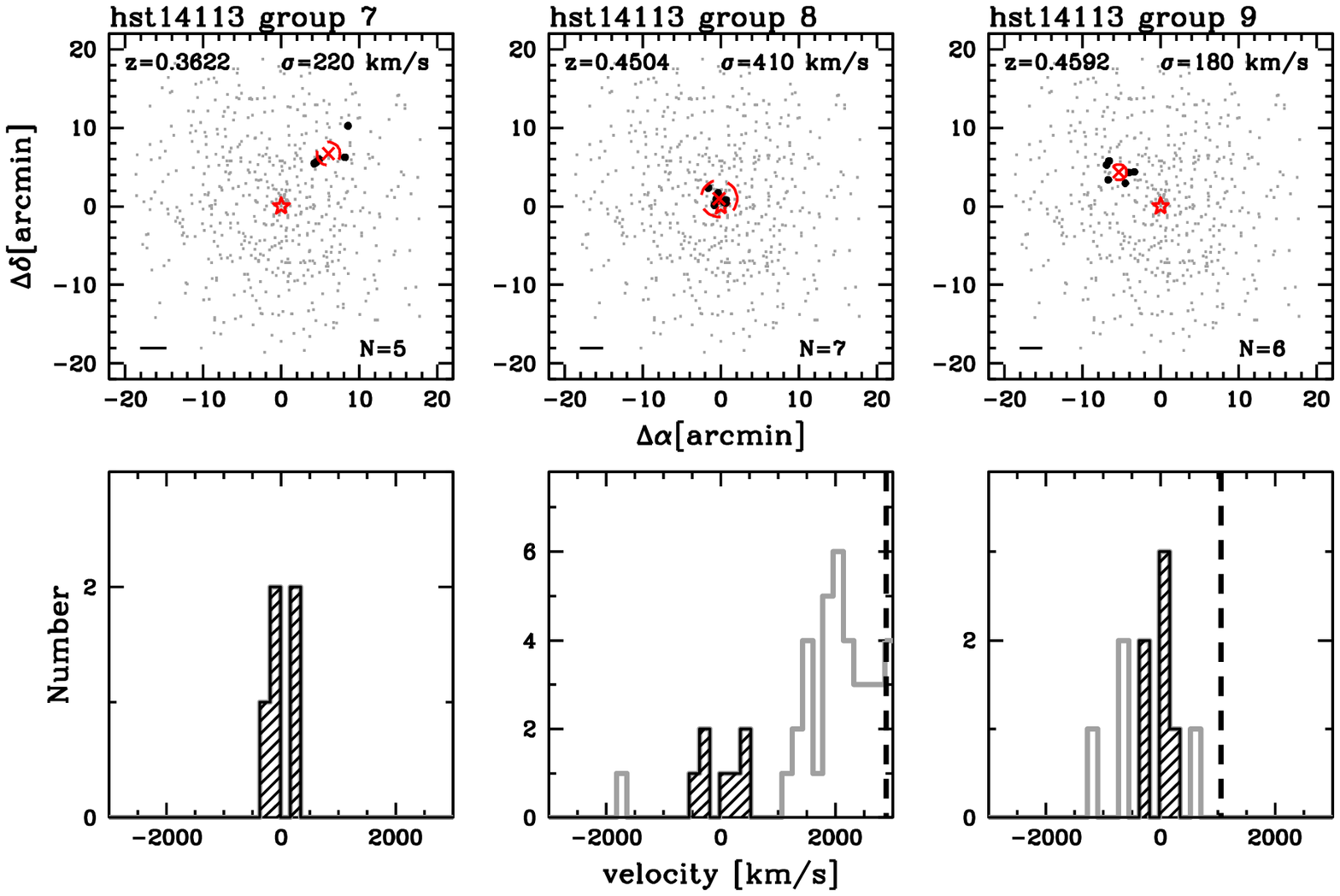}
\includegraphics[clip=true, width=18cm]{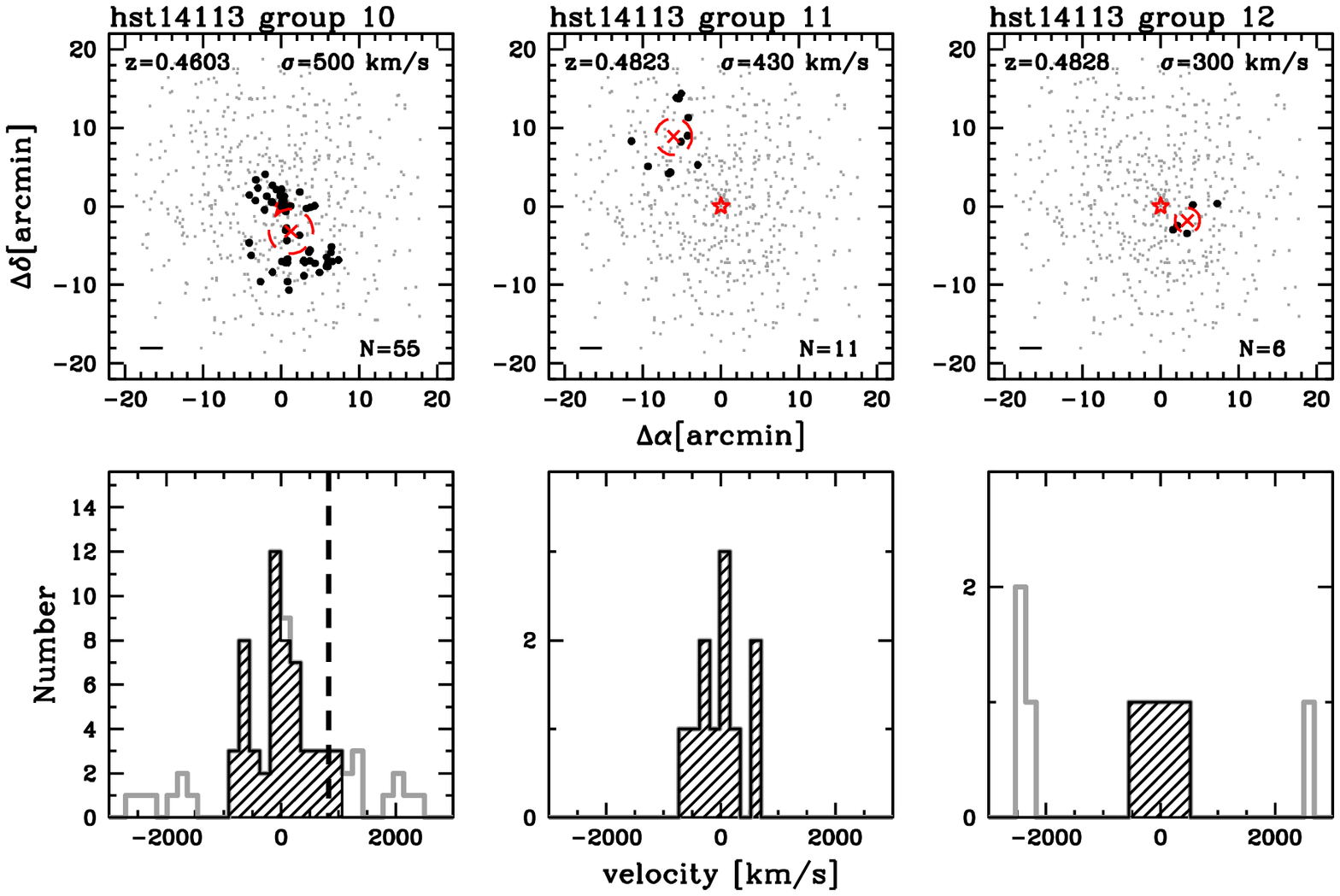}
\caption{Continued.}
\end{figure*}
\clearpage
\begin{figure*}
\ContinuedFloat
\includegraphics[clip=true, width=18cm]{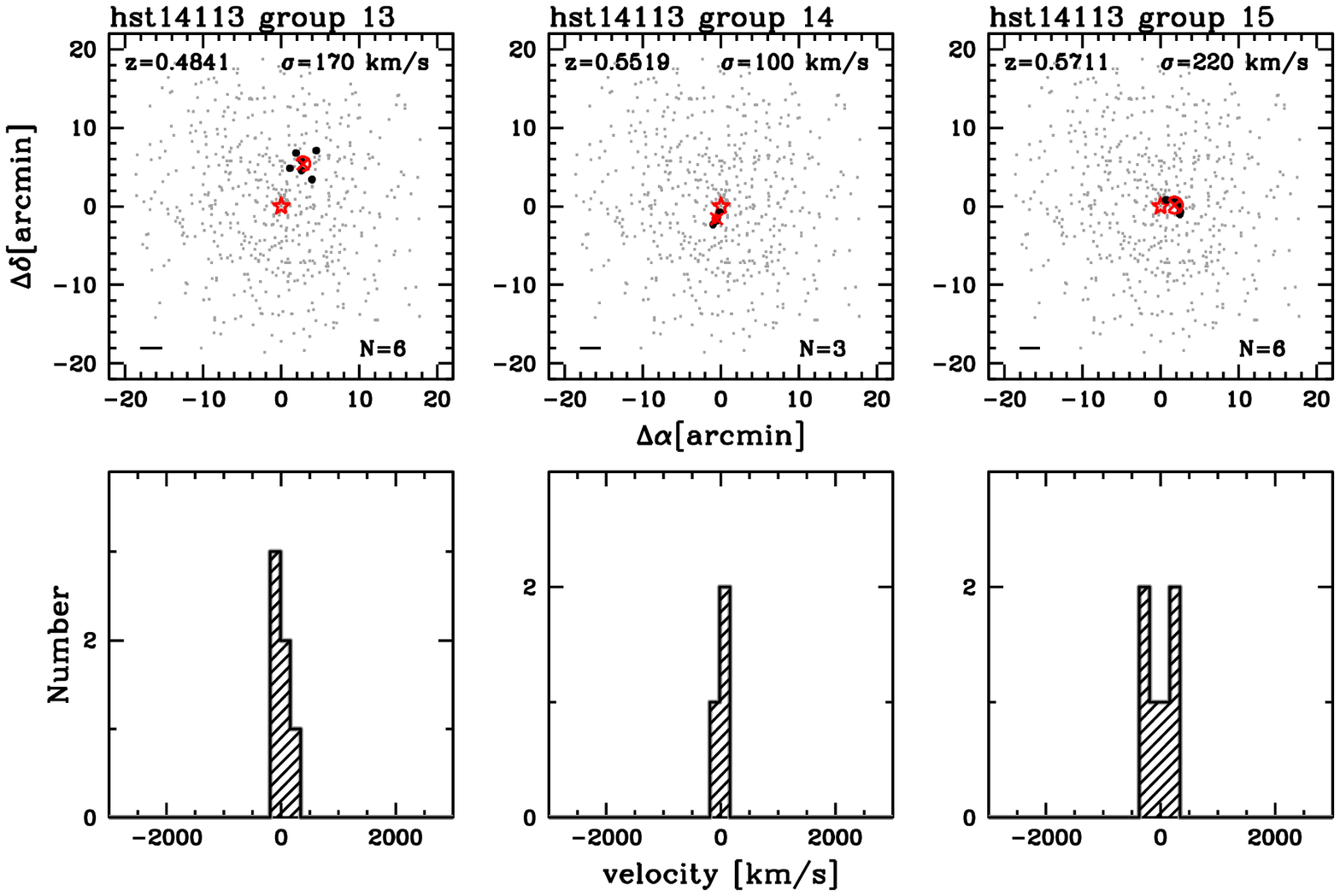}
\includegraphics[clip=true, width=18cm]{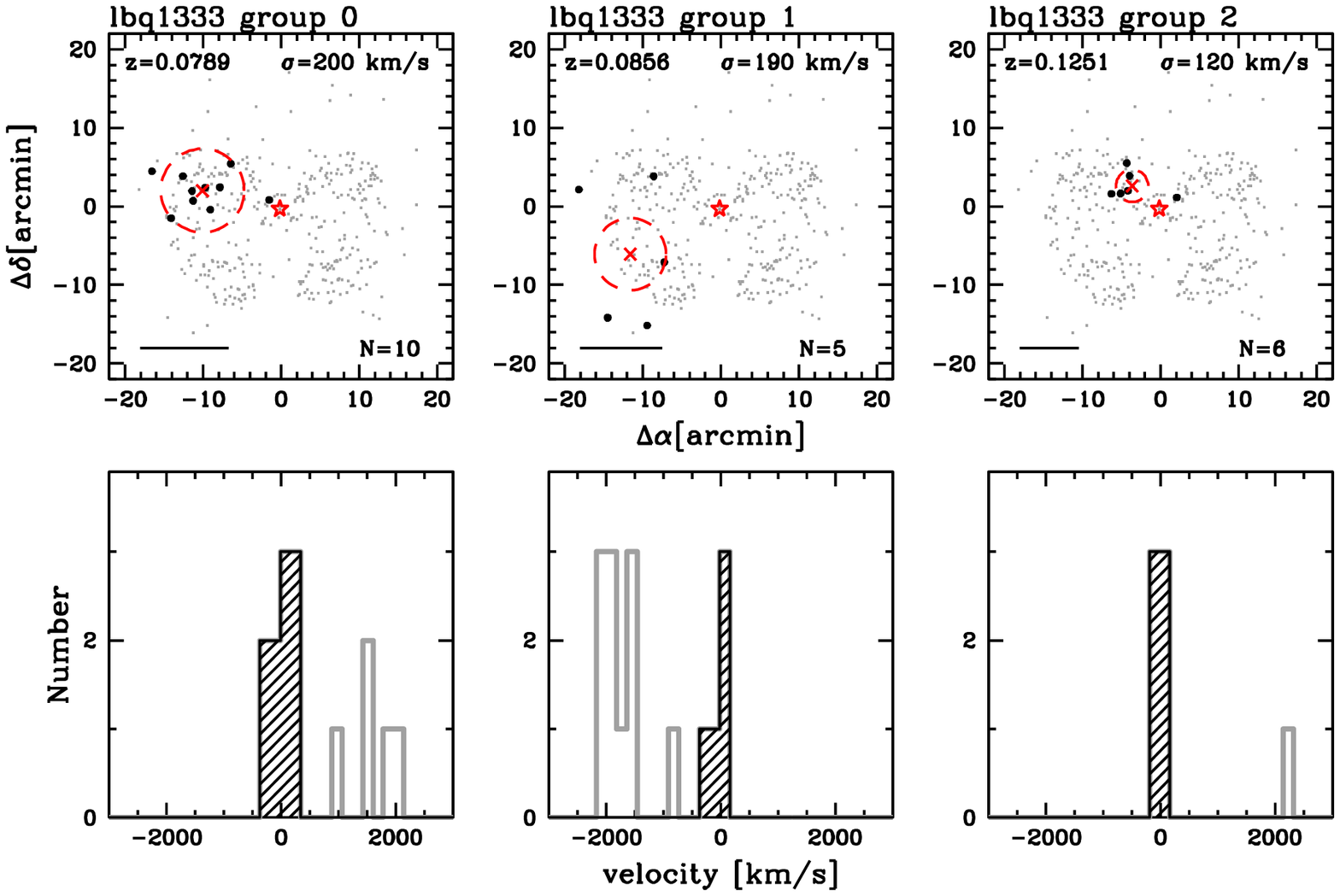}
\caption{Continued.}
\end{figure*}
\clearpage
\begin{figure*}
\ContinuedFloat
\includegraphics[clip=true, width=18cm]{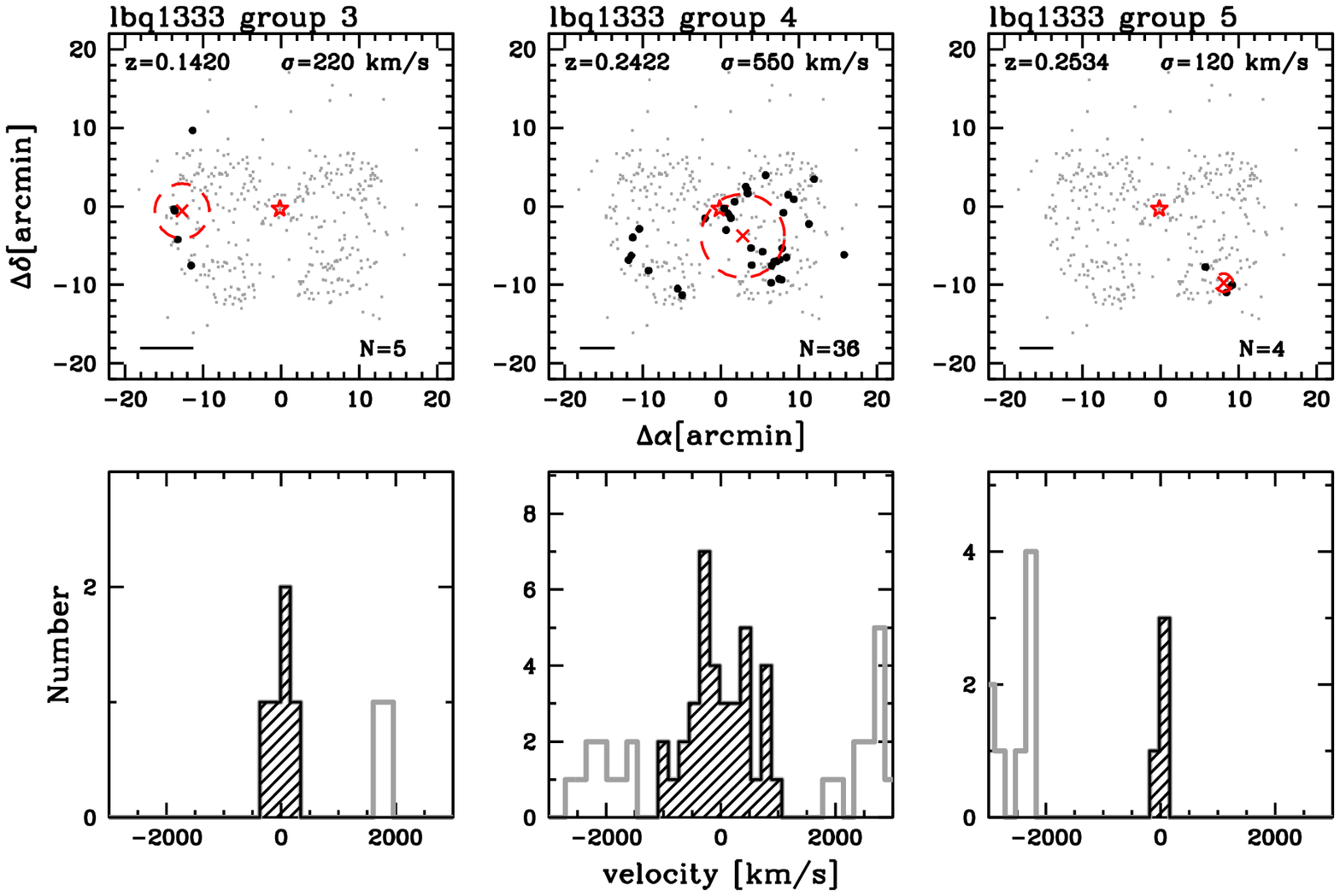}
\includegraphics[clip=true, width=18cm]{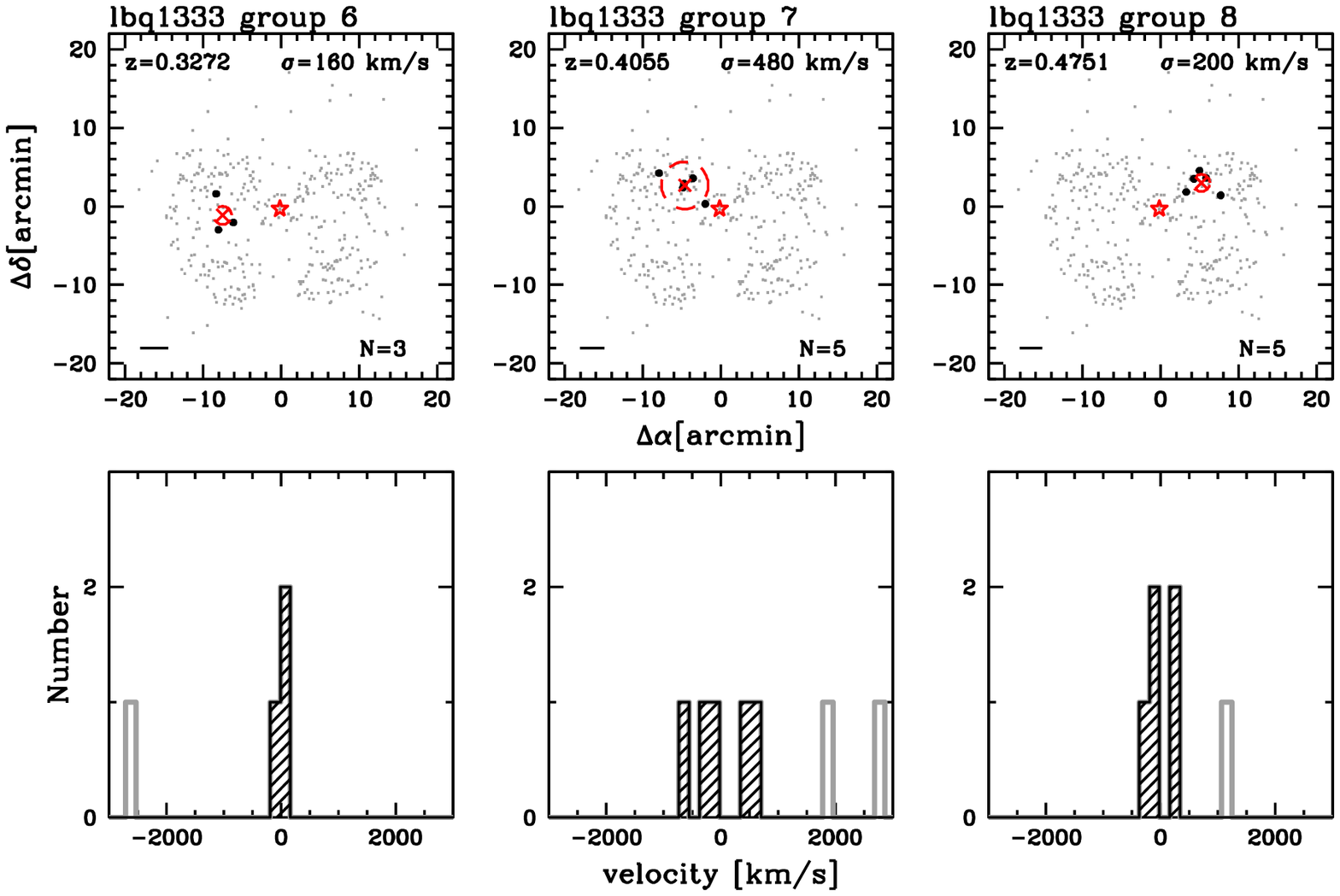}
\caption{Continued.}
\end{figure*}
\clearpage
\begin{figure*}
\ContinuedFloat
\includegraphics[clip=true, width=18cm]{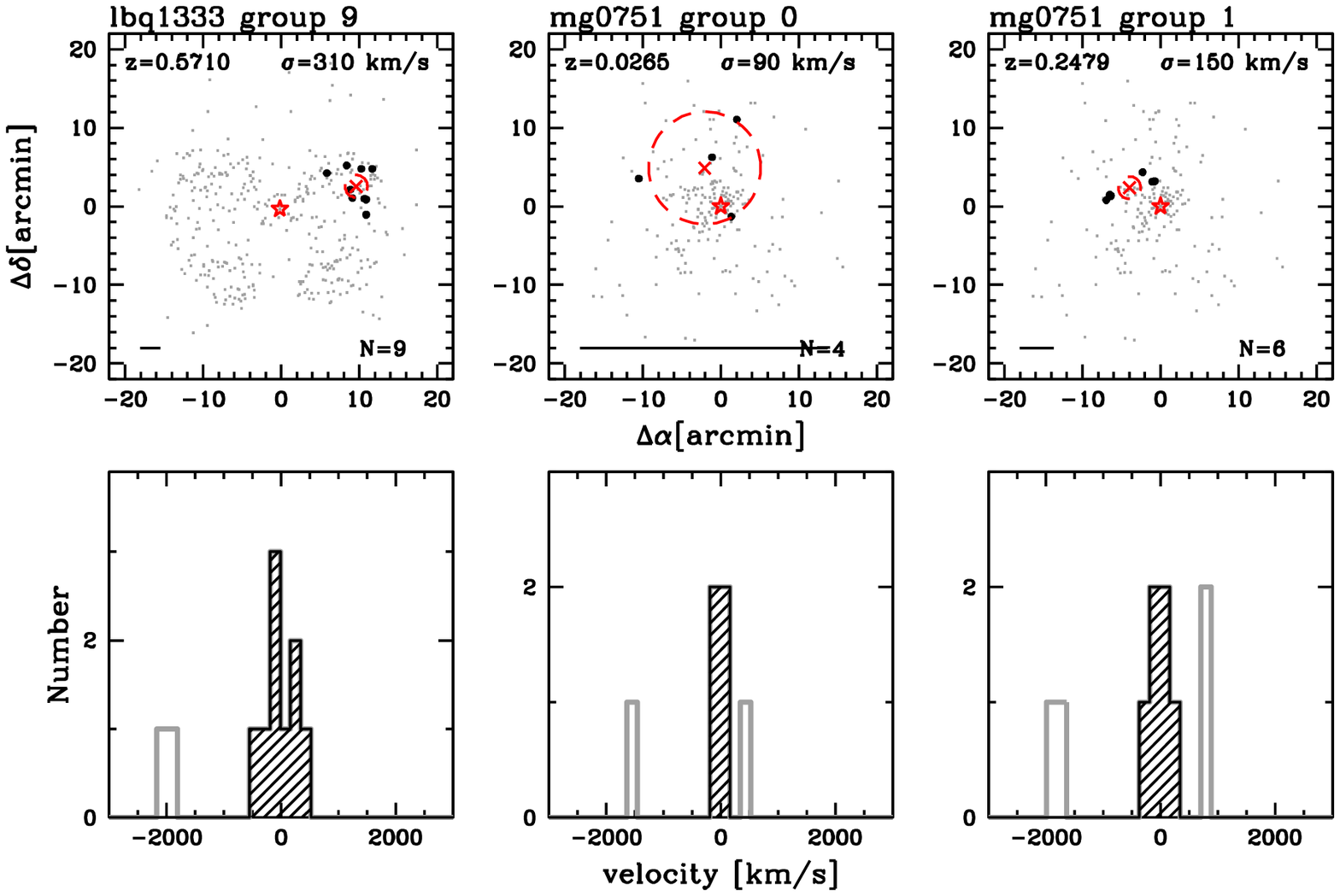}
\includegraphics[clip=true, width=18cm]{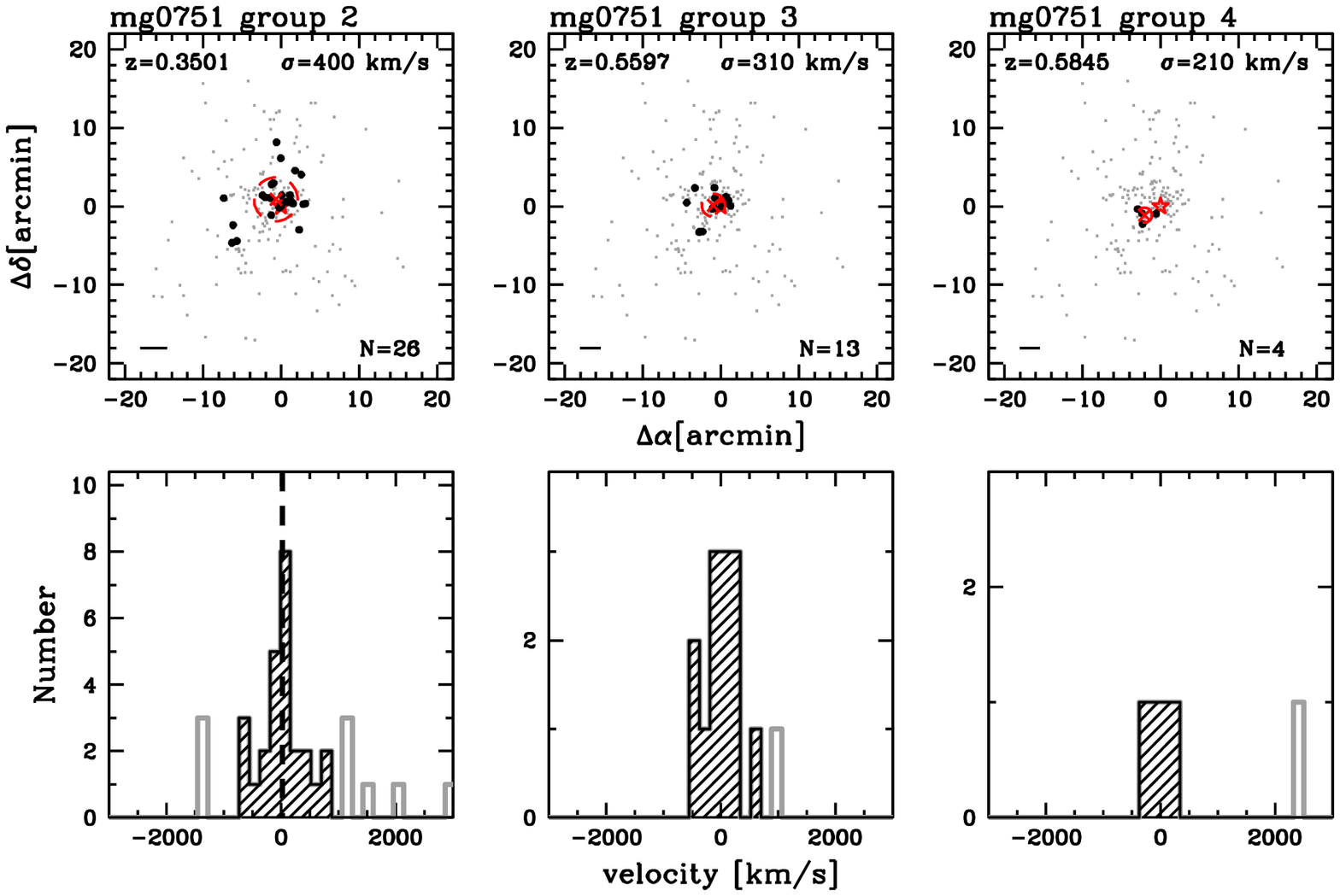}
\caption{Continued.}
\end{figure*}
\clearpage
\begin{figure*}
\ContinuedFloat
\includegraphics[clip=true, width=18cm]{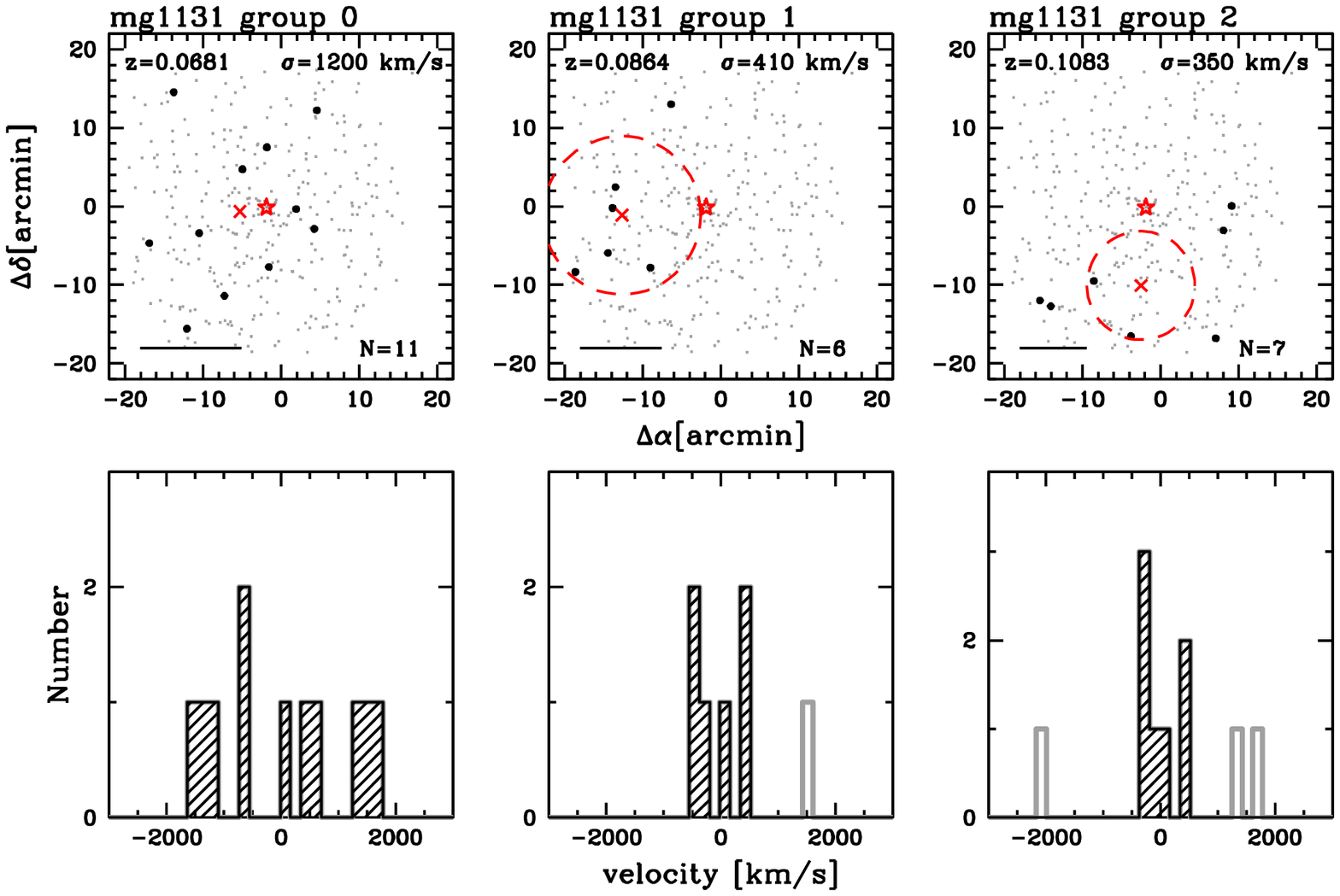}
\includegraphics[clip=true, width=18cm]{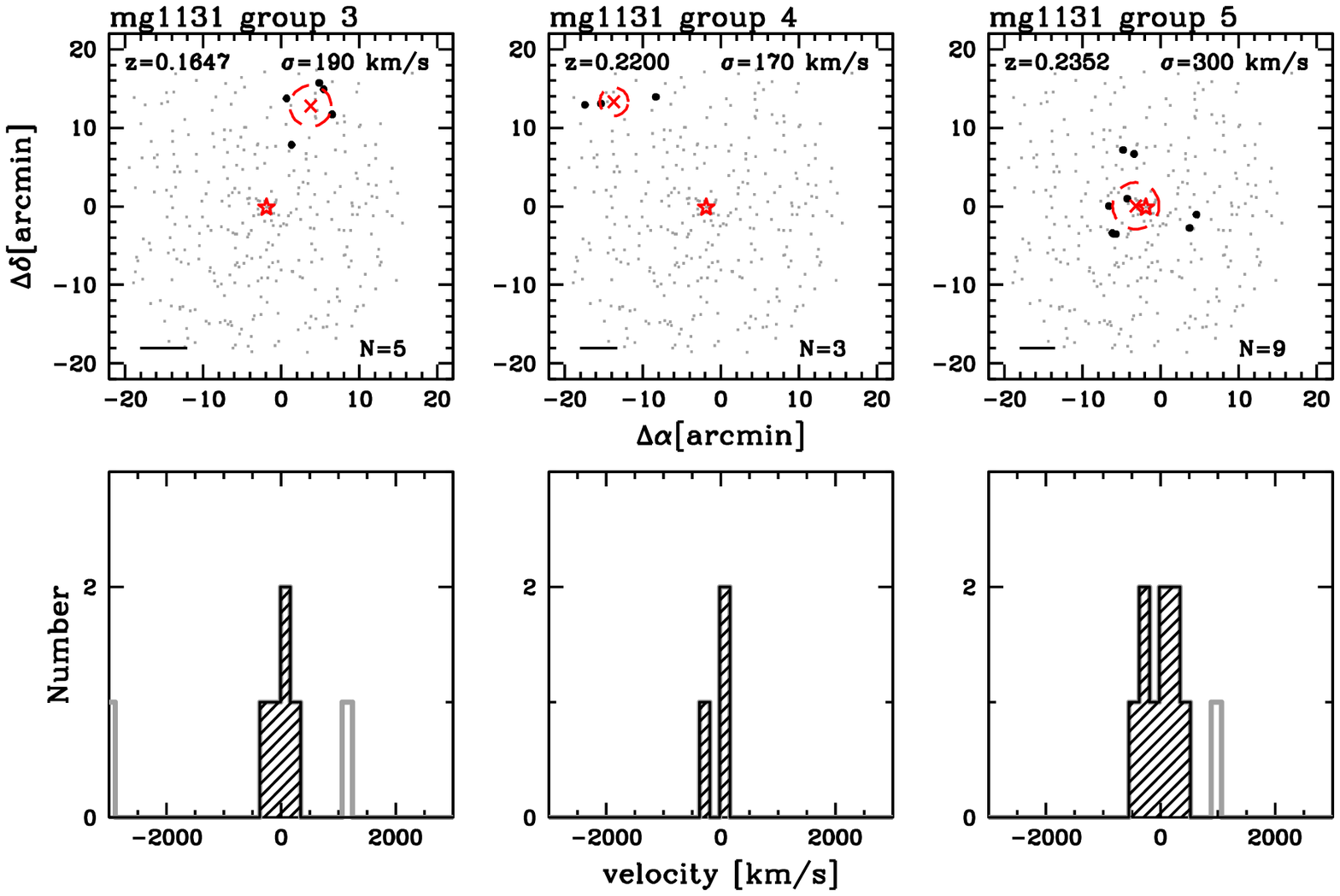}
\caption{Continued.}
\end{figure*}
\clearpage
\begin{figure*}
\ContinuedFloat
\includegraphics[clip=true, width=18cm]{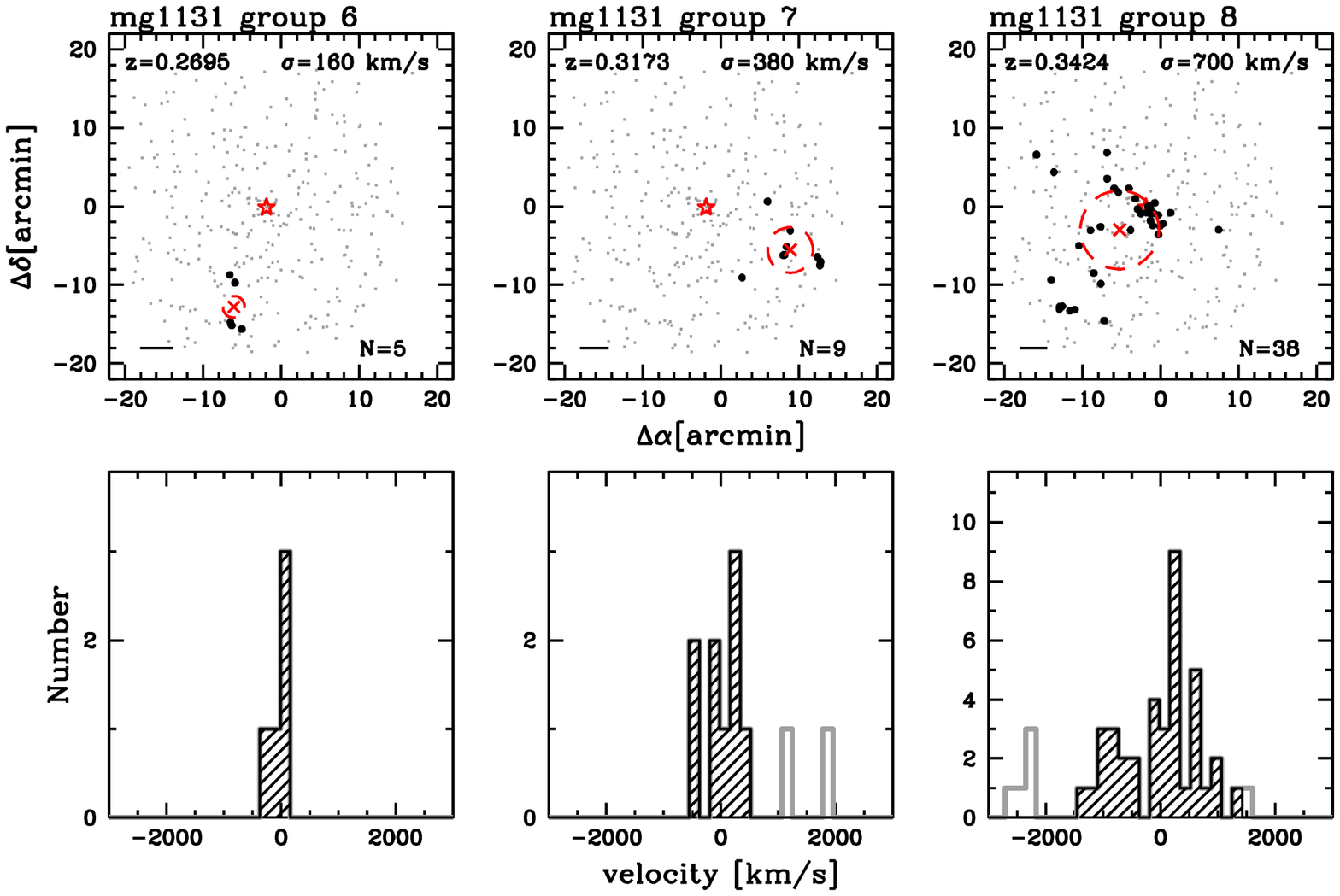}
\includegraphics[clip=true, width=18cm]{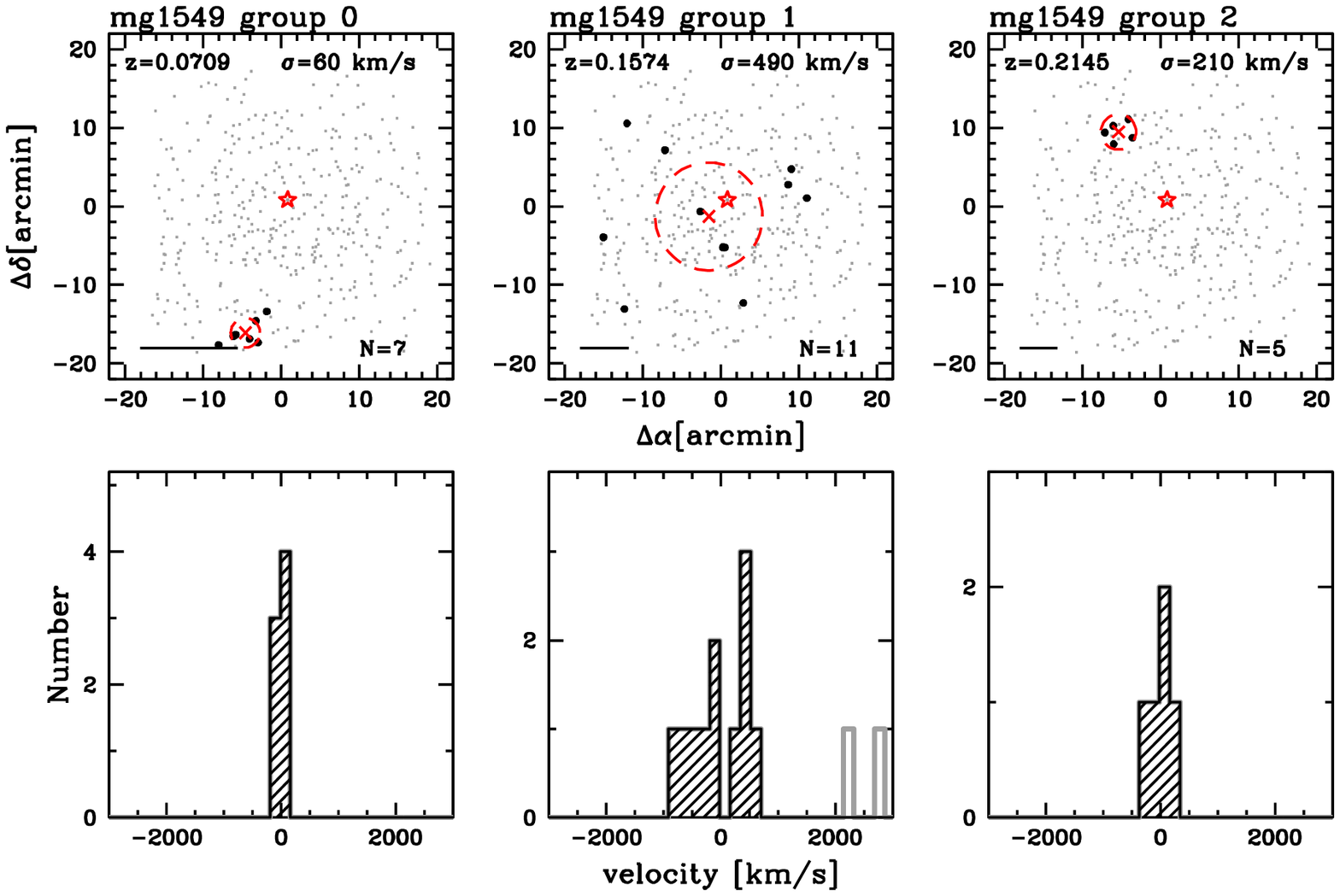}
\caption{Continued.}
\end{figure*}
\clearpage
\begin{figure*}
\ContinuedFloat
\includegraphics[clip=true, width=18cm]{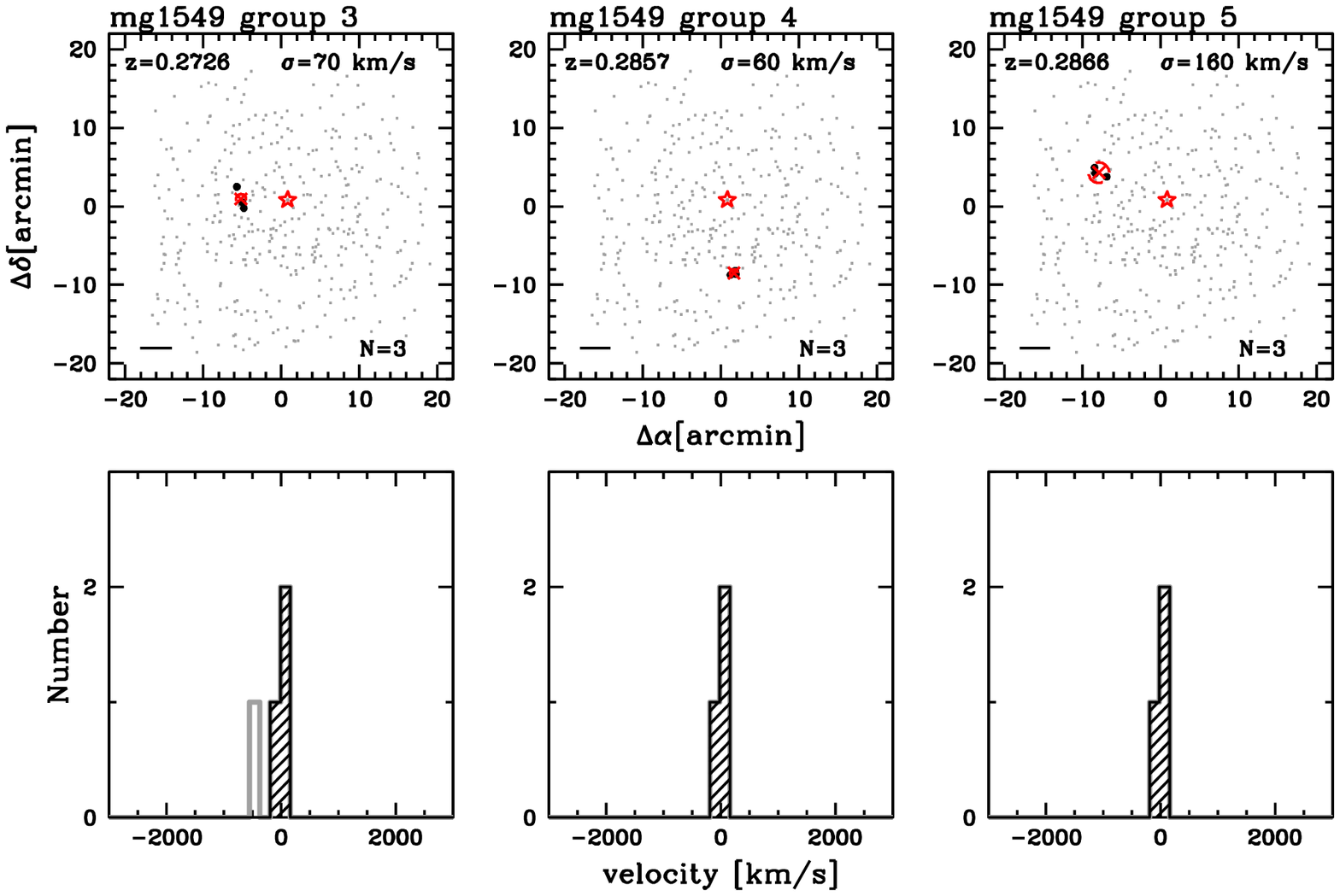}
\includegraphics[clip=true, width=18cm]{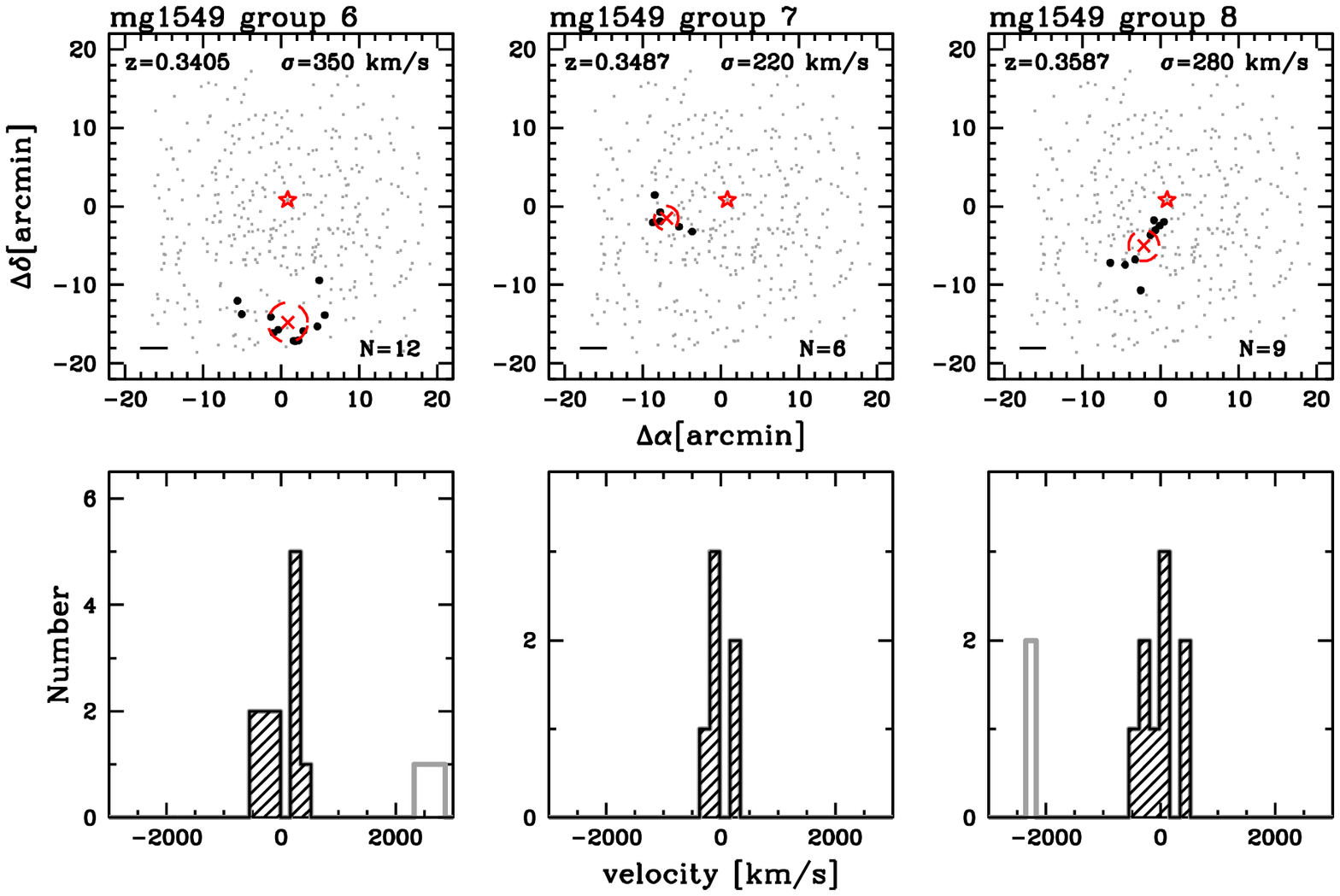}
\caption{Continued.}
\end{figure*}
\clearpage
\begin{figure*}
\ContinuedFloat
\includegraphics[clip=true, width=18cm]{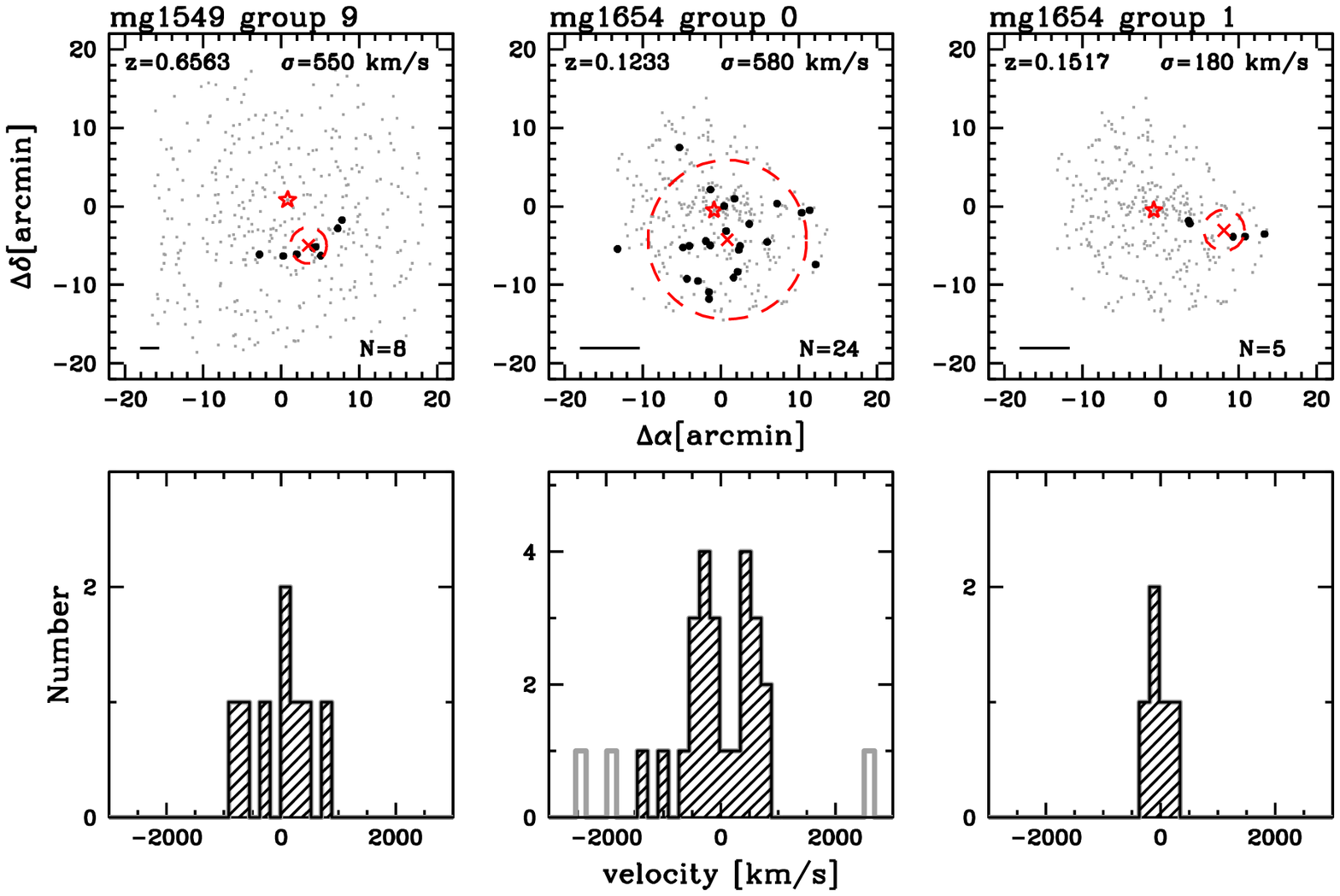}
\includegraphics[clip=true, width=18cm]{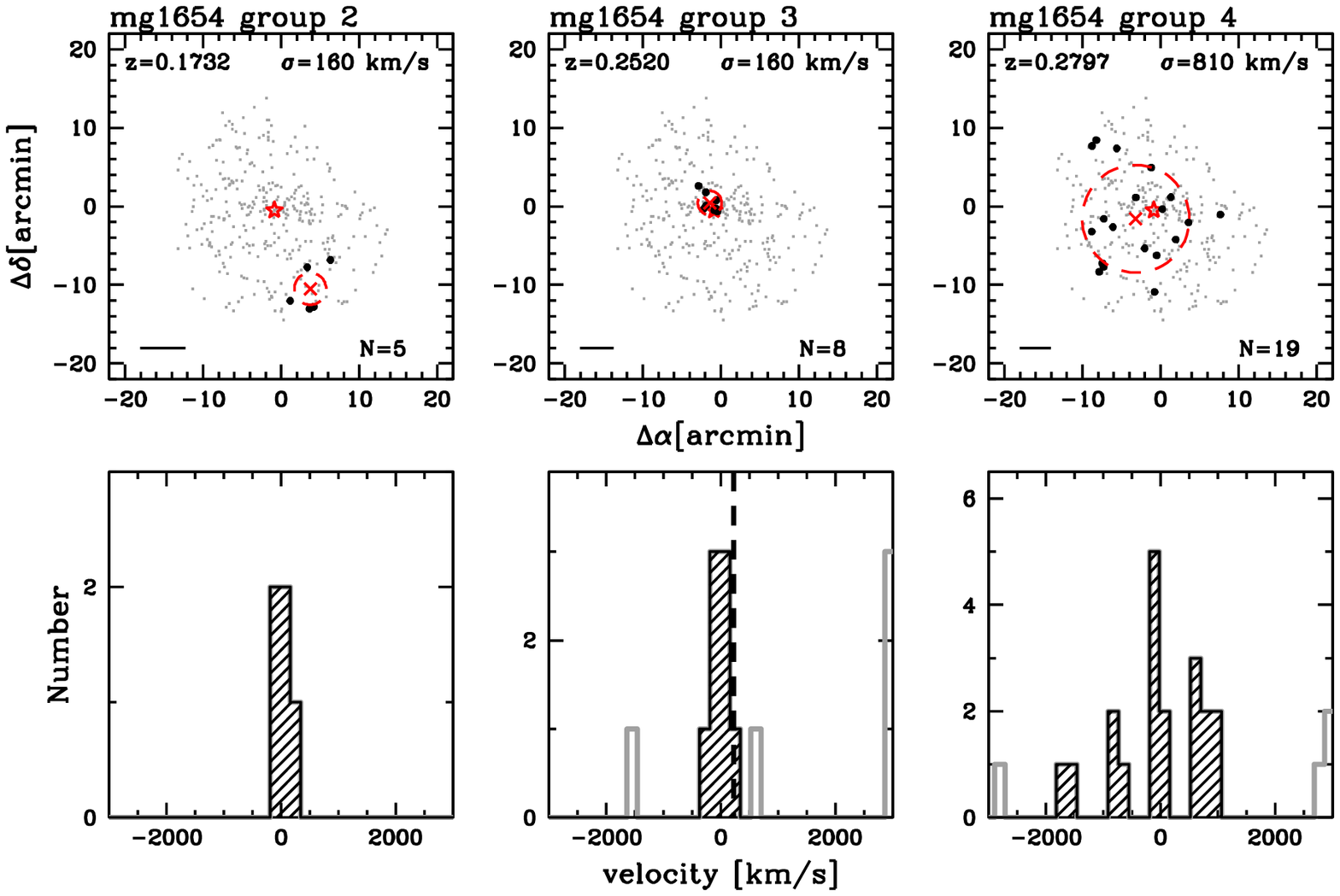}
\caption{Continued.}
\end{figure*}
\clearpage
\begin{figure*}
\ContinuedFloat
\includegraphics[clip=true, width=18cm]{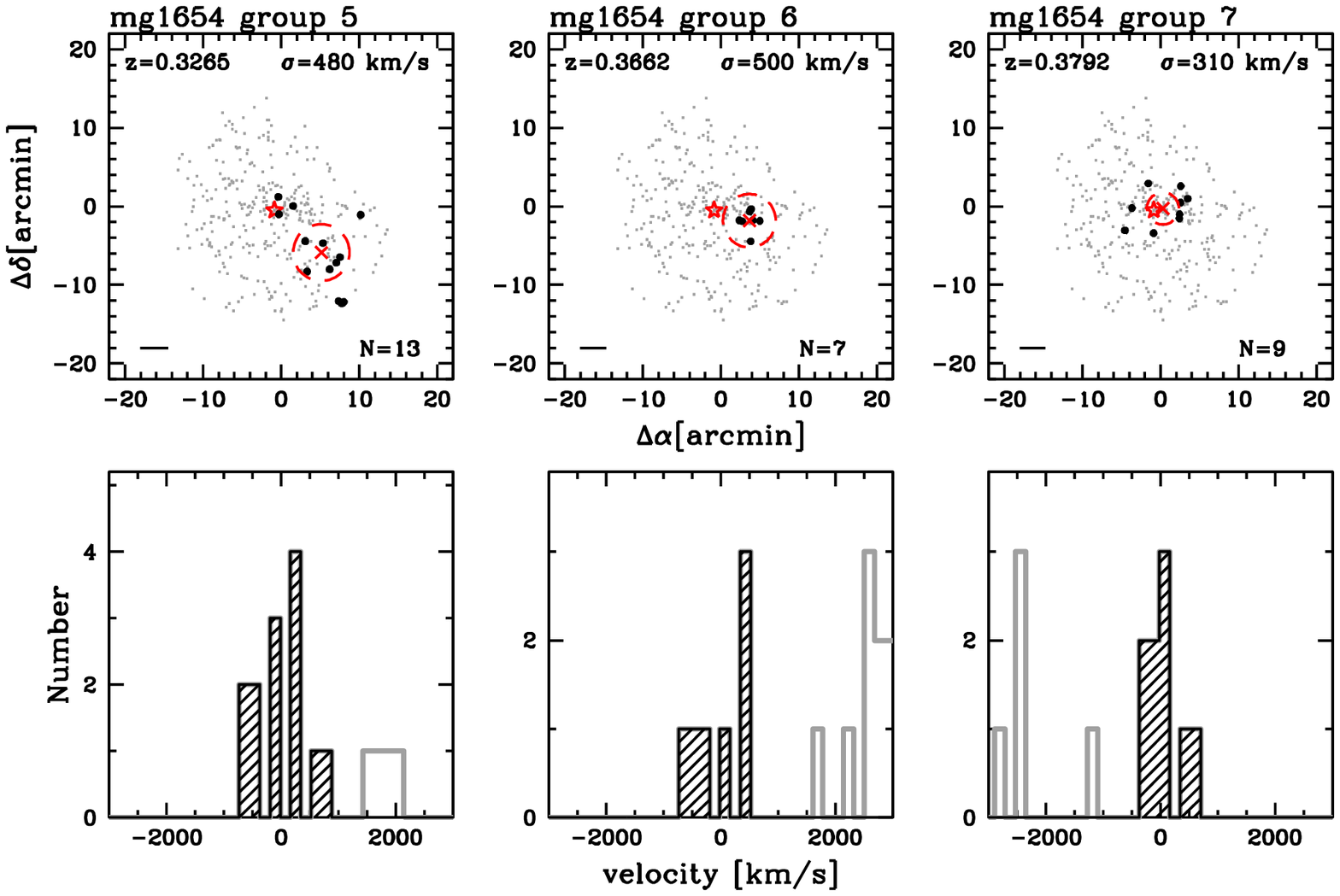}
\includegraphics[clip=true, width=18cm]{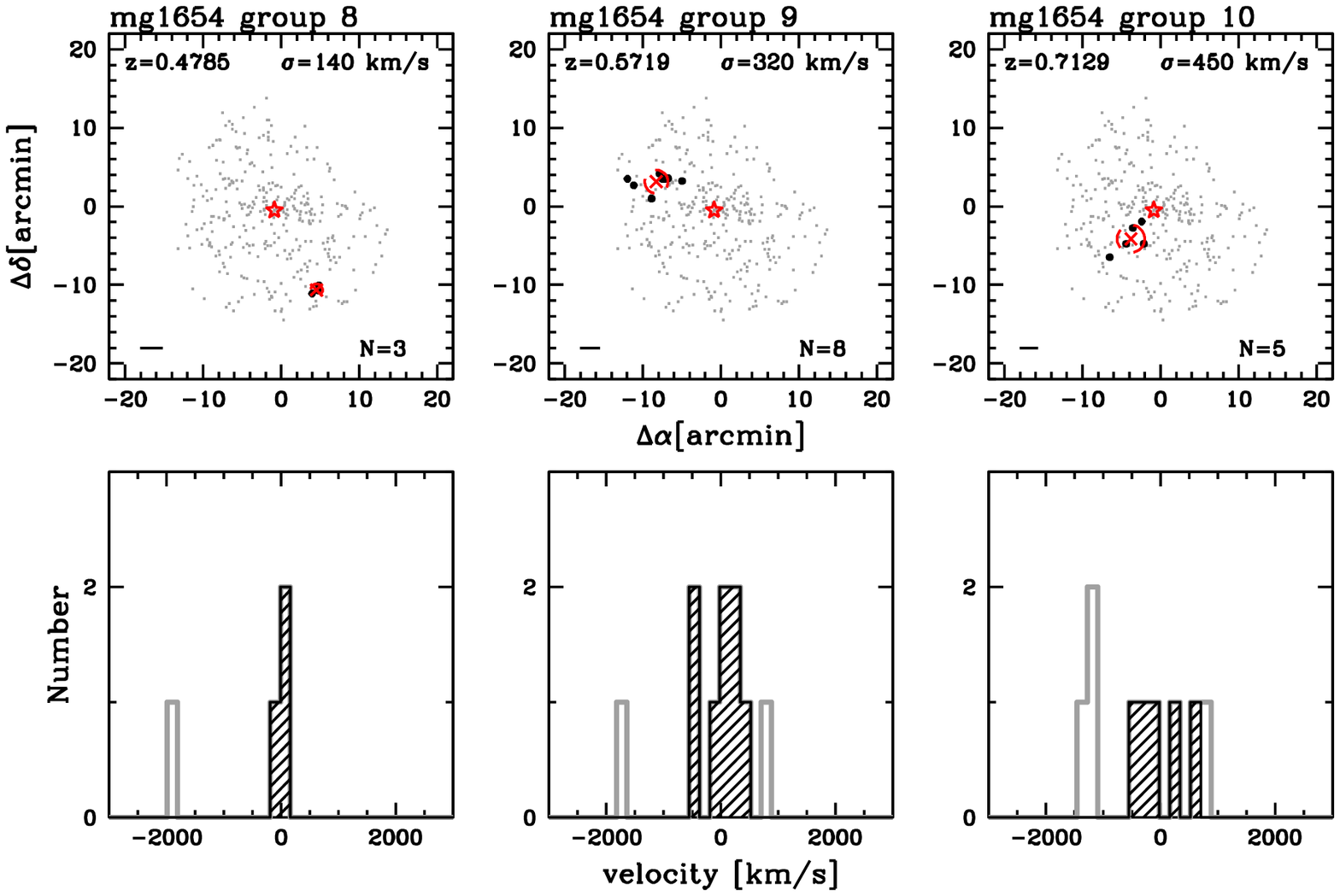}
\caption{Continued.}
\end{figure*}
\clearpage
\begin{figure*}
\ContinuedFloat
\includegraphics[clip=true, width=18cm]{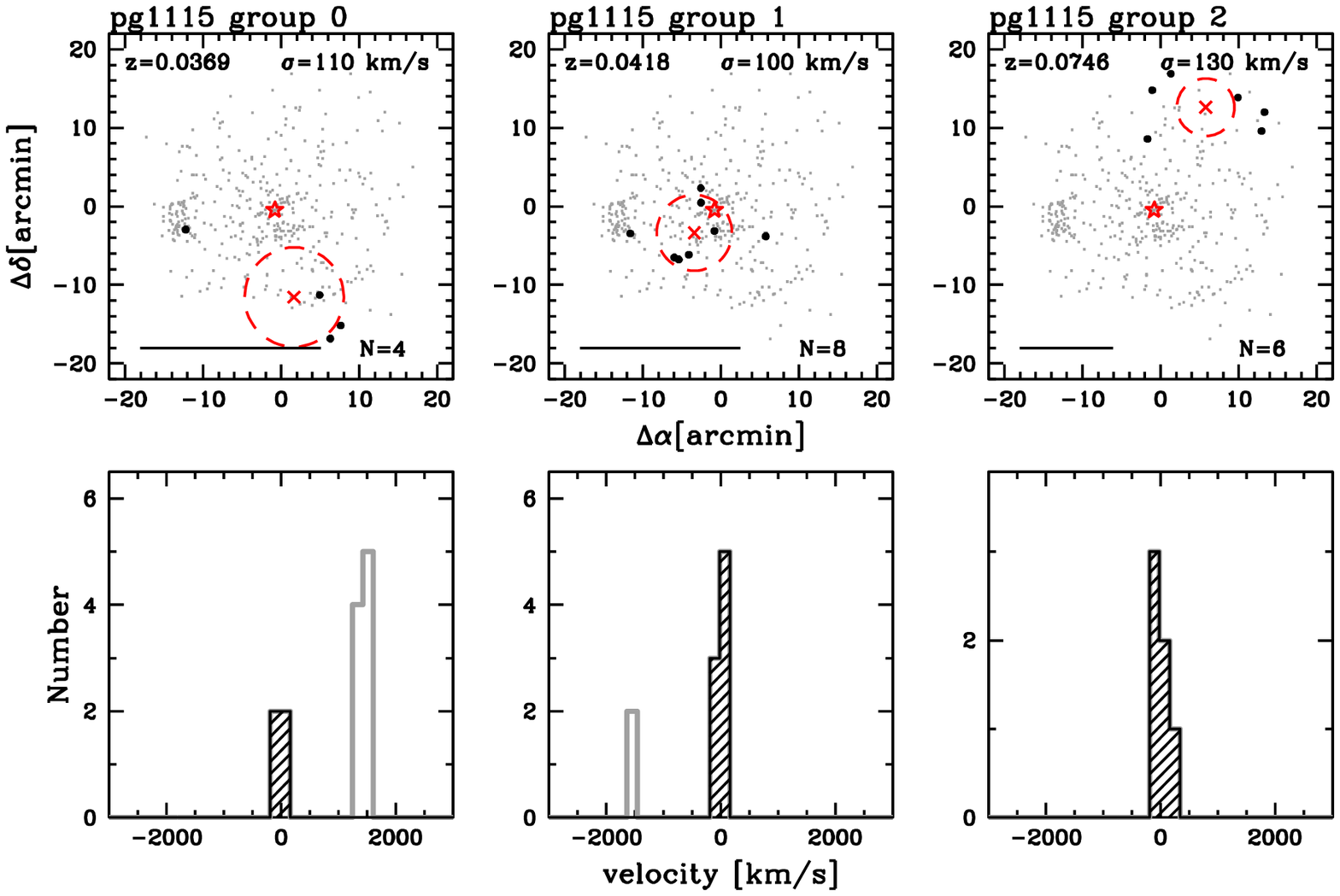}
\includegraphics[clip=true, width=18cm]{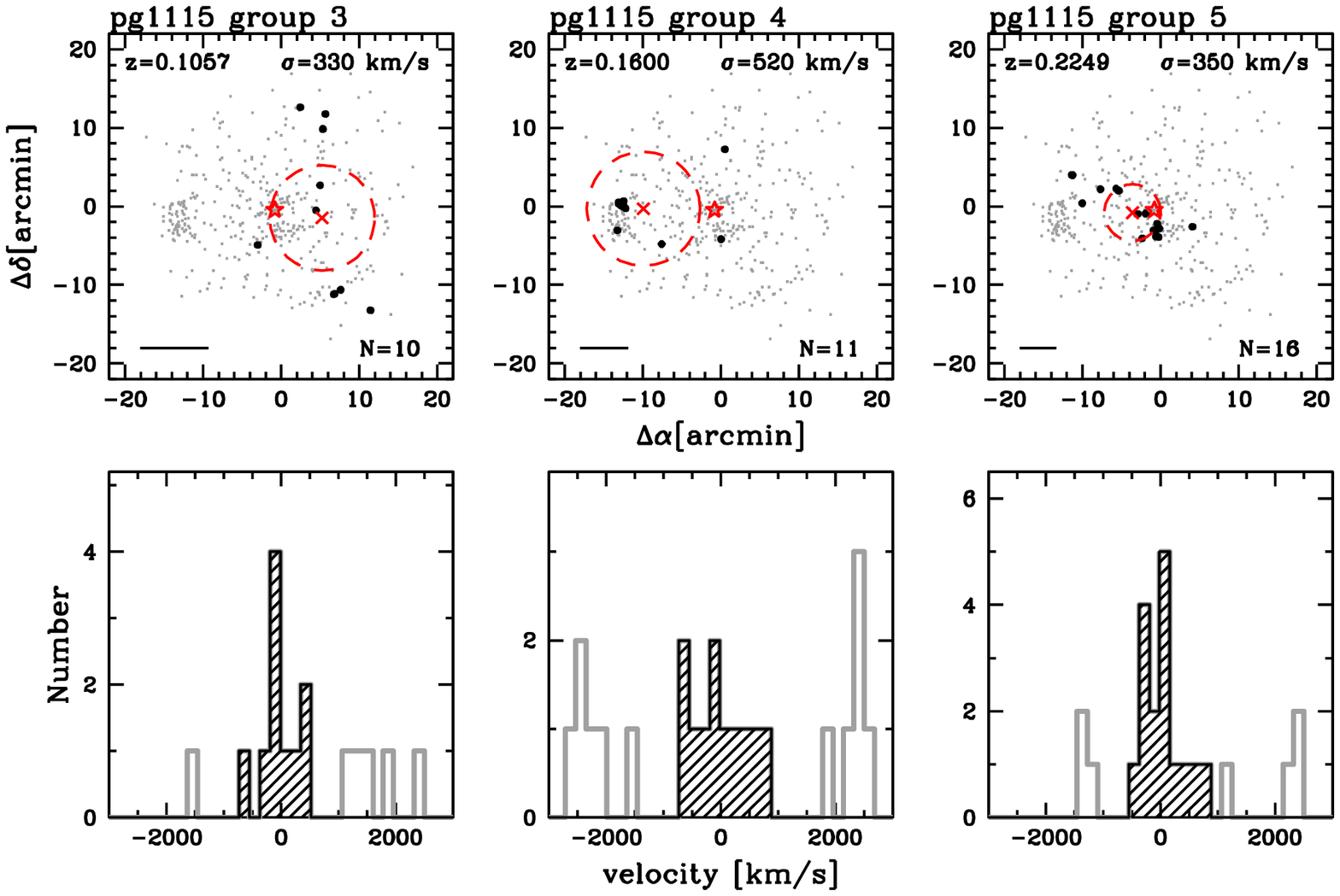}
\caption{Continued.}
\end{figure*}
\clearpage
\begin{figure*}
\ContinuedFloat
\includegraphics[clip=true, width=18cm]{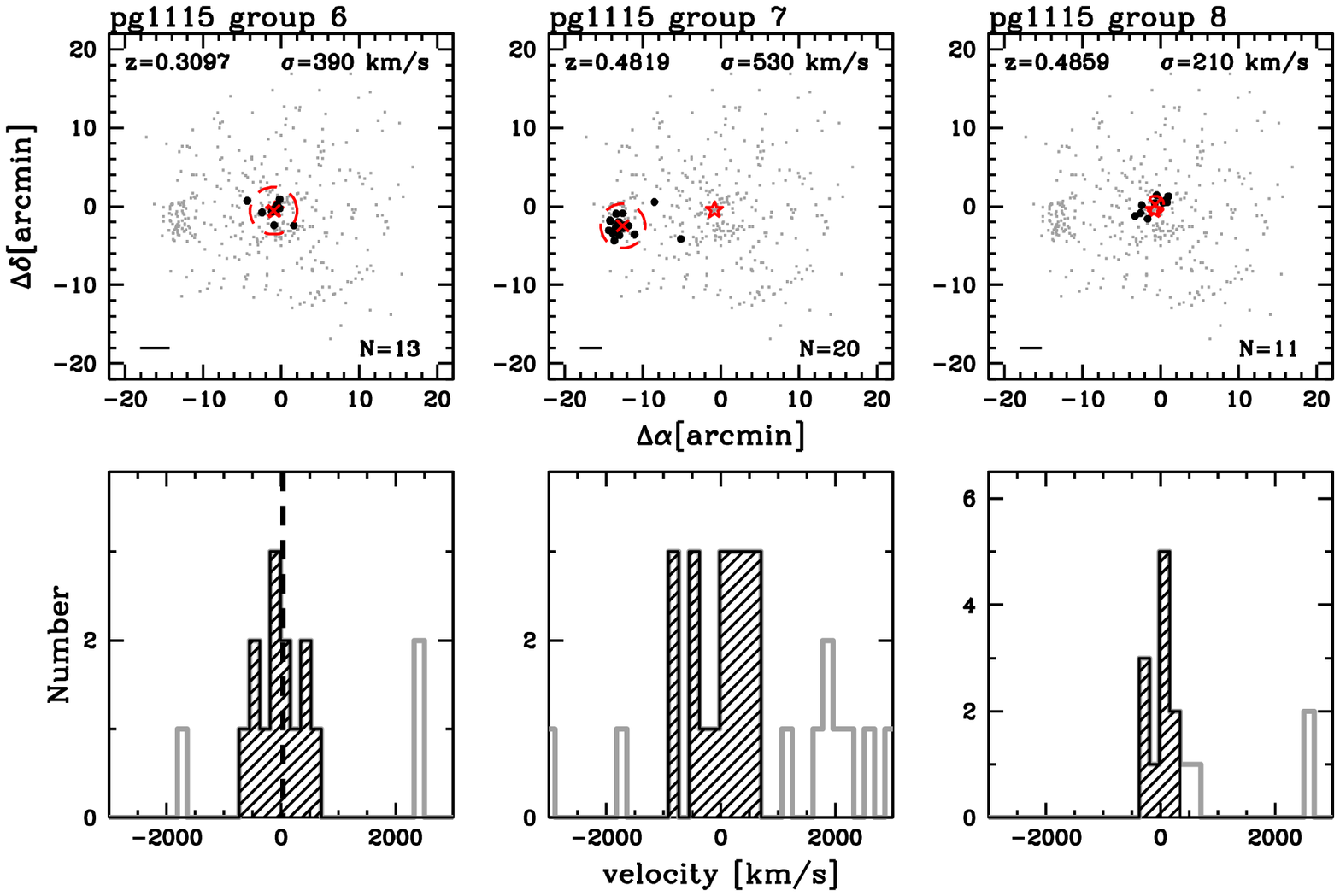}
\includegraphics[clip=true, width=18cm]{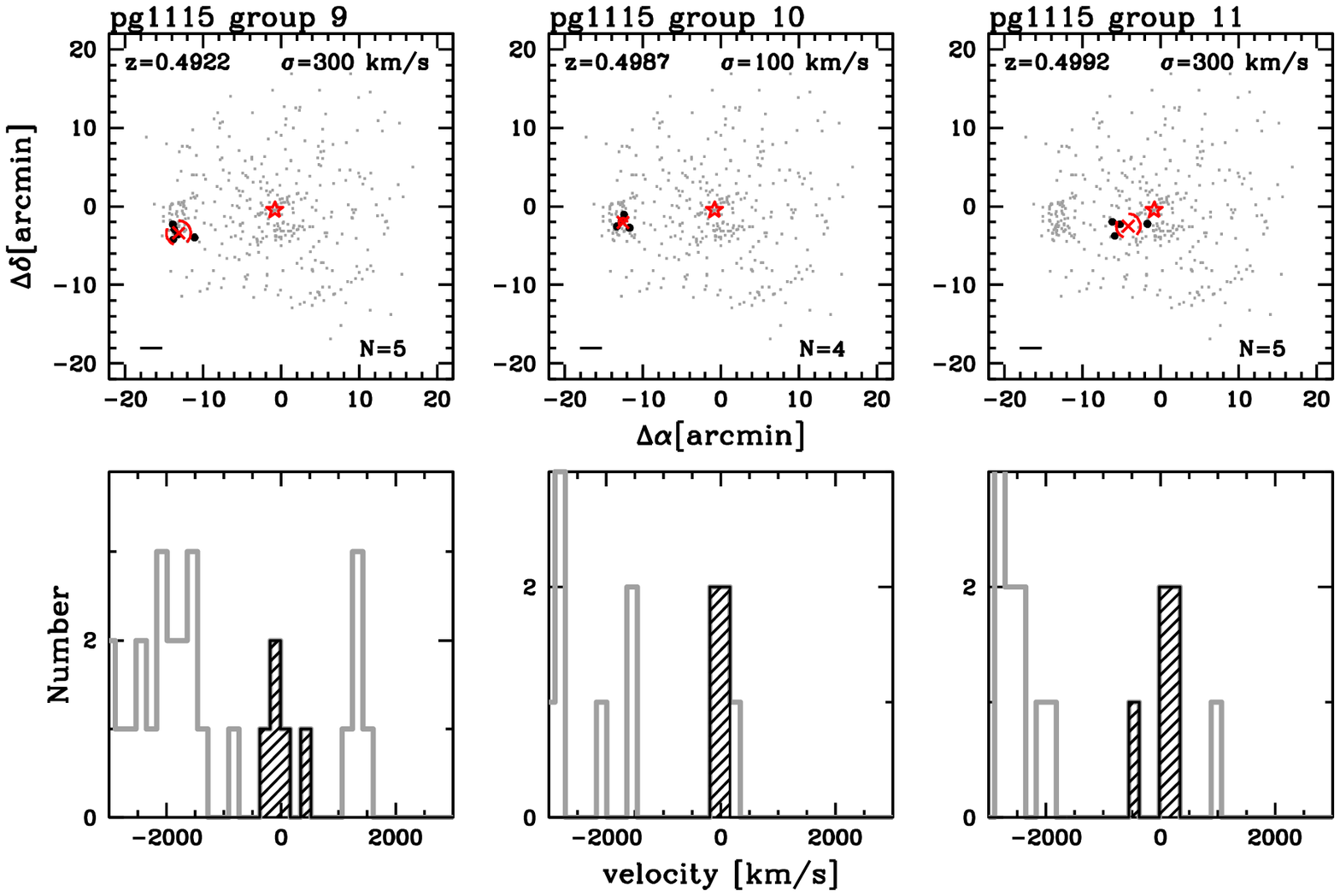}
\caption{Continued.}
\end{figure*}
\begin{figure*}
\ContinuedFloat
\includegraphics[clip=true, width=18cm]{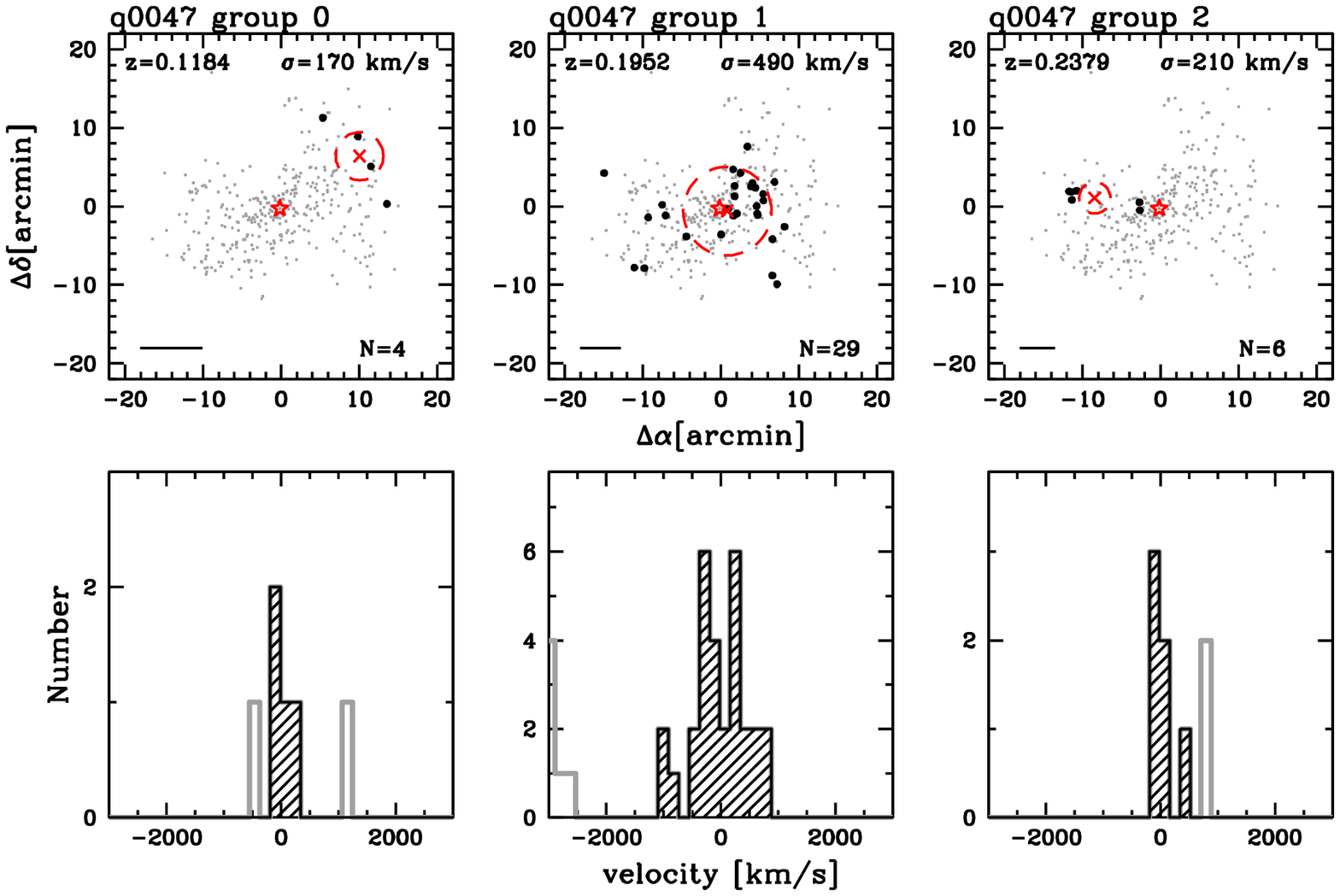}
\includegraphics[clip=true, width=18cm]{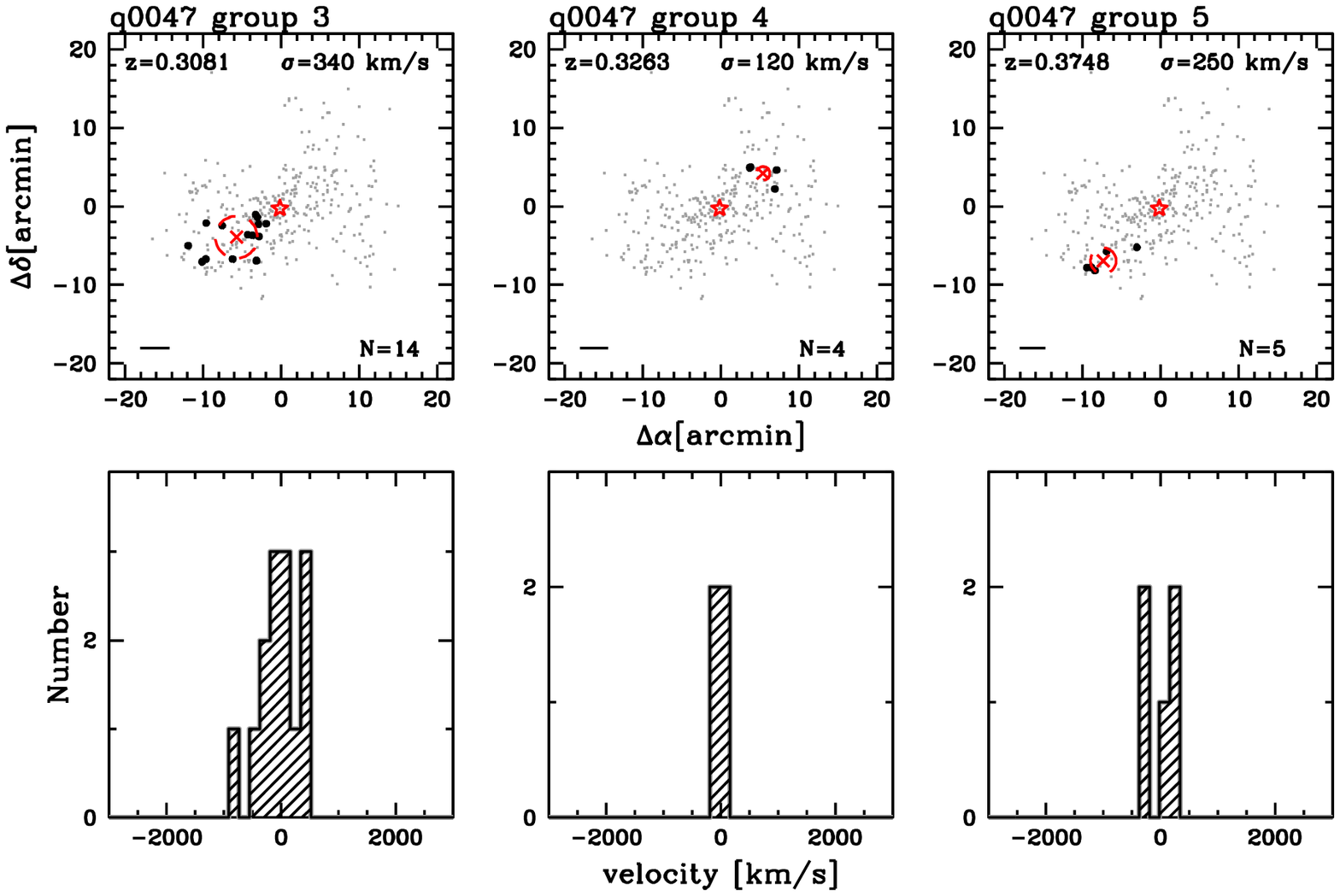}
\caption{Continued.}
\end{figure*}
\begin{figure*}
\ContinuedFloat
\includegraphics[clip=true, width=18cm]{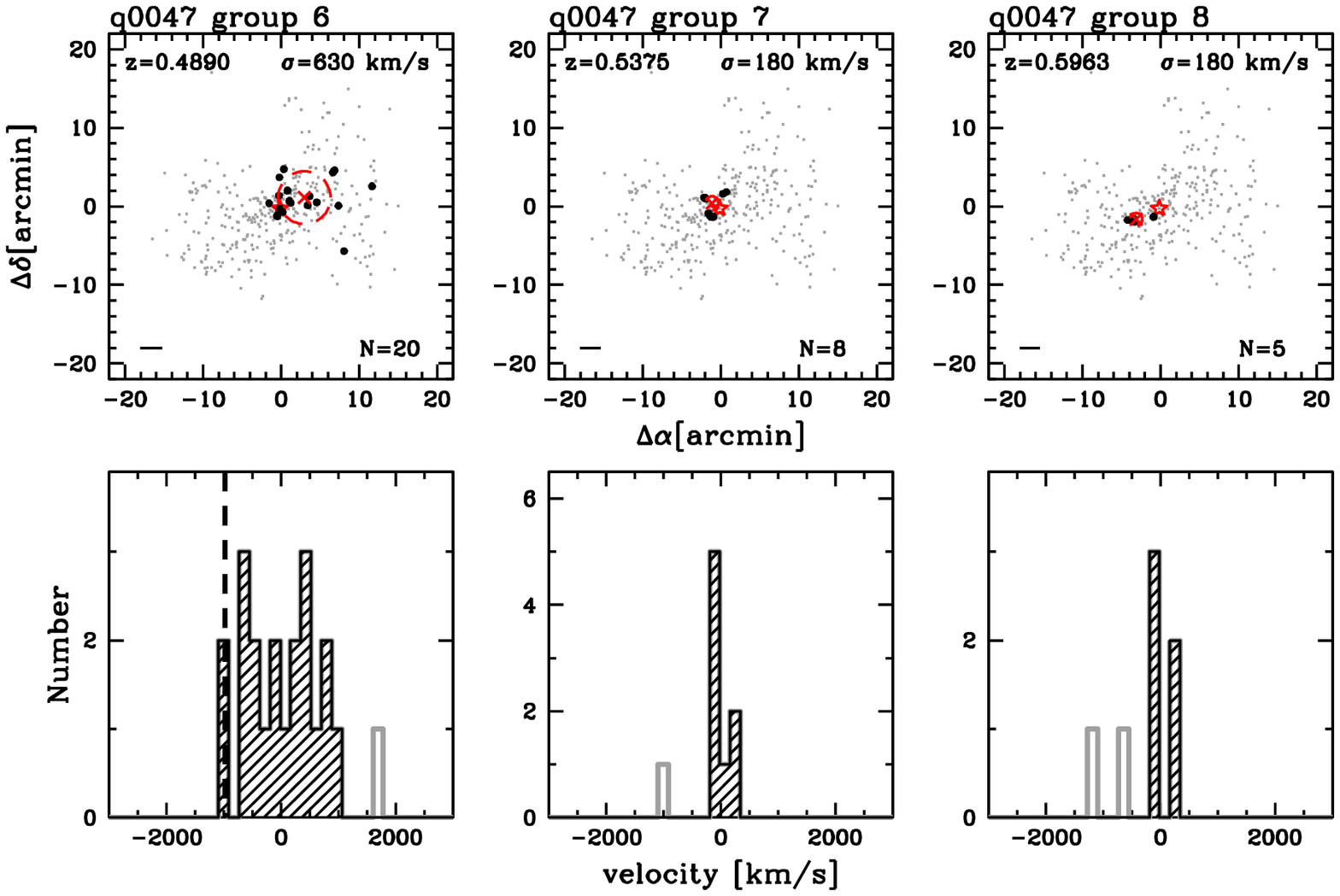}
\includegraphics[clip=true, width=18cm]{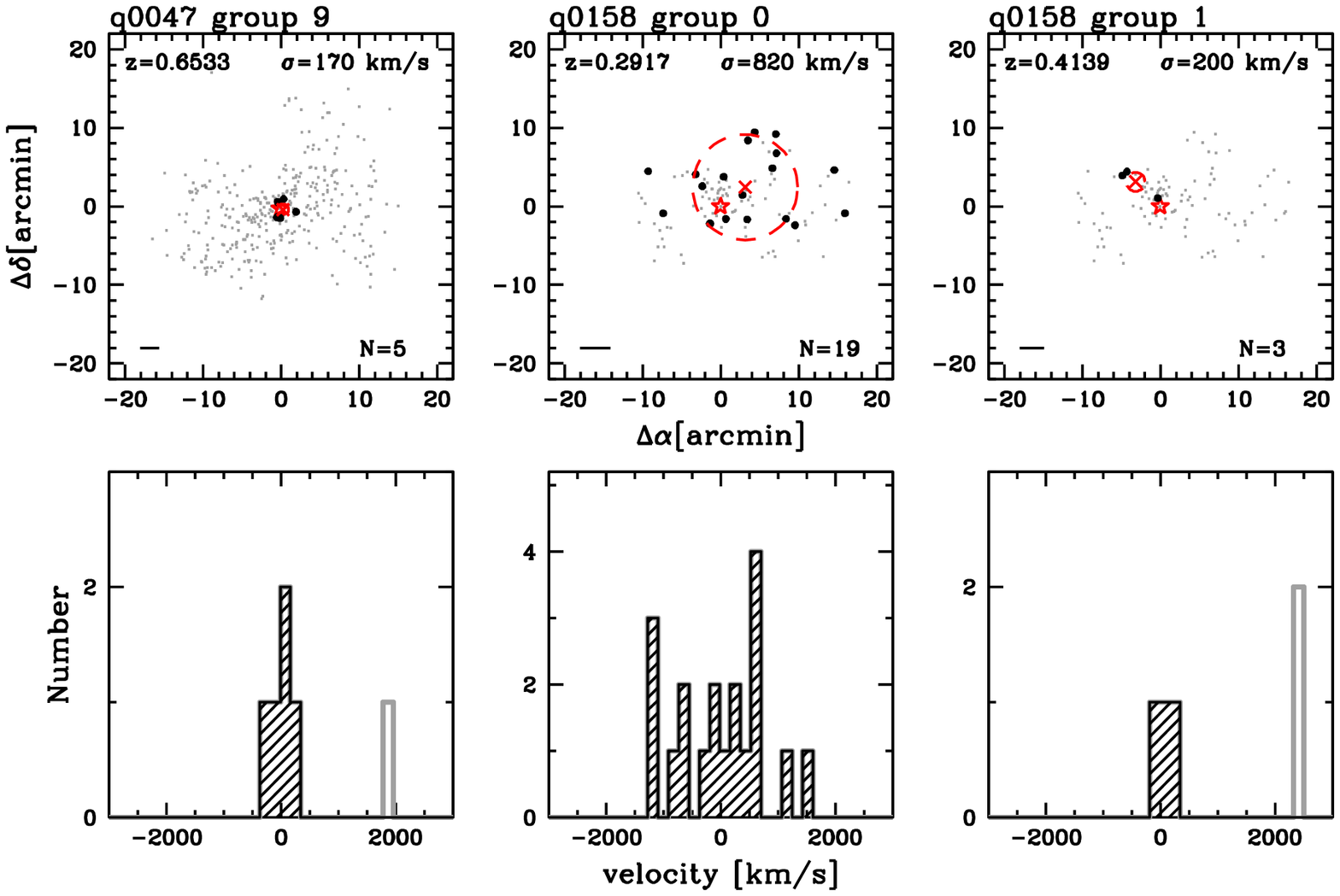}
\caption{Continued.}
\end{figure*}
\clearpage
\begin{figure*}
\ContinuedFloat
\includegraphics[clip=true, width=18cm]{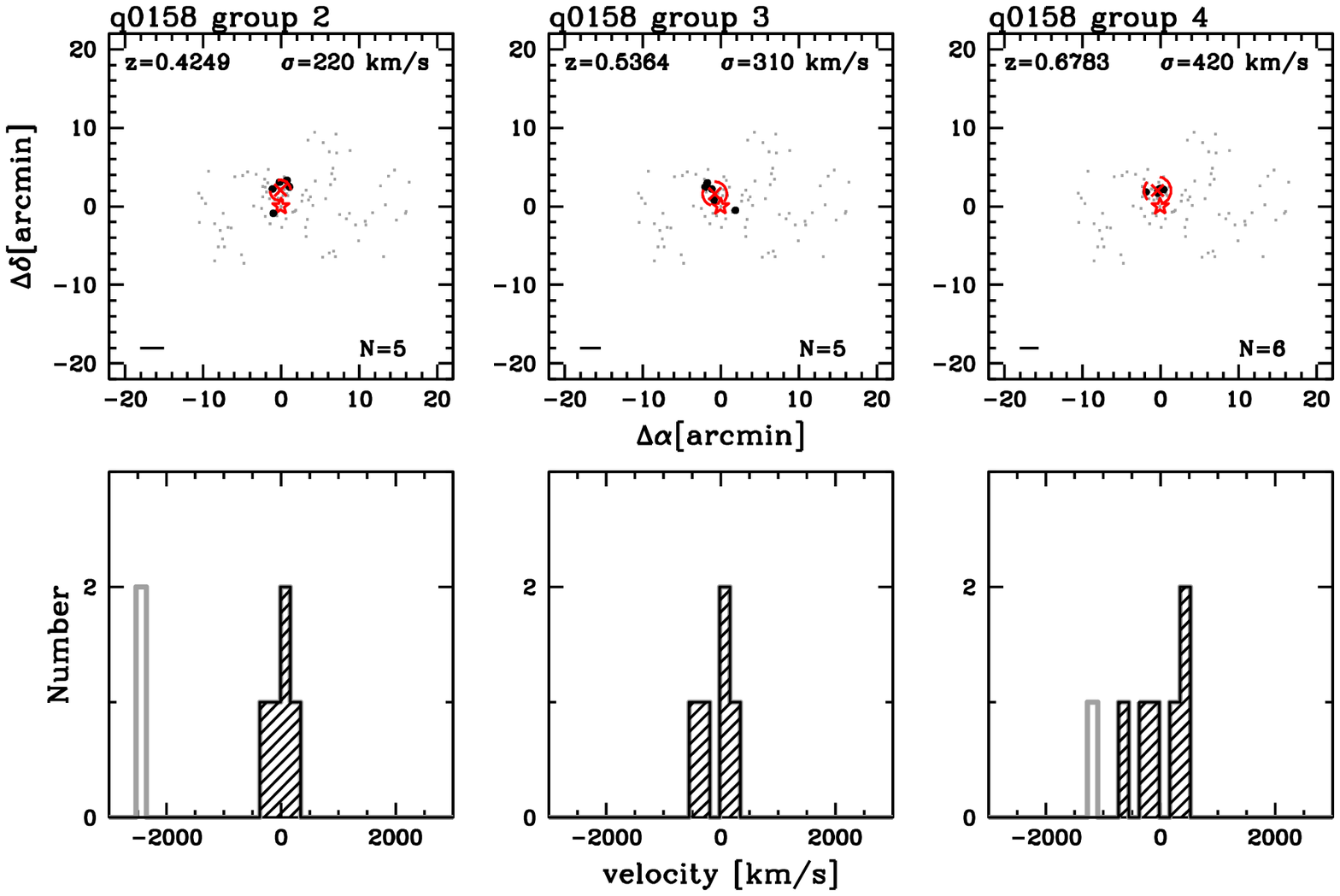}
\includegraphics[clip=true, width=18cm]{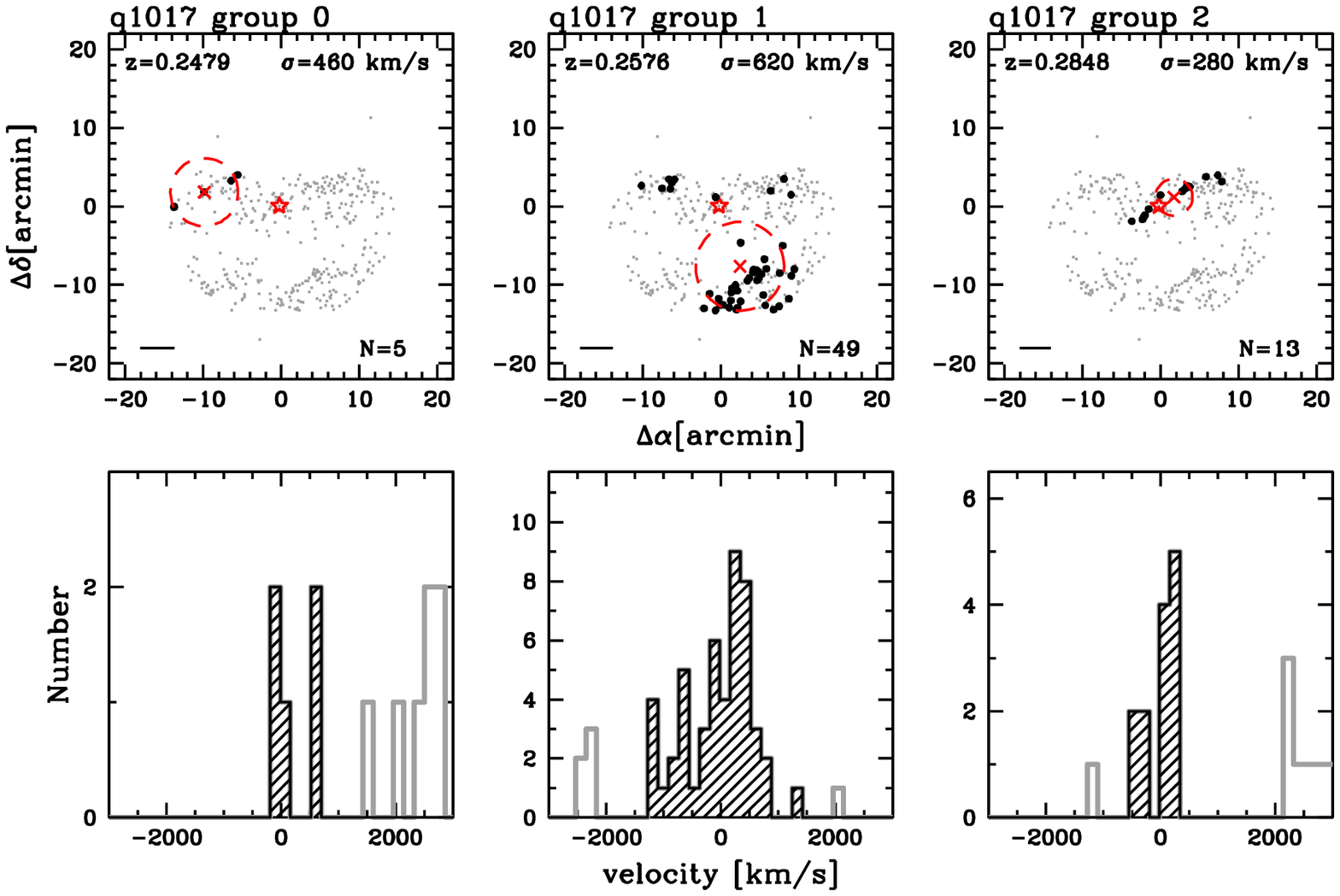}
\caption{Continued.}
\end{figure*}
\clearpage
\begin{figure*}
\ContinuedFloat
\includegraphics[clip=true, width=18cm]{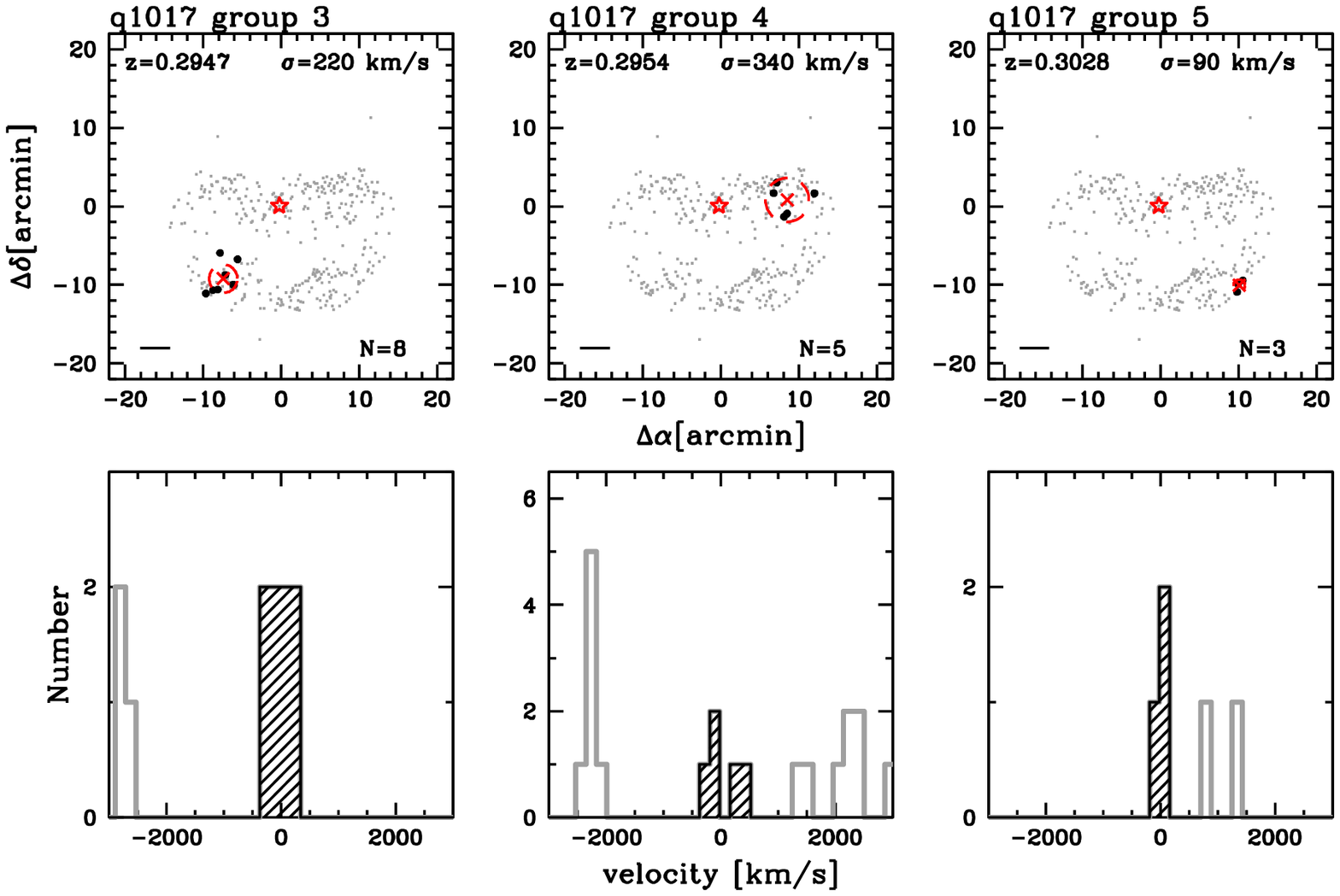}
\includegraphics[clip=true, width=18cm]{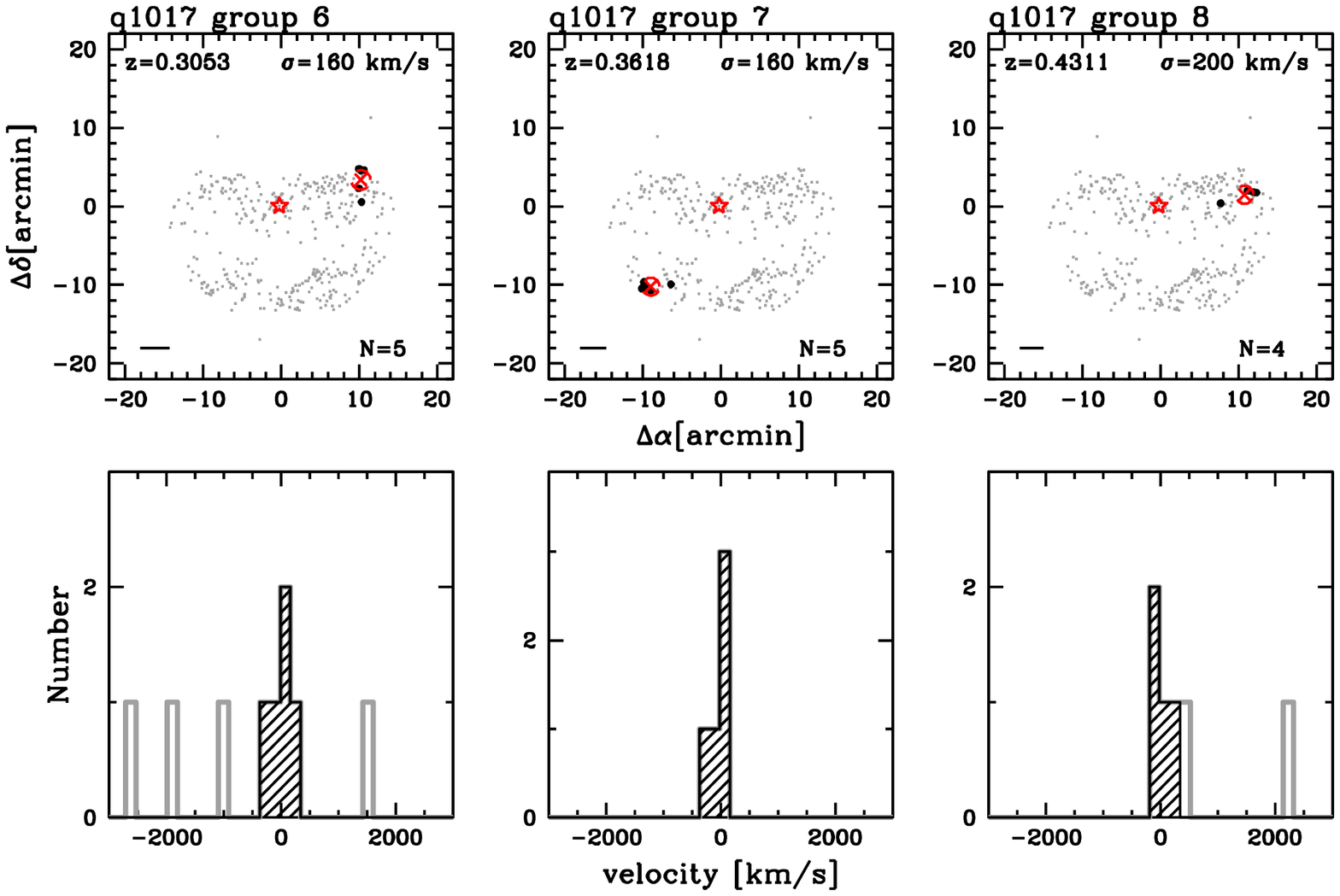}
\caption{Continued.}
\end{figure*}
\clearpage
\begin{figure*}
\ContinuedFloat
\includegraphics[clip=true, width=18cm]{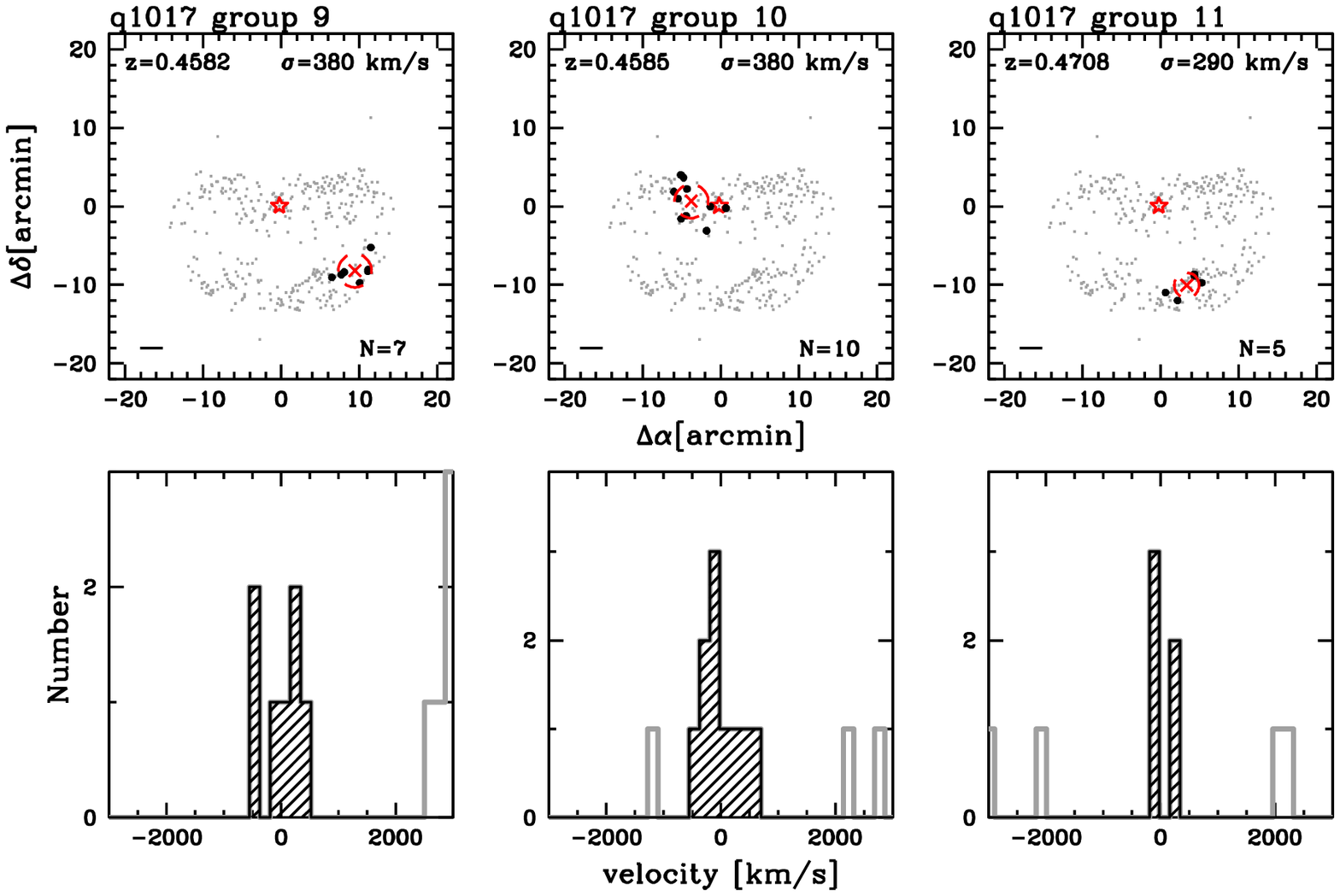}
\includegraphics[clip=true, width=18cm]{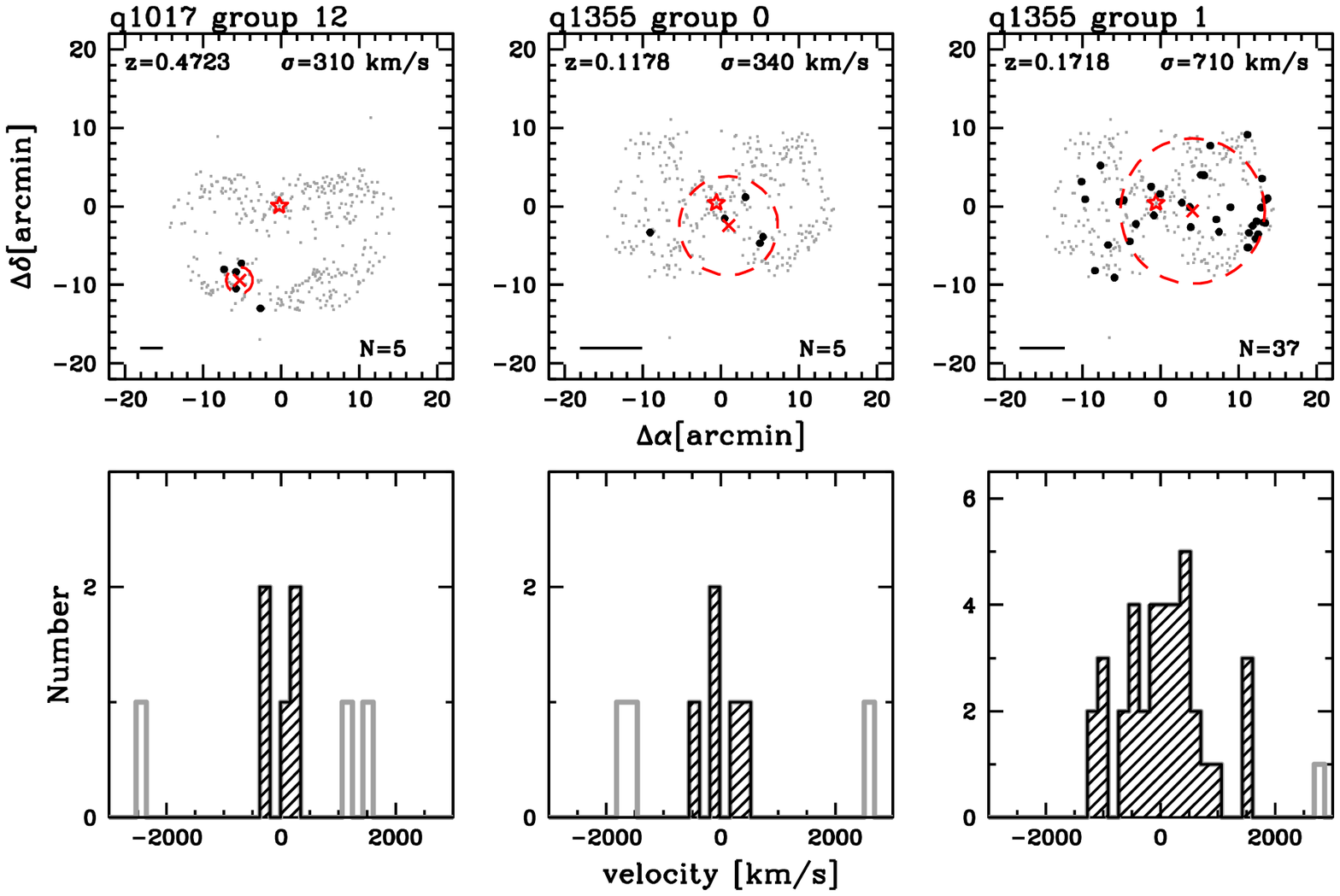}
\caption{Continued.}
\end{figure*}
\clearpage
\begin{figure*}
\ContinuedFloat
\includegraphics[clip=true, width=18cm]{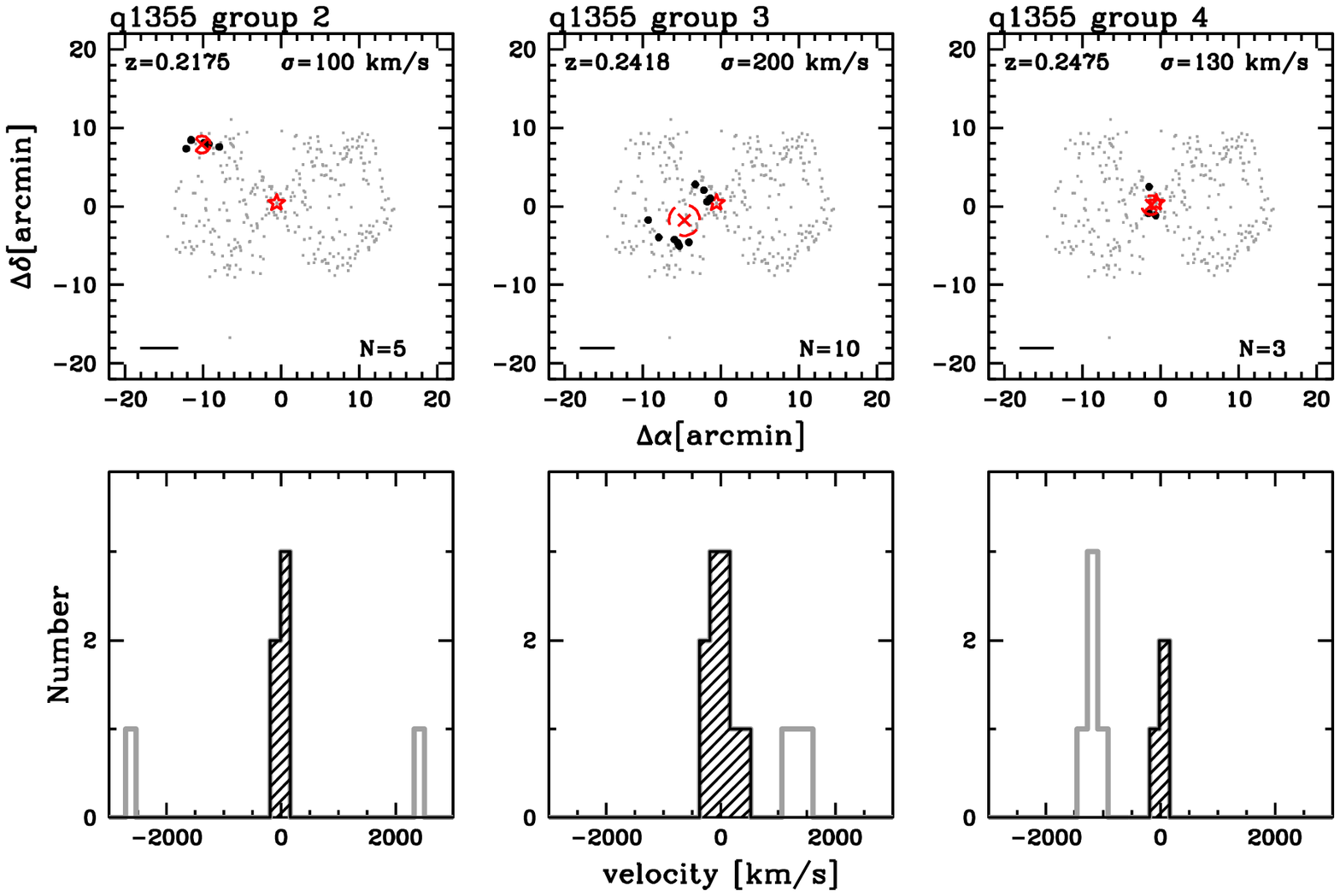}
\includegraphics[clip=true, width=18cm]{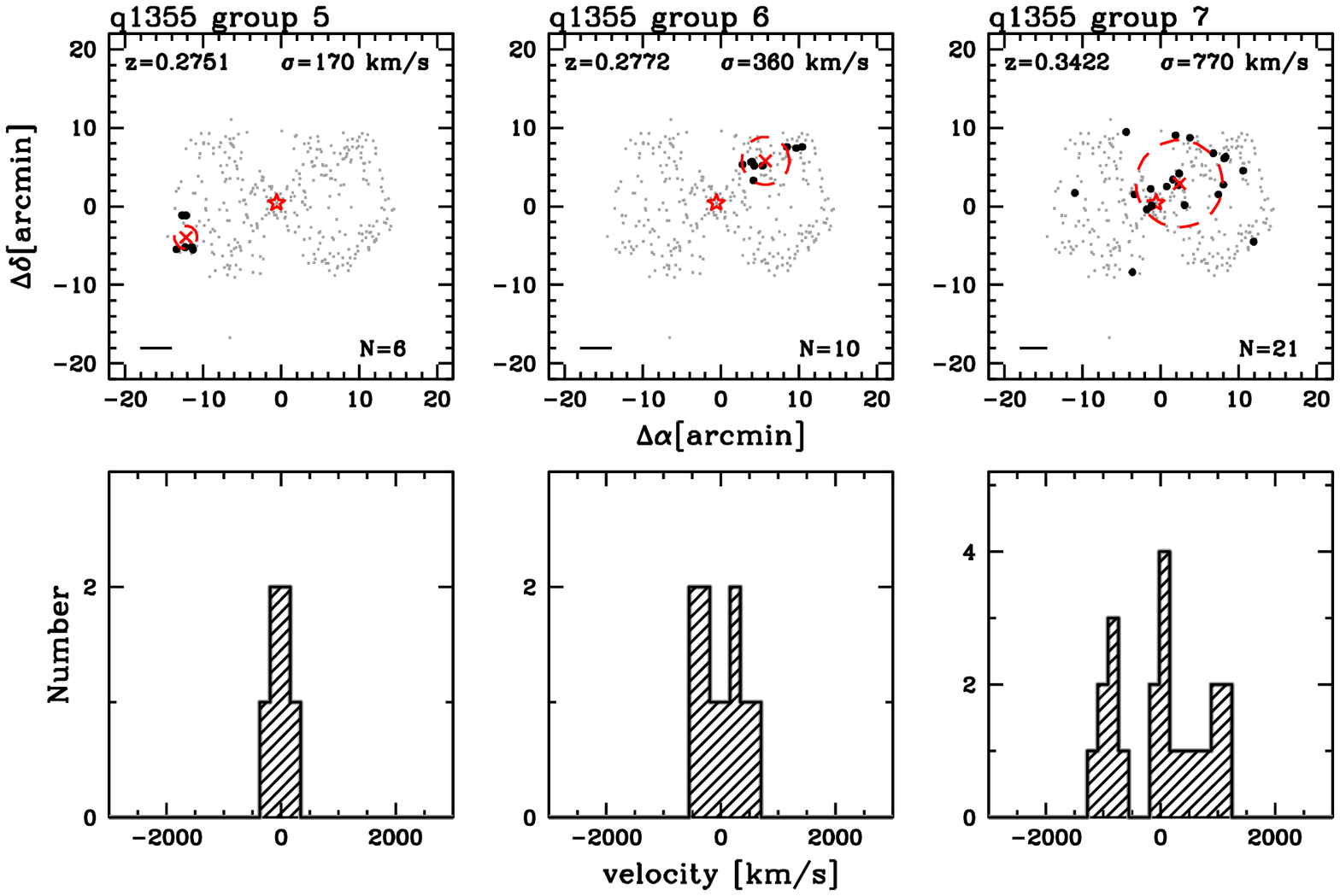}
\caption{Continued.}
\end{figure*}
\clearpage
\begin{figure*}
\ContinuedFloat
\includegraphics[clip=true, width=18cm]{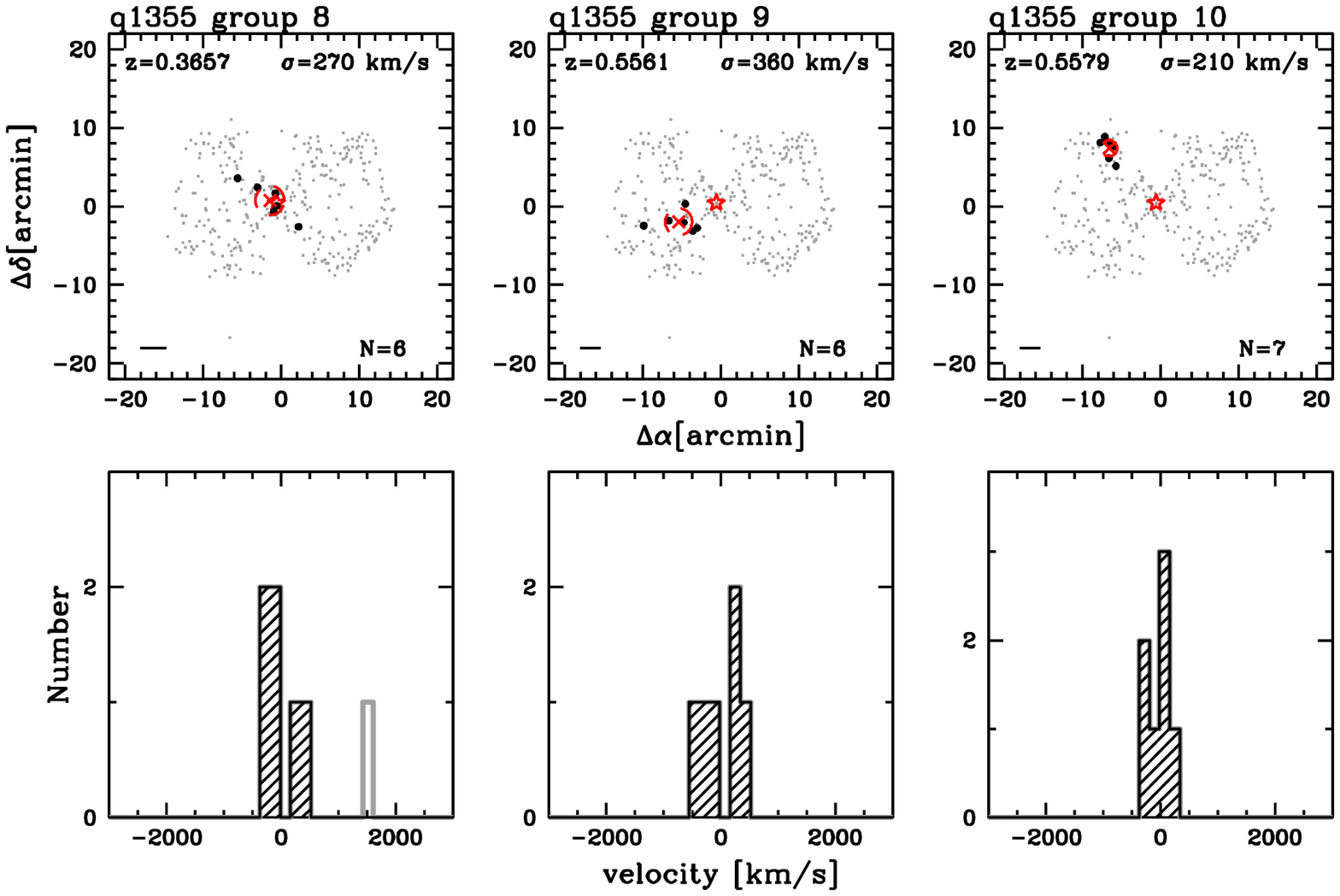}
\includegraphics[clip=true, width=18cm]{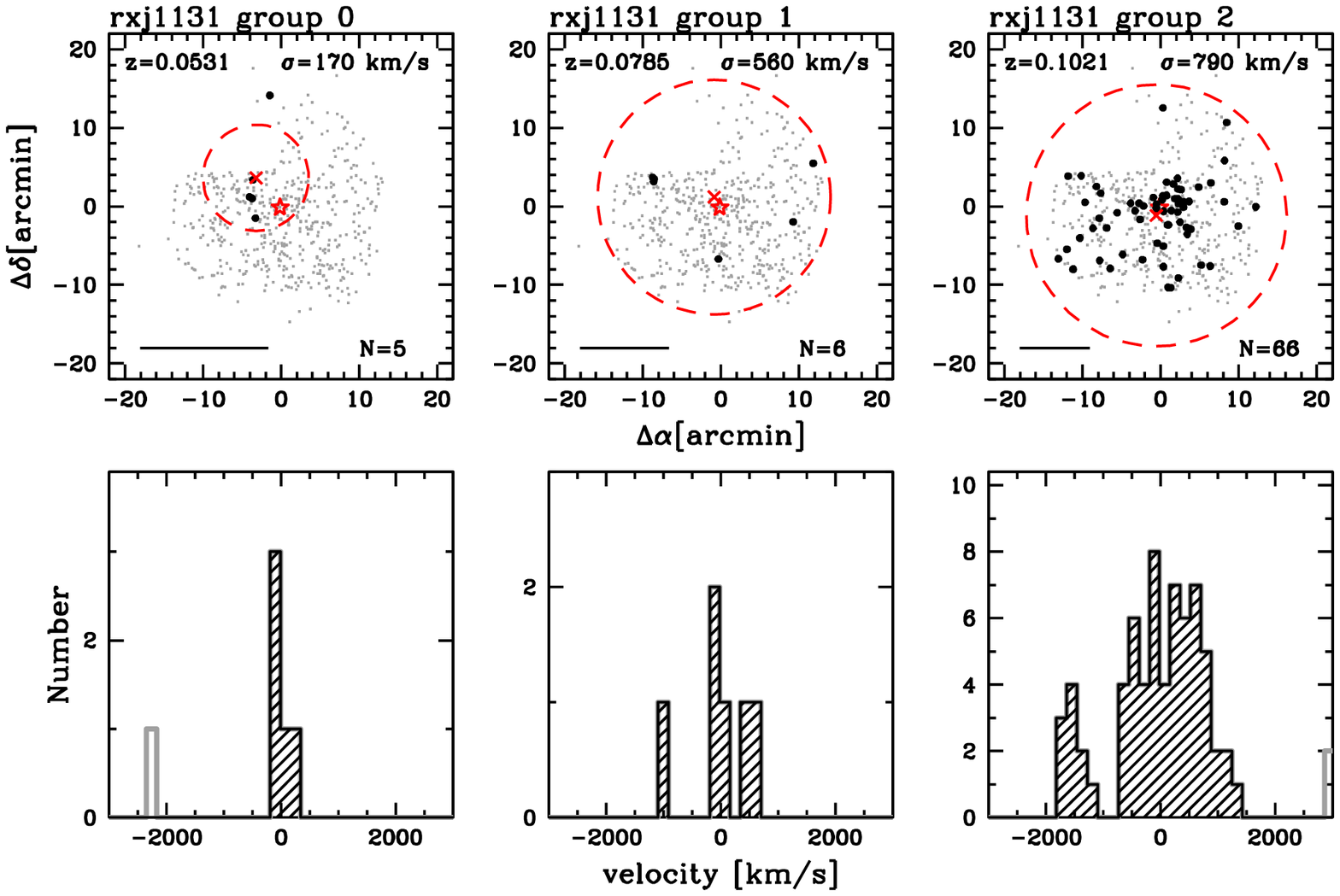}
\caption{Continued.}
\end{figure*}
\clearpage
\begin{figure*}
\ContinuedFloat
\includegraphics[clip=true, width=18cm]{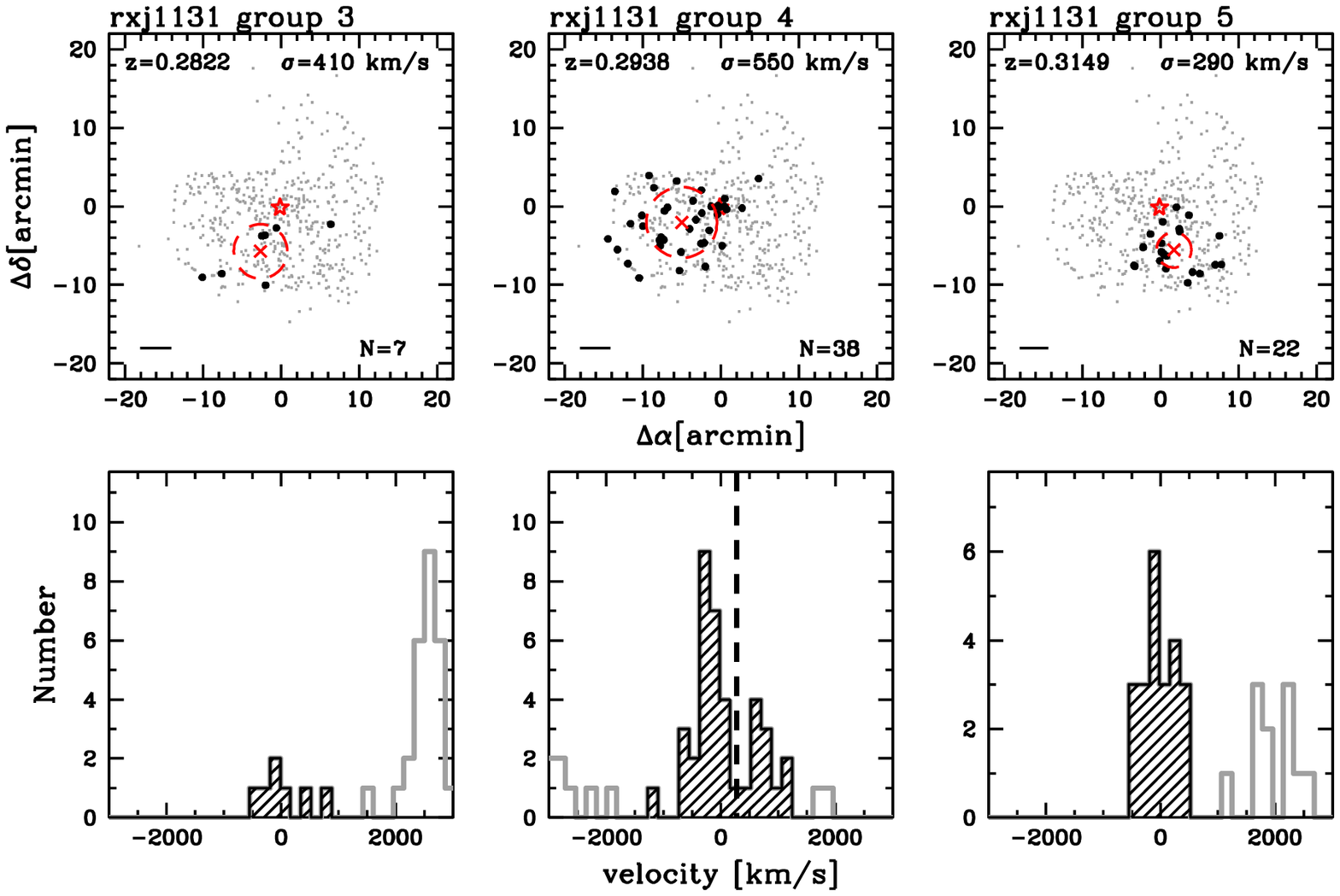}
\includegraphics[clip=true, width=18cm]{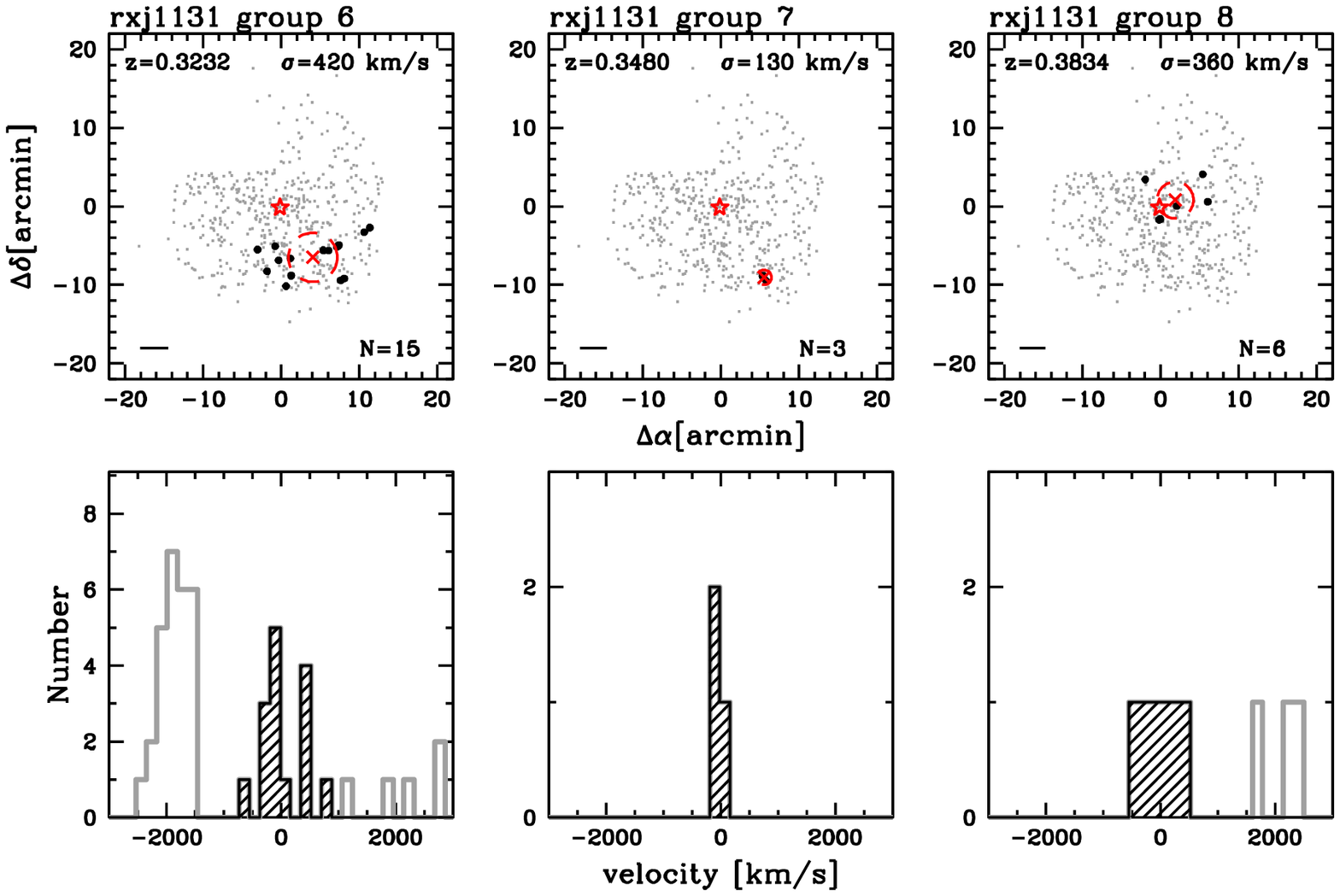}
\caption{Continued.}
\end{figure*}
\clearpage
\begin{figure*}
\ContinuedFloat
\includegraphics[clip=true, width=18cm]{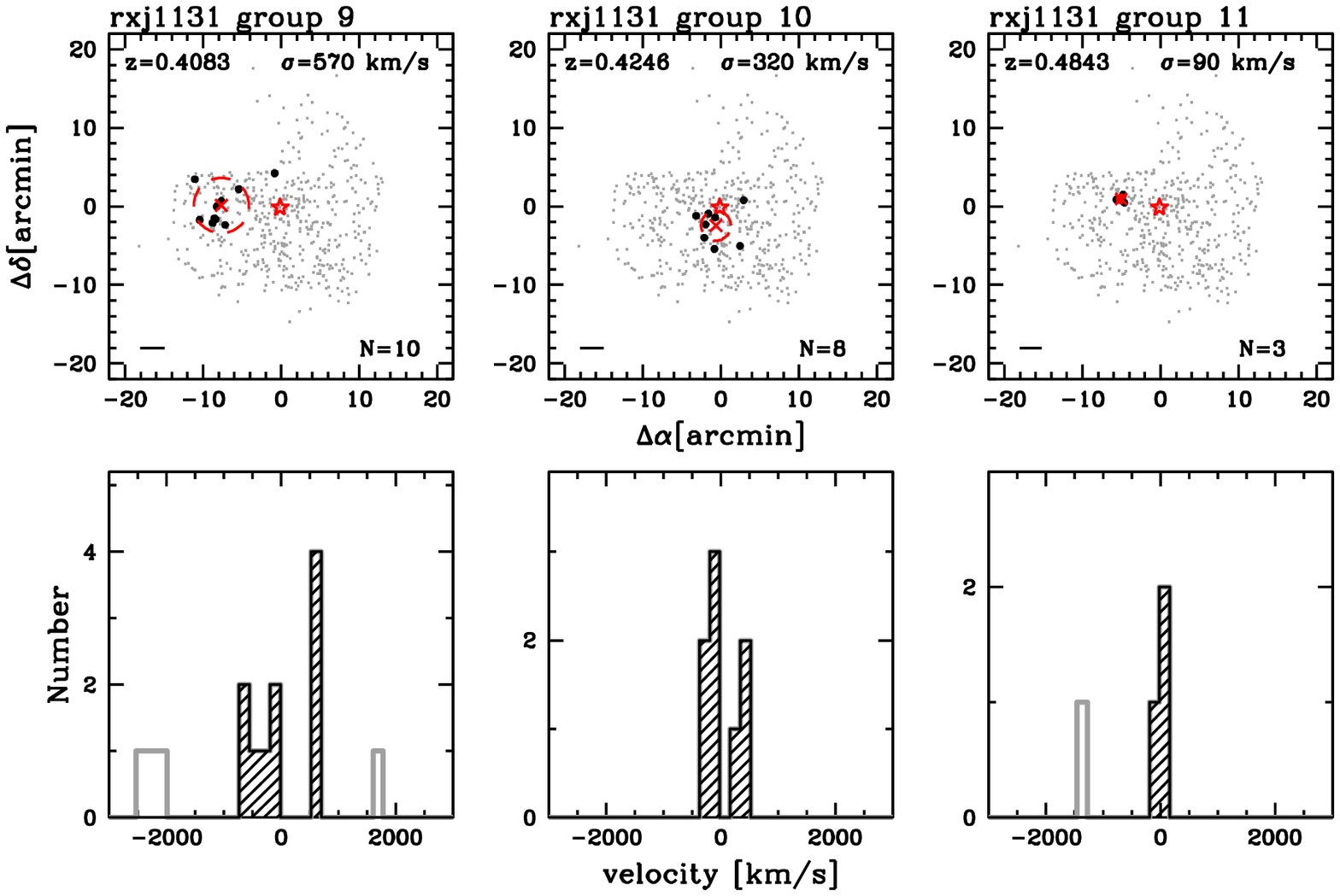}
\includegraphics[clip=true, width=18cm]{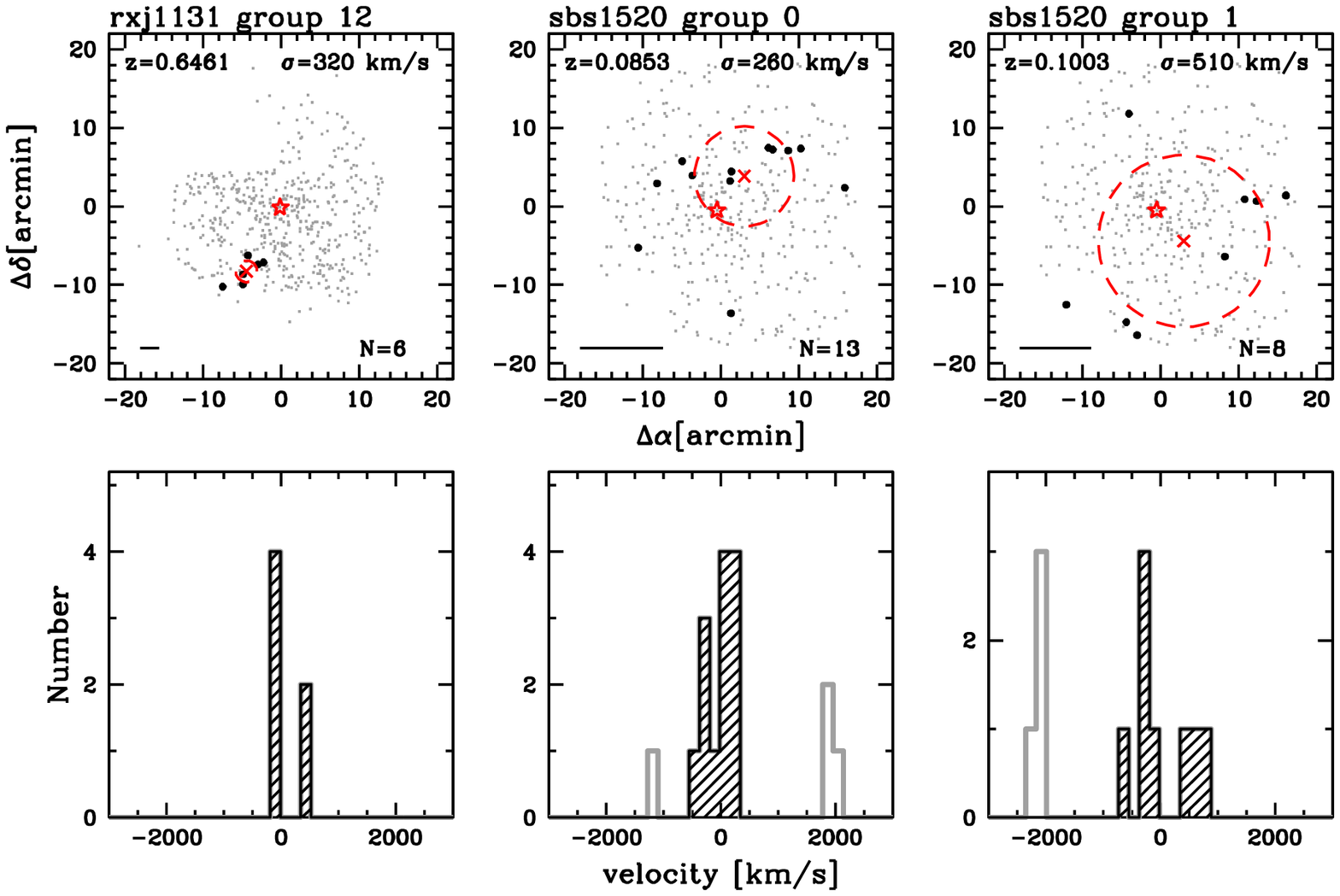}
\caption{Continued.}
\end{figure*}
\clearpage
\begin{figure*}
\ContinuedFloat
\includegraphics[clip=true, width=18cm]{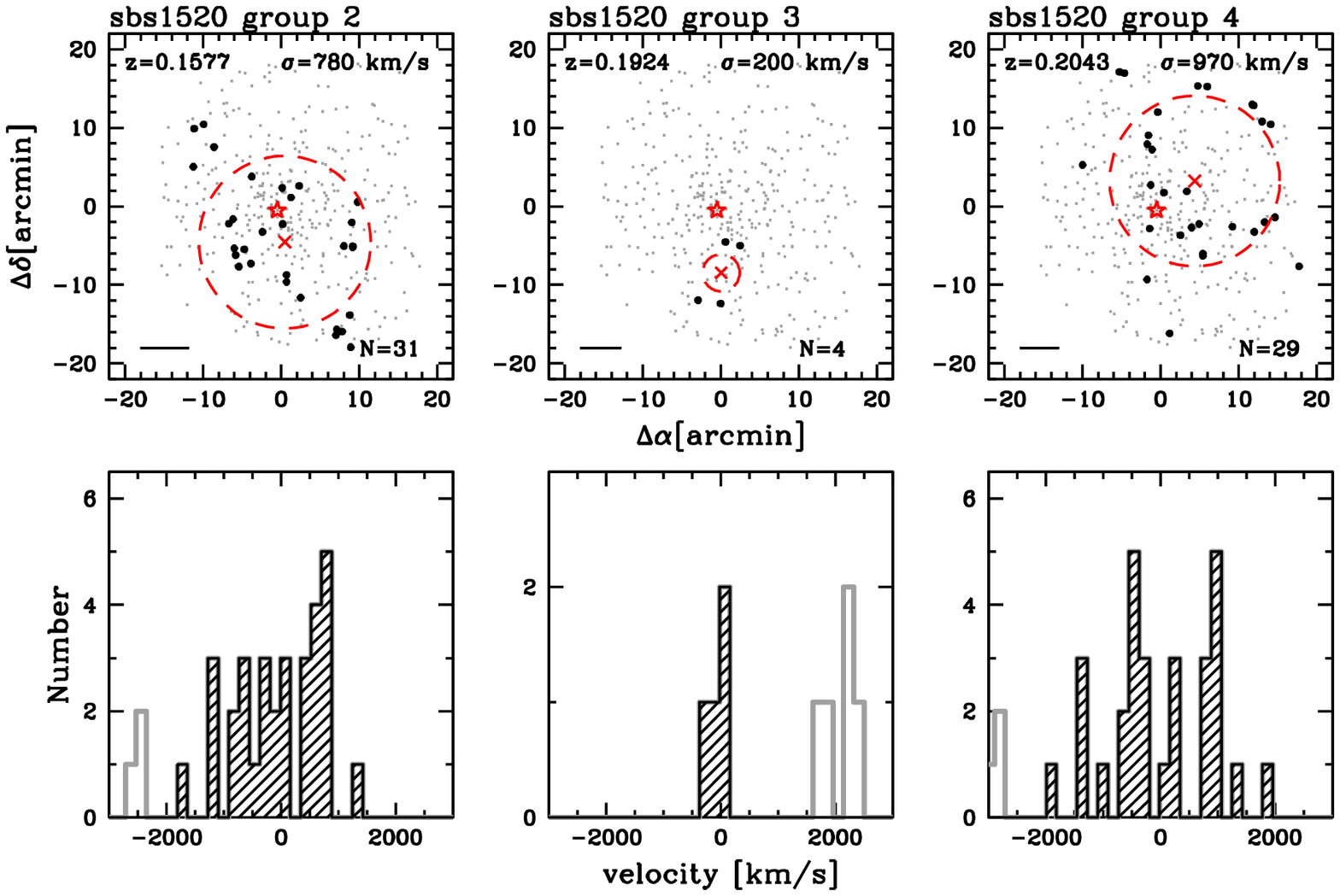}
\includegraphics[clip=true, width=18cm]{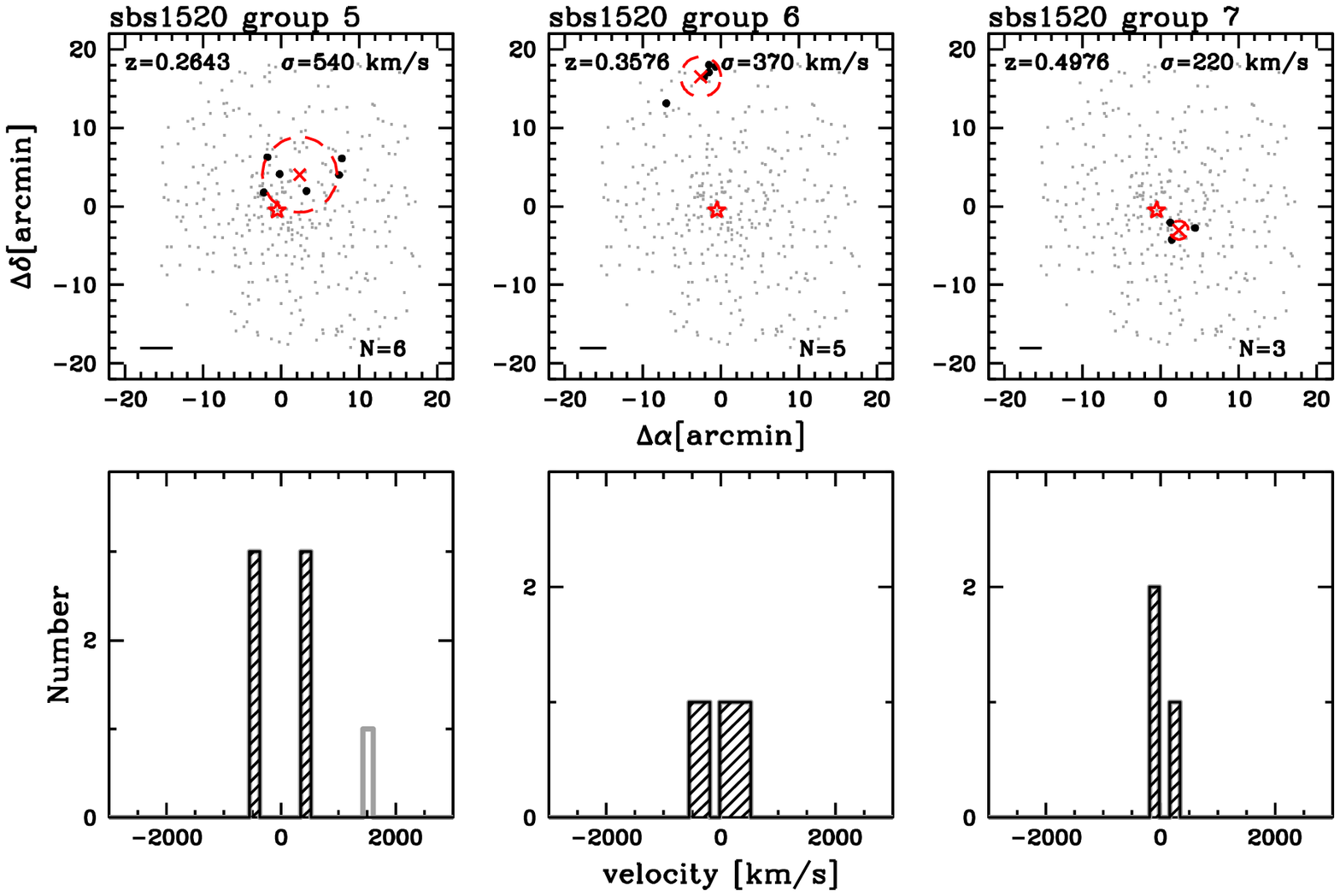}
\caption{Continued.}
\end{figure*}
\clearpage
\begin{figure*}
\ContinuedFloat
\includegraphics[clip=true, width=18cm]{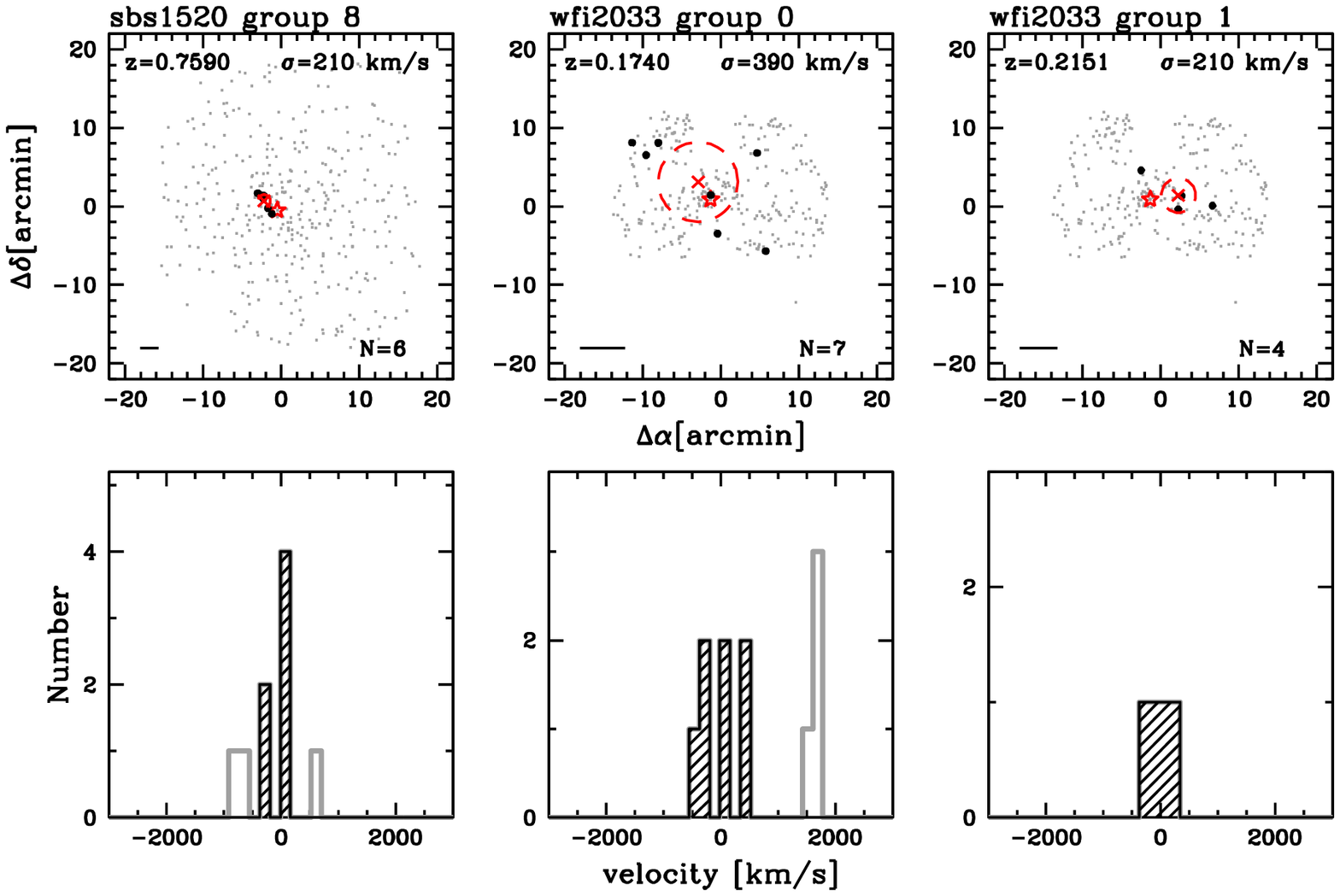}
\includegraphics[clip=true, width=18cm]{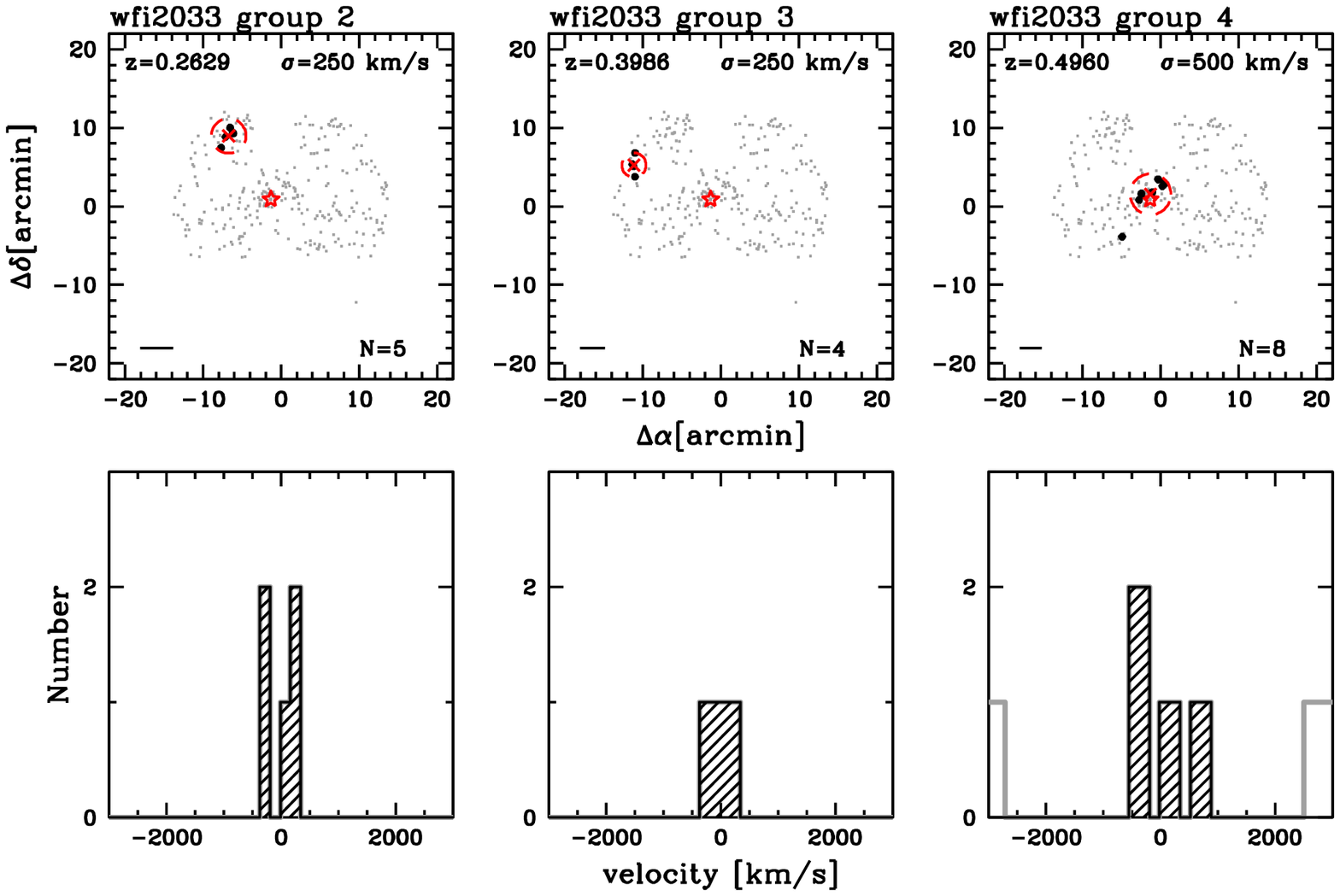}
\caption{Continued.}
\end{figure*}
\clearpage
\begin{figure*}
\ContinuedFloat
\includegraphics[clip=true, width=12cm]{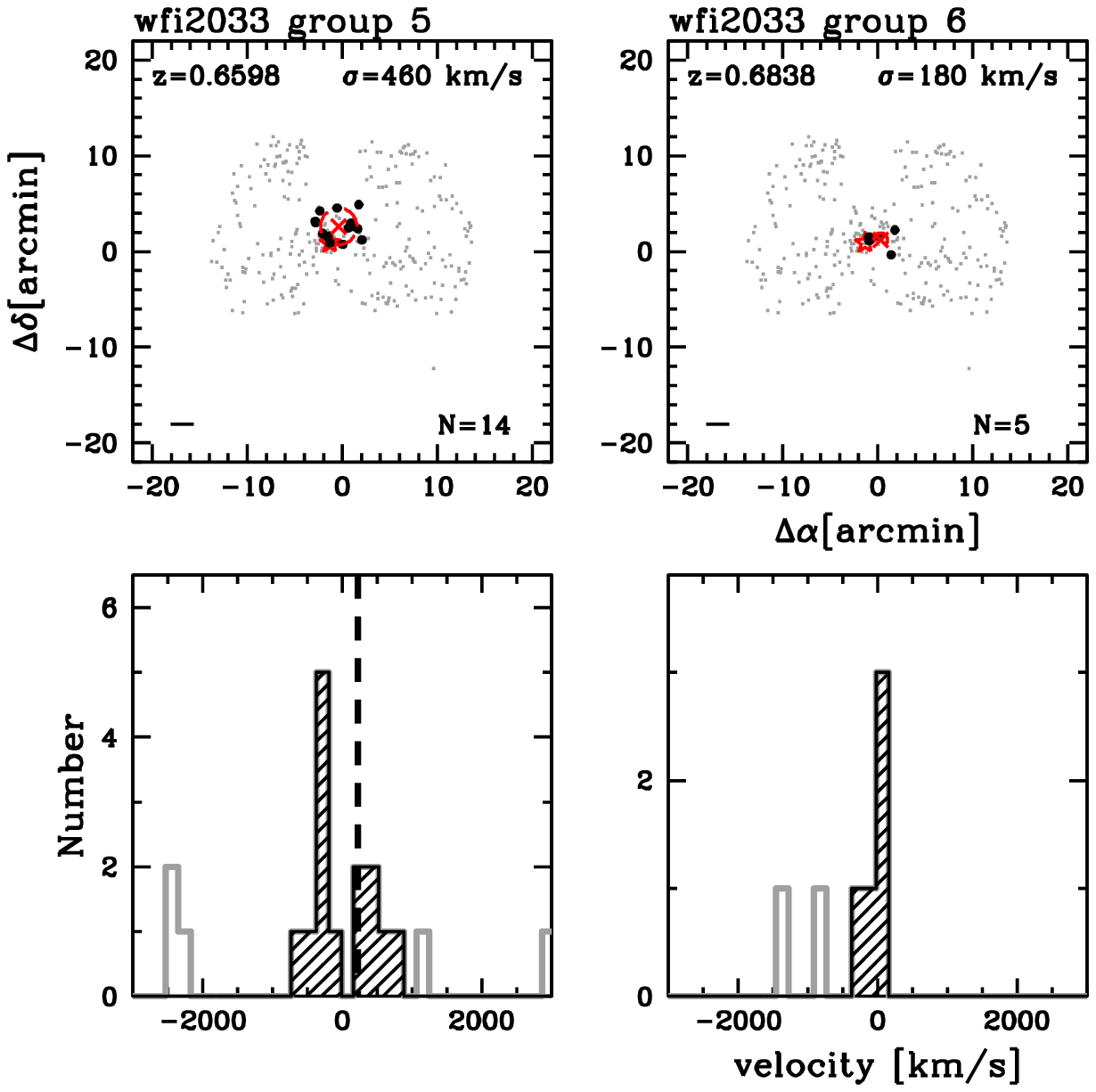}
\caption{Continued.}
\end{figure*}

\begin{figure*}
\includegraphics[clip=true, width=18cm]{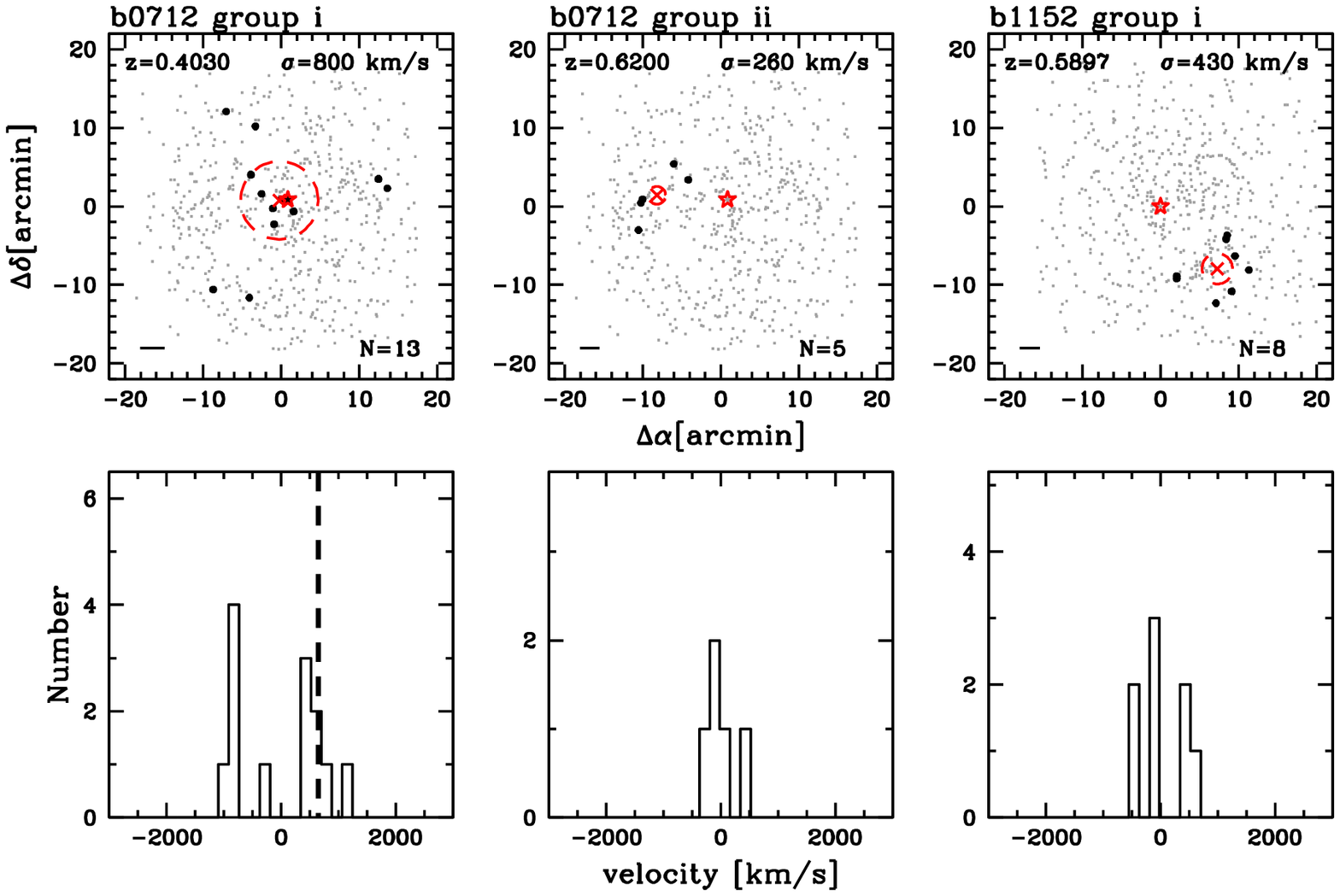}
\includegraphics[clip=true, width=18cm]{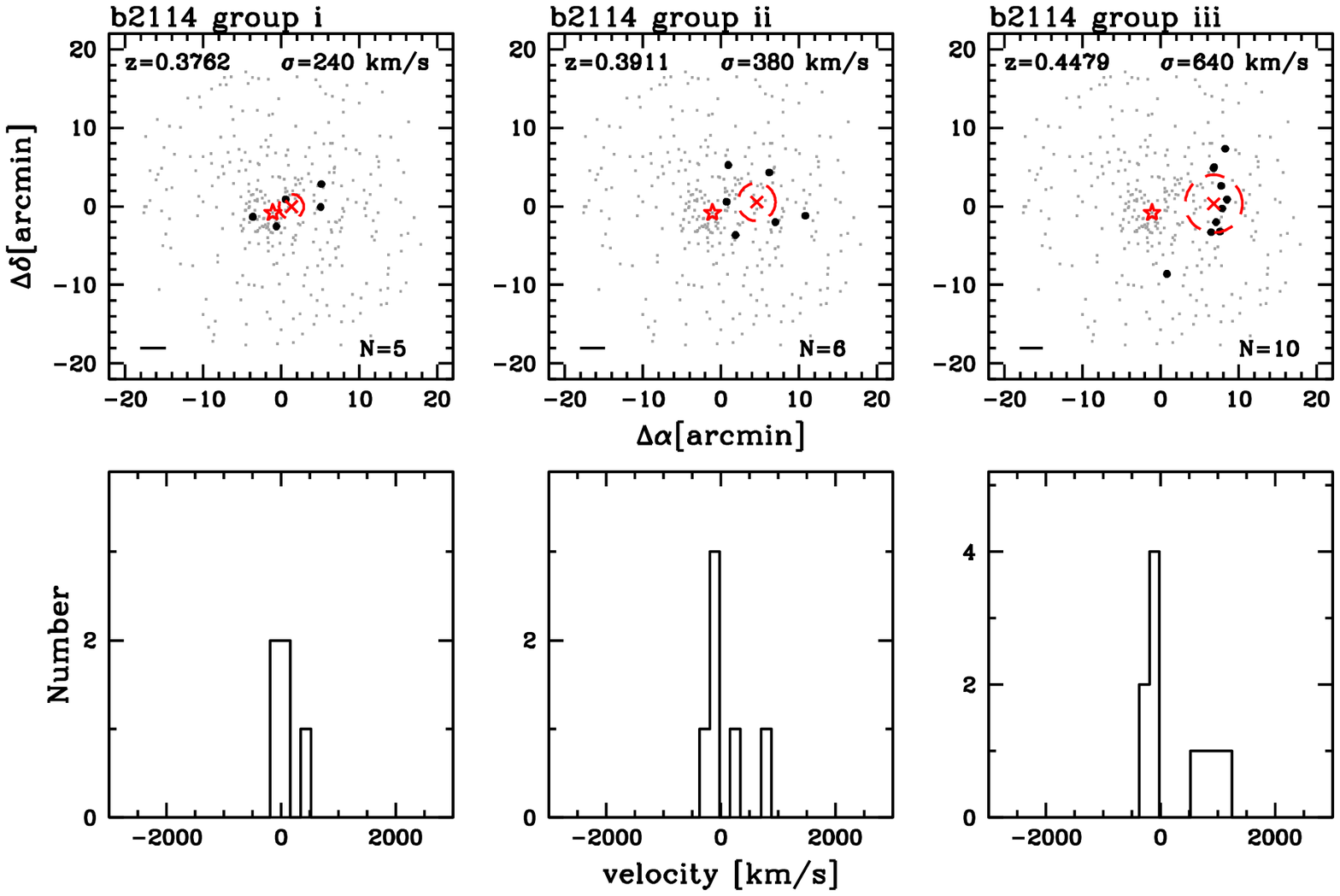}
\caption{As Figure \ref{fig:grpskyplotsandvelhists} but for the candidate groups.  There are no gray histograms, since we use no spatial criteria for original candidate group identification.}
\label{fig:grpskyplotsandvelhistscandgrps}
\end{figure*}
\clearpage
\begin{figure*}
\ContinuedFloat
\includegraphics[clip=true, width=18cm]{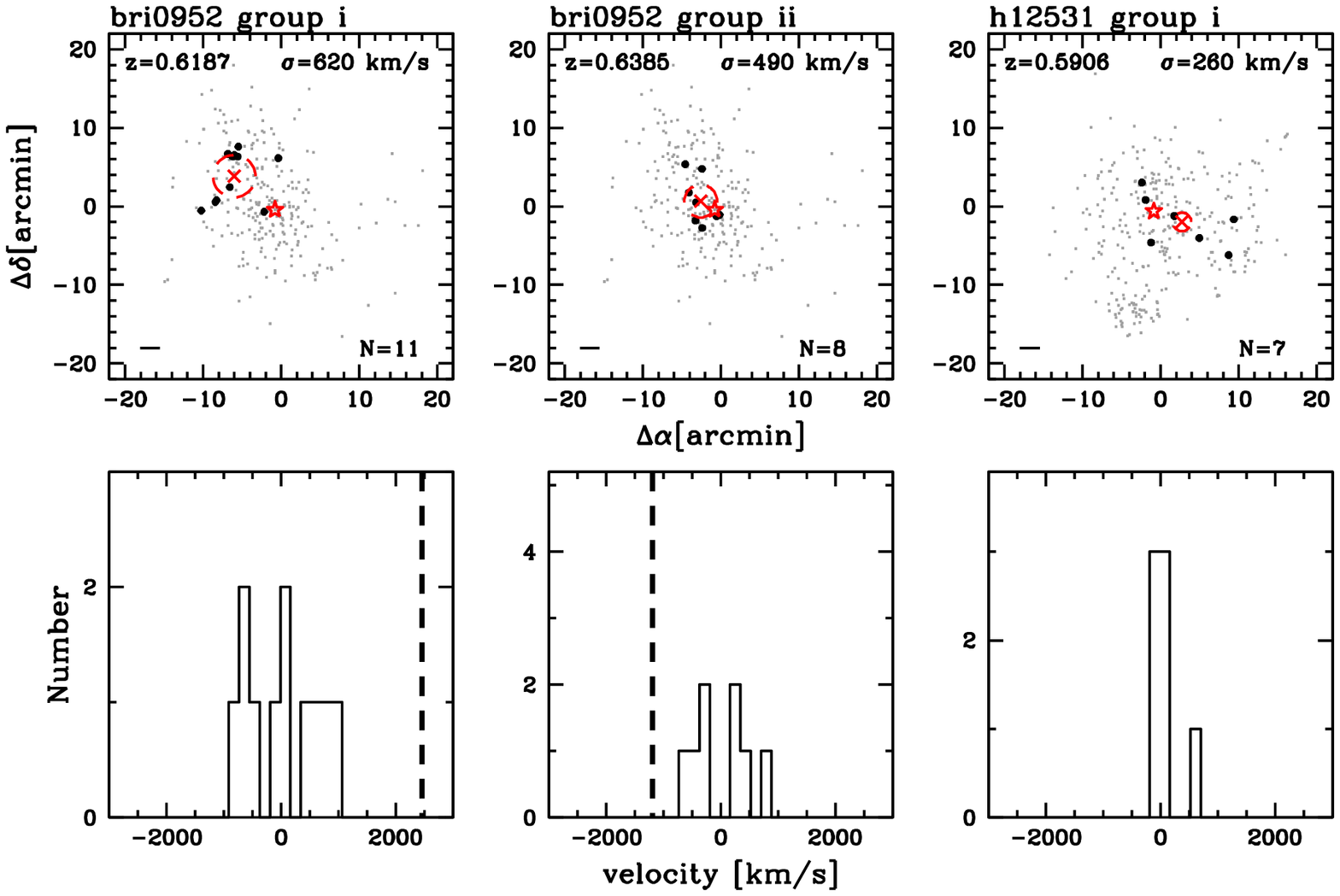}
\includegraphics[clip=true, width=18cm]{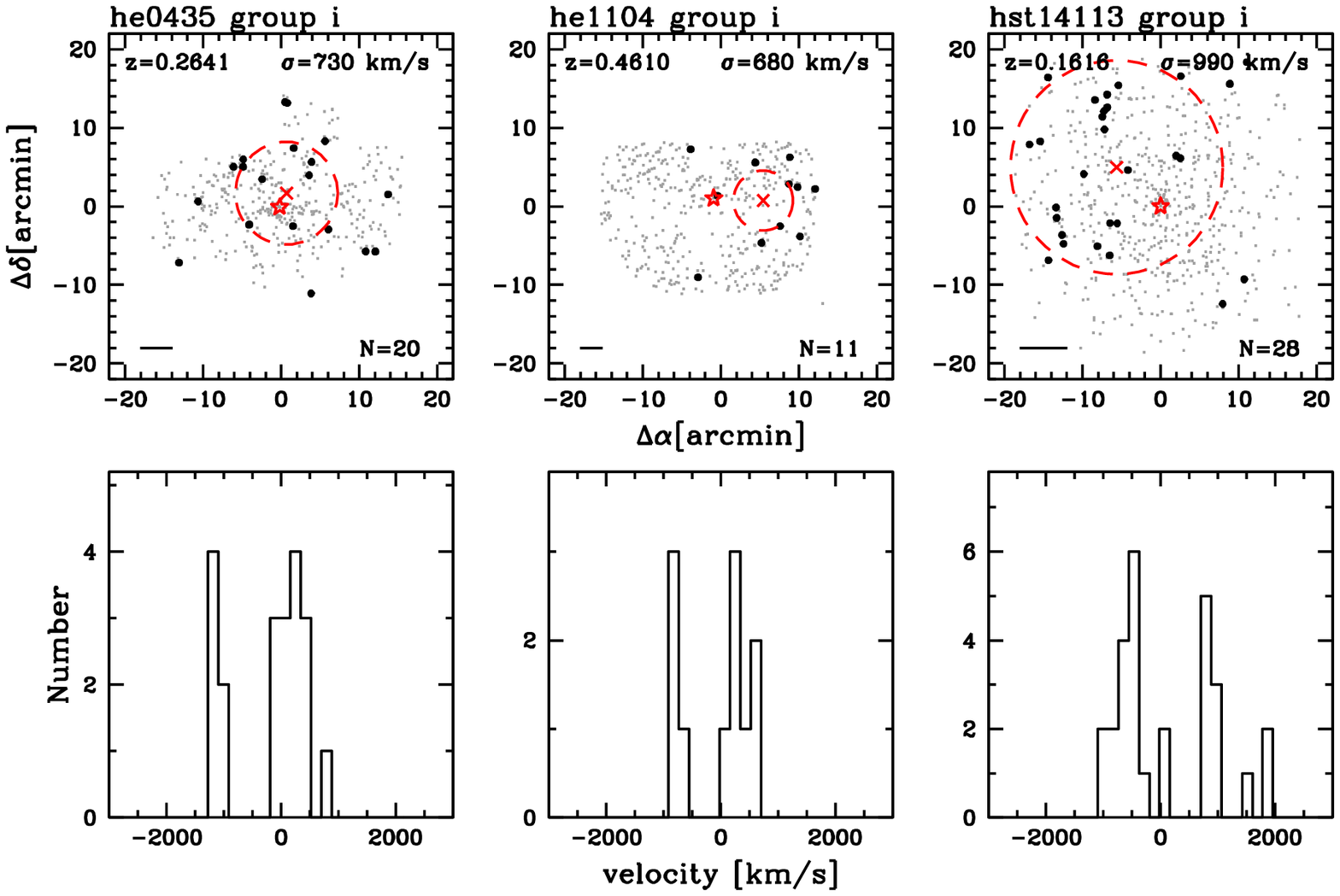}
\caption{Continued.}
\end{figure*}
\clearpage
\begin{figure*}
\ContinuedFloat
\includegraphics[clip=true, width=18cm]{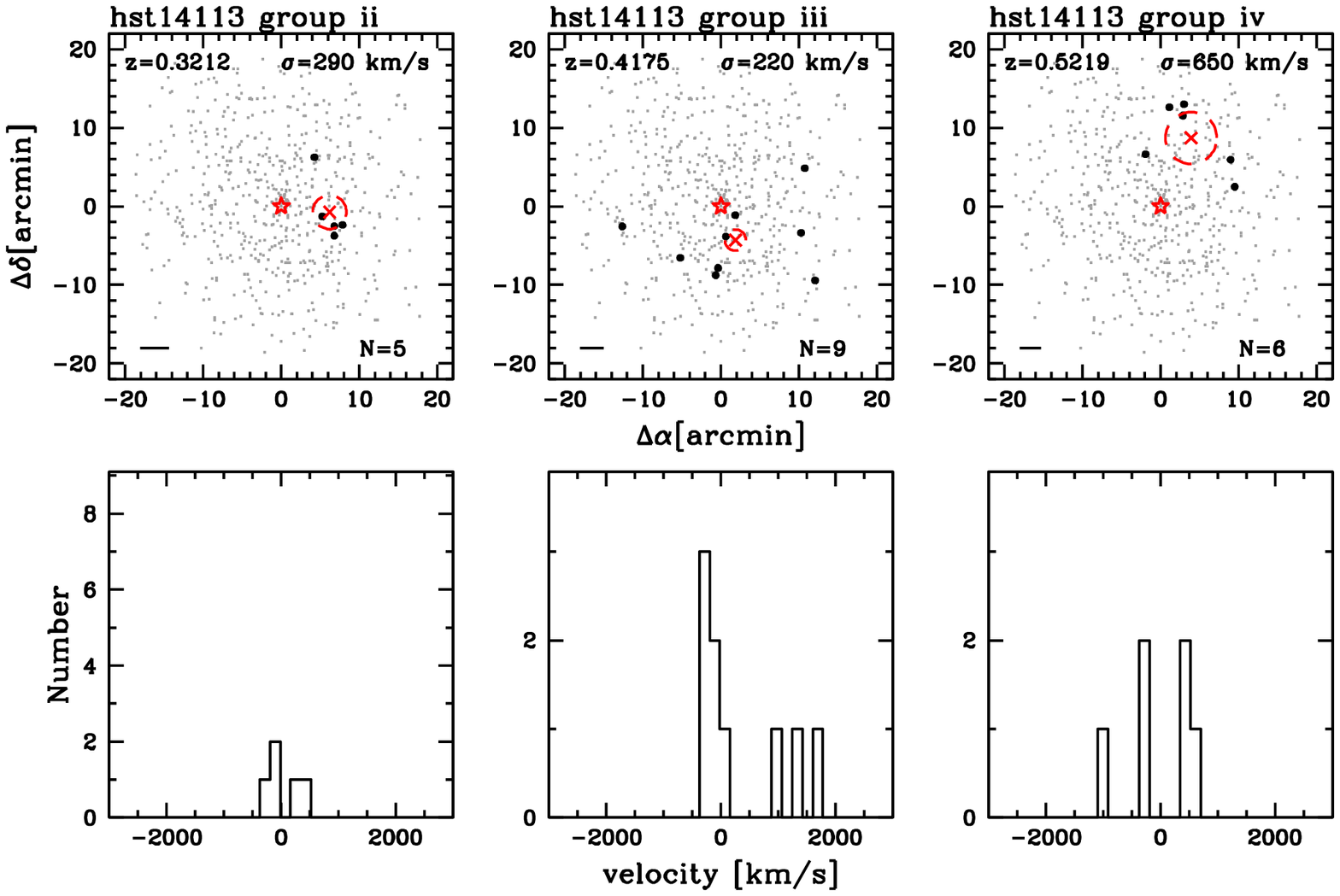}
\includegraphics[clip=true, width=18cm]{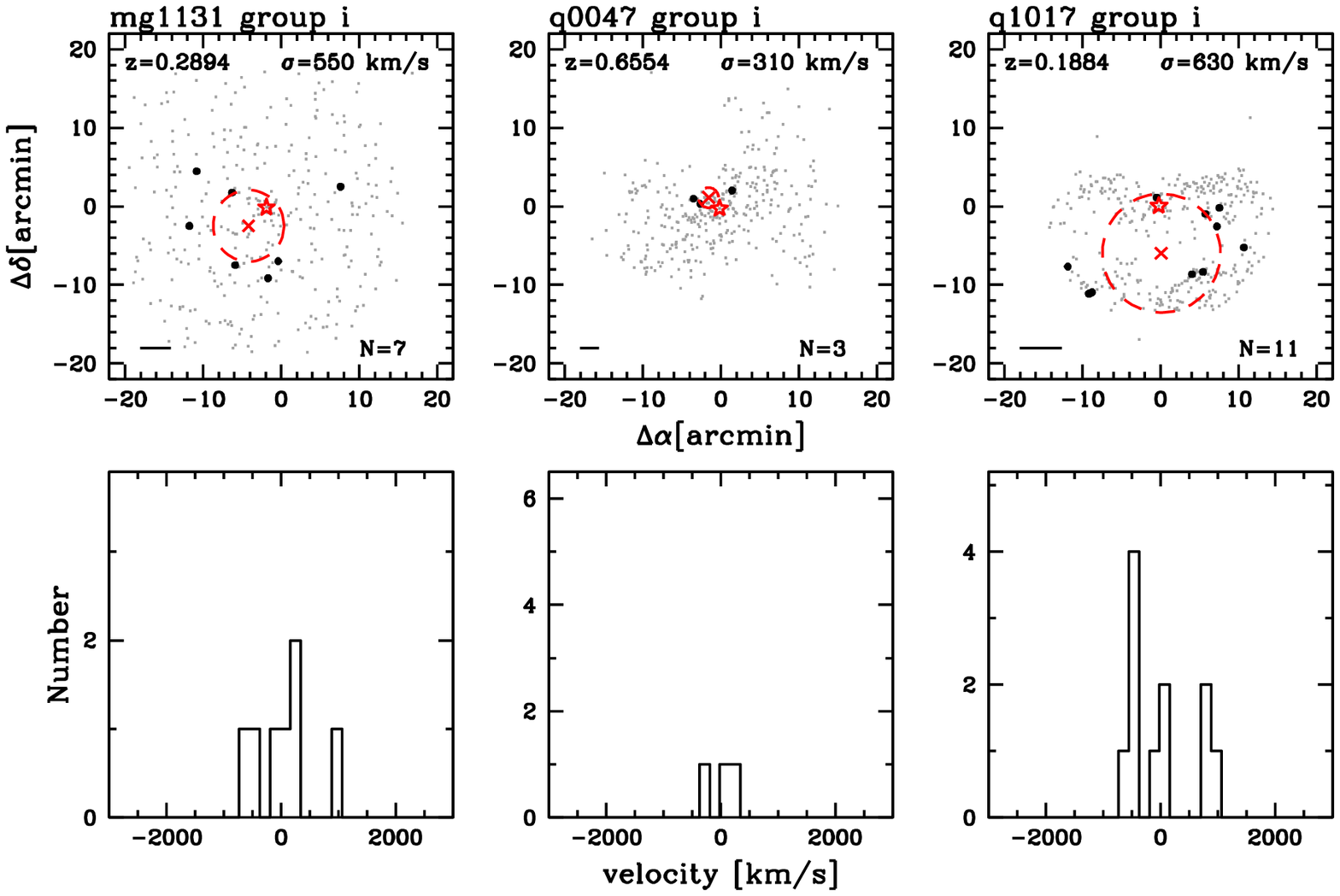}
\caption{Continued.}
\end{figure*}
\clearpage
\begin{figure*}
\ContinuedFloat
\includegraphics[clip=true, width=18cm]{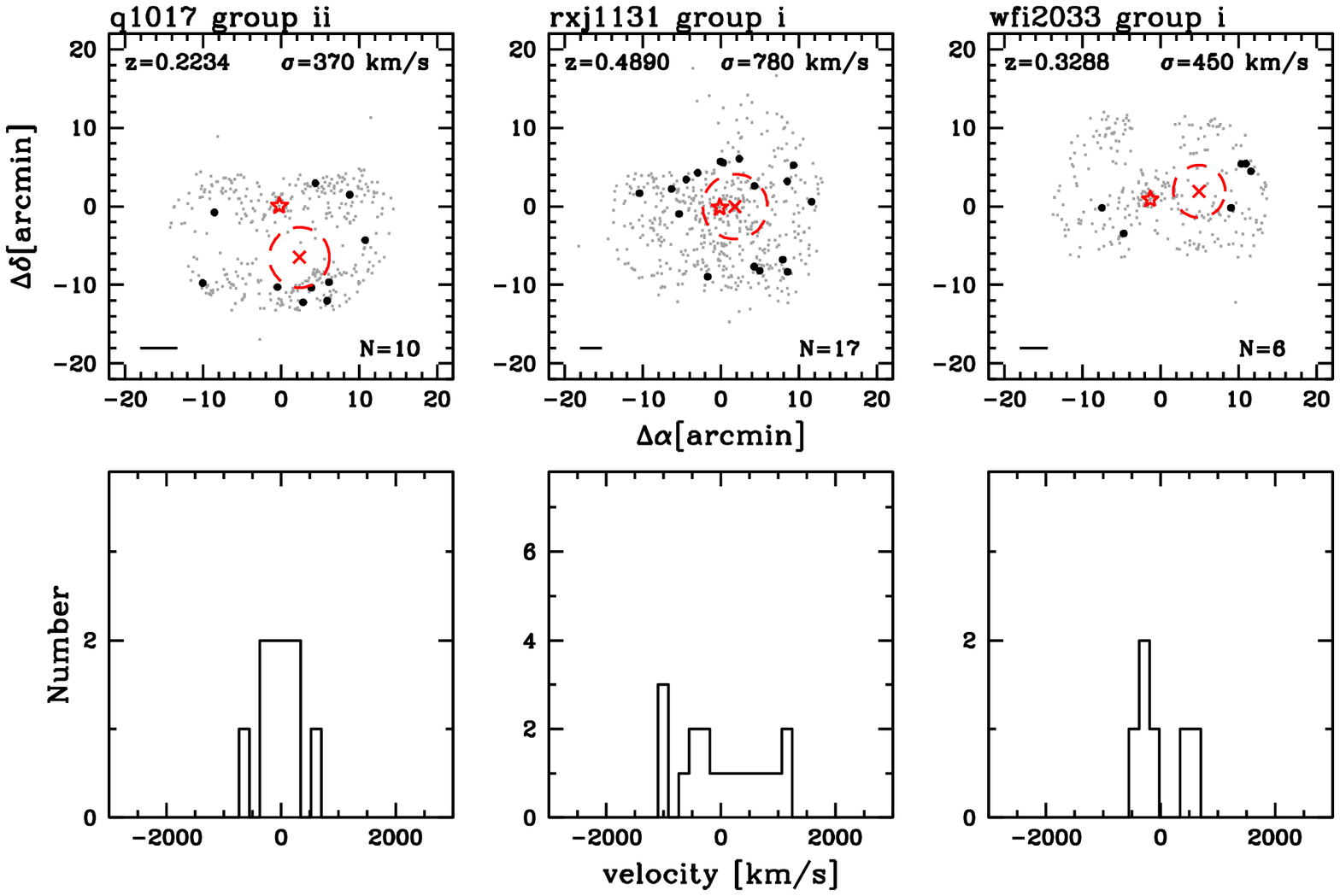}
\includegraphics[clip=true, width=12cm]{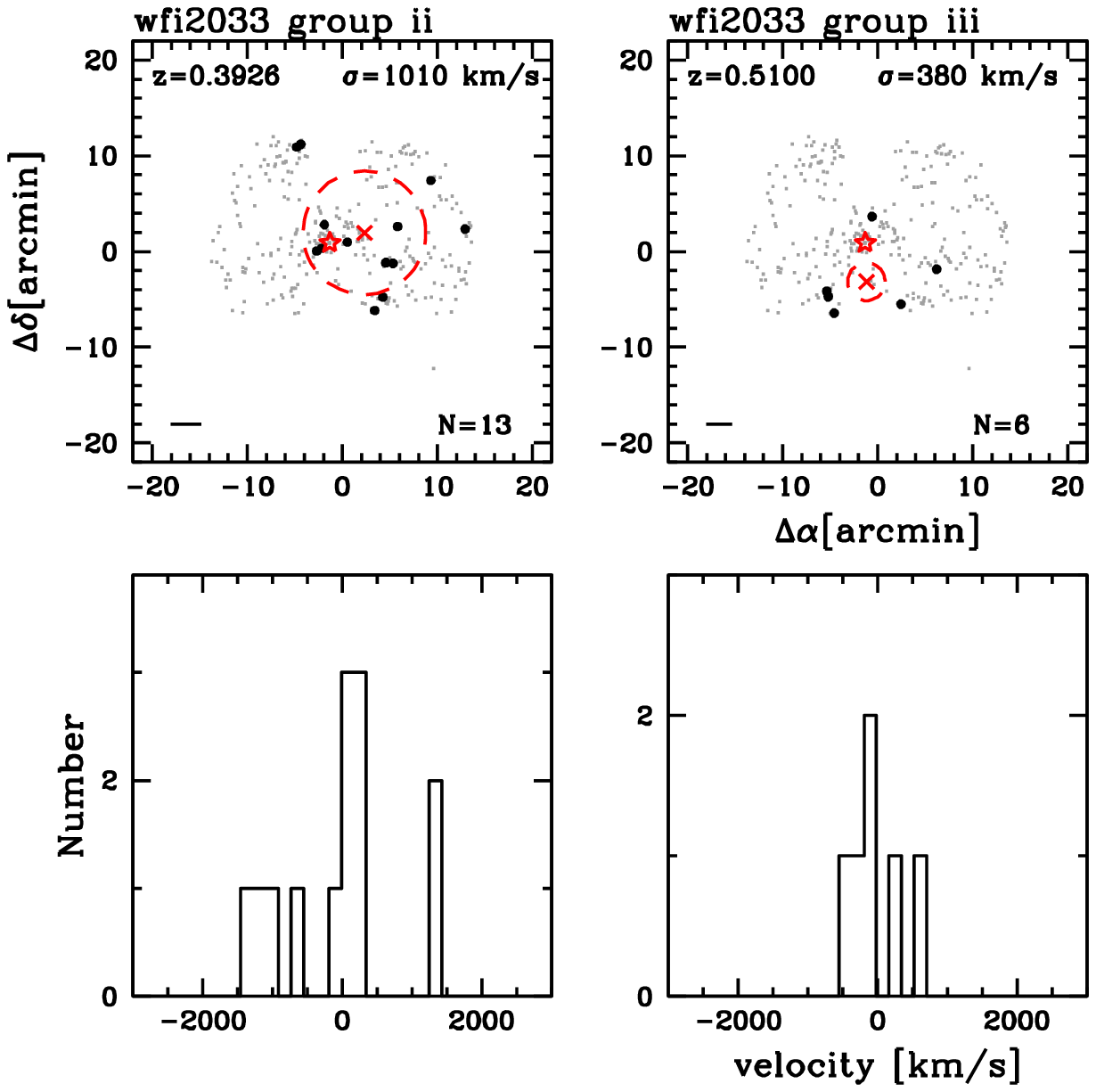}
\caption{Continued.}
\end{figure*}
\clearpage

\section{Group Members}
\label{appendix:membertables}

\begin{deluxetable*}{llllll}
\tablecaption{Algorithmic Group Members \label{table:alggrpgxys}}
\tablehead{
\colhead{Field} & \colhead{Group} & \colhead{$z_{gxy}$} & \colhead{$ez_{gxy}$} & \colhead{$RA_{gxy}$} & \colhead{$Dec_{gxy}$}  \\
\colhead{ } & \colhead{ } & \colhead{ } & \colhead{ } & \colhead{[deg]} & \colhead{[deg]}  
}

\startdata

   b0712 &   0 &    0.07849 &    0.00018 &    109.02408 &     46.84229 \\ 
   b0712 &   0 &    0.07773 &    0.00012 &    109.29465 &     46.89964 \\ 
   b0712 &   0 &    0.07797 &    0.00009 &    109.04527 &     46.85457 \\ 
   b0712 &   0 &    0.07865 &    0.00018 &    108.90497 &     46.89328 \\ 
   b0712 &   0 &    0.07866 &    0.00016 &    108.91337 &     46.91699 \\ 

\enddata
\tablecomments{Table \ref{table:alggrpgxys} is published in its entirety in the machine-readable format. A portion is shown here for guidance regarding its form and content.}

\end{deluxetable*}

\begin{deluxetable*}{llllll}
\tablecaption{Supergroup Members \label{table:supergrpgxys}}
\tablehead{
\colhead{Field} & \colhead{Group} & \colhead{$z_{gxy}$} & \colhead{$ez_{gxy}$} & \colhead{$RA_{gxy}$} & \colhead{$Dec_{gxy}$}  \\
\colhead{ } & \colhead{ } & \colhead{ } & \colhead{ } & \colhead{[deg]} & \colhead{[deg]}  
}

\startdata

   b0712 &   0 &    0.30306 &    0.00018 &    109.26192 &     47.09909 \\ 
   b0712 &   0 &    0.28987 &    0.00030 &    109.26274 &     47.11399 \\ 
   b0712 &   0 &    0.28954 &    0.00016 &    109.22754 &     47.18964 \\ 
   b0712 &   0 &    0.28460 &    0.00012 &    109.05324 &     47.01615 \\ 
   b0712 &   0 &    0.28980 &    0.00018 &    109.13190 &     47.06879 \\ 

\enddata
\tablecomments{Table \ref{table:supergrpgxys} is published in its entirety in the machine-readable format. A portion is shown here for guidance regarding its form and content.}

\end{deluxetable*}

\begin{deluxetable*}{llllll}
\tablecaption{Visually Identified Candidate Group Members \label{table:candgrpgxys}}
\tablehead{
\colhead{Field} & \colhead{Group} & \colhead{$z_{gxy}$} & \colhead{$ez_{gxy}$} & \colhead{$RA_{gxy}$} & \colhead{$Dec_{gxy}$}  \\
\colhead{ } & \colhead{ } & \colhead{ } & \colhead{ } & \colhead{[deg]} & \colhead{[deg]}  
}

\startdata

   b0712 &    i &    0.40473 &    0.00023 &    109.32741 &     47.17129 \\ 
   b0712 &    i &    0.40200 &    0.00030 &    108.78130 &     46.95694 \\ 
   b0712 &    i &    0.40554 &    0.00030 &    108.99552 &     47.14788 \\ 
   b0712 &    i &    0.40632 &    0.00018 &    109.29960 &     47.19134 \\ 
   b0712 &    i &    0.40489 &    0.00023 &    108.96837 &     47.12892 \\ 

\enddata
\tablecomments{Table \ref{table:candgrpgxys} is published in its entirety in the machine-readable format. A portion is shown here for guidance regarding its form and content.}

\end{deluxetable*}

\end{document}